\documentclass[preprint,12pt]{elsarticle}

\pdfoutput=1 

\usepackage{amssymb}
\usepackage{amsthm}
\usepackage{mathtools}

\usepackage{inputenc}

\usepackage{color}

\journal{Physics Reports}

\begin{document}

\begin{frontmatter}


%% \tnotetext[label1]{PITT-PACC 1515}
%% \fntext[label2]{PITT-PACC 1515}
\cortext[cor1]{Preprint numbers: PITT-PACC 1515;~~CERN-PH-TH-2015-259}
%\preprint{PITT-PACC 1515}

\title{Physics Opportunities of a 100 TeV Proton-Proton Collider}


\author[1]{Nima Arkani-Hamed}
\author[2]{Tao Han}
\author[3]{Michelangelo Mangano }
\author[4]{Lian-Tao Wang}

\address[1]{Institute for Advanced Study, 
Princeton, NJ, 08540, USA}
\address[2]{Department of Physics and Astronomy, Univ.~of Pittsburgh, 
Pittsburgh, PA 15260, USA \\
and Department of Physics, Tsinghua University, Beijing 100086, China}
\address[3]{Physics Department, TH Group, CERN, CH-1211 Genve 23, Switzerland }
\address[4]{Enrico Fermi Institute, Department of Physics, \\ and  Kavli Institute for Cosmological Physics, University of Chicago, \\ Chicago, IL 60637-1434, USA }

% \fntext[1]{IAS}
 
\begin{abstract}
The discovery of the Higgs boson at the LHC exposes some of the most
profound  mysteries
fundamental physics has encountered in decades, opening the door to
the next phase of
experimental exploration. More than ever, this will necessitate new
machines to push us deeper into the energy frontier.
In this article, we discuss the physics motivation and
present the physics potential of a proton-proton collider running at an energy significantly beyond that of the LHC and a luminosity comparable to that of the LHC. 100 TeV is used as a benchmark of the center of mass energy, with integrated luminosities of 3 ab$^{-1}-$30 ab$^{-1}$. 

\end{abstract}


\end{frontmatter}

\newpage

\tableofcontents

\newpage

%\chapter{Introduction}
%\label{Chapter:Introduction}

\newcommand{\ttbar}{t\bar{t}}
\newcommand{\iab}{ab$^{-1}$}
\newcommand{\ifb}{fb$^{-1}$}
\newcommand{\ipb}{pb$^{-1}$}
\newcommand{\lum}{cm$^{-2}$s$^{-1}$}
\newcommand{\tev}  {\ifmmode {\mathrm{TeV}} \else TeV\fi}

\section{Introduction}

\subsection{LHC, The Higgs Boson and Beyond}
With the discovery of the Higgs boson at the Large Hadron Collider (LHC) \cite{Aad:2012tfa,Chatrchyan:2012ufa}, fundamental
physics finds itself at one of the most exciting crossroads in its
history. The central questions today are the deepest ones that
have been posed in decades, related to the ultimate origin of the
elementary particles and even of space-time itself. Major new input
from experiments is needed for progress.

The LHC restarted in 2015 at higher energies, and will eventually
collect much more data. This will certainly significantly advance our
understanding. As we will discuss in this report, however, attacking some of the most 
profound theoretical questions of the 21st century, particularly ones associated with the largely mysterious Higgs particle, will 
necessitate another leap to higher energies. 
The future of fundamental physics on the $20-50$ year timescale
hinges on starting a huge new accelerator complex that can take us at
least one order of magnitude beyond the ultimate reach of the LHC.

There have been efforts in the community in planning the next step
beyond the LHC, which have intensified after the discovery of the
Higgs. Among the various options, a proton proton collider operating
at energies far beyond that of the LHC has emerged as an appealing
option, including the FCC-hh project promoted by CERN and the SppC
project promoted by IHEP in China. 100 TeV is typically used as a
benchmark energy for such a collider.

Many studies of the physics potential of an 100 TeV $pp$ collider have
been carried out in the recent past, and are continuously appearing in
the literature. 
%~\cite{Hook:2014rka,
%  Han:2014nja,Dawson:2014pea,Craig:2014lda,Curtin:2014jma,
%  Profumo:2014opa,Cohen:2014hxa,
%  Low:2014cba,Gori:2014oua,Khoze:2014kka,Ellis:2014kla,Acharya:2014pua,
%  Alves:2014cda,Cirelli:2014dsa,Altmannshofer:2014cla,diCortona:2014yua,
%  Bramante:2014tba,Curtin:2014cca}.
The results are still incomplete
and preliminary, many years of intensive work are still needed to
arrive at a complete description. At the same time, we  already
have a broad-brush picture of the physics capabilities of such a
machine. The studies have also highlighted a number of open questions
and future directions to explore. In this report we give a high-level summary
of the central scientific issues at stake, and draw on the studies
that have been carried out, to show that the leap in energy offered by
a 100 TeV $pp$ collider will allow us to robustly address some of the most important open questions in fundamental physics. 

Many of the most profound mysteries are intimately connected with the
Higgs particle, which is totally new, unlike anything we have seen
before. In many ways the Higgs is the simplest particle imaginable,
with no charge and no spin. This apparent simplicity is also
what makes it so beguiling. All other scalar particles we have seen have
been obviously composite, with a size close to their Compton radius.
The Higgs is not like this, appearing to be more point-like than
naturally expected on theoretical grounds.

The Higgs must also have a dynamical property we have never seen for
any of the other fundamental particles: it should be able to
interact not only with other particles, but also with itself! Indeed,
self-interaction is the most basic of all processes allowed by quantum
field theory, but spin and charge forbid point-like self-couplings for
all particles but the Higgs. The LHC will only scratch the surface of
this physics, but with the data from the 100 TeV collider we will be
able to unambiguously see and precisely measure the Higgs
self-interaction process, whose structure is deeply related to the
origin and mass of the Higgs itself.
\begin{figure}[h!]
\centering
\includegraphics[scale=0.4]{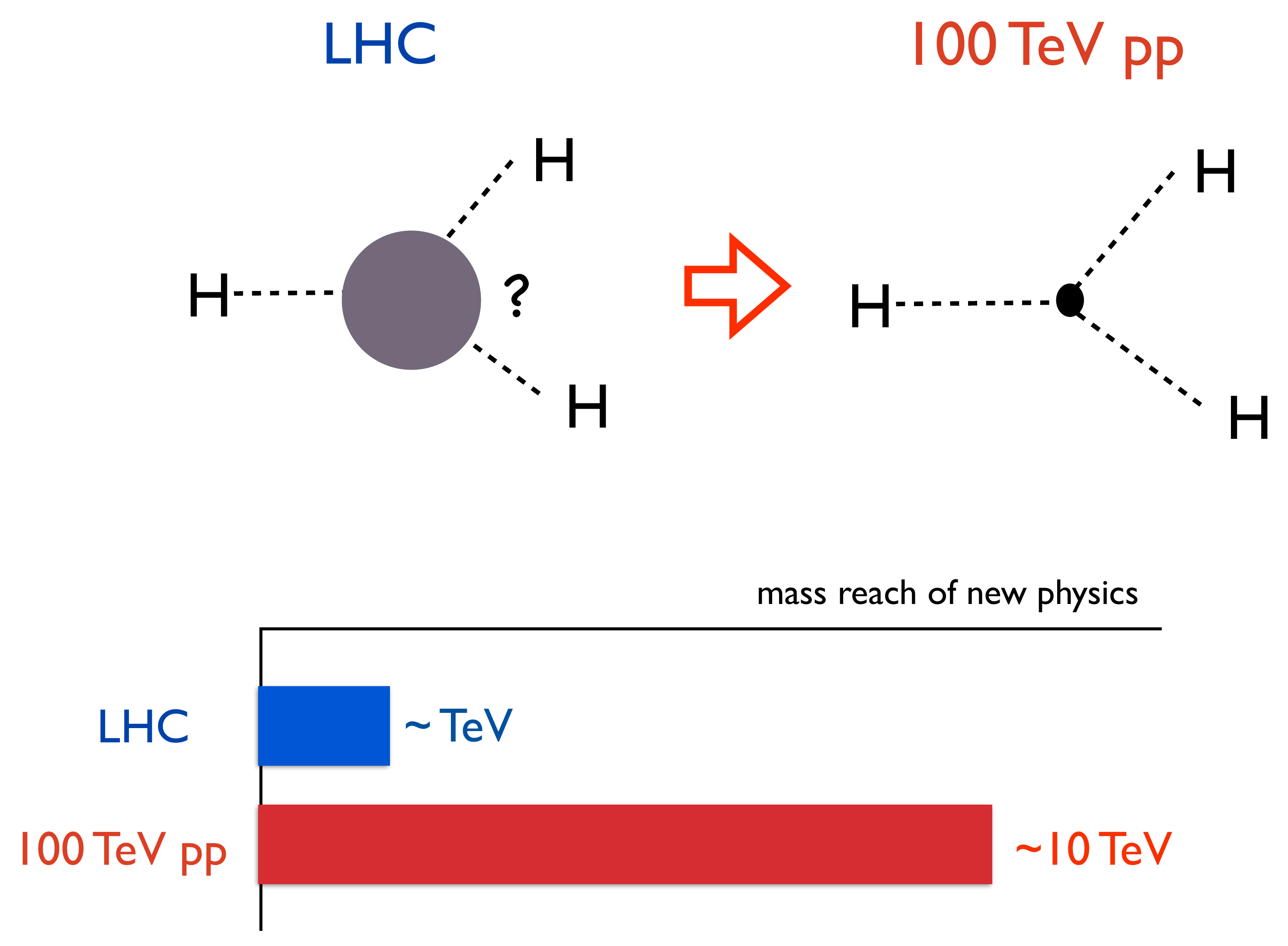}
\caption{A sketch of two of the major advances obtained by going to a 100 TeV
  $pp$ collider.  The 100 TeV $pp$ collider will see, for the first time,
  a fundamentally new dynamical process $-$ the self-interaction of an
  elementary particle $-$ uniquely associated with the Higgs. It will
  also improve the reach of the direct search of new physics
  particles by at least a factor of 5. }
\label{fig:main_theme}
\end{figure}
At an even more fundamental level, much of the excitement surrounding the
proposal of a 100 TeV $pp$ collider  stems from the bold leap into the completely uncharted new
territory that it offers, probing energy scales where we have long had
reasons to expect fundamental new physical principles at play.
The 100 TeV $pp$ collider will allow us to hunt for new
fundamental particles roughly an order of magnitude heavier than we can
possibly produce with the LHC, and new particles the LHC may produce
in small numbers will be produced with up to a thousand times higher
rate, giving us a new window into the quantum-mechanical vacuum of our
universe with a hundred-fold greater resolution than ever before.

These two points are sketched in Fig.~\ref{fig:main_theme}, and represent the major advances we will make by going to a 100 TeV $pp$ collider.

\subsection{New Colliders for a New Frontier}

Fundamental physics began with the twin revolutions of Relativity and
Quantum Mechanics. Much of the second half of the 20th century was occupied
with understanding the reconciliation of these principles within the
framework of quantum field theory, and identifying a specific quantum
field theory $-$ the Standard Model (SM) of particle physics $-$ describing
all particles and interactions we know of to date.

Theoretical consistency with relativity and quantum mechanics places
extremely strong constraints on theories of interacting massless
particles, almost completely dictating the possible menu of spins and
interactions.  At energies low enough compared to some fundamental
ultraviolet scale, physics is guaranteed to be described by Yang-Mills
theories and gravity coupled to particles of spin 0, 1/2, and also
possibly spin 3/2 with supersymmetry. The rigidity of this structure
is striking. Of course, most elementary particles are not massless,
but since the effects of mass are naively negligible at high energies,
these rules fix what physics at very high energies can look like, at
least until we hit the Planck scale where the usual notion of
space-time itself breaks down.

For particles with nontrivial spins, there is a jump in the number of
spin degrees of freedom between massless and massive particles.  For
instance, the massive $W$ and $Z$ bosons have spin one and three spin
degrees of freedom, but only two helicity degrees of freedom.  This
discontinuous difference between ``massless" and ``massive" obstructs
a smooth transition from the apparent complexity of low energy physics
to the simplicity of the high energy world whose structure is almost
entirely dictated by general principles.

Famously, in the SM, the addition of a single
particle $-$ the Higgs boson $-$ solves this problem, allowing us to
reassemble the degrees of freedom of massive particles at low energies
into the consistent high energy framework for massless particles.

The Higgs is certainly the simplest solution to the problem it
solves $-$ it is hard to imagine a simpler elementary particle, with no
spin or charge. But this simplicity is actually extremely surprising
and, in a literal sense, unprecedented, since we have never before
seen a point-like elementary particle of spin zero.  Indeed, violent
ultraviolet quantum fluctuations have the potential to generate huge
masses for elementary particles, but this does not happen for particles
with spin, where a change from ``massless'' to ``massive'' would change
the number of spin degrees of freedom discontinuously. However, the number of
spin degrees of freedom for massless and massive particles of spin
zero is the same, and so nothing shields the generation of huge scalar
masses, near the highest ultraviolet (UV) scales of the theory.

This logic is strongly supported from analogous phenomena in condensed
matter physics. Various materials can be engineered to be described by
non-trivial long-distance effective theories at very low
temperatures. Many of the key features of the SM, like
gauge fields and chiral fermions, can arise in a beautiful way as
emergent collective excitations of the system.  But interacting spin
zero particles like the Higgs are not seen: the only light scalars
that are ubiquitously present are Goldstone bosons $-$ like phonons
$-$ which are non-interacting at low energy. This makes sense because
the emergence of fermions and gauge fields can be robust and stable
against small variations in the detailed properties of the
material. Since this is not true for scalars, the only way to get
light scalars to emerge from a condensed matter system is to finely
adjust the microphysics of the material: for instance by putting it
under high pressure, looking for the thin slivers in parameter space
where a Higgs-like scalar becomes accidentally light. This expectation
has been borne out by recent experiments which do indeed fine-tune to
produce a particle resembling the Higgs boson of an (ungauged) $SO(3)
\to SO(2)$ symmetry breaking pattern \cite{Endres:2012wg}.

These good reasons for never having seen light scalars either in
particle physics or condensed matter systems make it all the more
remarkable to have finally found one with the Higgs! There is an irony
here: the development of the Higgs mechanism was greatly inspired by
the Landau-Ginzburg model of superconductivity.  However, the
Landau-Ginzburg model was never a real theory, only a phenomenological
model, and was replaced by BCS theory a few short years later. Many
theorists expected the same fate for the Higgs model of electroweak
symmetry breaking, with technicolor being the particle physics analog
of BCS theory. But it was the Higgs model that ended up being the
right answer in particle physics!

So while an oft-heard desire of particle physicists for many years has
been to find ``new physics'' beyond the Higgs, this is missing the
essential point: the Higgs itself represents ``new physics'' in a much
more profound way than any more complex discoveries would have
done. Its discovery closes the 20th century chapter of fundamental
physics while simultaneously kicking the door open to entirely new
questions that properly belong to the 21st century. These questions on
the table now are not about details, but are deeper and more
structural ones, leading back to the very foundations of quantum field
theory. It is striking that very similar questions are forced on us in
trying to reckon with the smallness of the cosmological constant and
the discovery of the accelerating expansion of the universe.

Obviously, the experimental future of the field will importantly
depend on results from the next run of the LHC. However, given what we
have already seen $-$ a light Higgs, but no evidence yet for physics
beyond the SM $-$ no matter what new physics the LHC does
or does not discover, building a complete
picture of the relevant physics will require new machines beyond the
LHC: not just for cleaning up details, but in order to answer the
big-picture questions that will set the direction of fundamental
physics for decades to come.

Let us begin by giving a lightning tour of the raw physics
capabilities of the 100 TeV $pp$ collider.  Thanks to the asymptotic
freedom and factorization theorem of QCD, hadronic collisions at high
energies can be calculable in perturbation theory, and we write the
production cross section of a final state $X$ as
\begin{eqnarray}
 && \sigma(pp\rightarrow X+ \text{anything}) =
   \int_{\tau_{0}}^{1} d\tau  \sum_{ij} \frac{d\mathcal{L}_{ij}}{d\tau}\  \hat{\sigma}(ij\rightarrow X),
    \label{factTheorem.EQ} \\
&&  \frac{d\mathcal{L}_{ij}}{d\tau}
= \frac{1}{1+\delta_{ij}}  \int^{1}_{\tau} \frac{d\xi}{\xi}
 \left[  f_{i/p}(\xi, Q_{f}^{2})f_{j/p}\left(\frac{\tau}{\xi},Q_{f}^{2} \right) + (i \leftrightarrow j) \right] ,
 \label{partonLumi.EQ}
\end{eqnarray}
where the parton luminosities $d\mathcal{L}_{ij}$ are given in
terms of the parton distribution functions (PDFs) $f_{i,j/p}$,
whose arguments are the fractions of momenta $(\xi,\tau/\xi)$ 
carried by the initial partons
$(i,j)$ and the parton factorization scale $Q_{f}$, and $\tau = \hat
s/s$, where $\sqrt{s}\ (\sqrt{\hat{s}})$ is the proton-proton beam
(parton-parton) 
center of mass (CM) energy.
$\hat{\sigma}$ is the partonic cross section for $ij \to
X$.

\begin{figure} [th]
\begin{center}
\includegraphics[width=0.48\textwidth]{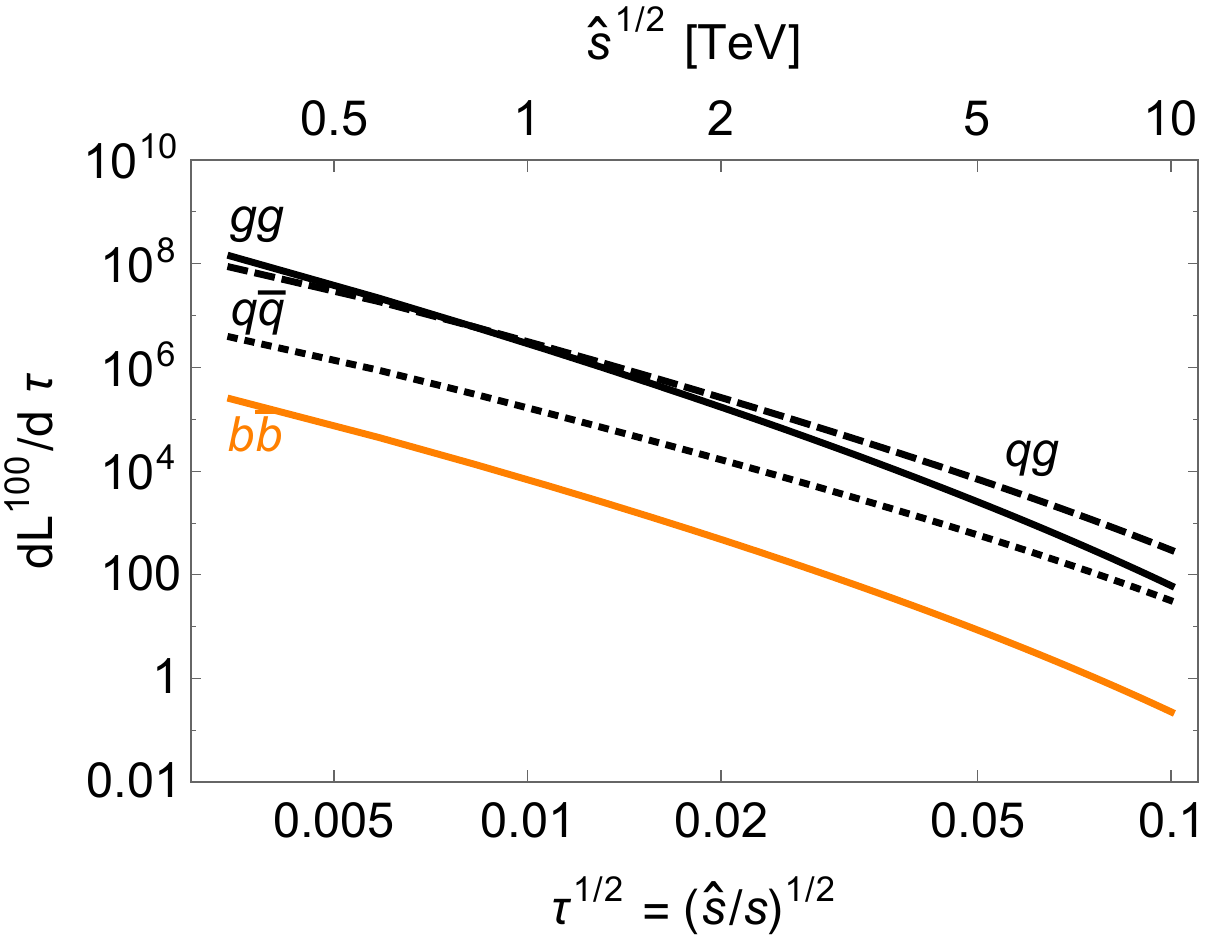} \quad
\includegraphics[width=0.48\textwidth]{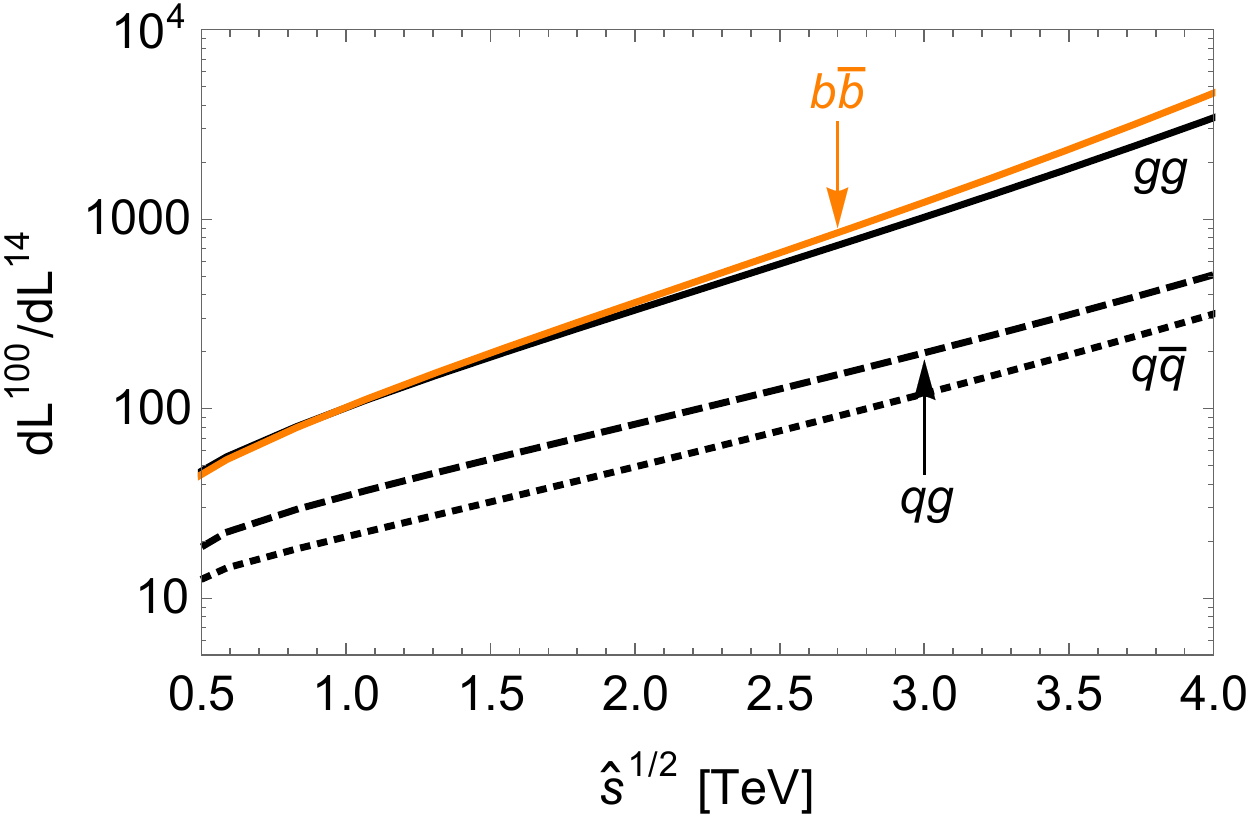}
\end{center}
\caption{Left panel: Parton luminosity densities at a 100 TeV $pp$ collider versus the average energy fraction $\sqrt\tau$ (lower scale) and the partonic
  CM energy $\sqrt{\hat s}$ (top scale); Right panel: luminosity ratios between 100 TeV and 14 TeV.  }
\label{fig:partons}
\end{figure}

Due to the rapid fall-off of parton luminosities at large $\tau$, 
the rate for processes that at a given $\sqrt{s}$ have a large value of
$\tau$ will increase dramatically when going to higher CM energies.
We illustrate this point in the left panel of
Fig.~\ref{fig:partons}, where we show the partonic luminosity densities versus
the average energy fraction $\sqrt\tau$ (lower scale) and the
partonic CM energy $\sqrt{\hat s}$ (top scale), and in the right panel the
luminosity ratios between 100 TeV and 14 TeV. We see the significant
increase of the partonic luminosities, by a factor ranging from $20-100$ at $\sqrt{\hat s}\approx 1$ TeV to  $300-5000$ at $\sqrt{\hat s}\approx 4$ TeV.
%ranging from $10^3 - 10^5$ depending on the partonic CM~energies.

Most importantly, the leap in energy at the 100 TeV $pp$ collider
gives a huge increase in the reach for new physics. A seven-fold
increase in CM energy relative to the LHC, with a
luminosity comparable to that of the LHC, increases the mass reach for
new particles significantly. For instance, the mass reach will be
extended by a factor of about five relative to the LHC for resonant
production of weakly or strongly interacting resonances, 
or by a factor of four for color-singlet pair production.
We illustrate this in detail in Sec.~\ref{sec:NP}.

\begin{figure}[h!]
\centering
\includegraphics[scale=0.5]{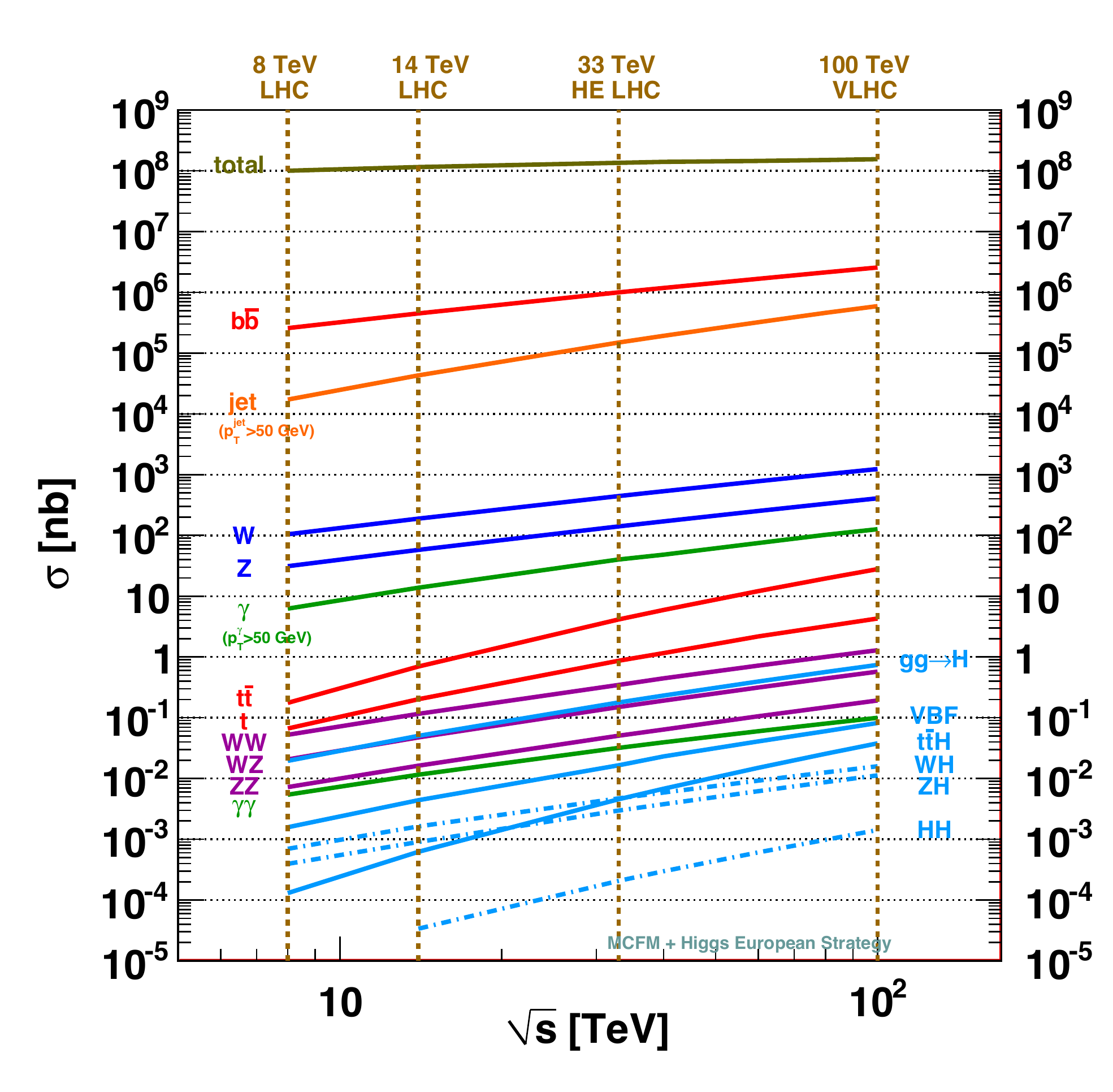}
\caption{Production rates of SM processes versus the $pp$ CM energy \cite{Campbell:2013qaa}.}
\label{fig:sm_rates}
\end{figure}

The huge increase in parton luminosity also leads to a substantial
enhancement of the production rates for the SM processes in going
from 14 to 100 TeV, as illustrated in Fig.~\ref{fig:sm_rates}
\cite{Campbell:2013qaa}. This will allow several extremely rare SM
processes to be potentially observable for the first time.

Measuring the triple Higgs coupling provides a direct probe of the
nature of the electroweak symmetry breaking, and this is best done at
the 100 TeV $pp$ collider, by looking for double-Higgs production, which
yields a sizable production cross section of $\cal{O}$(1 pb).
At the LHC, this process suffers from a low production rate and large
SM backgrounds. Moreover, one needs to disentangle different
contributions from different contributing diagrams.  At 100~TeV, this
process will however probe a SM Higgs self-coupling at about ten percent level
\cite{Yao:2013ika,He:2015spf,Barr:2014sga,Azatov:2015oxa}.
The 100 TeV $pp$ collder could also directly probe
the top Yukawa coupling, via $t\bar{t}H$ production, at the 1$\%$
level~\cite{Plehn:2015cta}.

Experiments at 100 TeV probe the SM in a regime where the electroweak
symmetry is effectively restored. A couple of new features are worth
noting (more details will be given in Section~\ref{sec:WZatHighE}).
First of all, in processes at the very high energies $\sqrt{\hat s} \gg
M_W$, EW gauge bosons are copiously produced by radiation. For $p_T$'s
approaching $\sim 10$ TeV, the electroweak Sudakov factor $4 \alpha_2
\log^2 (p_T^2/m_W^2) \sim 0.1$, and we have ``electroweak radiation"
in complete analogy with electromagnetic and gluon radiation.  For
instance, a $W$ or $Z$ gauge boson would be radiated off a light quark
with 10 TeV of energy with a probability of 10\% and off a gauge boson
with a probability of 20\%. 
These production rates are
one-to-two orders of magnitude higher than what we typically encounter
when considering the production of gauge boson in inclusive processes
at $\sqrt{\hat s}\sim M_W$: for example, as shown in
Fig.~\ref{fig:sm_rates}, the production rate of an additional $W$
boson in inclusive $W$ production is only at the per-mille level,
$\sigma(WW)/\sigma(W) \sim 10^{-3}$. 
This phenomenon
makes it possible to ``see'' traditionally invisible particles such as
neutrinos (or even weakly-interacting dark matter particles), through
electroweak radiation. This can be nicely illustrated by considering
the invisible decay of a $Z^\prime \to \nu \nu$. For heavy enough
$Z^\prime$'s, there is a significant rate for radiating $W,Z$'s off
the neutrinos.  The ratio $\Gamma(Z^\prime \to \nu
\bar{\nu})/\Gamma(Z^\prime \to \nu \bar{\nu} Z)$ only depends on the
mass of the $Z^\prime$, and so if this visible mode is abundant enough
we can directly determine the invisible rate (and thereby also
directly determine the $Z^\prime$ coupling to left-handed
leptons)~\cite{Hook:2014rka}.

Similarly to what happens for bottom quarks at the Tevatron, at
100~TeV one can expect processes where the energy involved is so large
that even a top quark can be considered as practically massless. In
this case, large logarithms of $\hat{s}/m_t^2$ can and must be
resummed, using a formalism similar to that adopted to describe the
bottom quark PDF at Tevatron or LHC energies. This will be discussed
in Section~\ref{sec:toppdf} and examples of relevant applications will be shown in Section~\ref{sec:NP}.
On a similar footing, as will be discussed in detail in
Section~\ref{sec:WZatHighE}, the electroweak gauge bosons, $W,Z$, may be
copiously radiated off the initial-state quarks, and can be treated as partons inside the proton.

At 100~TeV, the electroweak gauge bosons and the top
quark can be produced in highly boosted configurations.
Their decay products will be highly
collimated and thus form massive jets, $W,Z$- or top-jets. This new
phenomena will also naturally arise when new heavy particles are produced and
subsequently decay to the gauge bosons and top quarks.

\subsection{Luminosity of the 100 TeV $pp$ Collider}

In this discussion of physics opportunities, some brief comments about
the energy and luminosity requirements of the 100 TeV $pp$ collider are
in order.  The CM energy and luminosity of a proton proton
collider are crucial in determining its physics potential. Perhaps the
most obvious question is how energy and luminosity impact mass reach
for the production of new particles.  This review uses 100 TeV as a
benchmark, with integrated luminosities ranging from 3 ab$^{-1}$ to 30
ab$^{-1}$. The reach will roughly scale with the CM
energy, if other options are considered.

The question of required or target luminosity has been discussed in
detail recently \cite{Hinchliffe:2015qma}.  Larger integrated
luminosity leads to sensitivity to new physics with smaller signal
cross sections, which in turn enhances the new physics mass reach. To
be concrete, let us compare the reach of the LHC and a 100 TeV $pp$
collider for the production of massive particles with different
two-parton initial states, using estimates of reach based on scaling
of the parton luminosity.\footnote{For a useful tool to perform such
  estimates, see the link by Salam and Weiler at {\tt http://collider-reach.web.cern.ch/collider-reach/}.}
\begin{figure}[h!]
  \centering
  \includegraphics[scale=0.32]{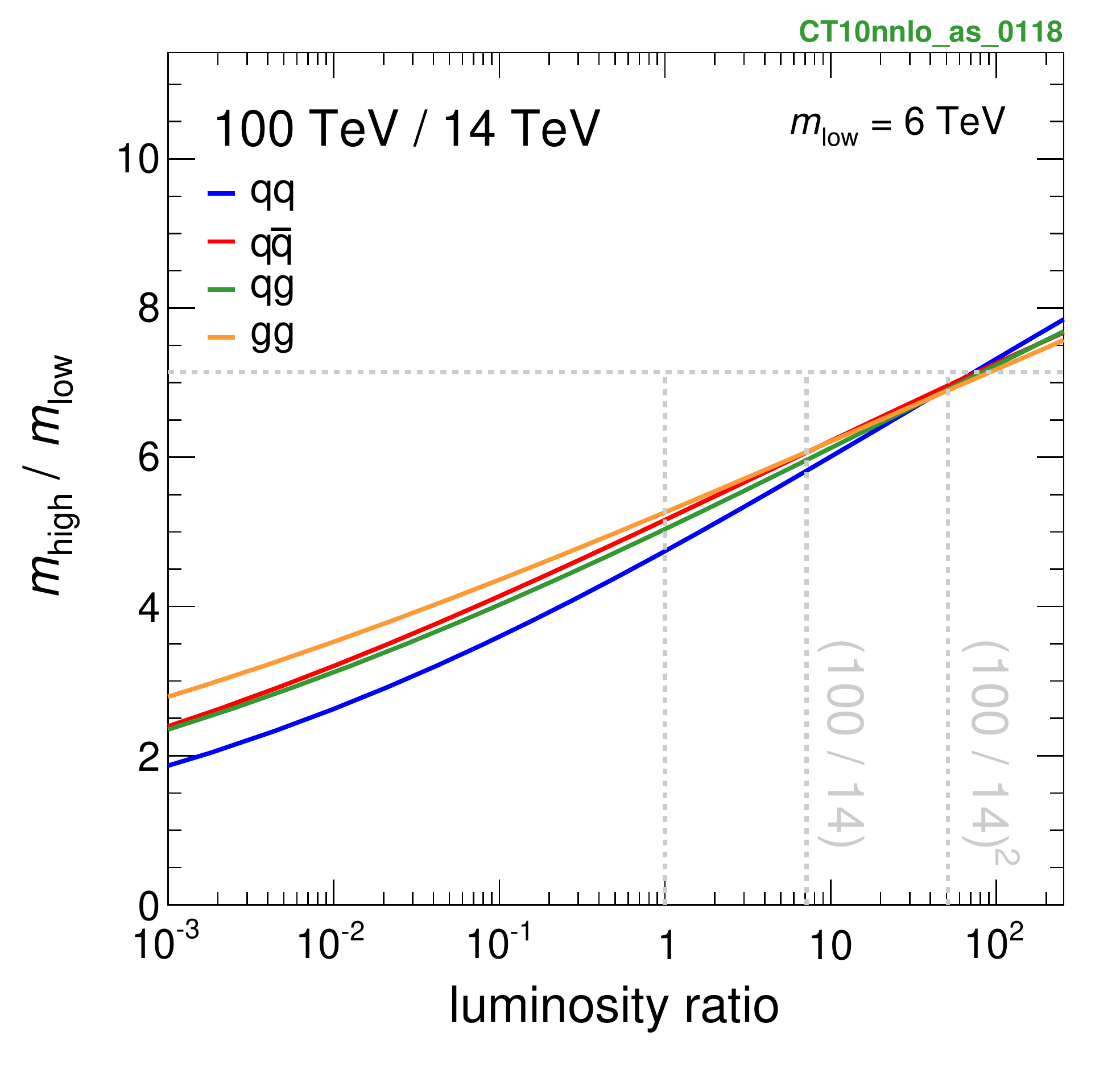} \quad
    \includegraphics[scale=0.32]{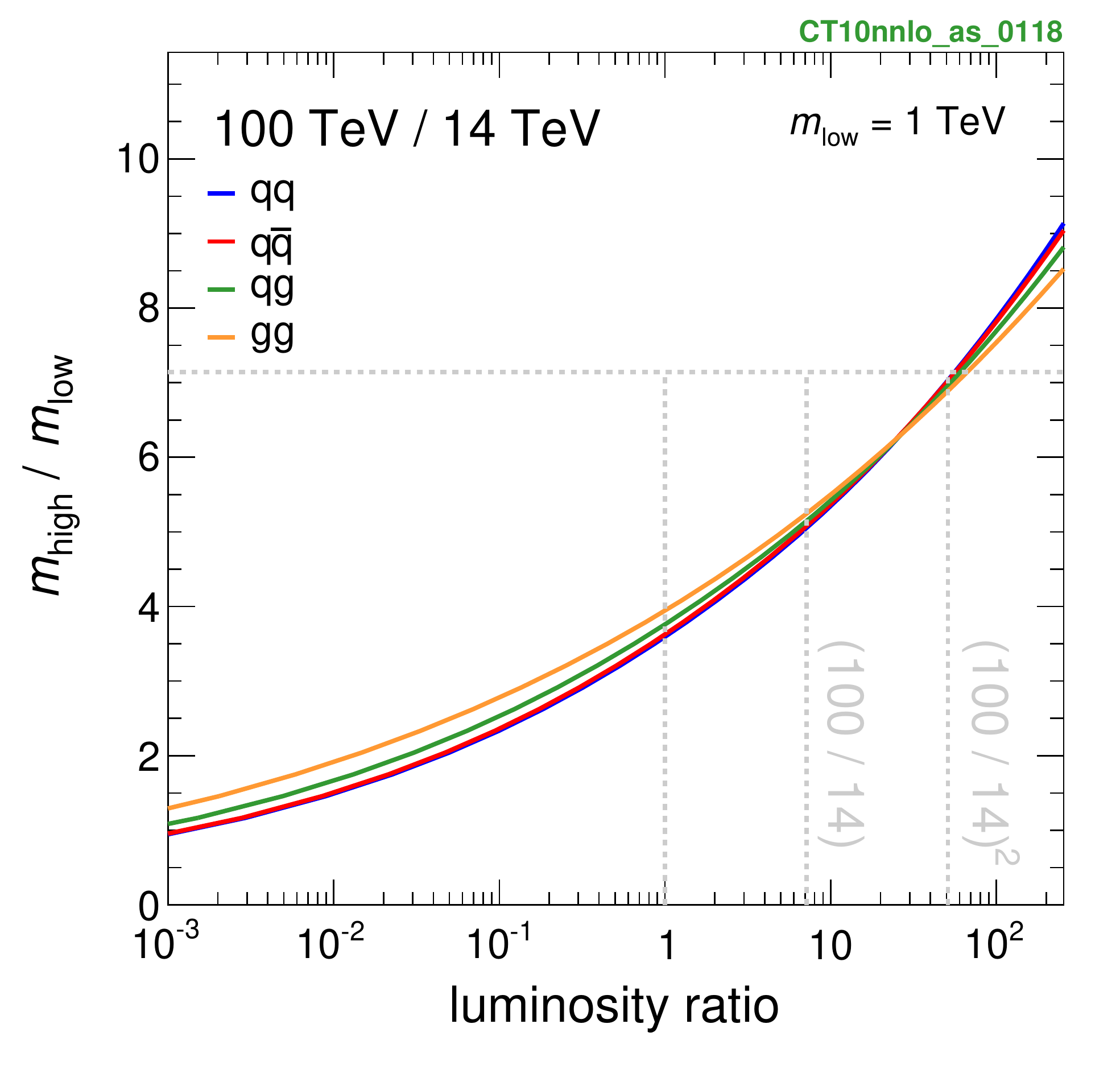}
  \caption{Ratio of the reach of new physics scale from the LHC and
    100 TeV $pp$ collider, shown as a function of the ratio of
    luminosity. New physics produced from different partonic initial
    states are considered.  The limit of LHC is assumed to be 6 TeV
    (left) and 1 TeV (right). }
  \label{fig:lumi_extrap}
\end{figure}

\begin{figure}[h!]
  \centering
  \includegraphics[scale=0.32]{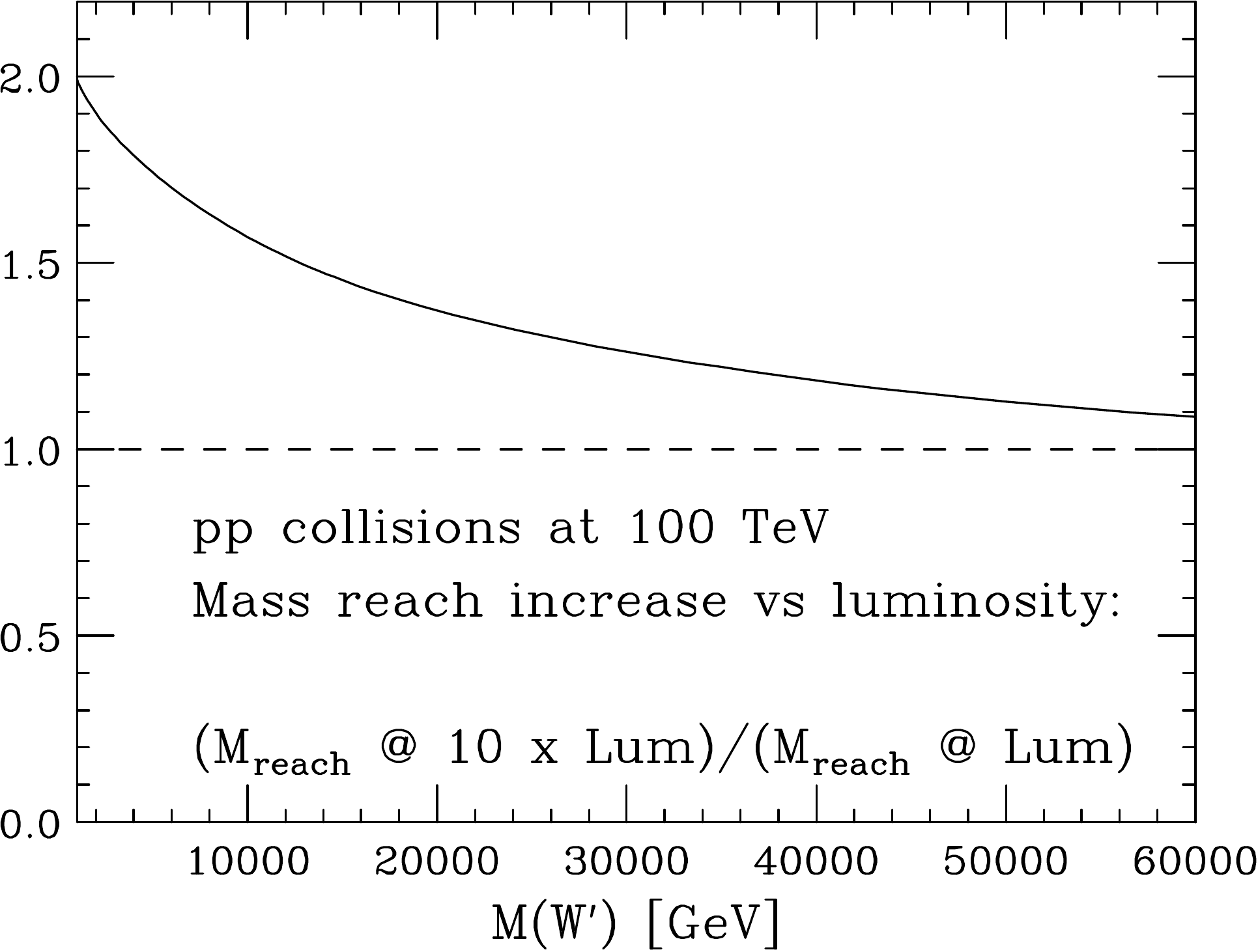} \quad
    \includegraphics[scale=0.32]{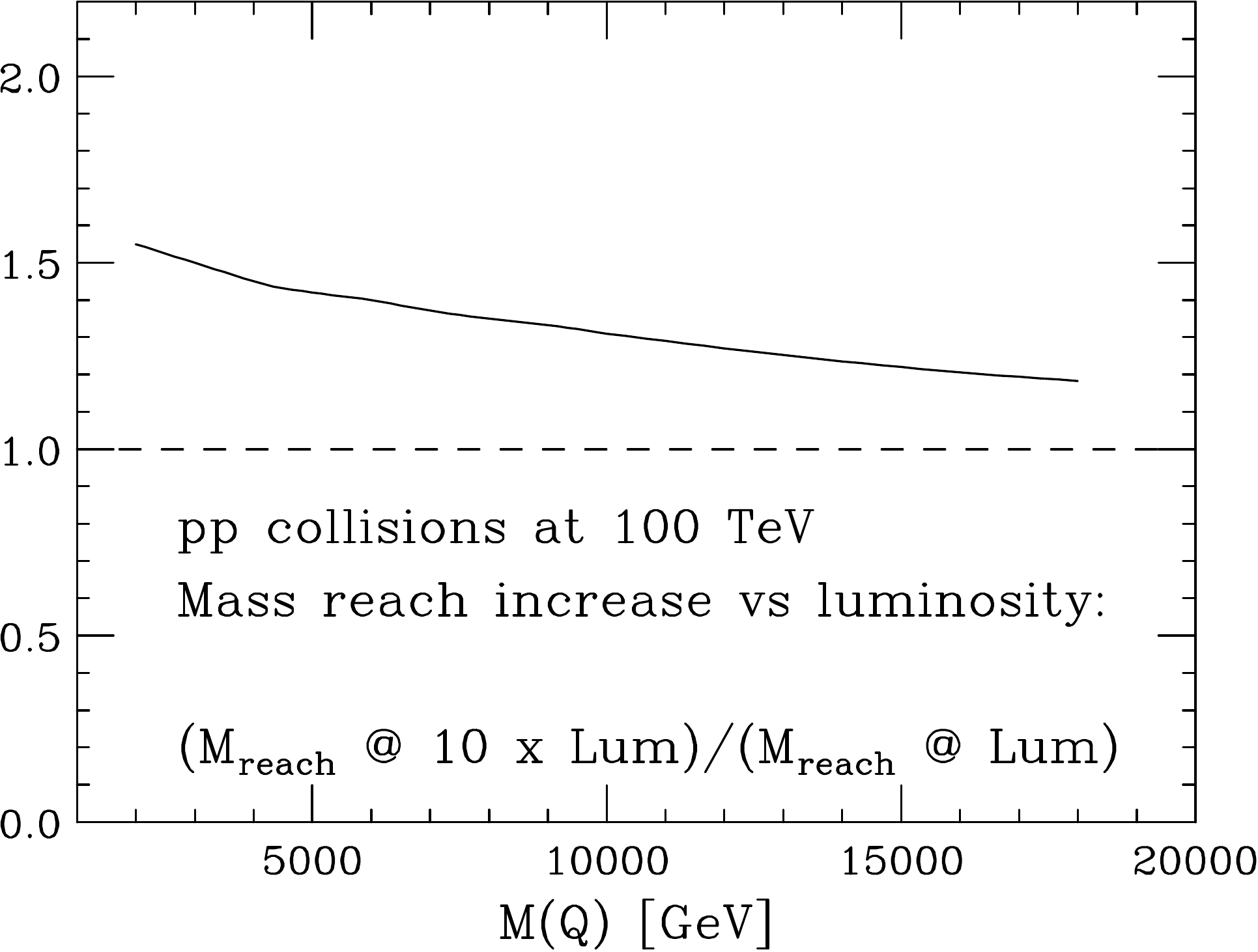}
  \caption{Two examples of the increase of the reaches of new physics
    with 10 times more luminosity. Left: Drell-Yan production of
    sequential $W^{\prime}$. Right: Pair production of heavy quark.}
  \label{fig:reachW_Q}
\end{figure}

First, we focus on the highest possible reach in mass. For the LHC, we
assume the reach for the scale of certain type of new physics is 6 TeV,
shown in the left panel of Fig.~\ref{fig:lumi_extrap}. This could be
the case for a 6 TeV $Z'$, or of a $\sim 3$~TeV gluino, pair produced.
While this is a crude estimate, it has been demonstrated to
be a reasonable approximation in a wide variety of examples and
suffices for our discussion here.  
To be more concrete, the increases
of the mass reach with 10 times more luminosity for two benchmark
examples, sequential $W^\prime$ and heavy quark, are shown in
Fig.~\ref{fig:reachW_Q}.
We see that the much larger increase in luminosity only gives us a
very modest gain in mass reach: this well-known fact is a direct
consequence of the steeply falling parton luminosity as a function of
parton CM energy. This is especially true when we consider
the highest reach in mass, which involves a regime where the parton
density falls off very fast and the ultimate reach is typically
limited by the production rate. In fact, in the luminosity range of 
$0.1-10^3$~ab$^{-1}$, the increase in mass reach is well approximated
by a logarithmic behavior, 
\begin{equation}
M(L)-M(L_0) \sim~7~\text{TeV} \, \log_{10}(L/L_0),
\end{equation}
with about 7~TeV increase in mass for a tenfold luminosity increase. 
The relative gain in mass reach therefore diminishes as the total luminosity is increased.  Even with just 3
ab$^{-1}$, the same as the target luminosity of the HL-LHC, the 100
TeV $pp$ collider can enhance the new physics reach by a factor of
5. This is a huge step and a large portion of the ratio of the CM 
energy $\sim 7$. Of course, given that partonic cross sections drop with the energy as 
\begin{equation}
\hat{\sigma} \sim \hat{s}^{-1}, 
\end{equation}
an increase of a factor of $\sim 50$ would be needed to
extend the reach by the full factor of $\sim 7$. Scaling violations in
the PDFs actually call for a slightly larger luminosity increase, as
discussed in detail in~\cite{Rizzo:2015yha}.

For the searches of lower mass particles, the parton density falls off
more slowly in the relevant regime and simple scaling suggests a larger
luminosity is necessary to achieve the same enhancement in the mass
reach, as demonstrated in the right panel of
Fig.~\ref{fig:lumi_extrap}. However, we note that making a sharp
statement in this case is much harder, since this is usually the case
with weak signals and large backgrounds. We need to identify particular
highly motivated cases to set the luminosity target. The most
important example is probably the measurement of the triple Higgs
coupling discussed earlier. The target here is to reach the 10$\%$
level accuracy, which is crucial in distinguishing qualitatively
different characters of the Higgs potential.
Preliminary studies of this process have been
performed \cite{Yao:2013ika,He:2015spf,Barr:2014sga,Azatov:2015oxa}. While
adopting sightly different assumptions about systematic uncertainties
and backgrounds, they nevertheless converge to a precision in the
range of $5-10\%$ for 30 ab$^{-1}$. Given the fundamental importance of
this question for setting an objective target for the luminosity,
future studies should be undertaken to settle it decisively.

Since typically a collider will start with a lower-than-nominal
luminosity, it is interesting a have a set of ``minimal'' luminosity goals.
\begin{figure}[h!]
  \centering
  \includegraphics[scale=0.45]{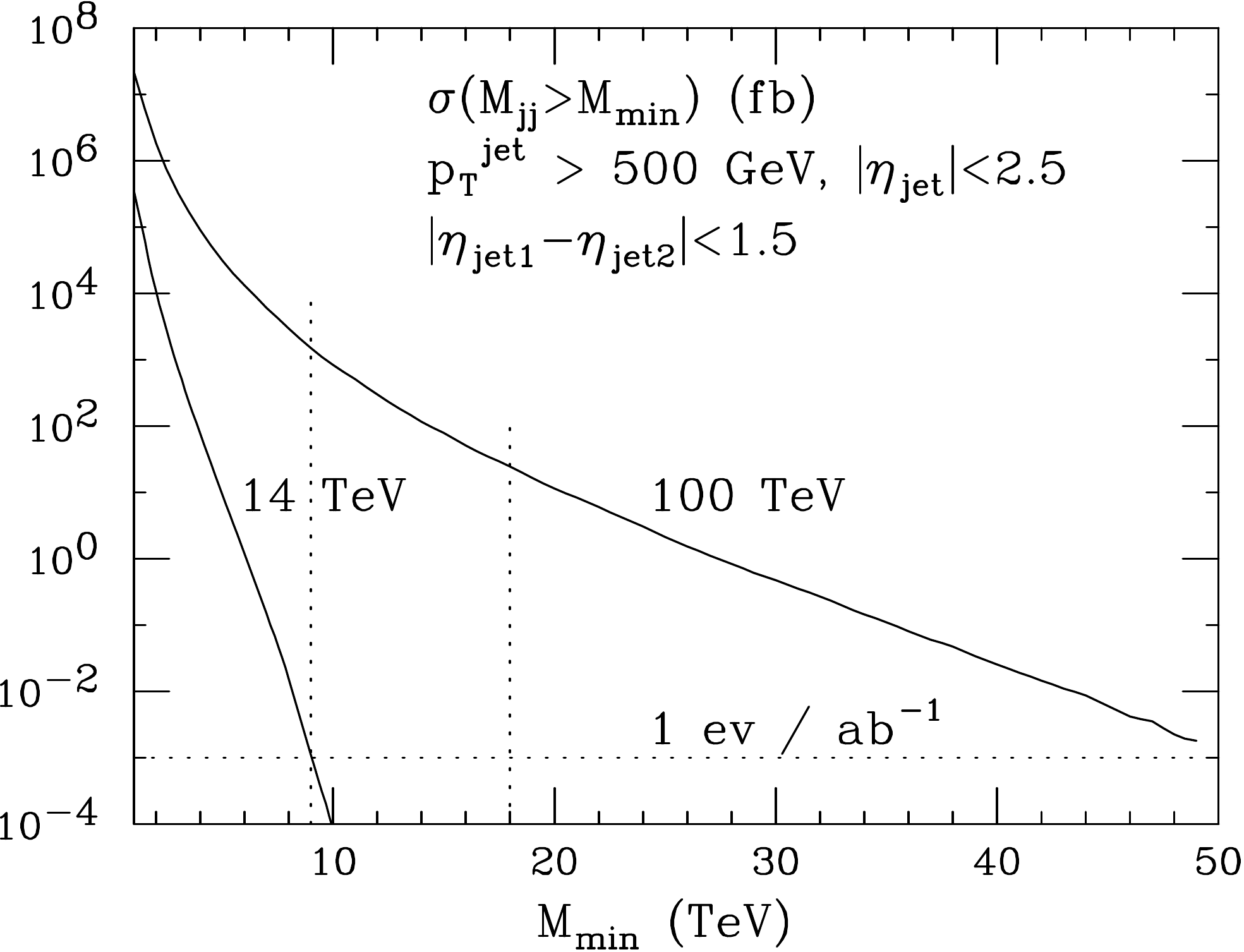}
  \caption{Cross sections for the production of dijet pairs
     with invariant mass $M_{jj}>M_{\mathrm{min}}$, at c.m.\ energies $\sqrt{s}=14$ and 100
     TeV. The jets are subject to the $p_T$ and $\eta$ cuts shown in
     the legend.}
  \label{fig:dijets}
\end{figure}
If we consider dijet production as a probe of the shortest distances,
we can extract a reference luminosity target from
Fig.~\ref{fig:dijets}, which shows the leading-order cross section to produce
central dijet pairs as a function of their invariant mass. The
LHC has a sensitivity at the level of 1 event per~\iab\ for dijet
masses above $\sim 9.5$~TeV.  At this mass, the 100 TeV cross section
is 6 orders of magnitude larger, which means that the HL-LHC
sensitivity can be recovered within 1~\ipb, i.e., in
less than a day of running at a luminosity of $10^{32}$~\lum. The
sensitivity to a mass range twice as large, 19~TeV, would require
50~\ipb, namely of the order of one month at $10^{32}$~\lum, and one
year of running at this luminosity would give us events with dijet mass well
above 25~TeV.

If we consider particles just outside the possible discovery reach of
the HL-LHC, which therefore the LHC could not have discovered, we find
the rate increases in the range of $10^4-10^5$ that we discussed
earlier, for $q\bar{q}$ and $gg$ production channels,
respectively. This means that integrated luminosities in the range of
$0.1-1$ \ifb\ are sufficient to push the discovery reach beyond what
the HL-LHC has already explored. This can be obtained in one year of
operations with initial luminosities as small as $2\times
10^{32}$~\lum.

Finally, we project in Fig.~\ref{fig:massreach_profile} the temporal
evolution of the extension of the discovery reach for various
luminosity scenarios, relative to the reach of 3~\iab\ at 14 TeV. The
left (right) plot shows results for a resonance whose couplings allow
discovery at HL-LHC up to 6~TeV (1~TeV). Once again, we notice that
the benefit of luminosity is more prominent at low mass than at
high mass. We also notice that, considering the multi-year span
of the programme, and assuming a progressive increase of the
luminosity integrated in a year, an early start at low luminosity does
not impact significantly the ultimate reach after several years of running. 

\begin{figure}[h!]
  \centering
  \includegraphics[scale=0.32]{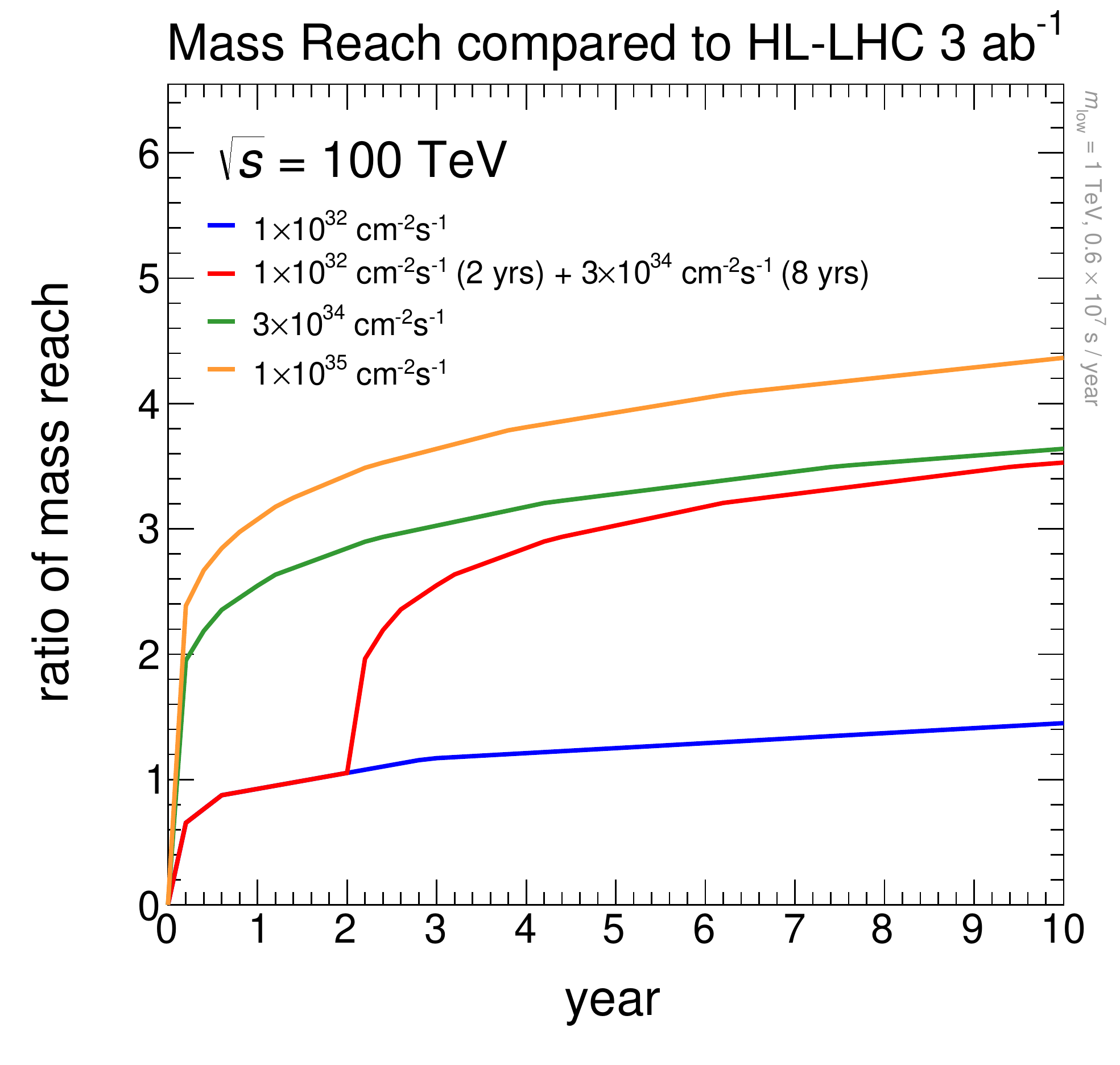} \quad
    \includegraphics[scale=0.32]{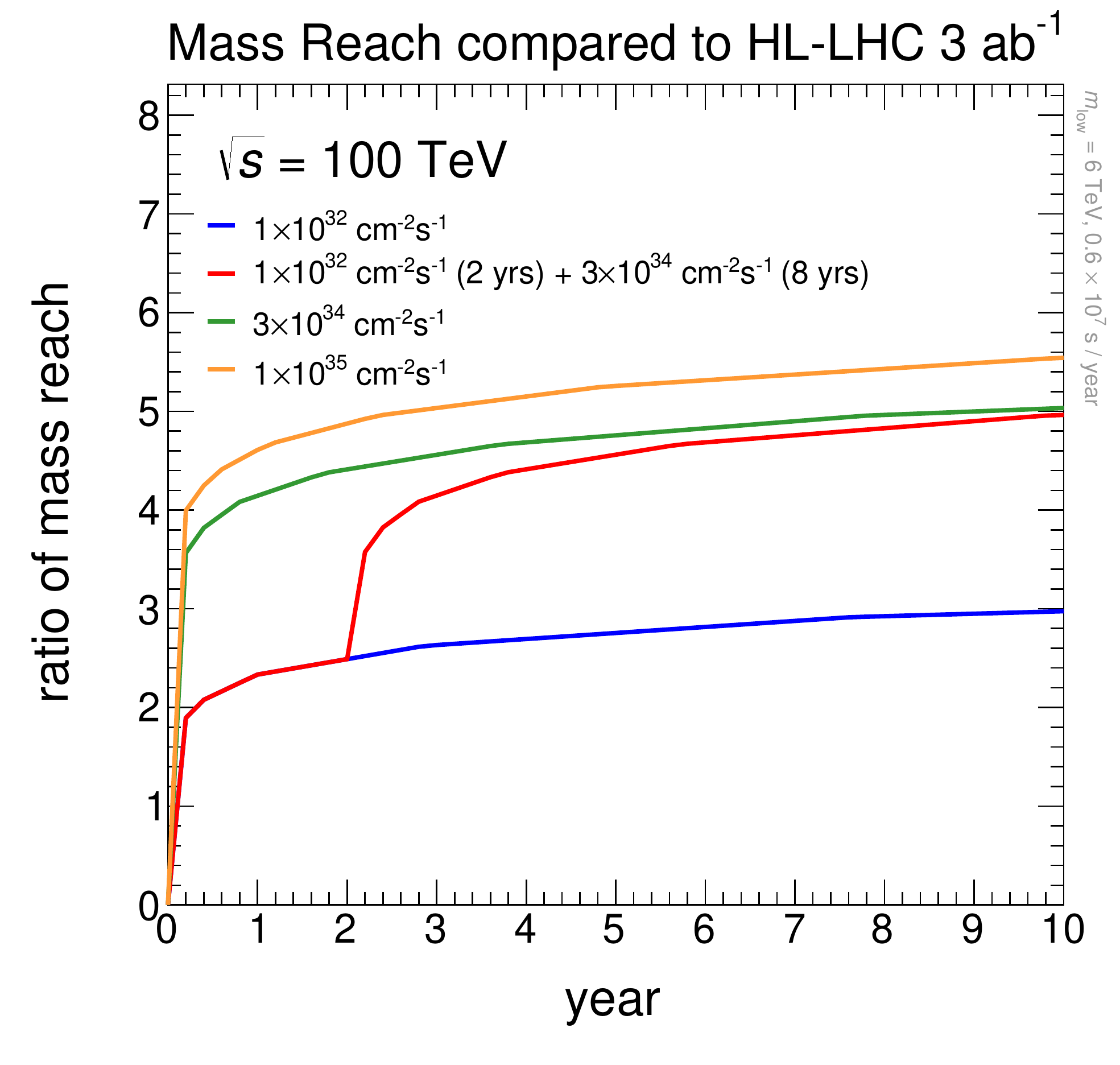}
  \caption{Evolution with time of the  mass reach at $\sqrt{s}=100~\tev$, relative to HL-LHC, under different luminosity scenarios (1 year counts for $6\times 10^6$ sec).
   The left (right) plot shows the mass increase for a ($q\bar{q}$)
   resonance with couplings enabling HL-LHC discovery at 6~TeV
   (1~TeV).}
  \label{fig:massreach_profile}
\end{figure}

The goal of an integrated luminosity in the range of $10-20$~\iab\ per
experiment, corresponding to an ultimate instantaneous luminosity
approaching $2\times 10^{35}$~\lum~\cite{benedikt}, seems therefore well-matched
to our current perspective on extending the discovery reach for new
phenomena at high mass scales, high-statistics studies of possible new
physics to be discovered at (HL)-LHC, and incisive studies of the
Higgs boson's properties.  Specific measurements may set more
aggressive luminosity goals, but we have not found generic arguments
to justify them. The needs of precision physics arising from new
physics scenarios to be discovered at the HL-LHC, to be suggested by
anomalies observed in $e^+e^-$ collisions at a future linear or circular
collider, or to be discovered at 100~TeV, may well drive the need for
even higher statistics. Such requirements will need to be established
on a case-by-case basis, and no general scaling law gives a robust
extrapolation from 14 TeV. Further work on \textit{ad hoc} scenarios,
particularly for low-mass phenomena and elusive signatures, is
therefore desirable.

For a large class of new-physics scenarios that may arise from the
LHC, less aggressive luminosity goals are acceptable as a compromise
between physics return and technical or experimental challenges. In
particular, even luminosities in the range of $10^{32}$~\lum\ are
enough to greatly extend the discovery reach of the 100~TeV collider
over that of the HL-LHC, or to enhance the precision in the
measurement of discoveries made at the HL-LHC.

We have given an overview of the impressive raw capabilities of the
100 TeV $pp$ collider. Of course, given that we can extrapolate the
SM alone to ultra-high energies, there is no guarantee
that this collider will see new particles. However, the production of
new particles has never been an aim in itself: the driving ambition of
our field has always been to uncover {\it new principles} of physics,
as they are needed. And as we have stressed, with the discovery of the
Higgs we are fortunate to find ourselves in an era where such
fundamentally new principles are called for, the character of which
will be illuminated by direct studies of the Higgs itself.
Nonetheless, in thinking about physics that may exist beyond the
Higgs, it is important to ask whether the reaches of the 100 TeV are
the right ones.
Our goal in the rest of this review is to address this issue, identifying  fundamental physics questions which are squarely within the cross-hairs of the 100 TeV $pp$ collider.

\section{The Electroweak Phase Transition}

\subsection{General Remarks}

For decades, particle physics has been driven by the question of what
breaks the electroweak symmetry.  With the discovery of the Higgs, we have
discovered the broad outlines of the answer to this question: the
symmetry breaking is associated with at least one weakly coupled
scalar field.  However, this gives us only a rough picture of the physics,
leaving a number of zeroth order questions wide open that must be
addressed experimentally, but cannot be definitively settled at the LHC.
These questions include what is the shape of the symmetry breaking
potential, and how is electroweak symmetry restored at high scales.

\hyphenation{pa-ra-me-tri-za-tion}

The SM picture for electroweak symmetry breaking follows
the Landau-Ginzburg parametrization of second-order phase transitions,
\begin{equation}
V(h) =  m_h^2 h^\dagger h + \frac{1}{2} \lambda (h^\dagger h)^2 , 
\end{equation}
with $m_h^2<0$ and $\lambda>0$. This is the simplest picture
theoretically, and the one we would expect on the grounds of effective
field theory, in which we include the leading relevant and marginal
operators to describe low energy physics.  On the other hand, as we
will review in more detail in our discussion of naturalness, this
picture is far from innocuous or ``obviously correct" --- for instance it
is precisely this starting point that leads to the all vexing
mysteries of the hierarchy problem!

The central scientific program directly continuing from the discovery
of the Higgs must thus explore whether this simplest parametrization
of electroweak symmetry breaking is actually the one realized in
Nature. And while we have discovered the Higgs, we are very far from
having confirmed this picture experimentally. As illustrated in Fig.~\ref{fig:EWPT_drawing}, the LHC will only probe the small, quadratic oscillations around the symmetry breaking vacuum, without giving us any idea of the global structure of the potential. For
example, the potential could trigger symmetry breaking by balancing a
negative quartic against a positive sextic \cite{Zhang:1992fs,Zhang:1994qb,Grojean:2004xa}, i.e.
\begin{equation}
\label{eq:h6_lagragian}
V(h) \rightarrow m_h^2 (h^\dagger h) + \frac{1}{2} \lambda (h^\dagger
h)^2 + \frac{1}{3! \Lambda^2} (h^\dagger h)^3 ,
\end{equation}
with $\lambda<0$. The potential might not even be well-approximated by
a polynomial function, and may instead be fundamentally non-analytic,
as in the early Coleman-Weinberg proposal for symmetry breaking~\cite{Coleman:1973jx}:
\begin{equation}
V(h) \rightarrow \frac{1}{2} \lambda (h^\dagger h)^2 {\rm log} \left[
\frac{(h^\dagger h)}{m^2} \right] .
\end{equation}

These possibilities are associated with totally different underlying
dynamics for electroweak symmetry breaking than the SM,
requiring new physics beyond the Higgs around the weak scale. They
also have radically different theoretical implications for
naturalness, the hierarchy problem and the structure of quantum field
theory.

\begin{figure} [h!]
\begin{center}
  \includegraphics[width=0.9\textwidth]{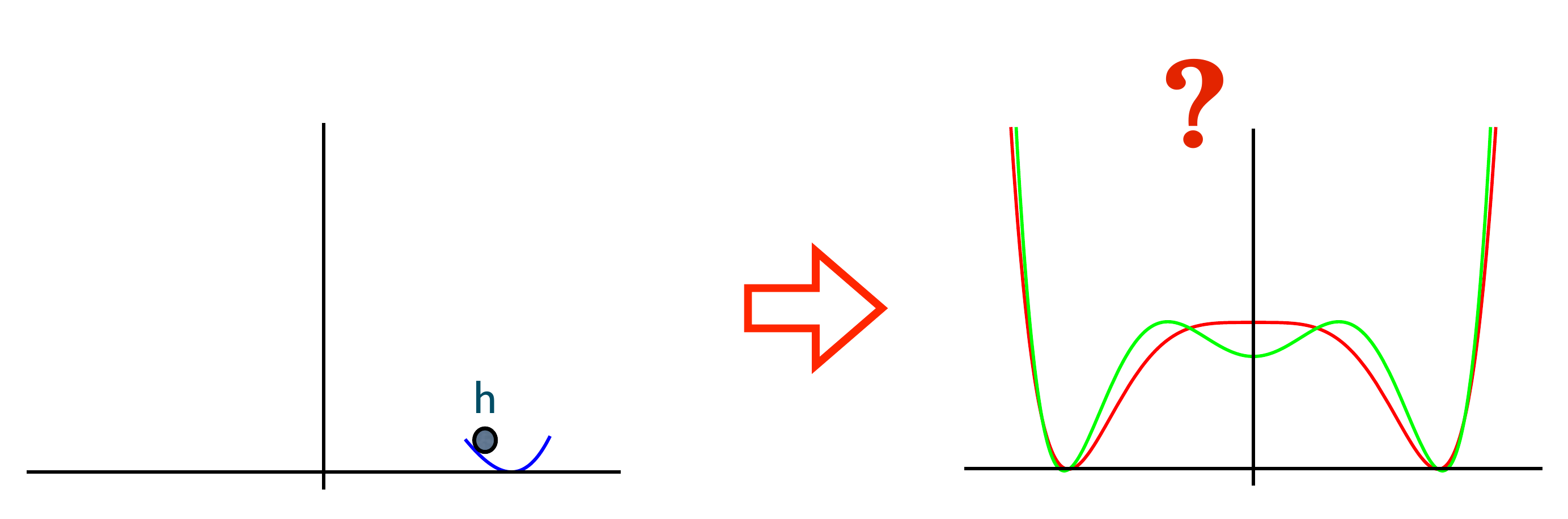}
\end{center}
\caption{Question of the nature of the electroweak phase transition.}
\label{fig:EWPT_drawing}
\end{figure}

The leading difference between these possibilities shows up in the
cubic Higgs self-coupling. In the SM, minimizing the potential gives
$v^2 = 2 |m_h|^2/\lambda$. Expanding around this minimum $h = (v
+H)/\sqrt{2}$ gives 
\begin{equation}
V(H) = \frac{1}{2}m_H^2 H^2 + \frac{1}{6} \lambda_{hhh} H^3 + \cdots,\ \ {\rm with}\ \ m_H^2 = \lambda v^2\ \  {\rm and}\ \  \lambda_{hhh}^{SM} = 3 (m_H^2/v). 
\end{equation}
Consider the example with the quartic balancing against a
sextic and, for the sake of simplicity to illustrate the point, let us
take the limit where the $m_h^2$ term in the potential can be neglected.
The potential is now minimized for $v^2 = 2 |\lambda| \Lambda^2$, and
we find 
\begin{equation}
m_H^2 = \lambda v^2,\ \  \lambda_{hhh} = 7 m_H^2/v = (7/3) \lambda_{hhh}^{SM},
\end{equation}
giving an $O(1)$ deviation in the cubic Higgs coupling relative to the
SM. In the case with the non-analytic $(h^\dagger h)^2$ log$(h^\dagger
h)$ potential, the cubic self-coupling is $\lambda_{hhh} = (5/3) \lambda_{hhh}^{SM}$.

Even larger departures from the
standard picture are possible --- we don't even know whether the dynamics
of symmetry breaking is well-approximated by a single light, weakly
coupled scalar, as there may be a number of light scalars,  and not all
of them need be weakly coupled! 

Understanding this physics is also directly relevant to one of the
most fundamental questions we can ask about {\it any} symmetry
breaking phenomenon, which is what is the order of the associated phase
transition.  Is the electroweak transition a cross-over, or might it
have been strongly first-order instead? And how do we attack this question experimentally? 
This question is another obvious next step
following the Higgs discovery: having understood what breaks
electroweak symmetry, we must now undertake an experimental program to
probe how electroweak symmetry is restored at high energies.

A first-order phase transition is strongly motivated by the
possibility of electroweak baryogenesis~\cite{Kuzmin:1985mm}. While the origin of the
baryon asymmetry is one of the most fascinating questions in physics,
it is frustratingly straightforward to build models for baryogenesis
at ultra-high energy scales, with no direct experimental consequences.
However, we are not forced to defer this physics to the deep
ultraviolet: as is well known, the dynamics of electroweak symmetry
breaking itself provides all the ingredients needed for
baryogenesis. At temperatures far above the weak scale, where
electroweak symmetry is restored, electroweak sphalerons are
unsuppressed, and violate baryon number. As the temperature cools to
near the electroweak transition, bubbles of the symmetry breaking
vacuum begin to appear. CP violating interactions between particles in
the thermal bath and the expanding bubble walls can generate a net
baryon number. If the phase transition is too gradual (second order),
then the Higgs vacuum expectation value (VEV) inside the bubbles turns on too slowly, so the
sphalerons are still active inside the bubble, killing the baryon
asymmetry generated in this way. However, if the transition is more sudden
(first order), the Higgs VEV inside the bubble right at the transition
is large, so the sphalerons inside the bubble are Boltzmann suppressed
and the baryon asymmetry can survive. This requires that 
\begin{equation}
\exp(-\Delta E_{sph}/T_c) < \exp(-10), 
\end{equation}
and can be translated to a rough criterion on the size of the Higgs expectation value at the
transition:
\begin{equation}
\frac{\langle h \rangle(T_c)}{T_c} >  0.6 - 1.6.
\label{eq:tc}
\end{equation}

In the SM with $m_H \approx 125$ GeV, the electroweak phase
transition is not strong enough to satisfy this condition. The CP
violation in the Cabibbo-Kobayashi-Maskawa (CKM) quark mixing matrix
is not large enough to generate the asymmetry.  Hence, in
order to make this beautiful idea work,  we have to go beyond the
SM. Getting the needed amount of CP violation is easy with
the addition of new particles and interactions near the weak scale,
without being in conflict with the stringent limits from electric
dipole moments for the electron and neutron. However, while we can
probe for new CP phases indirectly, by the continued search for
electric dipole moments, it is both difficult and highly-model
dependent to probe CP violation at colliders. On the other
hand, the physics needed for a sufficiently first-order phase
transition is a perfect target for future colliders. We will use the
requirement in Eq.~(\ref{eq:tc}) as our benchmark for probing an
``interestingly'' strong first order transition.

\subsection{Tests at the 100 TeV $pp$ Collider}

Colliders cannot replicate the high-temperature conditions of
the early universe at the electroweak scale. However, a 100 TeV collider can provide
an extremely powerful probe of  physics that could alter electroweak symmetry
breaking dynamics enough to make the phase transition first-order.
A large change in the structure of the Higgs potential
leads to an $O(1)$ deviation in the triple Higgs self-coupling
relative to the SM, which will be probed to about 10$\%$
level at a 100 TeV $pp$ collider. Furthermore, there must be additional particles
beyond the Higgs, with mass not too much heavier than the weak scale,
and relatively strongly coupled to the Higgs, in order to be able to
qualitatively change the order of the transition relative to the
minimal SM. While such particles can escape detection
at the LHC, they are a perfect target for 100 TeV colliders. Even in the
most difficult scenario in which the new particles only affect the phase transition at
loop-level, the combination of deviations in the Higgs triple coupling
and direct production of the new states at a 100 TeV collider covers most of the
allowed parameter space in the examples studied to date.

Of course, we are not claiming a ``no-lose'' theorem, and it may be possible to
engineer models that change the order of the phase transition while suppressing
the 100 TeV collider signals. However, such scenarios would appear to need some contrivance.
Our aim in this section is to show that a 100 TeV collider can robustly cover the space
of possibilities for simple models generating a first-order phase transition.

The simplest toy model for a first-order transition simply augments the
SM with a higher-dimension operator as in Eq.~(\ref{eq:h6_lagragian}) \cite{Zhang:1992fs,Zhang:1994qb,Grojean:2004xa,Bodeker:2004ws,Delaunay:2007wb}.
At leading order (which suffices for our purposes here) finite temperature
effects merely add the usual quadratic shift to the quadratic part
of the potential $m^2(T) \to m_h^2 + c T^2$ for a positive constant $c$
determined by the top Yukawa and gauge couplings.  A first-order phase
transition can be achieved if the quartic term is negative ($\lambda<0$).
As we saw earlier, in this example we have an
$O(1)$ deviation in the Higgs self-coupling, and this is a general
expectation for any theory where the first-order phase transition is
driven by a large change in the (zero-temperature) Higgs potential.

Purely by effective field theory rules, it is consistent to have a
theory where $(h^\dagger h)^3$ is the only dimension-6 operator at
leading order. It is amusing that this choice is even radiatively
stable at leading order: $(h^\dagger h)^3$ does not induce any of the
other dimension 6 operators involving the Higgs under 1-loop RG
evolution. 

In any reasonable UV completions we can expect other higher-dimension
operators in addition to $(h^\dagger h)^3$. While the UV physics may
preserve custodial $SU(2)$ and give suppressed contributions to the
precision electroweak operators, there is no symmetry distinction
between the $(h^\dagger h)^3$ operator and the operator $[\partial_\mu
  (h^\dagger h)]^2$, so they are expected to be generated as well, and
to affect the $ZZH$ couplings, which can be probed at the per-mille
level by $e^+ e^-$ Higgs
factories~\cite{Gomez-Ceballos:2013zzn,cepc_website,Fujii:2015jha,Linssen:2012hp}. However
it is the $(h^\dagger h)^3$ operator that is directly related to the
physics of the electroweak transition, which is most powerfully probed
at the 100 TeV collider.

We begin by considering the simplest example of a theory where these
couplings are generated at tree-level by integrating out a massive
singlet $S$ coupled to the Higgs. As we will see, this example
represents the ``easiest" case, where it is straightforward to get a
first-order phase transition, with large associated signals for 100 TeV colliders. Since this is an ``easy" case, we will use it
largely to illustrate the important physics points parametrically. We
will then move to the ``hard" case, where the order of the transition
is only affected at 1-loop.

The important interactions for this toy model are given by
\begin{equation}
m_h^2 h^\dagger h + \frac{\tilde{\lambda}}{2} (h^\dagger h)^2 + \frac{1}{2} m_S^2 S^2 +
a m_S S h^\dagger h + \frac{b}{3!} m_S S^3 + \frac{\kappa}{2} S^2 h^\dagger
h + \frac{1}{4!} \lambda_S S^4 .
\label{eq:singletmodel}
\end{equation}
The dimensionless couplings $a,b$ can be set to zero by a $Z_2$
symmetry under which $S \to - S$, but in the absence of such a symmetry they
should be present. We will concentrate on the limit where the $b m_S S^3$
interaction is negligible. Integrating $S$ out at tree-level gives rise to both the modified Higgs potential
as well the oblique Higgs operator as
\begin{equation}
m_h^2 h^\dagger h + \frac{\lambda}{2} (h^\dagger h)^2 + \frac{\kappa a^2}{2 m_S^2}
(h^\dagger h)^3 \, \, + \, \,  \frac{a^2}{2 m_S^2} (\partial_\mu
(h^\dagger h))^2 .
\end{equation}
Here $\lambda = \tilde{\lambda} - a^2$. Neglecting the $m_h^2$ term as above,
the first-order transition is driven with $\lambda<0$, $k>0$, and we can
determine the electroweak scale and Higgs masses as
\begin{equation}
v^2 = {4\over 3} {m_S^2 |\lambda| \over \kappa a^2},\quad m_H^2 = |\lambda| v^2.
\end{equation}
We can also find the shift in the $ZZH$ coupling as
\begin{equation}
\delta Z_H = {4\over 3} {a^2 v^2 \over m_S^2} = {4\over 3}{ |\lambda| \over \kappa}.
\end{equation}
In order to avoid an unwanted $O(1)$ shift to the $ZZH$ coupling,
we must have $\kappa \gg \lambda$. This is perfectly consistent since
$\lambda$ is highly perturbative. It is interesting that despite the
presence of a relatively strong coupling of the Higgs to a new massive
state, there are no difficulties whatsoever with large precision
electroweak corrections; this is closely related to the fact that the
$O(1)$ deviation in the Higgs cubic couplings associated with the
$(h^\dagger h)^3$ term does not radiatively induce precision
electroweak operators at one-loop. For the couplings to be self-consistently
perturbative, we must have 
\begin{equation}
\kappa^2/16 \pi^2 \lesssim |\lambda|\quad {\rm and}\quad a^4/16 \pi^2 \lesssim |\lambda|. 
\end{equation}
Since $\kappa$ cannot become too large, the correction
$\delta Z_H=(4/3) (|\lambda|/\kappa)$ cannot be too small and the singlet mass
$m_S = \sqrt{\frac{3 \kappa a^2}{4| \lambda|}} v$ cannot be too heavy, we thus find
\begin{equation}
\delta Z_H \gtrsim \frac{4}{3}\frac{\sqrt{|\lambda|}}{4 \pi} = 0.05,\quad m_S \lesssim \frac{\sqrt{3}}{2} 4 \pi v =2.7\ {\rm TeV}.
\end{equation}
A similar conclusion holds even if the $b m_S S^3$ term is included and dominates; the parametrics changes slightly and we find instead
\begin{equation}
\delta Z_H \gtrsim 4 \left(\frac{\sqrt{|\lambda|}}{4 \pi}\right)^{3/2} =
0.03, \quad m_S  \lesssim 2 \pi v \left(\frac{4 \pi}{\sqrt{|\lambda|}}\right)^{1/4} =3.4\ {\ \rm TeV}.
\end{equation}
Note that the bounds correspond to extreme limits of strong coupling, and it is most reasonable for the new couplings to be perturbative, so $m_S$ is most plausibly in the range of a few hundred GeV. 

These estimates quantify the intuitive expectation that any new physics giving
a first-order phase transition cannot be too heavy and too weakly
coupled to the Higgs. 

We also get an associated $O(1)$ deviation in the Higgs triple coupling,
and a singlet mass in the range of at most a few TeV, both of which are easily
accessible to a 100 TeV $pp$ collider. Since the singlet mixes significantly with the
Higgs, the singlet is produced just as heavy Higgs bosons would be, and the 
significant decays are $S \to HH, ZZ, W^+ W^-$ and $t \bar{t}$.
A rough estimate of the 100 TeV reach for $pp \to S \to HH$ in these
modes is shown in Fig.~\ref{fig:reach_Shh}. 
Here $c$ is a measure of the mixing between the singlet $S$ and the Higgs boson. We have $c \sim (a v)/m_S \sim (m_H/m_S)$, so this mixing is expected to be sizable.
\begin{figure} [h!]
\begin{center}
  \includegraphics[width=0.5\textwidth]{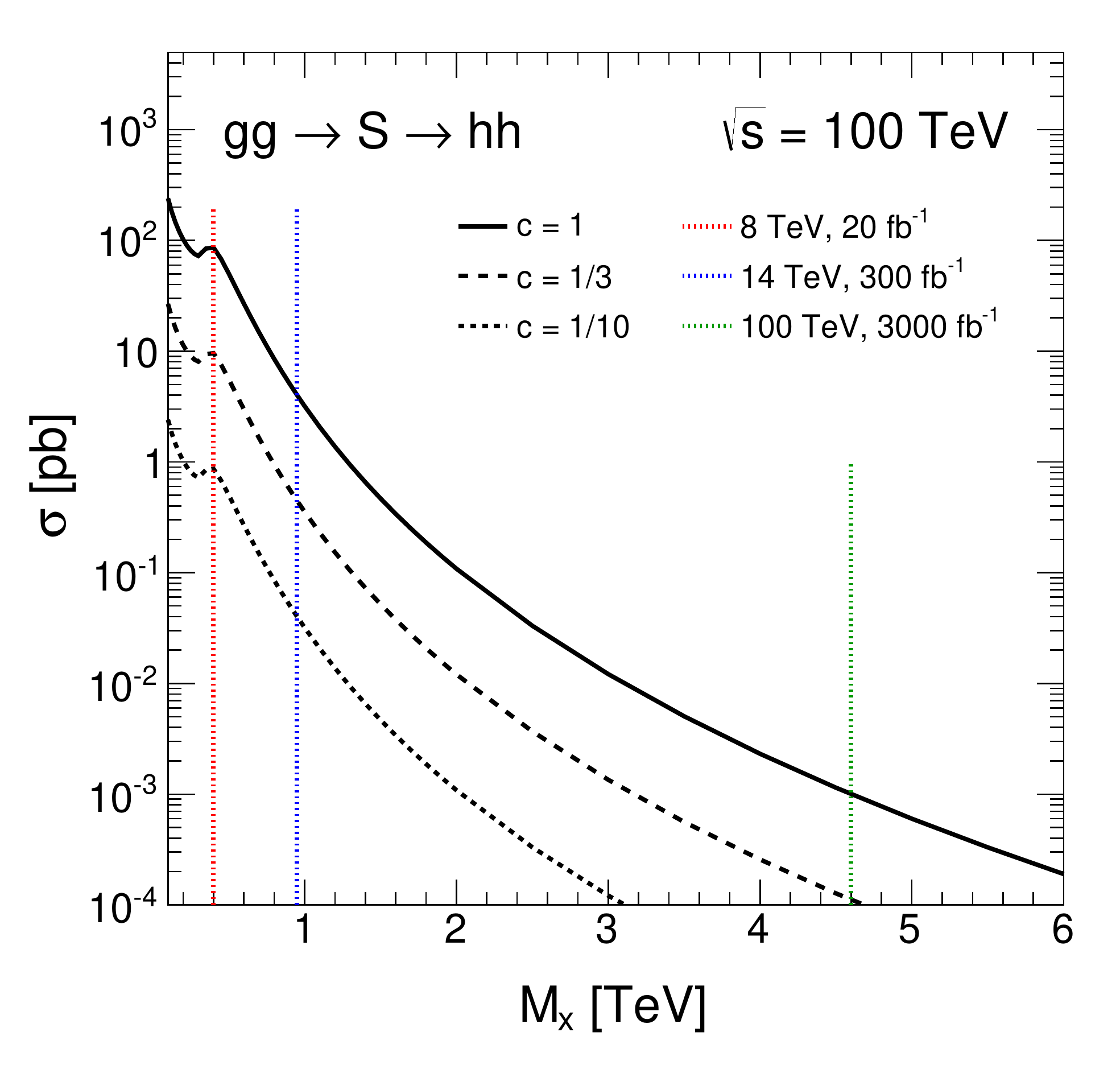}
\end{center}
\caption{Estimate of reach in the $gg \to S \to HH$ channel at HL-LHC
  and a 100 TeV collider extrapolating from an ATLAS
  search~\cite{Aad:2014yja}.  The reach for the $S$ scalar mass, shown
  by the vertical lines for different $pp$ CM energies,
  assume that one Higgs
  decays to $\bar{b}b$ and the other to $\gamma\gamma$, and refer to
  the case $c=1$.  }
\label{fig:reach_Shh}
\end{figure}

In the above analysis we have assumed that $m_S^2>0$, so that the
singlet is localized to the origin throughout the phase transition.
There is also a qualitatively different possibility with $m_S^2<0$.
Here, we can imagine that  it is really the phase transition for $S$
that dominates the physics, and drags the Higgs along with it, since
the effective Higgs mass term depends on $\langle S \rangle$ as
$m^2_{h, eff} = m_h^2 + a m_S \langle S \rangle + \kappa
\langle S \rangle^2$. The dynamics in the $S$-sector can make the $S$
phase transition strongly first-order, at a temperature $T_c \sim
\langle S \rangle$. Thus if we wish to have $\langle h \rangle/T_c
\sim 1$, we should have $\langle h \rangle \sim \langle S \rangle$.
This again gives us the obvious upper bound to the mass $m_S$, $m_S <
4 \pi \langle S \rangle \sim 4 \pi v \sim 2$ TeV, and $S$ 
accessible to direct production at a 100 TeV collider. 

Having discussed the ``easy" cases for new physics giving a
first-order electroweak phase transition, let us consider what appears
to be the most difficult possible case, where a first-order
electroweak phase transition is driven entirely by radiative effects,
coupling the Higgs to SM singlet fields. This case is
realized in our singlet model, if we further impose a $Z_2$ symmetry
so that $a,b=0$. This makes $S$ exactly stable, and it
could indeed be a component of Dark Matter. However this aspect is not
relevant to our discussion; we may always assume a minuscule amount of
$Z_2$ breaking giving a small $a,b$ which allow
$S$ to decay on cosmological timescales.

As with our tree-level example, there are two qualitatively
different cases to consider. When $m_S^2>0$, the role of the
singlet is to give a large deformation to the Higgs potential at
1-loop, enabling a first-order phase transition directly in the Higgs
direction. This will require $\kappa$ to be large, which can be
accomplished within a consistent weak-coupling approximation. In this
case we expect a large correction to the zero-temperature Higgs
potential and so an $O(1)$ deviation in the Higgs triple coupling. On
the other hand, if $m_S^2<0$, we can have a two-step phase transition,
where a first-order transition in $S$ forces a first-order transition
for $h$.

A detailed analysis of the model parameter space allowing a strong
first-order phase transition has recently been given in
\cite{Curtin:2014jma}, as shown in Fig.~\ref{fig:parameter_Z2}.
\begin{figure} [h!]
\begin{center}
  \includegraphics[width=0.55\textwidth]{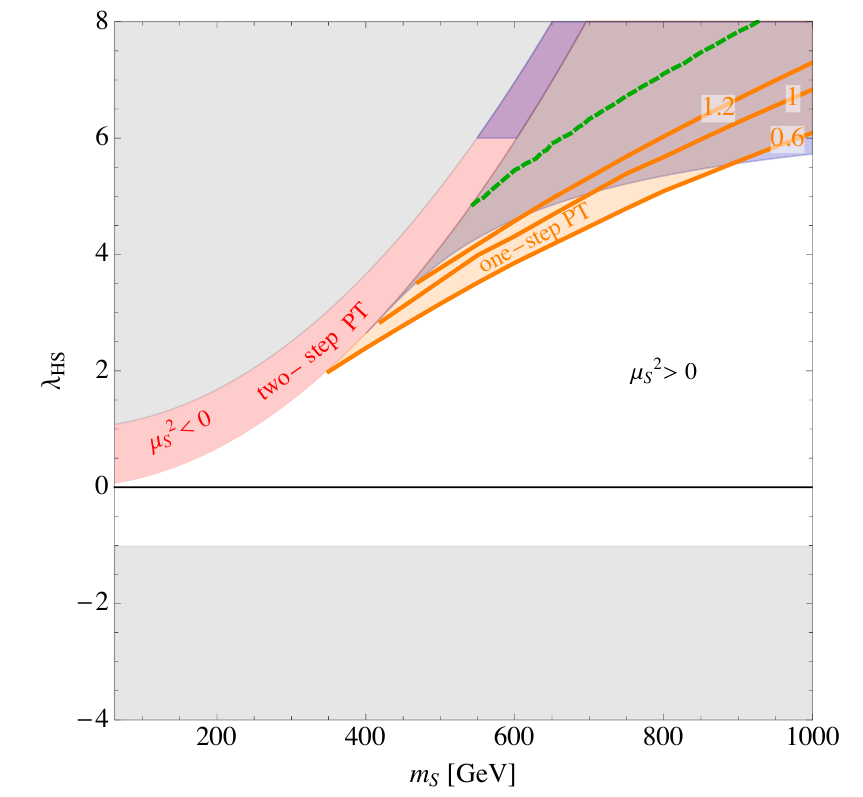}
\end{center}
\caption{Parameter space with first order phase transition in the
$Z_2$ model \cite{Curtin:2014jma}.
Red shaded region: for $m_S^2<0$ ( $m_S^2$ is denoted as $\mu_S^2$ in this figure~\cite{Curtin:2014jma}), it is possible to choose $\lambda_S=\kappa/2$ (in Eq.~(\ref{eq:singletmodel})) to get tree-induced two-step first-order electroweak
phase transition. Orange contours: value of $v_c/T_c$ for $m_S^2 > 0$.
The orange shaded region indicates $v_c/T_c > 0.6$, where a one-step transition can be sufficiently first-order for electroweak baryogenesis. Above the green dashed line, singlet loop corrections generate a barrier between $h = 0$ and $h = v$ even at zero temperature, but results in the dark shaded region might not be reliable.}
\label{fig:parameter_Z2}
\end{figure}
The two-step transition operates for smaller values of the singlet
masses and couplings, while larger masses and couplings can give rise
to the modified Higgs potential giving the one-step transition along
the Higgs direction.

In all cases, the singlet $S$ is lighter than $\sim 1$ TeV, and so
certainly kinematically accessible to a 100 TeV collider. In this worst-case
scenario, since $S$ only couples in pairs to the SM via
the Higgs, as long as $m_S > m_H/2$ we must produce it via off-shell
Higgses.  Furthermore, if $S$ is collider-stable, we are looking for
missing energy signals very much like standard invisible Higgs decay
searches, the main difference being the much smaller, non-resonant
$SS$ production cross-section. The dominant channels for $SS$
production are in Vector-Boson-Fusion (VBF) $qq \to qq SS$, as well as
in associated production $qq \to V SS$ for $V=W^{\pm},Z$.
\begin{figure} [h!]
\begin{center}
\quad
  \includegraphics[width=0.56\textwidth]{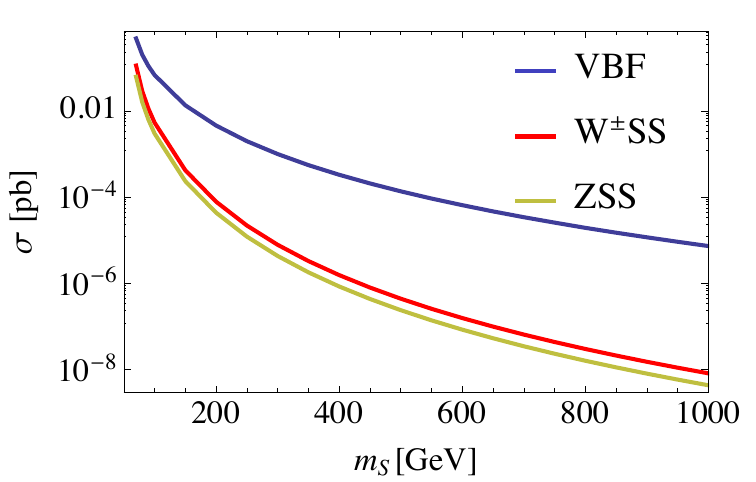}
\hfil
  \includegraphics[width=0.4\textwidth]{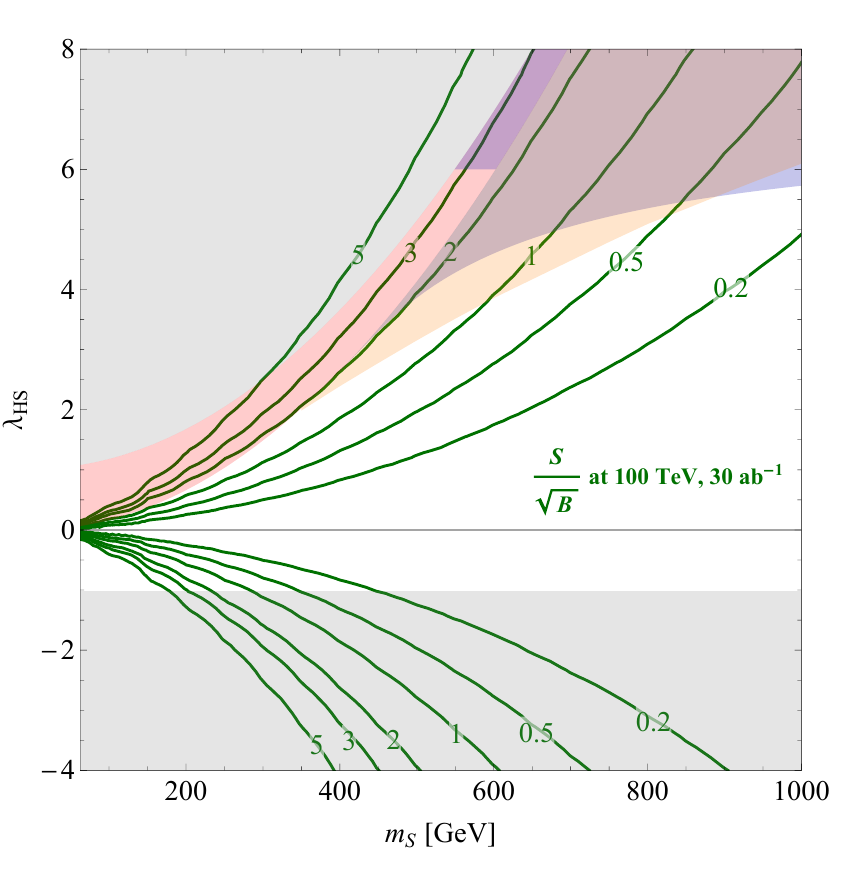}
\end{center}
\caption{
Left: Production rate for the VBF process at a 100 TeV collder. Right: $S/\sqrt{B}$ of VBF process at the 100 TeV $pp$ collider for an integrated luminosity of 30 ab$^{-1}$ \cite{Curtin:2014jma}.}
\label{fig:VBF_singlet_rate}
\end{figure}
The cross sections for these processes at a 100 TeV collider are  shown in the
right panel of Fig.~\ref{fig:VBF_singlet_rate}.
\quad
These cross sections are very small, between $10^{-2} \to
10^{-4}$~pb.  There is also a large background, in the VBF
production of $Z \to \nu \bar\nu$, which is $\sim 10^3$~pb at 100 TeV. The
authors of \cite{Curtin:2014jma} imposed a simple set of cuts to isolate the
signal,  demanding exactly two forward jets with $p^T_{1,2}>40$ GeV
and $\eta_{1,2}<5$, a missing energy cut $\slash\!\!\!\!{E}_T>150$~ GeV, jet
separation $|\eta_1 - \eta_2| > 3.5$ and $|\eta_{1,2}|>1.8$, and
$M_{jj}>$800 GeV, while rejecting leptons with $|\eta|<2.5$ and
$p_T>$15 GeV. The contours for $S/\sqrt{B}$ in the $(m_S,\kappa)$ plane
resulting from their analysis are shown in the right panel of
Fig.~\ref{fig:VBF_singlet_rate}.

Already this simple analysis suggests that the entire region of the
two-step transition can be probed by direct $SS$ production at a 100 TeV collider. Note that this is a rough first pass at studying this signal,
and one may expect to do significantly better. The main limiting
factor is the huge $Z \to \nu \bar\nu$ background, but this may be
measured directly from the data in the familiar way, from the
kinematically identical $Z \to \ell^+\ell^- $ process, which should give a
sharp handle on the systematics.
In the part of parameter space giving the one-step transition, the
direct production of $SS$ is swamped by the $Z \to \nu \bar\nu$
background. However, this is exactly the case in which we expect an
$O(1)$ deviation to the Higgs cubic coupling, as shown in the left
panel of  Fig.~\ref{fig:HiggsCoupling_Z2}.
\begin{figure} [h!]
\begin{center}
\includegraphics[width=0.42\textwidth]{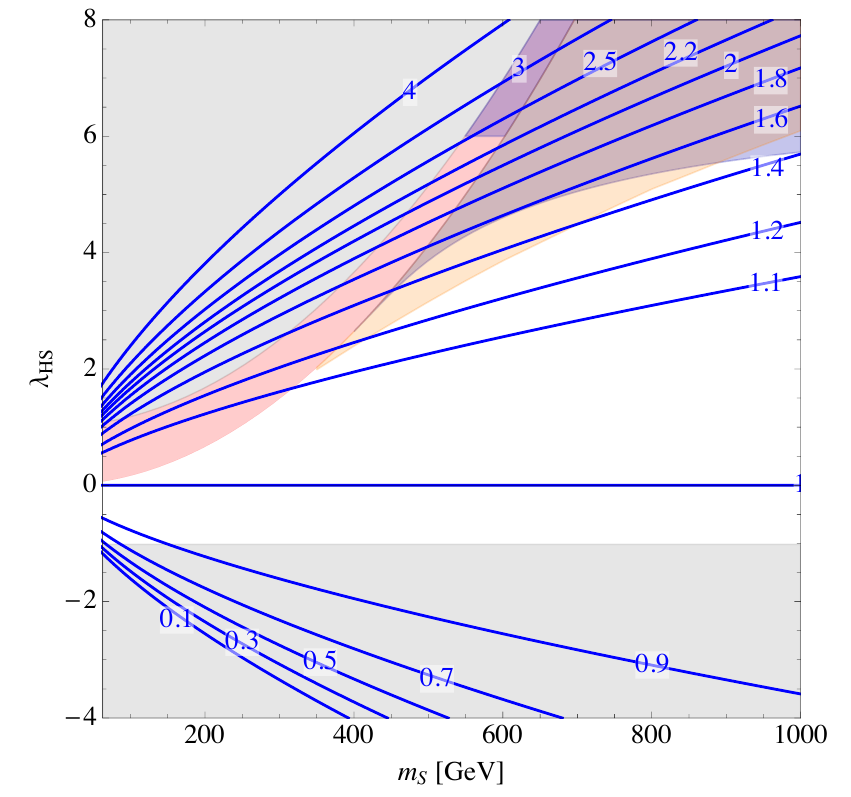}
\quad
\includegraphics[width=0.42\textwidth]{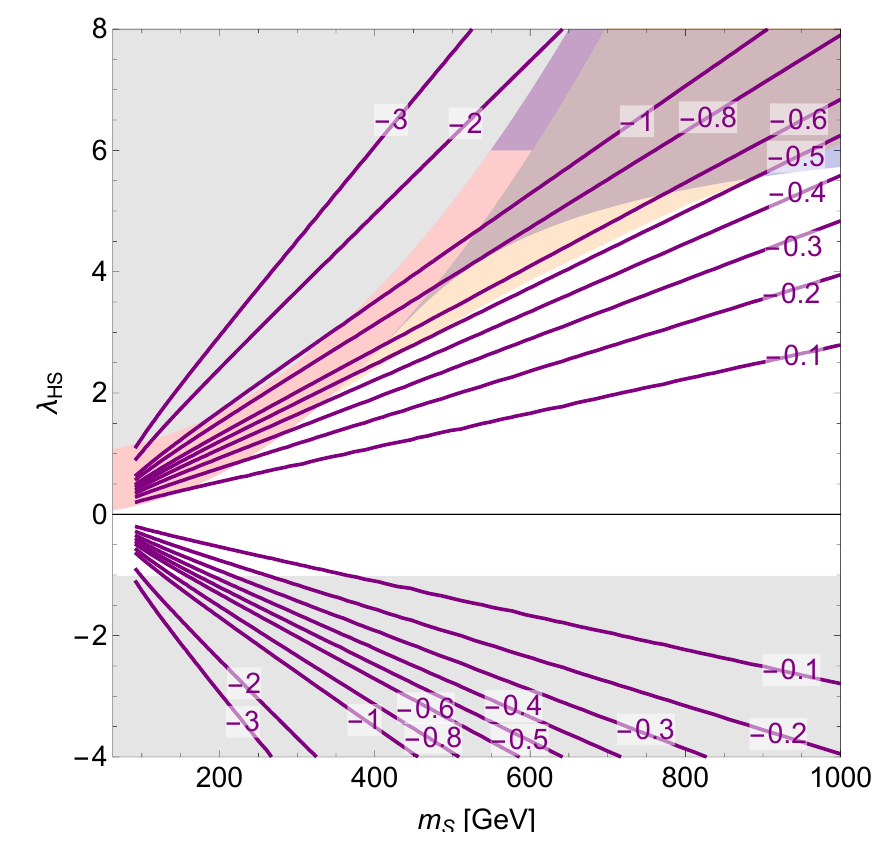}
\end{center}
\caption{Left: Shift in triple Higgs coupling in the $Z_2$ singlet model. Right: Percentage shift in the $e^+ e^- \rightarrow ZH$ cross section, which is directly proportional to $\delta Z_H$. }
\label{fig:HiggsCoupling_Z2}
\end{figure}
We see that even pushing to the limit of $\langle h \rangle/T_c \sim 0.6$, we must have a deviation in the triple Higgs coupling of at least
20\%, which is visible at a 100 TeV collider.

\begin{figure} [h!]
\begin{center}
\includegraphics[width=0.5\textwidth]{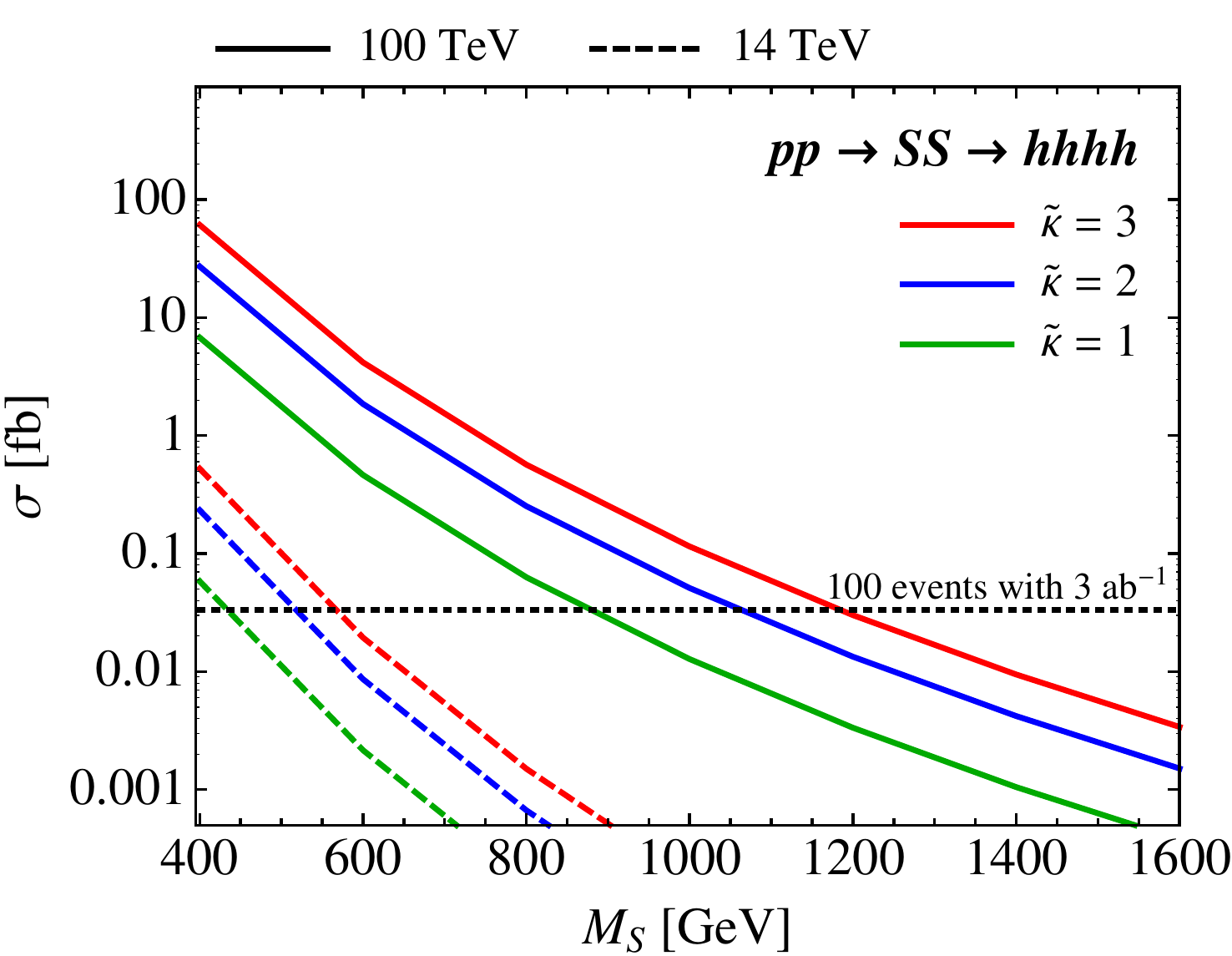}
\end{center}
\caption{Rate of process $pp \to SS \to HHHH$ at the LHC and an 100 TeV $pp$ collider. }
\label{fig:pairS}
\end{figure}

We conclude that, even in this very worst case scenario, a 100 TeV pp collider
allows us to probe the physics giving us a first-order phase
transition. Needless to
say, even small modifications from this worst-case scenario can make
detection much easier.  For instance, if the $Z_2$ symmetry is broken
by an even tiny amount so that $a > 10^{-10}$, then $S$ will decay as
$S \to HH$ inside the detector. Direct $S$ production will be much
easier to see, giving a spectacular signal $pp \to SS \to HHHH$. This
should allow a 100 TeV $pp$ collider to cover the allowed range of
$m_S$ up to 1 TeV. While a detailed study is left for future work, an
estimate of the reach for producing 100 events is shown in
Fig.~\ref{fig:pairS}.  Note that, while at fixed mass the cross
section at 100 TeV is at least $\sim 100$ times larger than at the LHC, the
mass reach is $\sim$ 2.5 times greater, compared to the typical factor
of $\sim 5$ we are accustomed to.  This is because both the production
and decay vertices of the off-shell Higgs are suppressed by factors of
$(v/E)$ at high energies, and the cross-section scales as $v^4/E^6$
rather than the usual $1/E^2$. These suppressions would be absent with
more physical Higgses in the final state. It would be interesting to
see whether such final states with large Higgs multiplicity can be seen at
a 100 TeV $pp$ collider.

We have seen in our simple examples something we expect to hold more
generally for models that drive a first-order phase transition: there
should be large signals at a 100 TeV collider, either through the
direct production of new states, or via an $O(1)$ deviation in the
cubic Higgs self-coupling. Probing the electroweak transition does not
need a $10^3$ TeV $pp$ collider; 100 TeV $pp$ collisions are just right to
robustly probe this physics.

\section{Naturalness of the Electroweak Scale}
\label{sec:natural}

The notion of naturalness, as introduced by Ken Wilson and Gerard 't
Hooft in the late 1970's, is deeply connected to our understanding of
the structure of effective field theory, strongly supported by
analogies with condensed matter physics. Naturalness has been the
dominant force driving our thinking about physics beyond the Standard
Model for the past four decades, suggesting a rich spectrum of new
physics at the weak scale.

However, there have also been reasons to question this doctrine
throughout this period.  Most glaringly, naturalness seems to fail
spectacularly for the cosmological constant, though this involves
mysteries of gravity and cosmology that may not be relevant for
particle physics.  Within particle physics, there have also been a
number of counter-indications to naturalness, from the lack of
indirect signals that might have been induced by new physics at the
weak scale in low energy flavor and CP violation to the absence of new
states going back to LEP and the Tevatron. The absence of new physics
at LHC Run 1 continues this trend and appears to put naturalness under
further pressure. Settling the ultimate fate of naturalness is perhaps
the most profound theoretical question of our time that is amenable to
experimental tests, and will largely dictate the future development
of fundamental physics in this century.

We will begin with a brief overview of this set of ideas to put them
in context and elucidate their importance. As we will see, on top of
what we learn from LHC14, a 100 TeV collider is certain to play a
decisive role in unraveling this physics.

\subsection{On the Mass of the Higgs Boson}

A good place to begin a discussion of naturalness is to look at the name of the Standard Model itself, which has an apt moniker, since it gives us a {\it model}, rather than a deeper {\it theory}, for electroweak symmetry breaking. This is most obviously seen by the fact that $m_h^2$ is a parameter of the theory; its value is not predicted, but must be taken from experiment. Even the most qualitative property of the Higgs potential --- the negative sign of $m_h^2$, leading to symmetry breaking --- is not predicted. The SM allows us to {\it model and parametrize} symmetry breaking, but it certainly does not give us a real {\it understanding} of its origin.

The famous quadratically divergent radiative corrections to the Higgs
mass, dominantly from the top quark at 1-loop
\begin{equation}
\delta m_h^2 \sim \frac{3 y_t^2}{8 \pi^2} \; \Lambda_{\rm UV}^2 \sim
(0.3\; \Lambda_{\rm UV})^2 
\end{equation}
is one indication of the fact that the Higgs mass parameter cannot be computed in the SM. Note that purely within the SM, nothing obliges us to think about ``UV sensitivity", ``fine-tuning" or the ``hierarchy problem" --- since there is no computation of the Higgs mass in the SM, these notions are not precise. There is no well-defined computation of the Higgs mass to complain about ``fine-tuning", and there is certainly no theoretical inconsistency with taking the value of the weak scale from experiment.
However, we will immediately confront these issues in attempting to find a real theory where we can actually {\it calculate} the Higgs mass.

What should such a theory look like? Especially over the past century, we have been driven by the reductionist paradigm, in which explanations for mysterious low energy phenomena are to be found in a more fundamental high energy theory. Following this tradition, there should be an UV scale $\Lambda_h$, above which we find the theory in which the Higgs mass becomes calculable. Unlike the SM, in this theory there will be a concrete formula for the Higgs mass, which should take the form
\begin{equation}
m_h^2 = a \Lambda_h^2 + b \frac{3 \lambda_t^2}{8 \pi^2} \Lambda_h^2 + \ldots
\end{equation}
with $a,b,\ldots$ dimensionless constants that are calculable in the theory.

There are then two possibilities: (A) $\Lambda_h \sim m_h$ with $a,b, \ldots$ of $O(1)$. In this case we say the physics is ``natural", and the physics at the scale $\Lambda_h$ gives a complete account of electroweak symmetry breaking. Otherwise (B) $\Lambda_h \gg m_h$; this entails an extreme correlation between deep UV and IR physics. While such a correlation is a logical possibility, we have never seen anything like this before, anywhere else in physics.

Let us illustrate these possibilities with a concrete example, to show how the naturalness issues are forced upon us as soon as we find a theory in which the Higgs mass becomes calculable. Let us start with a toy model of a light scalar $\Phi$ with mass $m_\Phi^2$,  which is in the adjoint representation of an $SU(2)$ gauge group with coupling $g$.  This mass $m_\Phi^2$ is incalculable just as the Higgs mass is incalculable in the SM. But there is a simple UV completion where it can be unambiguously computed: consider a five-dimensional gauge theory with gauge coupling $g_5$ compactified on a circle of radius $R$. The gauge field is obviously massless in the UV, but at energies much smaller than $1/R$ and at tree-level, we have a massless four-dimensional gauge field with coupling $g^2 = g_5^2/R$, and a massless scalar in the adjoint representation. The scalar will pick up a mass at 1-loop, and the radiative corrections in the full theory are  calculable:
\begin{equation}
m_\Phi^2 = \frac{3 \zeta(3)}{\pi^2} \times \frac{3 g_4^2}{4 \pi^2} \times \frac{1}{R^2}.
\end{equation}

Now, $1/R$ also sets the mass of the new states in the theory --- the Kaluza-Klein excitation of the gauge boson. Thus, in this UV completion where the scalar mass becomes calculable, it is simply impossible to keep the scalar much lighter than the new KK states: there must be ``new physics" in the model, parametrically at exactly the energy scale predicted from the classic back-of-the-envelope estimates following from the quadratic divergence in the low energy theory, cut-off at the scale $1/R$.

Simple variants of this model, where the extra dimension is an
interval, are used in various guises of realistic theories for the
Higgs as a pseudo-Nambu-Goldstone boson, interpreted in either
extra-dimensional or four-dimensional terms. Of course the realistic
theories include a top quark with an adjustable Yukawa coupling
$y_t$. Once again, the Higgs mass can be completely calculated as
\begin{equation}
m_h^2 = [a (g^2/8 \pi^2) - b (3 y_t^2/8 \pi^2)] \times 1/R^2, 
\end{equation}
where $a,b>0$ are calculable, and the masses of the KK excitations of the
gauge fields and fermions are also calculable multiples of $1/R$. Even
the signs of these contributions are fixed; remarkably, one can
compute that when the top Yukawa is large, electroweak symmetry is
necessarily broken. This beautifully explains one qualitative fact ---
why is electroweak symmetry broken? --- as a consequence of the
seemingly unrelated qualitative fact that the top Yukawa is larger
than gauge couplings.

Note that, in this UV completion, it is possible to make the Higgs
much lighter than the KK excitation set by $1/R$, but only if the
couplings $g$ and $y_t$ happen to be adjusted to be extremely close to
a particular ratio.  Absent supersymmetry, there is nothing relating
these couplings, and indeed they vary with scale, so the needed
coincidence at just the right scale would be completely accidental. To
say this more vividly, if as theorists we wished to simulate this
model on a computer, we would have to very delicately move around in
parameter space in order to make the scalar very light, giving an
operational meaning to ``fine-tuning" in a concrete calculation. Of
course it is logically possible that if such a model were realized in
nature, the couplings would happen to be arranged in just a way as to
yield a light scalar. But then the explanation for the generation of
the weak scale would be deferred to the higher-energy theory, which
predicts the seemingly random choices of $y_t$ and $g$ needed to make
this happen, entailing the extreme correlation between UV and IR
physics we alluded to.

If we discount the possibility of extreme UV/IR correlations, this logic predicts that light scalars with non-derivative gauge and Yukawa interactions can never be ``lonely'' --- they must always be within a weak-coupling loop factor of heavier new physics. This conclusion has been borne out in all examples we have seen in Nature to date. For instance the charged pion is just an electromagnetic loop factor lighter than the $\rho$ meson. And we have a nice understanding for the striking absence of non-derivatively coupled scalars in condensed matter systems.

There is of course a famous example from condensed matter physics,
however, where we do see light scalars, and where the word
``fine-tuning'' has direct experimental relevance.  This is the
Landau-Ginzburg description of a system very close to a second-order
phase transition, say in a metal.  Within the reductionist paradigm,
one might naively imagine that the detailed microphysics of the
material would provide the explanation for the lightness of the scalar
field in this system.  However, this assumption is incorrect, because,
in this system, it is not the physics of the material itself that
controls the mass of the scalar, but rather the fact that the system
is coupled to an external heat bath with a temperature that can be
dialed by an experimentalist.  In this example, the experimentalist
must ``fine-tune'' the temperature to make the scalar very light.
However, from the point of view of an observer within the material
itself, the reductionist paradigm breaks down, since the explanation
for macroscopic phenomena is not simply given by specifying the
microphysics of the system, but also crucially depends on the presence
of a ``multiverse'' outside it. The much discussed picture of an
enormous landscape of vacua, populated by eternal inflation, is one
possible analog of this scenario for particle physics.

Given the experimental observation of a light elementary Higgs scalar,
we are confronted with three qualitatively different possibilities: if
the reductionist paradigm continues to be the correct guide---as it
has been for centuries---we must either discover physics to make the
Higgs mass natural, or we must allow a possibility we have never seen
before, that of an extreme correlation between the physics of the deep
UV and IR. Alternatively, we must acknowledge the failure of the
reductionist paradigm altogether, and admit that the explanation for
the lightness of the Higgs is not to be found in our
microphysics. {\it Any} of  these three conclusions would have
monumental implications for the future of fundamental physics.

\subsection{Natural Theories and the Tests at the 100 TeV $pp$ Collider}

The most conservative possibility is that naturalness holds. Even this conservative possibility involves major extensions to our picture of physics. Only a few theoretical possibilities for solving the hierarchy problem have emerged over the past few decades, starting from the early proposals of technicolor \cite{Weinberg:1975gm,Susskind:1978ms} and variants with the Higgs as a pseudo-Goldstone boson \cite{Kaplan:1983fs,ArkaniHamed:2001nc,Agashe:2004rs}; the supersymmetric SM \cite{Dimopoulos:1981zb}; and the proposals of large \cite{ArkaniHamed:1998rs} and warped \cite{Randall:1999ee} extra dimensions, the latter of which are in fact holographically dual \cite{Maldacena:1997re} to versions of technicolor and composite Higgs models. Technicolor was in many ways the most conservative,  simplest and most beautiful of these possibilities, but has been conclusively ruled out by the discovery of a light Higgs. Supersymmetry remains the best studied and most attractive possibility, especially given the striking success of supersymmetric gauge-coupling unification, precise at the percent level \cite{Dimopoulos:1981zb,Langacker:1991an}.

However, with the continued absence of both indirect and direct
evidence for new physics to date, it is also conceivable that we will
come to see that naturalness is not a good guide to TeV scale physics,
as it has perhaps already been seen to fail for the cosmological
constant. The two alternatives to naturalness represent much more
radical paradigm changes; it is true that without further positive
clues from experiment we will not know which of the options is
correct, but being forced into either of these directions would be an
epochal shift, akin to the move away from the aether triggered by the
null result of aether-drift experiments over a century ago.

Given the magnitude of the stakes involved, it is vital to get a clear
verdict on naturalness from experiment, and a 100 TeV collider will be
necessary to make this happen.  To this end, we will be maximally
conservative, and with a few exceptions will operate under the
assumption that LHC14 sees no evidence for physics beyond the SM. Let
us recall why this would be surprising from the usual perspective of
naturalness. Consider the top-loop contribution to the Higgs mass
\begin{equation}
\delta m_h^2 \sim (3 \lambda_t^2/8 \pi^2) \Lambda_{\rm UV}^2 \sim (0.3
\Lambda_{\rm UV})^2. 
\end{equation}
Asking for $\delta m_h^2$ not to be larger than $m_h^2$
tells us that there must be some new state lighter than $\sim 400$
GeV, related to the top by some new symmetry that allows it to cancel
the UV sensitivity. The couplings of this new state must be determined
by $\lambda_t$; in addition, since the ``3'' in the expression for
$m_h^2$ arises from the number of colors, the simplest possibility is
that the ``top-partner" is also colored. This is what happens in most
well-studied natural theories. The top partner in supersymmetric
theories is the (colored) stop, while the fermionic top-partners in
Little Higgs and composite Higgs theories are also colored. This is
the way in which naturalness predicted a bonanza of new physics for
the LHC, since {\it colored} 400 GeV particles could have been
copiously produced even at LHC8.

Of course this has not happened, and if the LHC continues to see nothing but the Higgs, any colored top partners will be pushed to being heavier than $\sim 1$ TeV, indicating a level of fine-tuning of typically a few percent for electroweak symmetry breaking.  As a canonical example, consider the case of supersymmetric theories, in which stop loops generate a contribution to the Higgs mass, logarithmically enhanced starting from the scale $\Lambda_{\rm X}$ where supersymmetry (SUSY) breaking is first communicated to the minimal supersymmetric SM (MSSM). This leads to a rough measure for the degree of fine-tuning, $\Delta^{-1}$~\cite{Barbieri:1987fn}, as
\begin{equation}
\Delta^{-1} \sim 10^{-2} \left(\frac{1\, {\rm TeV}}{m_{\tilde{t}}}\right)^2 \left(\frac{5}{{\rm log}(\Lambda_{\rm X} / {\rm TeV})} \right) ,
\end{equation}
and $m_{\tilde{t}} \sim 1$ TeV is tuned at the percent level.

The question then becomes, how bad is percent-level tuning? Certainly it seems qualitatively different than the $10^{-30}$ levels of tuning usually discussed if $\Lambda_{UV}$ is close to the Planck scale.  Furthermore, we have seen accidents of order a few percent elsewhere in physics, ranging from the surprisingly large nucleon-nucleon scattering length, to the accident that the moon can nearly perfectly eclipse the sun! We do not tend to associate deep significance with these accidents at this level. Thus, while a failure to discover colored top-partners would be a major blow to the most popular natural theories of the weak scale, given the relative ubiquity of percent-level tunings in physics, it would perhaps not be a completely decisive blow. It is logically possible that the LHC might have just been a bit unlucky, and the new states could be a little heavier,  slightly beyond its reach. The 100 TeV collider will then play a critical role to settle the issue. If the new particles are indeed ``just around the corner", then 100 TeV collisions will produce them in enormous abundance. On the other hand, the 100 TeV reach for colored top partners will be able to discover them up to masses about 5 times higher than the LHC, pushing the fine-tuning to the $10^{-3}-10^{-4}$ level, a degree we have never seen before anywhere else in particle physics.

But one may justifiably ask: if the LHC sees nothing beyond the Higgs,
does not this already kill the possibility of a completely natural
theory for electroweak symmetry breaking? Would the only role of
future colliders be to further clinch an already clear case? The
answer to this question is an emphatic ``No". What is true is that in
all the natural theories for the weak scale developed over twenty
years ago, we might have already expected to see new colored
top-partners at the LHC. However this does not prove that the idea of
naturalness itself is wrong, only that the particular natural
scenarios theorists invented through the 1990's are not realized in
Nature. As we have emphasized already, this is not a new surprise
delivered to us by the LHC, since there were already indirect
indications that these theories could not be fully natural going back
to the absence of indirect signals for new physics in low energy
experiments and at LEP. Motivated by these considerations, in the mid
2000's new classes of natural theories of EWSB were developed, where
the top partners are not colored, but are charged under mirror gauge
groups. These includes variations on the ``Twin Higgs"
\cite{Chacko:2005pe}, which realizes this idea with the Higgs as a
pseudo-Goldstone boson, and ``Folded SUSY" \cite{Burdman:2006tz},
where supersymmetry is ultimately responsible for the naturally light
scalars. These ``color-neutral natural" theories are much less
constrained by LHC searches, and indeed, {\it completely natural}
regions of parameter space for these theories could be completely
missed by the LHC. They provide an existence proof that the idea of
naturalness can survive the LHC era entirely unscathed, and there may
be further ideas along these lines that have yet to be unearthed. Thus
no new physics at the LHC will not decide the fate of naturalness, the final verdict awaits $pp$ collisions at 100 TeV.

We begin with a discussion of more conventional theories with colored
top partners, and, for concreteness, we will discuss these issues
mostly in the context of supersymmetric theories; a detailed
investigation of other scenarios is left for future studies.

If the MSSM is just mildly tuned we should be able to produce all the
superpartners at a 100 TeV collider. The reach for stops in particular
will be critical; any gain in mass reach relative to the LHC is
squared in the measure of tuning.  Another interesting possibility is
minimally split supersymmetry
\cite{Wells:2003tf,ArkaniHamed:2004fb,Giudice:2004tc,ArkaniHamed:2006mb,Arvanitaki:2012ps,ArkaniHamed:2012gw}. Here,
the spectrum has one-loop splitting between the gauginos (and perhaps
higgsinos) compared to the scalars, as typically happens in the
simplest models of SUSY breaking. The gauginos/higgsinos are at the
TeV scale for reasons of dark matter, while the scalars have a mass
$m_S \sim 10^2-10^3$ TeV, entailing a ``meso-tuning" of $O(10^{-6})$
for electroweak symmetry breaking, while preserving gauge coupling
unification and removing all flavor and CP difficulties of the MSSM.
The usual SUSY boundary conditions for the Higgs quartic coupling is
then easily compatible with the observed $m_h = 125$ GeV for heavy
scalar masses in this range. Interestingly, this tells us that the
gluino cannot get heavier than $\sim 20$ TeV, quite apart from any
constraints on the electroweak part of the spectrum from dark
matter. So for mini-split SUSY, a 100 TeV collider should be able to
produce the gluino, and the electroweak-inos as well.

At a 100 TeV $pp$ collider, we expect an significant improvement of the mass reach of the  superpartners beyond those of the LHC. This is a direct consequence of the large increase of the production rates, shown in Fig.~\ref{fig:SUSY}, resulting from the increase of the center of mass energy.  

\begin{figure}[h!]
\begin{center}
  \includegraphics[width=0.5\textwidth]{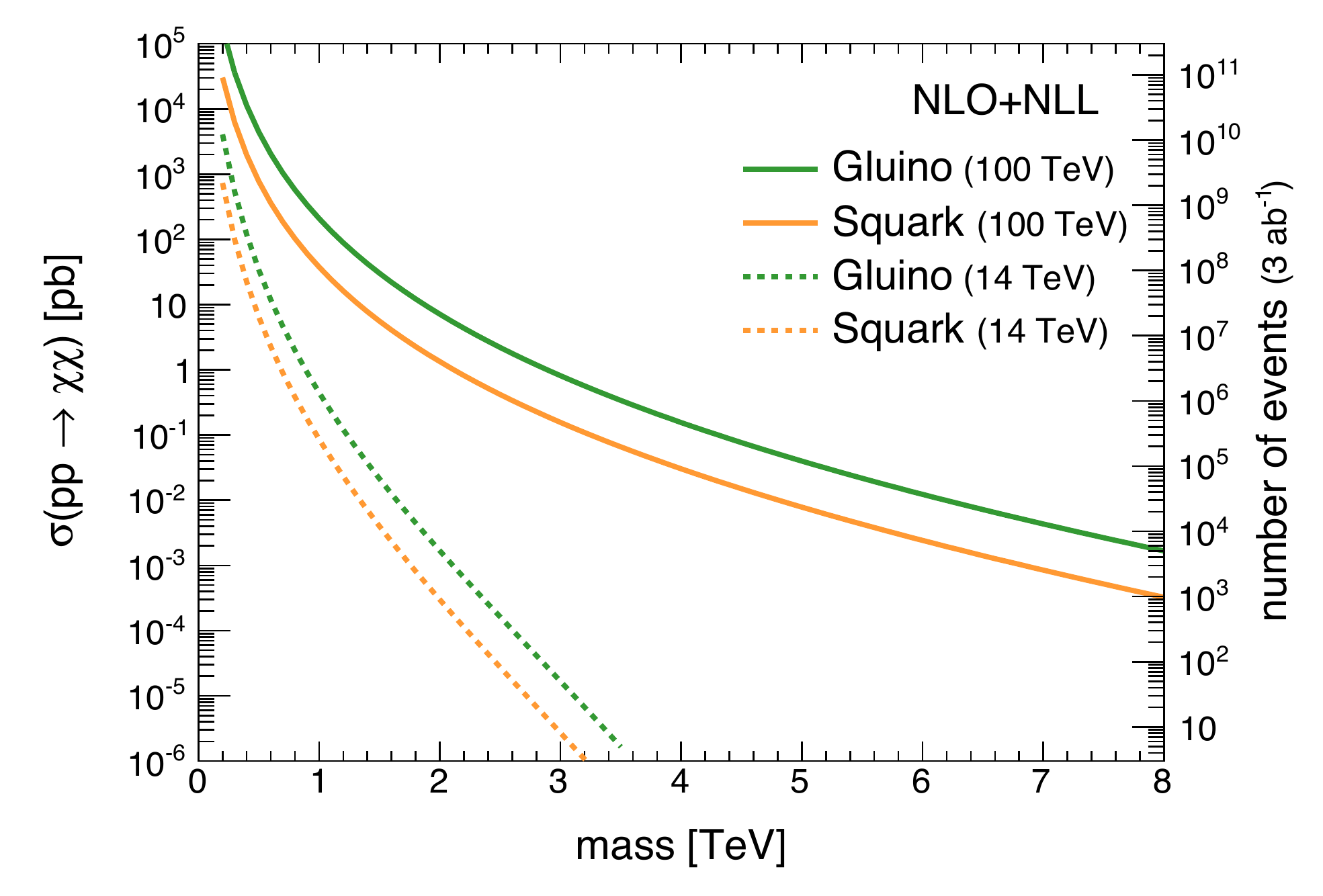} 
  \includegraphics[width=0.48\textwidth]{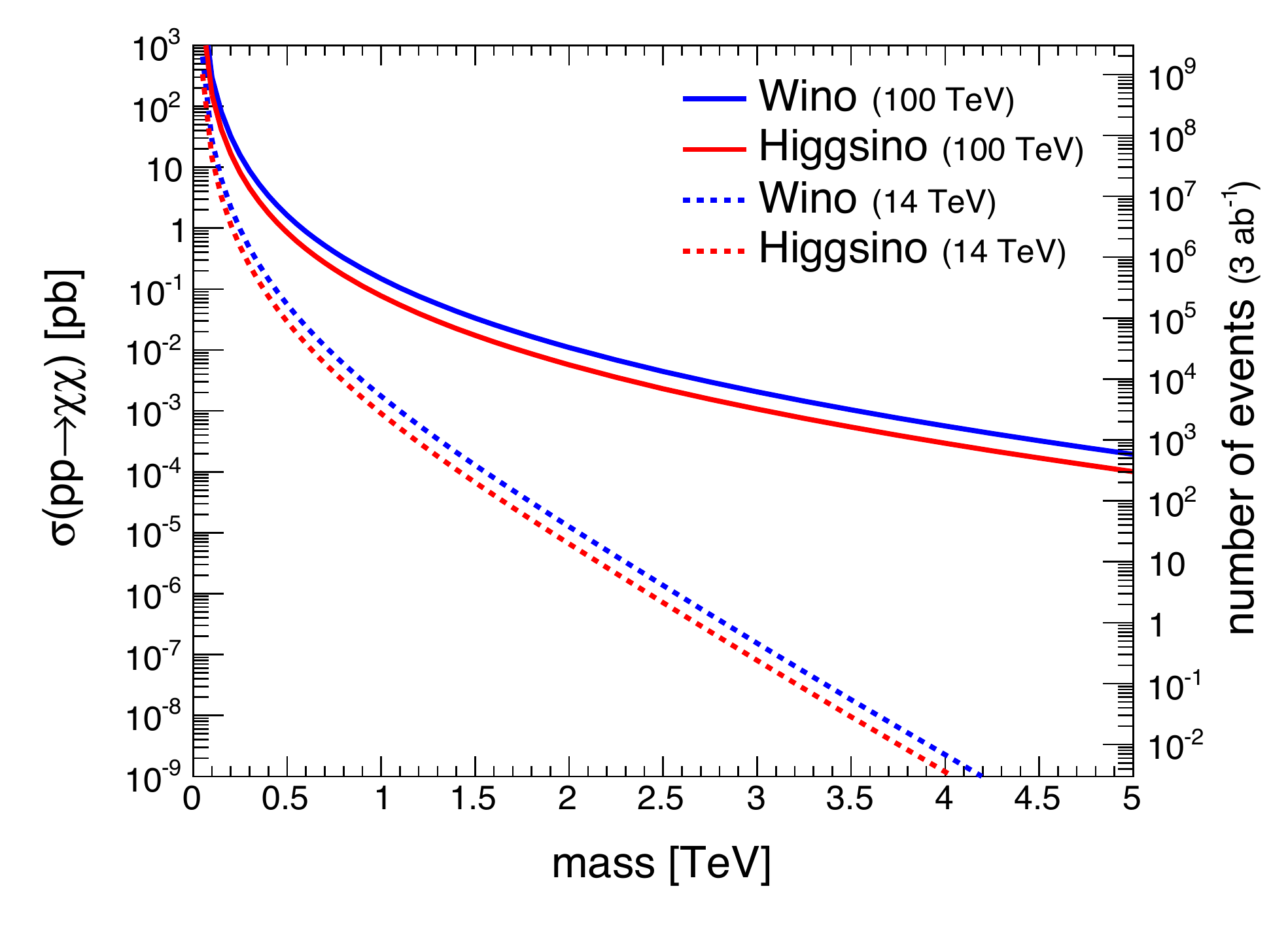}
\end{center}
\caption{Pair production of SUSY particles for (a) gluino and squark,
  and (b) higgsino and wino  at 14 and 100~TeV, see also Ref.~\cite{Borschensky:2014cia}}
\label{fig:SUSY}
\end{figure}

An investigation of the SUSY reach for 100 TeV colliders was carried out in \cite{Cohen:2013xda} for a number simplified models of SUSY production and decay, covering most of the qualitatively interesting scenarios. We summarize their findings here, referring to \cite{Cohen:2013xda} for details of their analysis.

The first simplified model is that of gluino pair production, with
gluinos decaying to neutralino + light flavors, $\tilde{g} \to q
\bar{q} \tilde{\chi}^0$. This process will dominate if the squarks are
heavier than the gluino, and is particularly well-motivated in the
case of split SUSY. The reach is obviously most powerful if there is a
large splitting between the gluino and neutralino masses, and is shown
in the left panel of Fig.~\ref{fig:gluino_reach}, comparing also to  a 33 TeV $pp$ collider 
and to the LHC at 14 TeV. The 100 TeV discovery reach goes up to
$m_{\tilde{g}} = 11$ TeV, about 5 times the reach of LHC at 14 TeV.
\begin{figure} [h!]
\begin{center}
  \includegraphics[width=0.3\textwidth]{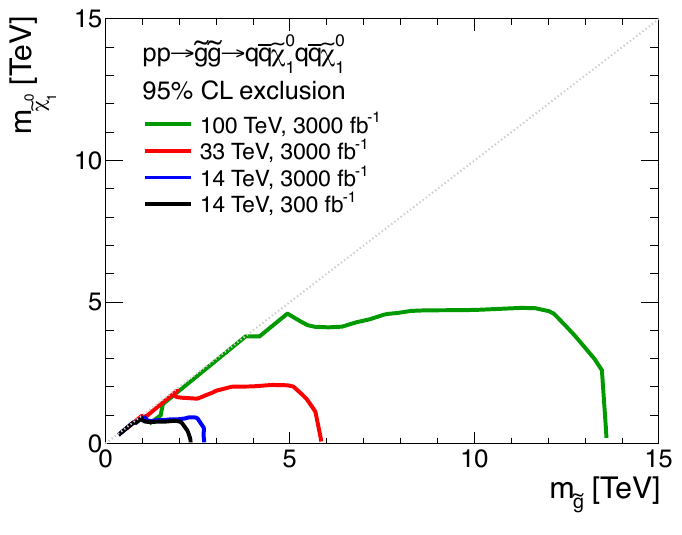}
  \includegraphics[width=0.35\textwidth]{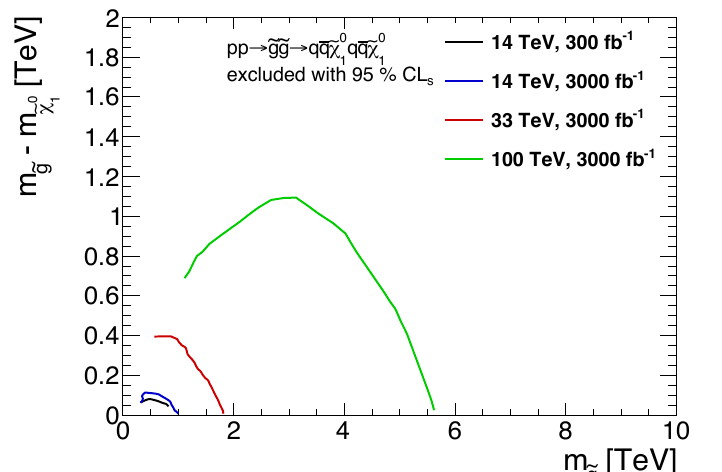}
  \includegraphics[width=0.3\textwidth]{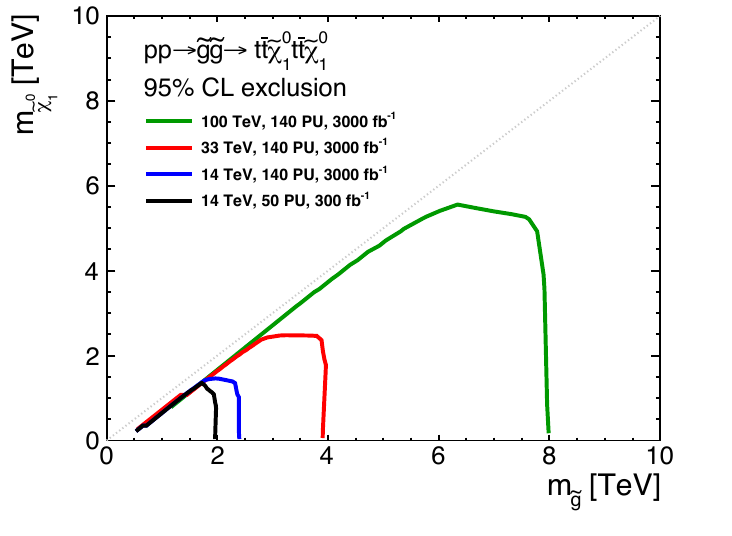} \\
  \includegraphics[width=0.3\textwidth]{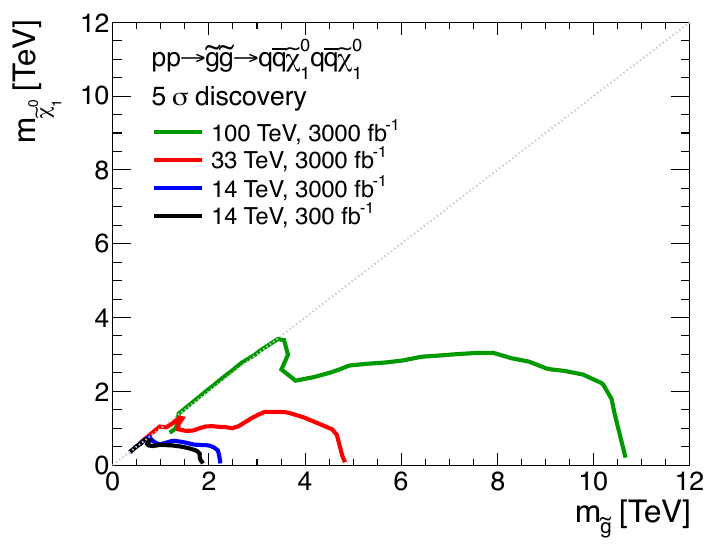}
  \includegraphics[width=0.35\textwidth]{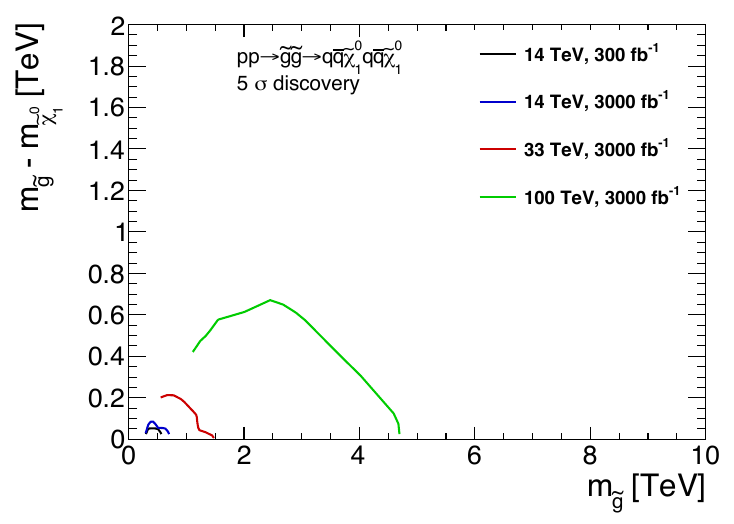}
  \includegraphics[width=0.3\textwidth]{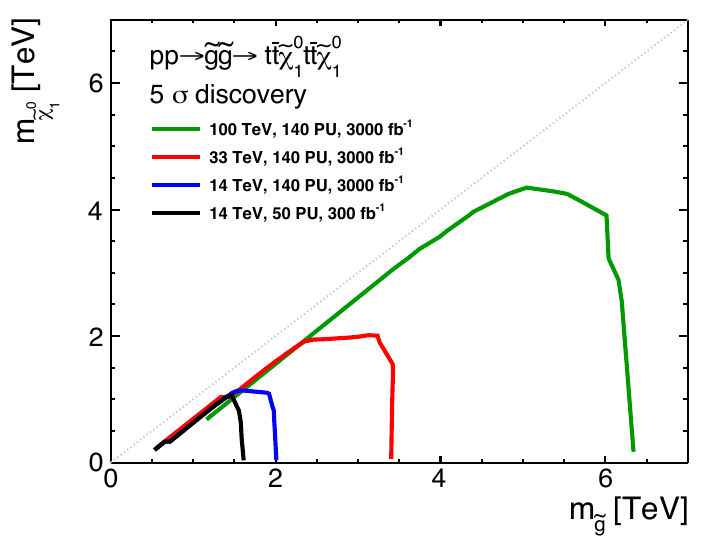}
\end{center}
\caption{Reach for gluino at 100 TeV $pp$ collider for separated (left column) and compressed (center column) spectrum. The reach for gluino decay dominated by $\tilde{g} \to t \bar{t} \chi_0$ is shown in the right column. The $95 \% $ exclusion reach and $5 \sigma$ discovery potential are shown in the top and bottom rows, respectively.}
\label{fig:gluino_reach}
\end{figure}
If the gluino and neutralino are instead relatively degenerate, the decay products will be too soft to see and one will have to rely on the emission of initial or final state radiation to tag the events. The mass reach still goes up to an impressive $\sim 5$ TeV in this case, as shown in the middle panel of Fig.~\ref{fig:gluino_reach}.
It is also interesting to consider that gluinos decay dominantly to
top quarks and the neutralino: $\tilde{g} \to t \bar{t}
\tilde{\chi}^0$. This can easily arise from top-down theories, since
stops are typically driven to be lighter than the first two
generations of squarks, under RG evolution, and is again particularly
well-motivated in split SUSY. In this case, the reach is shown in the
right panel of Fig.~\ref{fig:gluino_reach}.

A much more challenging case with smallest production cross-section for colored particles, is the pair production of the first-two generation squarks, which are taken to be degenerate, followed by the decay to the lightest neutralino as $\tilde{q} \to q \tilde{\chi}^0$. The gluino is taken to be much heavier than the scalars. It is not easy to realize such a scenario from a top-down point of view, since a heavy gluino will quickly drag up the squarks under RG evolution. The reach for the case with the squarks significantly split from the neutralino is shown in the left panel of Fig.~\ref{fig:squark_reach}, while the case with more nearly degenerate squarks and neutralino is shown in the middle panel of Fig.~\ref{fig:squark_reach}.
\begin{figure} [h!]
\begin{center}
  \includegraphics[width=0.3\textwidth]{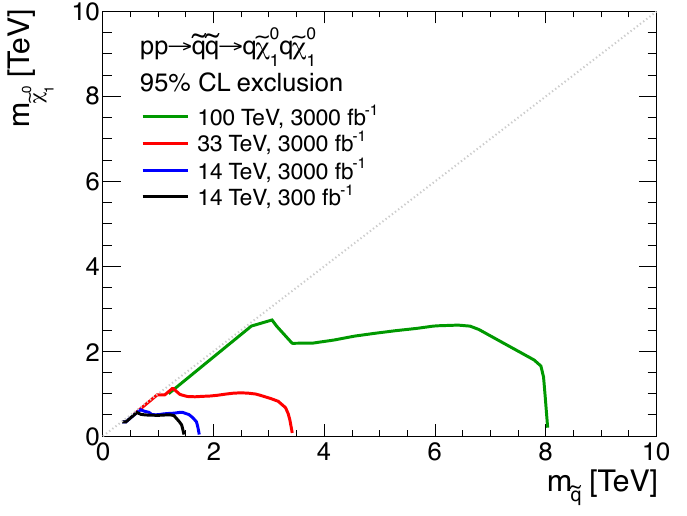} \quad
  \includegraphics[width=0.3\textwidth]{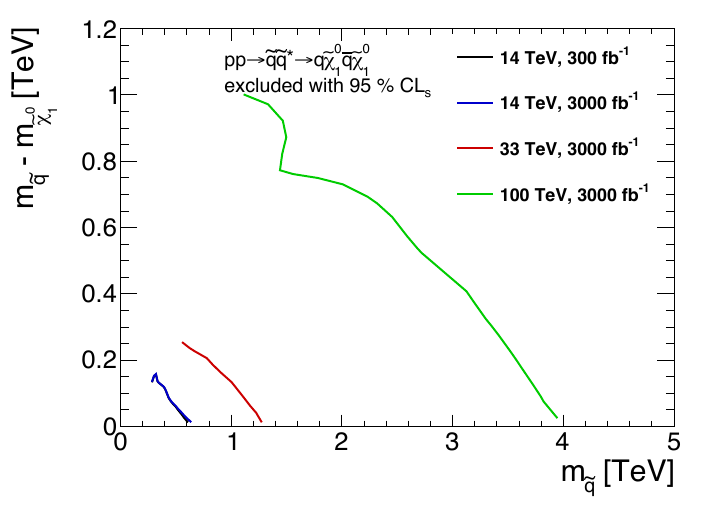}
  \includegraphics[width=0.3\textwidth]{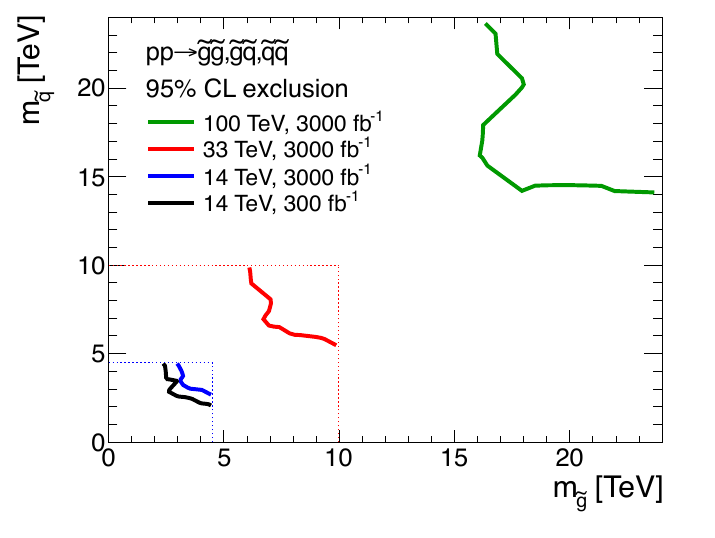} \\
  \includegraphics[width=0.3\textwidth]{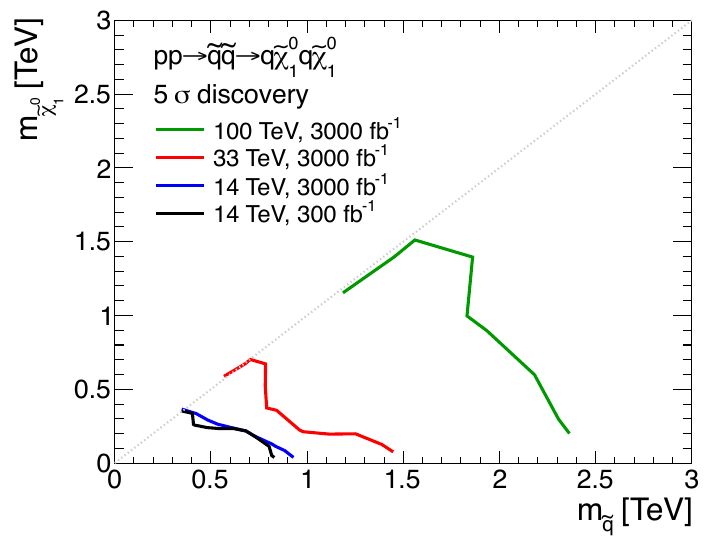} \quad
  \includegraphics[width=0.3\textwidth]{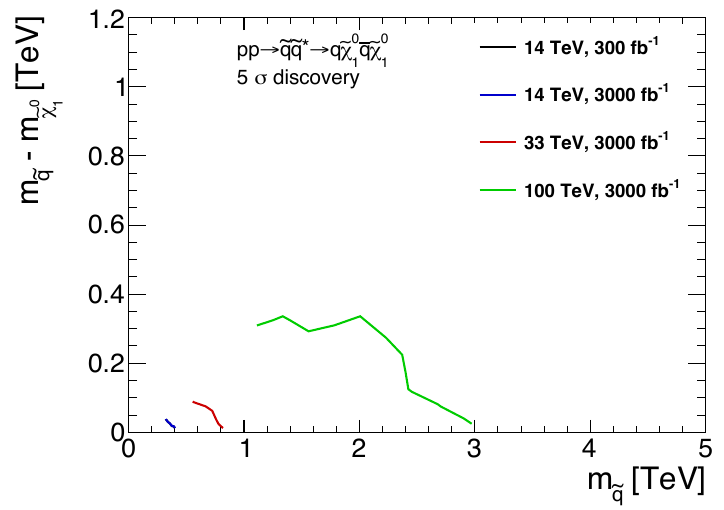}
  \includegraphics[width=0.3\textwidth]{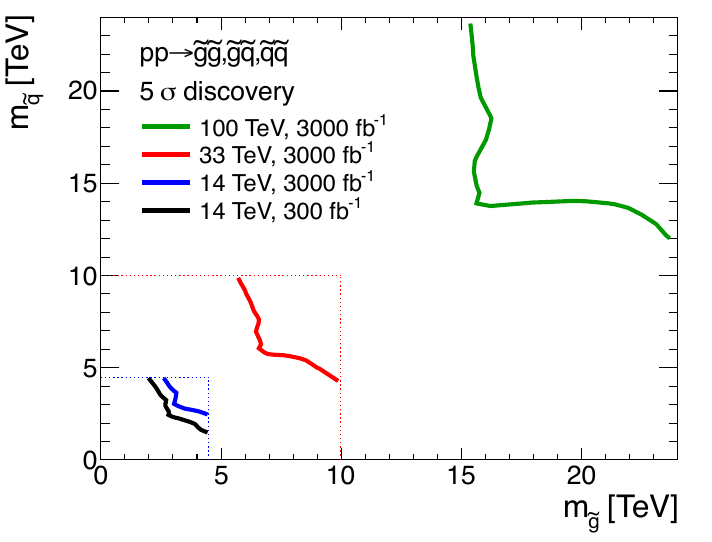} \\
\end{center}
\caption{Reach of squark with separated (left column) and compressed (middle column) spectrum. The reach in the case of gluino and squark with comparable mass is shown in the right column. The $95 \% $ exclusion reach and $5 \sigma$ discovery potential are shown in the top and bottom rows, respectively. }
\label{fig:squark_reach}
\end{figure}

Since the production cross-section is so small, it is difficult to see the signal over the background, so the improvement of the discovery reach at a 100 TeV collider relative to the LHC is not as pronounced here as in the previous examples. Nonetheless the exclusion reach is very impressive in both cases, again representing a factor $\sim 5$ improvement relative to the LHC.

The final simplified model is closest to a ``typical" supersymmetric
spectrum, where both the gluino and first-two generation squarks are
light enough to be produced at a 100 TeV collider, via pair-production
of $\tilde{g} \tilde{g}, \tilde{q} \tilde{q}$, and also associated
production $\tilde{g} \tilde{q}$. If the gluino is heavier than the
squark, it decays to the squark and neutralino as $\tilde{g} \to
\tilde{q} \tilde{\chi}^0$, while if the gluino is lighter than the
squark, it decays to light flavors + neutralino as $\tilde{g} \to q
\bar{q} \tilde{\chi}^0$, and similarly for the squark, which decays as
$\tilde{q} \to q \tilde{g}$ if heavier than the gluino, and $\tilde{q}
\to q \tilde{\chi}^0$ if lighter than the gluino. The neutralino is
taken to be much lighter than the gluinos and squarks. The 100 TeV
reach is shown in right panel of Fig.~\ref{fig:squark_reach}.  This
shows an amazing reach up to $m_{\tilde{g}}, m_{\tilde{q}} \sim 15$
TeV.

We now turn to the 100 TeV reach for stops, which will probe masses up
to the $5-10$~TeV range, pushing the fine-tuning measure to the
$10^{-4}$ level. It is interesting to note that with moderately large
tan$\beta$, stops in the $5-10$ TeV range can also be easily
responsible for pushing the Higgs mass up to $125$ GeV. To be
conservative, we look at the simplified model with all particles but
the stop and the lightest neutralino decoupled, considering the QCD
production of $\tilde{t} \tilde{t}^*$, followed by $\tilde{t} \to t
\tilde{\chi}^0$.  The same search is of course being carried out at
the LHC, but an interesting novelty arises in 100 TeV collisions. With
heavy enough stops, the top quarks produced in the decay are so highly
boosted, that it becomes more difficult to identify the individual top
decay decays products as compared to the LHC. Thus simply scaling up
the LHC analysis to 100 TeV is suboptimal, and identifying highly
boosted tops becomes an important challenge for 100 TeV detectors. It
is possible to use a strategy less dependent on unknown detector
response: when a highly boosted top decays hadronically, the muons
from the resulting $b$ decays will be collinear with the top jet; thus
requiring a lepton inside a jet can be used to effectively tag the
boosted tops~\cite{Cohen:2014hxa,Aguilar-Saavedra:2014iga}.

The 100 TeV reach for direct stop production is shown in the left
panel of Fig.~\ref{fig:stop_reach}, for the two usual cases of a
separated and compressed spectrum:
\begin{figure} [h!]
\begin{center}
  \includegraphics[width=0.45\textwidth]{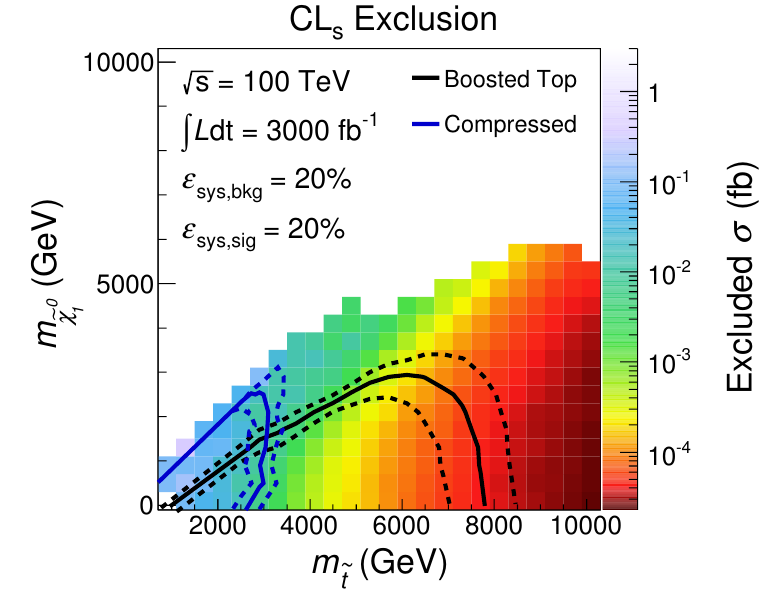} \quad
  \includegraphics[width=0.35\textwidth]{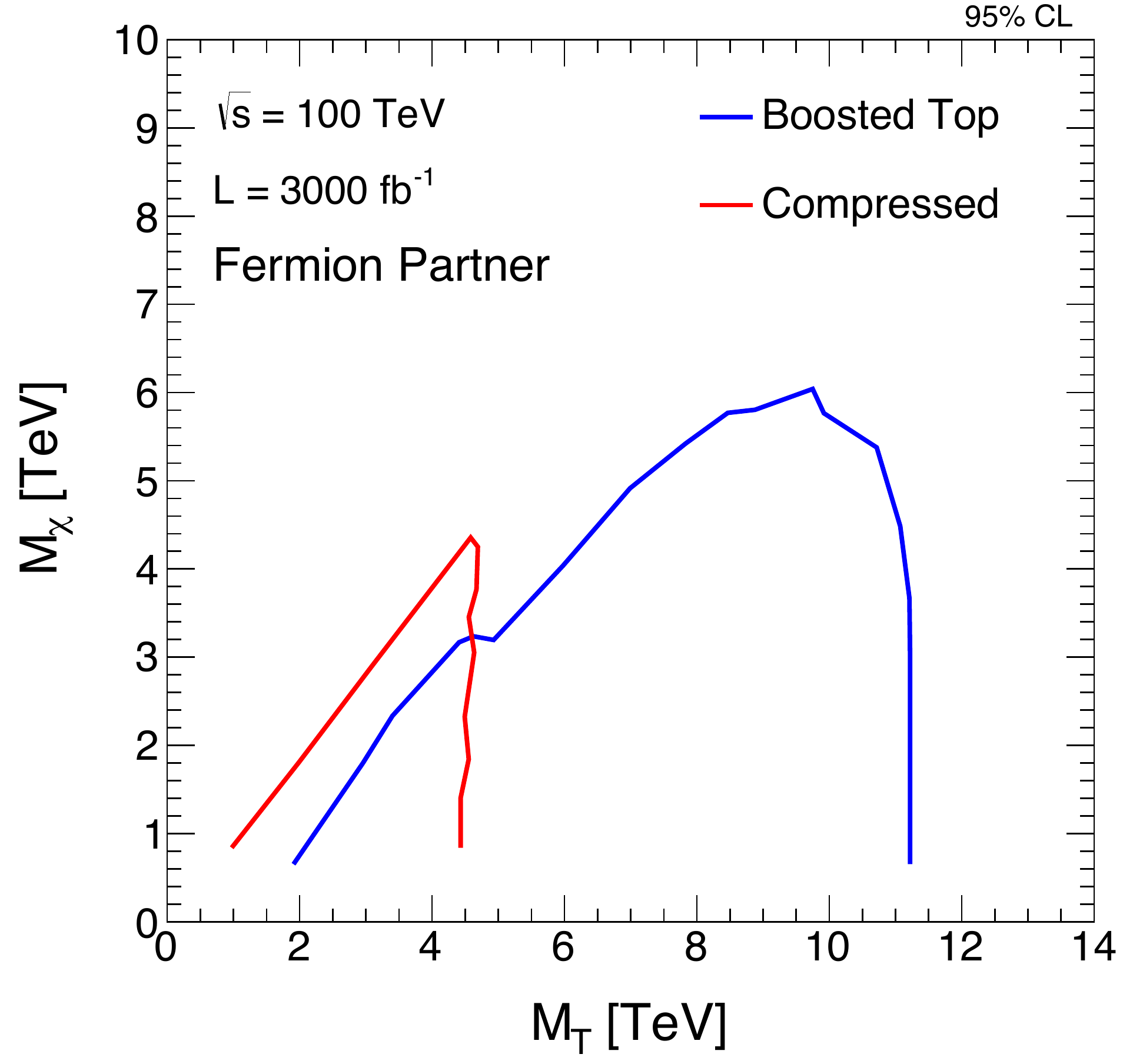} \\
    \includegraphics[width=0.45\textwidth]{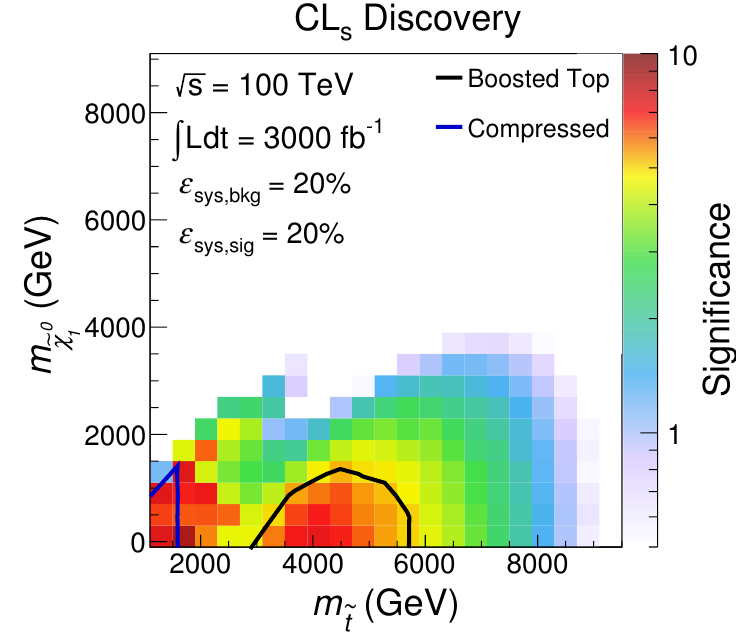} \quad
  \includegraphics[width=0.35\textwidth]{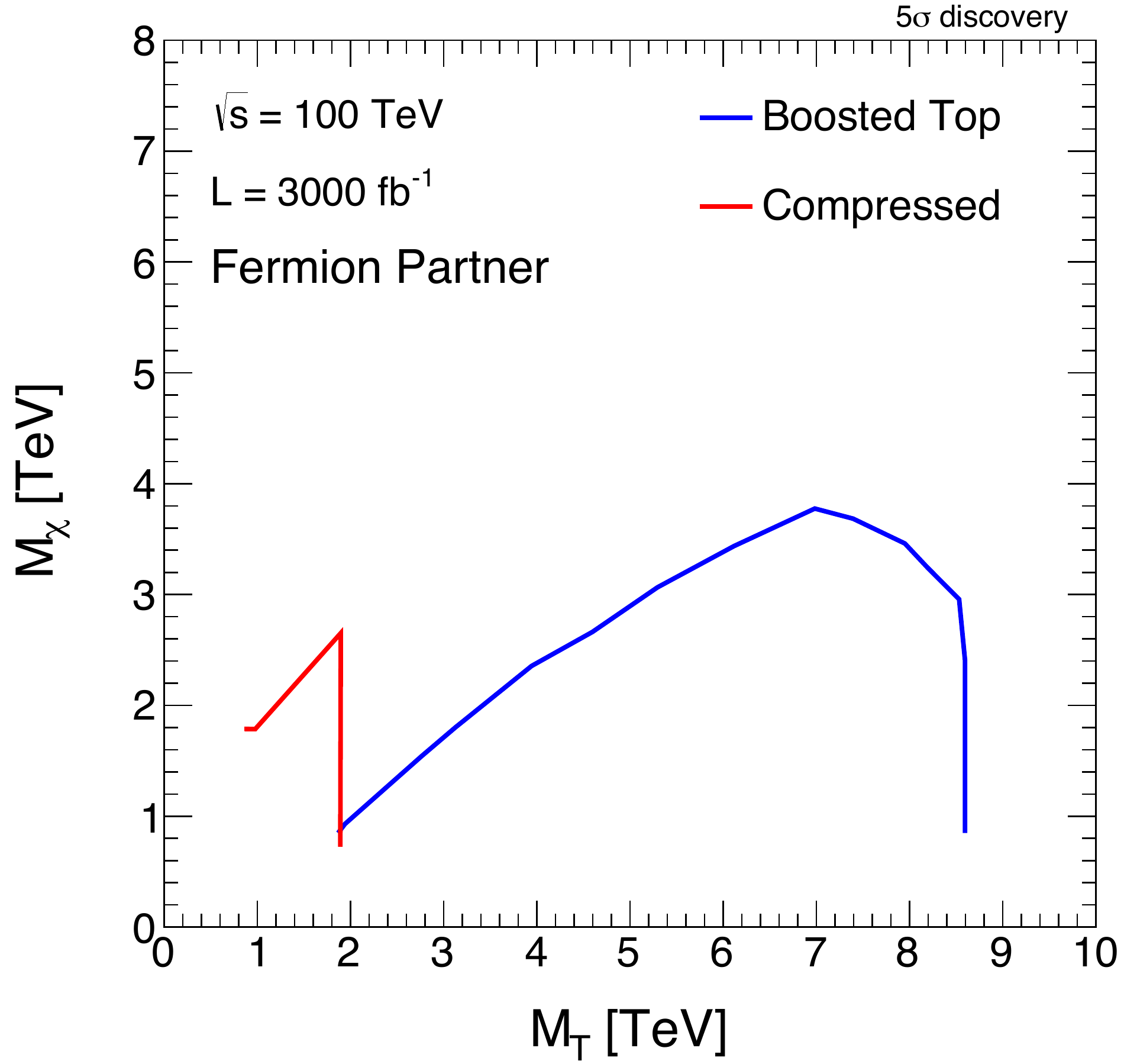}
  \end{center}
\caption{Reach for stops (left column) and fermonic top partners (right column) at a 100 TeV $pp$ collider.}
\label{fig:stop_reach}
\end{figure}
The gain from using the boosted top tagging can be clearly seen. Stops can be discovered (excluded) up to $\sim$ 6 (8) TeV with this method.

Similar reaches are possible for non-supersymmetric theories. For instance, in composite Higgs models, we have fermionic top partners $T^\prime$. Depending on whether we have the ``T-parity" analog of R-parity, these may decay to tops + missing energy, or via $T^\prime \to t Z, T^\prime \to t h$. A dedicated projection to the reach for these models at 100 TeV collider has not yet been done. However, in the case with T-parity, the signal is very similar to that of the stop. Therefore, we can use the stop reach and get a rough estimate of the reach of $T^\prime$ by matching the production rate and mass splitting. The result is  shown in the right panel of  Fig.~\ref{fig:stop_reach}.

All of this discussion has assumed no signals for new physics at the
LHC. In the more optimistic case that LHC {\it does} produce, e.g., superpartners, the need to proceed to the higher energies of 100
TeV collisions is even more urgent, for two obvious reasons. First,
given that we have not seen any superpartners at LHC8, while LHC14
could be powerful enough to discover them, it is unlikely to produce
them in high enough numbers for the more detailed study needed to
ascertain what the particles are trying to tell us about TeV scale
physics. As a simple example, consider a gluino with mass of 1.5 TeV,
just at the LHC Run 1 limit. Roughly $10^4$ of these particles will be
produced through the LHC14 program, certainly enough to be able to
claim a discovery, but not much else. The  careful examination of its
properties, necessary to even hope for a zeroth order claim that
supersymmetry has been discovered, will need a 100 TeV collider,
producing $\sim 10^7-10^8$ gluinos of the same mass. Second, the fact
that we have not seen any new physics at LHC Run 1 also makes it very
unlikely that the entire spectrum of new states will be produced at
LHC14. Consider the example of ``natural SUSY", where the stops and
gluinos are light. At the same time, the first two generations should plausibly be heavier than
$\sim 5$ TeV, enough to eliminate their dangerous contribution to
electric dipole moments. But they cannot get too heavy, as they induce a logarithmically enhanced negative mass for the (light) third-generation squarks \cite{Dimopoulos:1995mi,ArkaniHamed:1997ab}, and so cannot be pushed higher than at most $\sim 30$ TeV.  Finding these heavier scalars will be critical for a zeroth-order understanding of the spectrum, which entangles the physics of flavor and supersymmetry breaking in a fascinating way. While these scalars are well outside the reach of the LHC, they will be accessible to a 100 TeV collider. The most powerful production channel is the associated production of the gluino and first-two generation squarks, as shown in Fig.~\ref{fig:associate_reach}.
\begin{figure} [h!]
\begin{center}
  \includegraphics[width=0.47\textwidth]{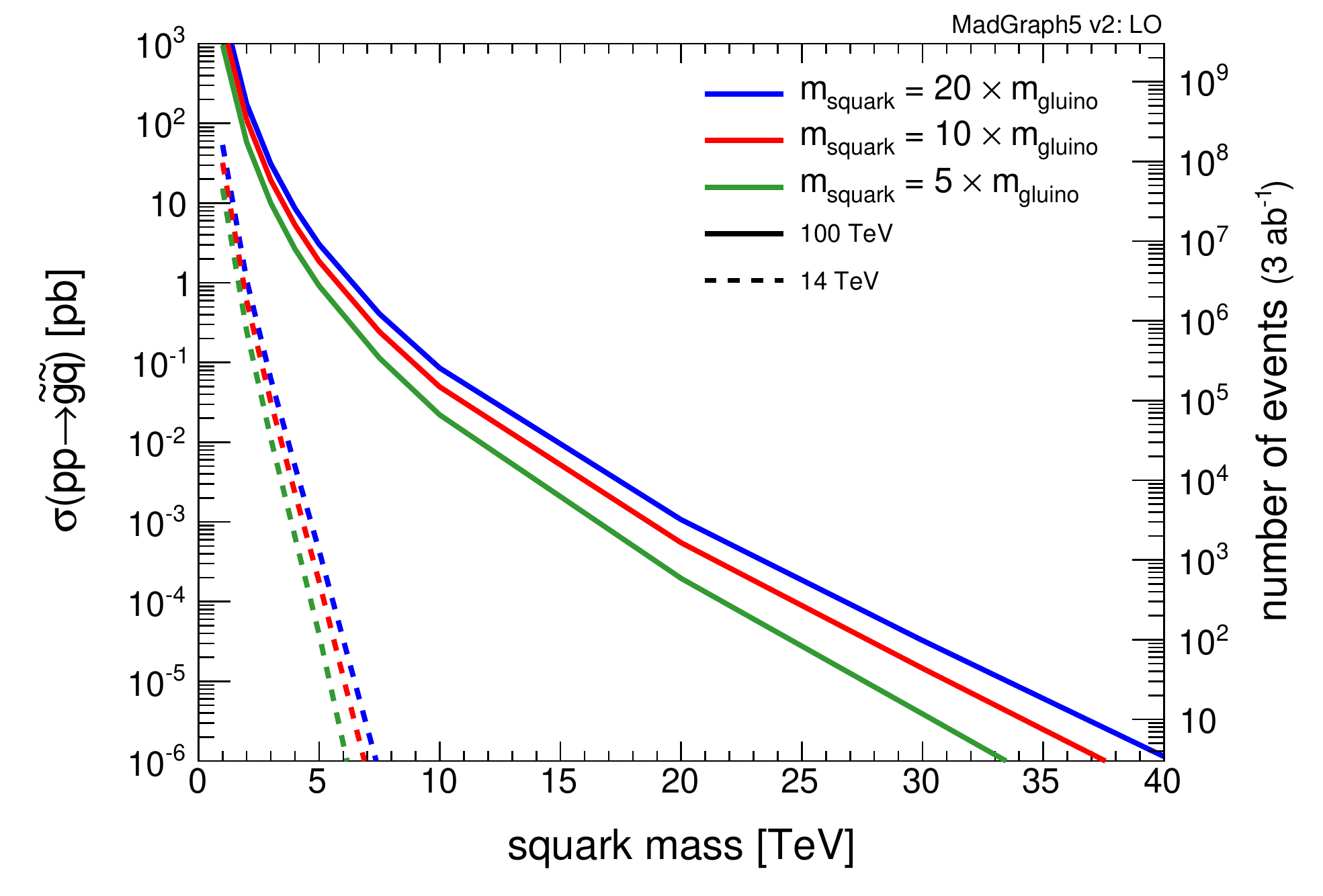}
  \includegraphics[width=0.47\textwidth]{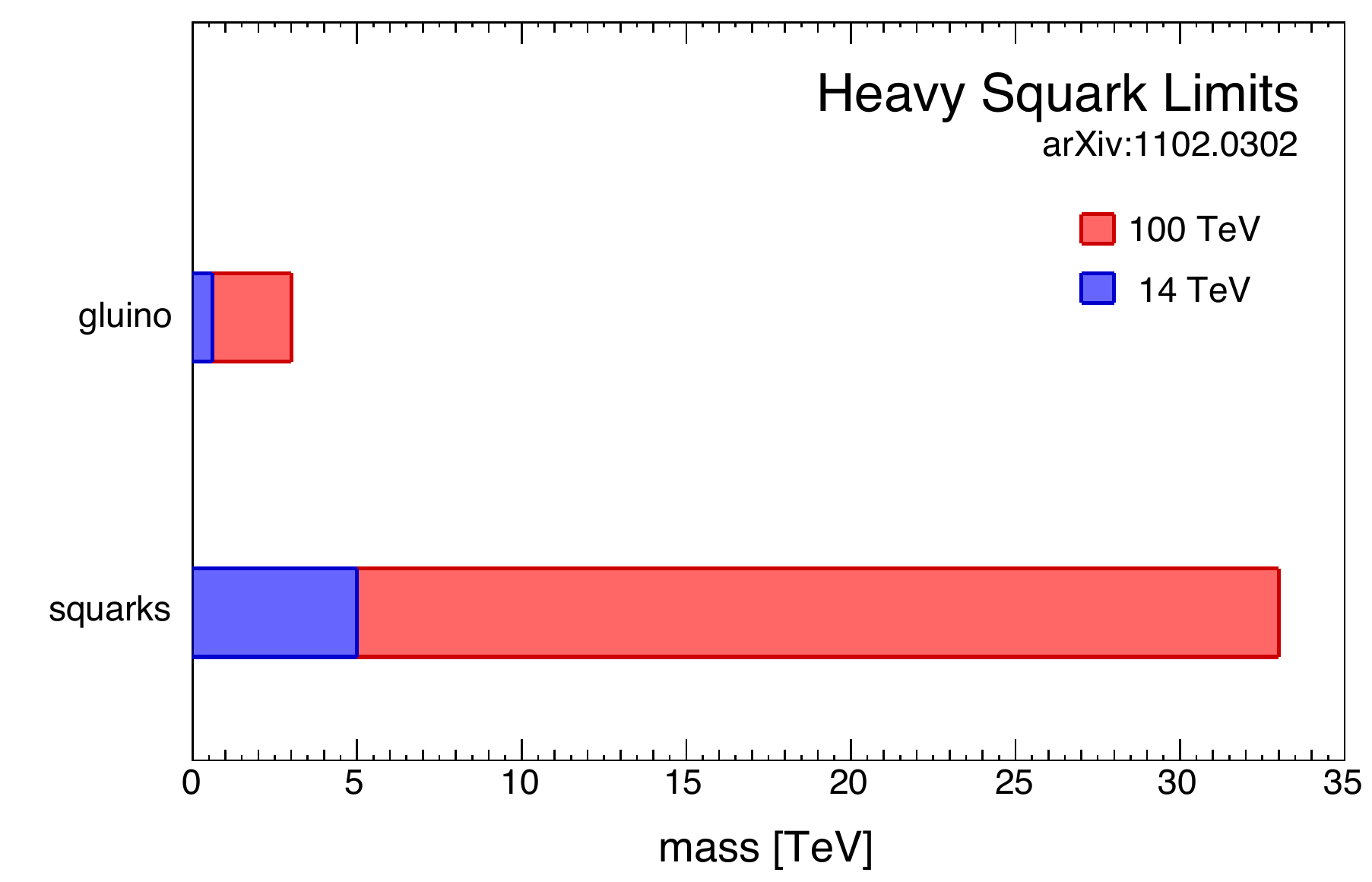}
\end{center}
\caption{Cross section (left panel) and reach (right panel) for a heavy squark produced in association with a light gluino at the 100 TeV $pp$ collider. }
\label{fig:associate_reach}
\end{figure}
The reach for squarks goes up to an incredible $\sim 35$ TeV, covering the entire range of masses for the first-two generation scalars of natural SUSY.

The supersymmetric implementations of neutral naturalness do not generate this large oblique Higgs operator at tree-level. In the simplest cases, the top partners are scalars like the stop, but charged under a mirror SU(3), with six states in total. We can parametrize all the interesting possibilities from the bottom up: we imagine that there is some number $N_\phi$ of new scalars $\phi_I$, and a quartic interaction with the Higgs
\begin{equation}
\frac{1}{2} c_\phi (\phi_I \phi_I) h^\dagger h .
\end{equation}
Some or all of the global symmetries acting on the $\phi_I$ might be gauged, either by the SM electroweak interactions, or mirror interactions. There must be an underlying symmetry that relates $c_\phi$ to the top Yukawa coupling, so as to guarantee
\begin{equation}
c_\phi \times N = \lambda_t^2 \times 6 \to c_\phi = \frac{6 \lambda_t^2}{N}.
\end{equation}

It is also possible to directly produce the $\phi_I$ at the 100 TeV $pp$ collider, again the discussion is analagous to the production of the $S$ singlets in our discussion of the electroweak phase transition. There, the phase-transition requirement forced $S$ to be light enough and sufficiently strongly coupled, for the $\phi_I$ naturalness plays the same role. For simplicity the $\phi_I$ are taken to be degenerate. The signals is just as we had before, vector-boson fusion production of the $\phi_I$, which escape the detector (or decay invisibly). The 100 TeV reach is shown in Fig.~\ref{fig:Higgs_portal_sppc}, along with the effective $|c_\phi|$  associated with the case $N=6$.
\begin{figure} [h!]
\begin{center}
  \includegraphics[width=0.65\textwidth]{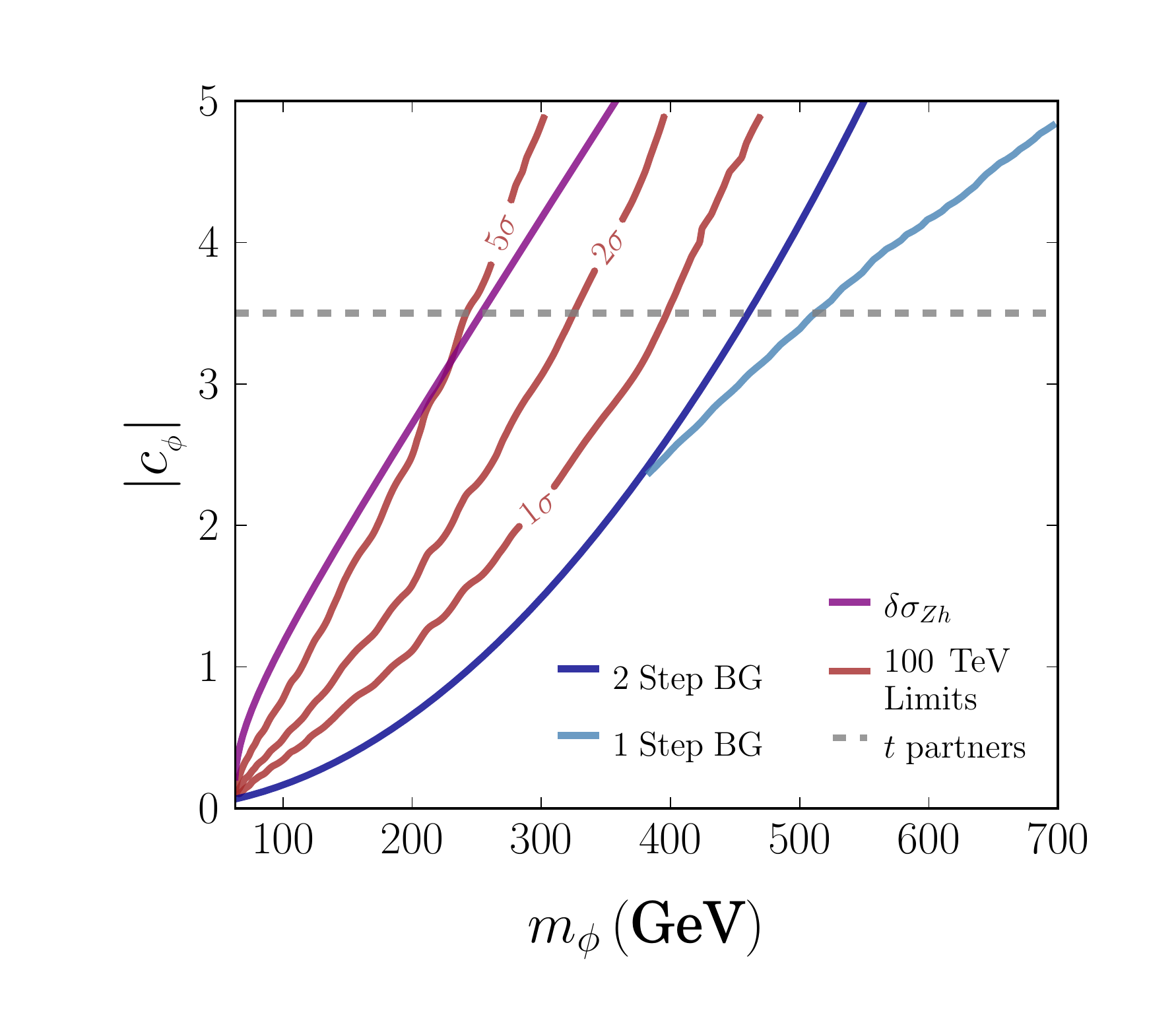}
\end{center}
\caption{The reach for neutral top partners produced through Higgs portal at a 100 TeV $pp$ collider \cite{Craig:2014lda}. 
The lines labelled as ``1 Step BG" and ``2 Step BG" correspond to two scenarios in which the Higgs portal interactions with top partners can make the electroweak physics transition first order. The horizontal dashed line denotes the coupling needed to cancel top quadratic divergence.}
\label{fig:Higgs_portal_sppc}
\end{figure}

Thus a 100 TeV collider has a reach for $5\sigma$ discovery up to $\sim 250$ GeV, and a $2\sigma$ exclusion up to $\sim 350$ GeV, pushing to the boundaries of the natural region. 

In all examples studied so far, we have seen that a 100 TeV collider can decisively settle whether fully natural theories of electroweak symmetry breaking are realized in Nature. While we have avoided using the language of ``no-lose theorems" to discuss the physics opportunities of this machine, it is possible that more detailed future studies can actually formulate a sensible ``no-lose theorem" for probing naturalness at a 100 TeV collider; preliminary examples along these lines can be found in the recent works \cite{Curtin:2015bka}.

\section{Dark Matter} %%%%%%%%%%%%%%%%%%%%%%%%%%%%%%%%%%%%%%%%%%%%%%%

The existence of cold dark matter is one of the most direct and powerful pieces of evidence for physics beyond the Standard Model. There are a huge range of possibilities for what the dark matter might be, since for any mass we can simply adjust the number density to get the needed energy density today, with $\Omega_{DM} h^2 \sim 0.1$. Even if the new particle physics is completely specified, the main uncertainty is cosmological: what determines the abundance of the new particles in the early universe?

\subsection{WIMP Dark Matter}

Weakly Interacting Massive Particles (WIMPs) remain the best motivated and well-studied possibility for dark matter by giving a clear answer to this question: the dark matter particles interact with the SM and are thermalized in the early universe. Assuming a standard cosmological history, the present abundance of dark matter can be unambiguously computed once the underlying particle physics is fixed, in much the same way as the abundance of light elements is predicted in big bang nucleosynthesis.

The relic abundance of dark matter particles is set by their annihilation cross section in the early universe 
~\cite{Lee:1977ua,Goldberg:1983nd,Steigman:2012nb}  
\begin{equation}
\Omega h^2 = 0.11 \times \left(\frac{\langle \sigma v \rangle_{{\rm freeze}}}{2.2 \times 10^{-26}~{\rm cm}^3/{\rm s}}\right)^{-1} ,
\end{equation}
with $\sigma \propto g_{\rm eff}^4/M_{\rm DM}^2$.  Therefore, to avoid overclosure,  the limit on the dark matter mass is
\begin{equation}
M_{\rm DM} < 1.8~{\rm TeV} \left(\frac{g^2_{\rm eff}}{0.3}\right) .
\end{equation}
As has been long appreciated, it is quite remarkable that the TeV scale emerges so naturally in this way, assuming dark matter couplings comparable in strength to the electroweak gauge interactions. This gives a strong, direct argument for new physics at the TeV scale, independent of any theoretical notions of naturalness.

Compellingly, dark matter often falls out of theories of physics
beyond the SM without being put in by hand. Indeed, if the SM is
augmented by new physics, not even necessarily close to the weak
scale, but far beneath the GUT scale, the interactions with new states
should respect baryon and lepton number to a very high degree. Since
all SM particles are neutral under the discrete symmetry
$(-1)^{B+L+2S}$, any new particles that are odd under this symmetry
will be exactly stable. This is the reason for the ubiquitous presence
of dark matter candidates in BSM physics. It is thus quite plausible
that the dark matter is just one part of a more complete sector of
TeV-scale physics; this has long been a canonical expectation, with
the dark matter identified as e.g. the lightest neutralino in a theory
with TeV-scale supersymmetry. The dominant SUSY processes at hadron
colliders are of course the production of colored particles---the
squarks and gluinos---which then decay, often in a long cascade of
processes, to SM particles and the lightest supersymmetric particle (LSP), resulting in the well known missing energy signals at hadron colliders. This indirect production of dark matter dominates, by far, the direct production of dark matter particles through electroweak processes.

However, as emphasized in our discussion of naturalness, it is also
worth preparing for the possibility of a much more sparse spectrum of
new particles at the TeV scale. Indeed, if the idea of naturalness
fails even slightly, the motivation for a very rich set of new states
at the hundreds-of-GeV scale evaporates, while the motivation for WIMP
dark matter at the TeV scale still remains. This is for instance part
of the philosophy leading to models of split SUSY: in the minimal
incarnation, the scalars and the second Higgs doublet of the MSSM are
pushed to $\sim 10^2-10^3$ TeV, but the gauginos (and perhaps the
higgsinos) are much lighter, protected by an $R$-symmetry.  The
scalars are not so heavy as to obviate the need for $R$-parity, so the
LSP is still stable, and must be set at the TeV scale in order not to
overclose the universe, thereby making up some or perhaps all of the
dark matter.

In exploring dark matter at colliders, therefore, it is most prudent to first look for direct production of dark matter, rather than dark matter arising in the decay products of other states that may not be accessible. We will therefore explore the reach of a 100 TeV collider for the production of new states with only electroweak quantum numbers, which also certainly give the simplest possible picture for what the dark matter could be. The simplest case of all would be a single new state: a real triplet or vector-like doublet adds the fewest possible number of degrees of freedom to the SM, and no new interactions, so the only free parameters are the particle masses. We can be slightly more general and allow for the presence of additional singlet states. Including just singlets, doublets, and triplets gives a minimal ``module" for dark matter, which we will consider, described by the Lagrangian
\begin{align}
  \begin{split}
\Delta L = & \;\; M_1 \tilde{B} \tilde{B} + M_2 \tilde{W} \tilde{W} + \mu \tilde{H}_u \tilde{H}_d \\
           & + \sqrt{2} \kappa_1 h^\dagger \tilde{W} \tilde{H}_u + \sqrt{2} \kappa_2 h \tilde{W} \tilde{H}_d + \frac{\kappa^\prime_1}{\sqrt{2}} h^\dagger \tilde{B} \tilde{H}_u + \frac{\kappa^\prime_2}{\sqrt{2}} h \tilde{B}\tilde{H}_d .
  \end{split}
\end{align}
Since the quantum numbers are the same as binos ($\tilde{B}$), winos ($\tilde{W}$), and higgsinos ($\tilde{H}_{\rm u,d}$) of supersymmetric theories, we will use this notation and language in referring to these states, as ``charginos", ``neutralinos", ``the LSP", and so on.  Much of our analysis is, however, free of supersymmetric assumptions: supersymmetry only relates the new Yukawa couplings to the SM gauge couplings as $\kappa \sim g$ and $\kappa^\prime \sim g^\prime$, but this won't play an essential role in most of our discussion.

Given this spectrum of electroweak states, we can consider two obvious limits. One of these states can be significantly lighter than the others; if it is also significantly heavier than $M_Z$, then the dark matter is close to being a ``pure" electroweak state, so we can have a ``pure wino" or ``pure higgsino" (a ``pure bino" has no interactions at leading order and so is not relevant to our discussion). Alternately, the lightest state can be a significant admixture of different electroweak states.

For both the higgsino and wino, the electroweak multiplet contains
charged and neutral states that would be degenerate in the absence of
electroweak symmetry breaking; however, a small splitting between
these states arises after electroweak symmetry is broken.  There is a
calculable radiative correction to the splitting, that can be thought
of as the difference between the ``electrostatic" energy of the photon
and $Z$ fields for the charged and neutral components, giving 
\begin{equation}
\Delta m \sim \alpha_{\rm EM} M_Z.  
\end{equation}
This irreducible splitting is $\Delta m
= 166$ MeV for winos~\cite{Ibe:2012sx} and $\Delta m = 355$ MeV for
higgsinos~\cite{Thomas:1998wy}. Further splittings can also arise from
UV effects, by integrating out heavier particles (for instance the
heavier electroweak states). 
For the higgsino, the leading dimension-5 operator 
generates a splitting between the charged and neutral states as
\begin{equation}
(\kappa^2/M) (h^\dagger \tilde{H}_u)(h \tilde{H}_d)\quad \Rightarrow \quad 
\Delta m \sim M_Z^2/M. 
\end{equation}
 For the wino, the leading dimension-5
operator does not split the two states, and we have to go to the
dimension-7 operator, which generates an even smaller splitting as
\begin{equation}
(\kappa^4/M^3) (h^\dagger \tilde{W} h)^2 
\quad \Rightarrow \quad 
\Delta m \sim M_Z^4/M^3.
\end{equation}
Comparing the radiative and UV splittings, if there is just an
$\mathcal{O}(1)$ difference in mass between the wino and the rest of
the states, the UV splittings become much smaller than the radiative
splitting.

Since the wino and higgsino have sizable electroweak gauge interactions, they annihilate very efficiently; this is why their masses have to be pushed to $1-3$ TeV to be thermal relics. By contrast, the bino has no electroweak couplings at all. Therefore it is interesting to consider the dark matter as having a sizable admixture of bino together with wino or higgsino. Since the mixing between the states arises through electroweak symmetry breaking, in the limit where the masses $M_1,M_2, \mu$ are large compared to $M_Z$, the mixing angles will be very small, suppressed by powers of $(M_Z/M)$, unless some pair of the diagonal masses are close to degenerate, as with the case of ``well-tempered" neutralinos~\cite{ArkaniHamed:2006mb}.  For the case where the bino/higgsino are nearly degenerate, the mixing terms are parametrically $\sim M_Z$,
and this also sets the size of the splitting between the charged and neutral states, which can be typically $\sim 20-50$ GeV. For the bino/wino case, the mixing terms are parametrically $\sim M_Z^2/M$, and we expect somewhat smaller splittings.

Thermal relic pure winos and higgsinos must have a mass of 3.1 TeV and 1.1 TeV respectively to account for all the dark matter. At smaller masses they can still account for a significant fraction of the dark matter, for instance a 2 TeV wino can account for half of the dark matter. Mixed dark matter can be lighter but masses around $\sim 500$ GeV are typical.

The direct detection of pure winos and higgsinos is extremely challenging. The leading dark matter-nucleon interaction arises at 1-loop, and gives rise to a tiny spin-independent cross section \cite{Hisano:2010fy,Hill:2013hoa}
\begin{eqnarray}
\nonumber
\sigma_{\rm SI} 
 \left\{  
\begin{array}{ll}
\approx 10^{-47}\ {\rm cm}^2 \qquad   {\rm  for\  winos}, & \\
\leq 10^{-48}\ {\rm cm}^2 \qquad   {\rm  for\  higgsinos }. &  
\end{array}
\right.
\end{eqnarray}
%$\sigma_{\rm SI}=10^{-47}$ cm$^2$ for winos and $\sigma_{\rm SI} \leq 10^{-48}$ cm$^2$ for higgsinos~
These cross sections are just at the border of the irreducible neutrino scattering floor for direct detection experiments, and with TeV masses the rates are also too low to be seen in any of the planned experiments. Mixed dark matter is a much more promising target for direct detection, and is already tested by current limits, but a sizable region of parameter space continues to be viable.

We can also consider indirect detection of high energy particles resulting from dark matter annihilation near the center of our galaxy. Of course predictions for indirect detection rates are fraught with astrophysical uncertainties, and it is difficult to get robust limits in this way. Nonetheless, pure winos are constrained in an interesting way, since their annihilation cross section has a significant Sommerfeld enhancement~\cite{Hisano:2004ds}. The absence of any signals in the HESS experiment for high energy gamma photons from the galactic center~\cite{Abramowski:2013ax} sets limits on the fraction of dark matter a wino of a given mass can comprise.  A 3 TeV wino making up all the dark matter is excluded for a standard NFW dark matter distribution, though it is allowed for more ``cored" profiles~\cite{Cohen:2013ama,Fan:2013faa,Hryczuk:2014hpa,Ovanesyan:2014fwa,Baumgart:2014saa}. The current limits are summarized in Fig.~\ref{fig:winofraction}.
\begin{figure} [h!]
\begin{center}
  \includegraphics[width=0.9\textwidth]{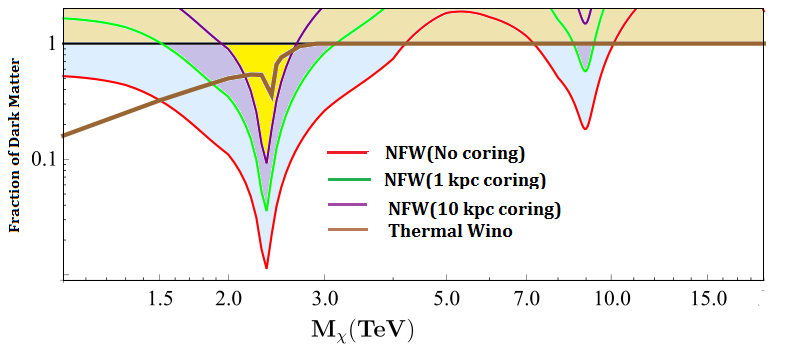}
\end{center}
\caption{Exclusion plot for an NFW profile with the wino making up only some fraction of the  dark matter~\cite{Baumgart:2014saa}.}
\label{fig:winofraction}
\end{figure}

Future indirect detection experiments, such as CTA, could move the wino bounds down to 1 TeV, subject to the same astrophysical uncertainties. But we can see that thermal relic winos making up an $\mathcal{O}(1)$ fraction of dark matter are certainly still consistent. For both pure higgsinos as well as mixed dark matter, the annihilation is not significantly Sommerfeld enhanced, and there are no interesting limits  from indirect detection,

It is striking that the very simplest models of dark matter---pure winos and higgsinos---could be completely inaccessible to direct detection experiments, while astrophysical uncertainties make it hard to interpret indirect detection limits. We are left with directly producing the dark matter at accelerators.  Relic winos and higgsinos forming a significant component of dark matter, which have masses in few TeV scale, are hopelessly out of reach for direct production at the LHC, which has an ultimate reach up to $\sim 300-400$ GeV for pure wino and $\sim 200$ GeV for pure higgsino production.  Moreover, only a fraction of the parameter space for mixed dark matter is accessible to direct production at the LHC.

\subsection{WIMP Dark Matter at the 100 TeV $pp$ Collider}

As we will see shortly, however, the huge increase of rate at 100~TeV
will allow a much larger range of the relevant parameter space to be
explored. The most basic process we will first consider is dark matter
pair production. Since the dark matter escapes the detector without
leaving a trace, we need to look for additional hard radiation of SM
particles from the process---quarks or gluons, photons, $W/Z$'s, and
Higgses.  Of these, the ``monojet" channel where a quark or gluon is
radiated typically gives the best sensitivity. For mixed states we can
have a mass splitting $m_{\chi^\pm} - m_{\chi^0} \sim 20-50$ GeV
between the chargino and neutralino states.  In this case, in addition
to a hard jet, it is possible to search for low $p_T$ leptons
resulting from a chargino or neutralino, which decay to the LSP and
leptons. We call this the soft lepton channel. On the other hand, when
the lightest state is pure, the radiative mass splitting is tiny and
the decay length is long, leaving a striking signature of a high-$p_T$
charged track abruptly ending when the chargino decays to the LSP and
very soft, likely undetected, SM particles. We include this
disappearing-tracks search in our considerations as well.

\noindent{\bf Monojets}:  %%%%%%%%%%%%%%%%%%%%%%%%%%%
Our first analysis looks for a single hard jet produced in association with a pair of dark matter particles, the classic monojet plus missing energy search. Monojet searches for dark matter and large extra dimensions have been carried out both at the Tevatron and the LHC.
The backgrounds for this channel include SM processes with a hard jet and neutrinos.
 Processes with leptons also comprise part of the background because
 leptons can fail to be tagged if they are outside the detector
 acceptance, not isolated, or too soft. This is a very challenging
 channel with the uncertainty dominated by the background's systematics.

\noindent{\bf Soft Leptons}: %%%%%%%%%%%%%%%%%%%%%%%%%%% In the case
of mixed dark matter, where we have splittings of $\Delta m \sim
20-50$ GeV, the heavier states can also be pair produced, and decay to
the dark matter via off-shell gauge bosons, which then decay
hadronically or into low-$p_T$ leptons. The hadronic decays are
difficult to extricate from the noisy hadronic environment, but it is
possible to tag the soft leptons. This is different from the standard
multilepton searches where there are both more and harder leptons.  It
has been noted that triggering on a hard jet, as in the monojet
search, is advantageous in a soft lepton search.

\noindent{\bf Disappearing Tracks}: %%%%%%%%%%%%%%%%%%%%%%%%%%
The third analysis leverages the near degeneracy of charginos and the LSP for pure electroweak states.  Due to the tiny mass splitting, the dominant decay $\chi^{\pm} \to \pi^{\pm} \chi^0$ can have a long enough lifetime -- $c\tau \sim 6$ cm for winos -- to leave a track in the inner detector.  This chargino track disappears within the inner detector when it decays to a neutralino and soft pion.  This is a promising search channel with no obvious physics background.  Searches can also be done when the charginos have a shorter or longer lifetime and look for displaced vertices and stable charged massive particles, respectively.

\noindent{\bf Multi-Lepton} %%%%%%%%%%%%%%%%%%%%%%%%%%% 
Finally when
one moves away from the compressed region of parameter space, any mass
splitting between the next-to-lightest supersymmetric particle
(NLSP) and the LSP can be generated and it is most natural to cast
limits in the NLSP-LSP mass plane.  For these searches, there are
multiple leptons from the NLSP-to-LSP decays whose energies scale with
the NLSP-LSP splitting.  They are energetic enough that the hard jet
required for triggering in the soft lepton search is unnecessary.
These searches can be categorized by the particular combination of
leptons for which they are looking.  Here we consider the three lepton
($3\ell$), the opposite-sign di-lepton (OSDL), and the same-sign
di-lepton (SSDL) signatures, although the $3\ell$ is always the most
sensitive.  Multi-lepton searches are based on the observation that
while the signal has large mass splittings and heavy invisible
particles, the background has neither and so has harder back-to-back
jets, with leptons, than the signal.

As the optimal search strategy strongly depends on the splittings, it
would be interesting to look at the overlap and transitions between
the approaches discussed above.  This more detailed analysis deserves
focus in future studies.

\subsubsection{Pure Wino} %%%%%%%%%%%%%%%%%%%%%%%%%%%%%%%%%%%%%%%%%%%%%%%

The pure wino has nearly degenerate charged and neutral states. The pair production of the chargino proceeds via Drell-Yan production through an $s$-channel $Z/\gamma^*$, while the production of a chargino/neutralino proceeds through an $s$-channel $W$. The charginos decay to the neutralino and a soft pion.

The mass reach in the monojet channel for a pure wino is shown in Fig.~\ref{fig:wino}.
\begin{figure} [t]
\begin{center}
  \includegraphics[scale=0.32]{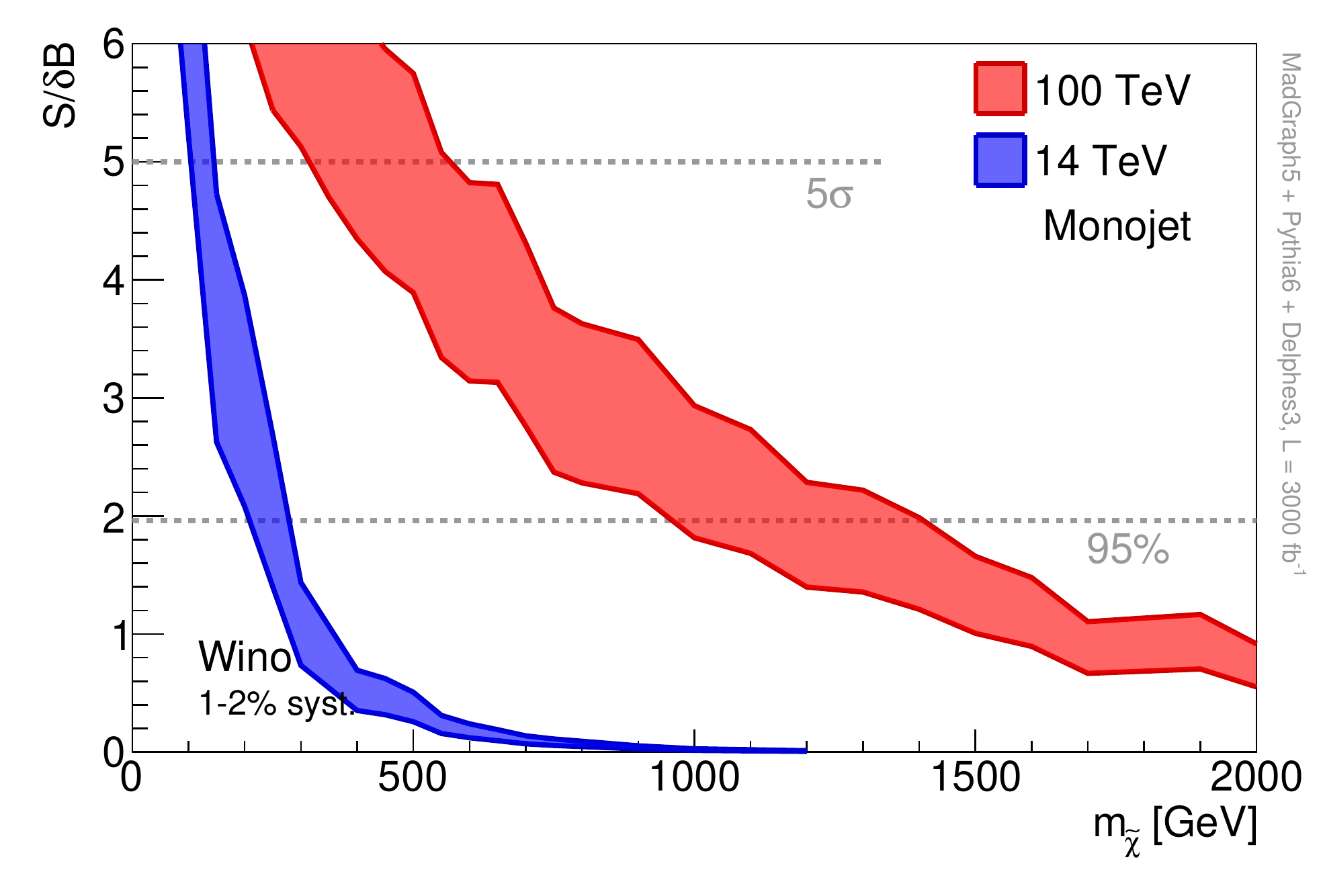}
    \includegraphics[scale=0.32]{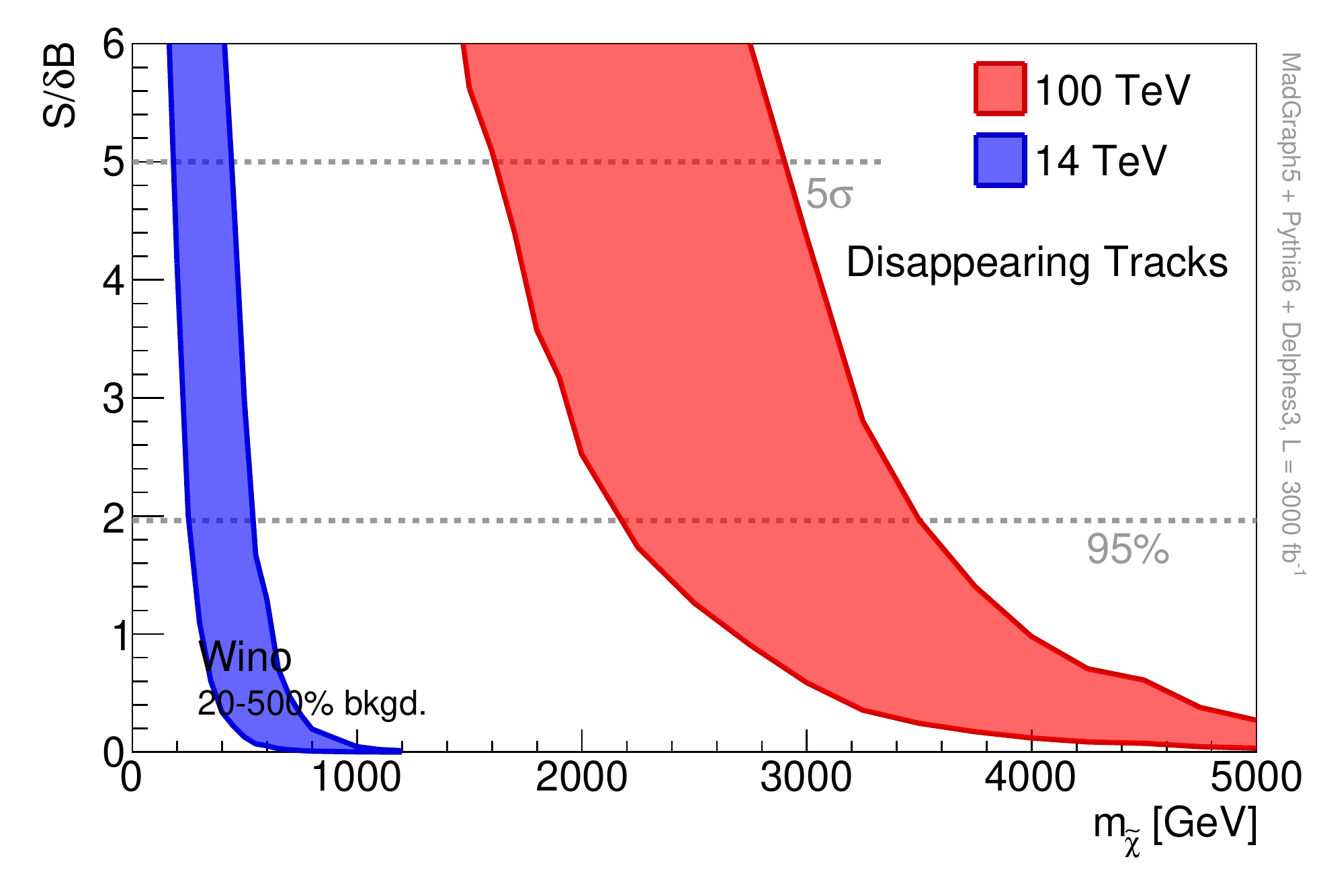}
\end{center}
\caption{Left: The mass reach for the pure wino in the monojet channel
  with $\mathcal{L}=3~\rm{ab}^{-1}$ for the 14 TeV LHC (blue) and
  at 100~TeV (red).  The bands are generated by varying the background
  systematics between $1-2 \%$ and the signal systematic uncertainty
  is set to $10 \%$~\cite{Low:2014cba}. Right: The mass reach in the
  pure wino scenario in the disappearing track channel with
  $\mathcal{L}=3~\rm{ab}^{-1}$ for the 14 TeV LHC (blue) and at
  100~TeV (red).  The bands are generated by varying the background
  normalization between $20-500 \%$~\cite{Low:2014cba}.}
\label{fig:wino}
\end{figure}
The dominant uncertainty in the reach comes from the systematics
 of the background, which is varied between $1-2$\%, generating
the bands in the plot.
Naively scaling by total event rates the systematics from current ATLAS studies
\cite{ATLAS:2012zim} (see Ref.~\cite{CMS:rwa} for the CMS study) 
would yield $0.5$\% for $3~\rm{ab}^{-1}$, but this is
clearly overly optimistic. Choosing the systematic error $ \sim 1-2$\%
as we have done may also be optimistic, but it sets a reasonable
benchmark, and underscores that minimizing these systematics should be
a crucial factor taken into account in the design of the 100~TeV
detectors. Given the same integrated luminosity, the monojet search
increases the reach relative to the LHC by nearly a factor of 5, as
shown in the left panel of Fig.~\ref{fig:wino} .

Due to the tiny mass splitting $\Delta m = 166$ MeV between the chargino and the neutralino, the decay lifetime can be long. The resulting disappearing track is a very distinctive signal in this case.
Since the dominant background for a disappearing track search would be mis-measured low-$p_T$ tracks, we cannot accurately predict the backgrounds in the not-yet-designed 100~TeV detectors. Nonetheless, we can calibrate against the present ATLAS searches for disappearing tracks \cite{Aad:2013yna} (see Ref.~\cite{CMS:2014gxa} for the CMS search).  
For example, we can require that $d^{{\rm track}}>30$ cm, with tens of signal events passing all cuts. The resulting mass reach is shown in the right panel of Fig.~\ref{fig:wino},
and the bands result from varying the background normalization upwards and downwards by a factor of 5.
The disappearing tracks could be extremely powerful, with the potential to both convincingly rule out, or discover, thermal wino dark matter.

\subsubsection{Pure Higgsino} %%%%%%%%%%%%%%%%%%%%%%%%%%%%%%%%%%%%%%%%%%%%%%%

Pure higgsinos are also produced through $s$-channel $Z$'s and $W$'s, and the analysis is similar to the pure wino case. The reach of the monojet search is shown in the left panel of Fig.~\ref{fig:higgsino}.
\begin{figure}[h!]
  \begin{center}
  \includegraphics[scale=0.32]{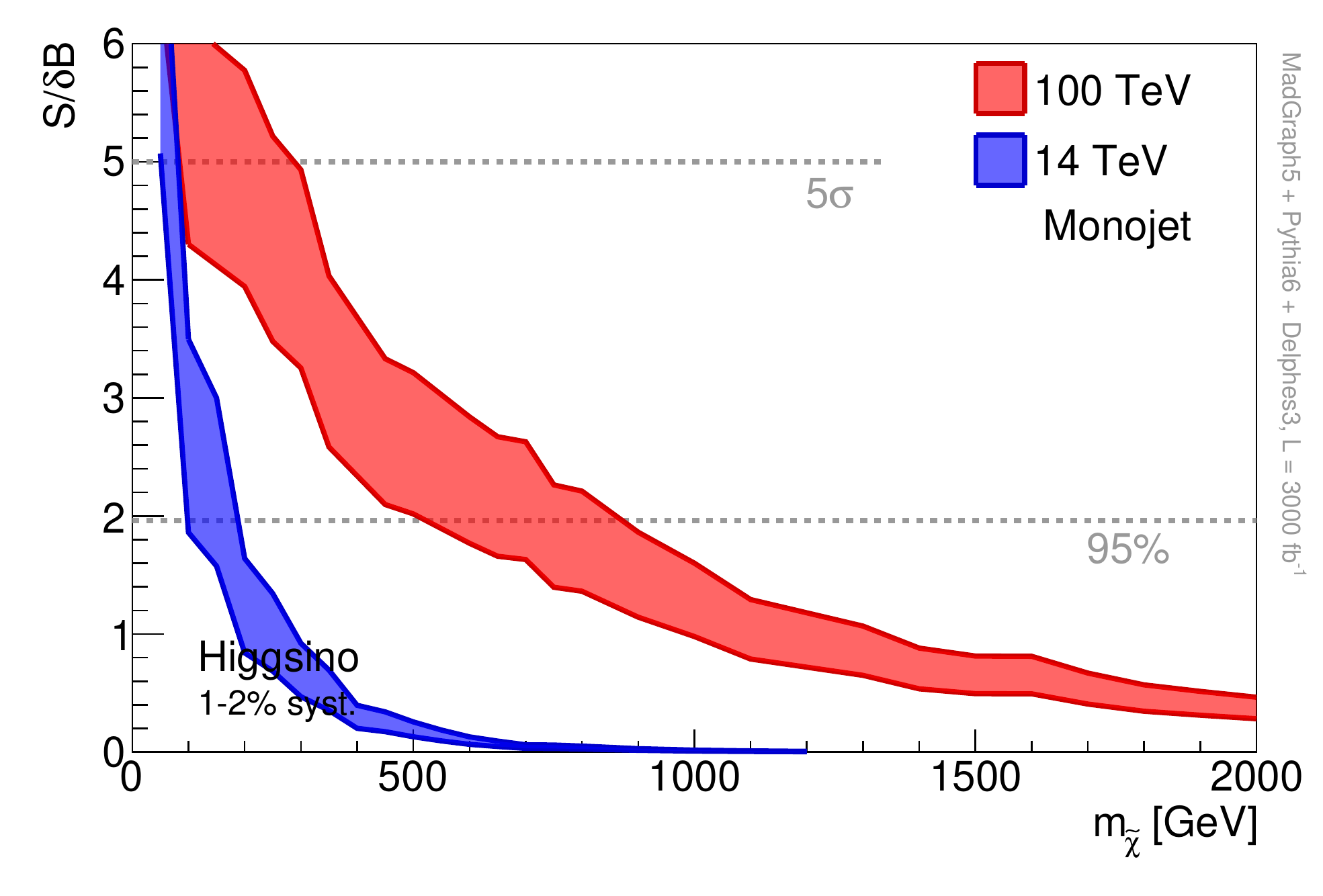}
  \includegraphics[scale=0.32]{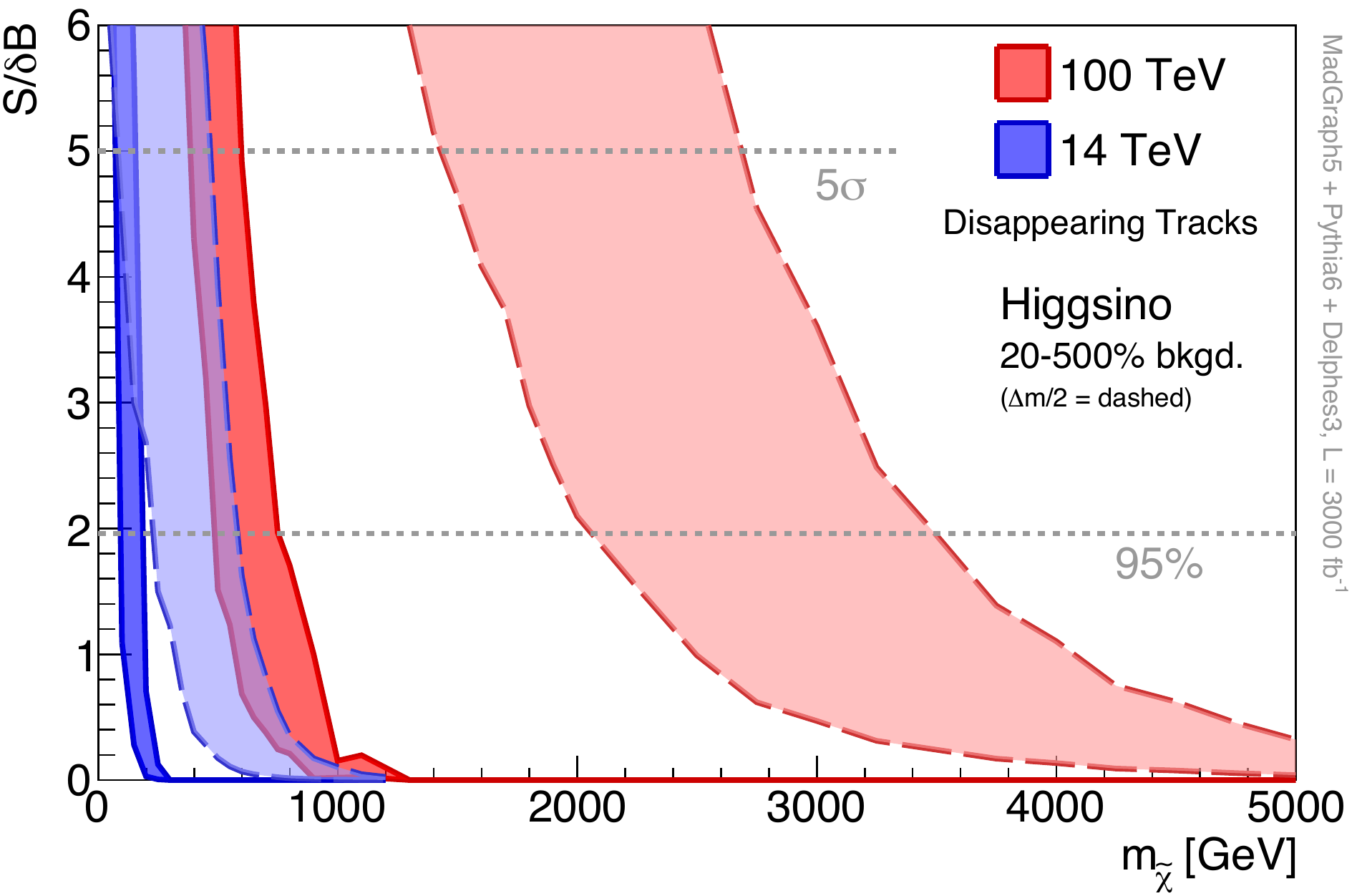}
  \end{center}
  \caption{Left: The mass reach for the pure higgsino in the monojet
    channel with $\mathcal{L}=3~\rm{ab}^{-1}$ for the 14 TeV LHC
    (blue) and at 100~TeV (red).  The bands are generated by varying
    the background systematics between $1-2 \%$ and the signal
    systematic uncertainty is set to $10
    \%$~\cite{Low:2014cba}. Right: The mass reach for the pure
    higgsino in the disappearing-track channel with
    $\mathcal{L}=3~\rm{ab}^{-1}$ for the 14 TeV LHC (blue) and at
    100~TeV (red).  The bands are generated by varying the background
    normalization between $20-500 \%$~\cite{Low:2014cba}.}
  \label{fig:higgsino}
\end{figure}
As for winos, the search improves by nearly a factor of 5 in mass relative to the LHC; the weaker reach relative to winos is due to the smaller production cross section. With optimistic systematics, higgsinos can be excluded up to 800 GeV.

We can next look at the disappearing-tracks search. If the splitting between the states is purely radiative, the lifetime for the higgsino is much shorter than for the wino, since the lifetime scales as $\tau \propto \Delta m^{-5}$.
This makes the disappearing-track search less effective than the monojet search for higgsinos; the reach is shown in the right panel (solid contour) of  Fig.~\ref{fig:higgsino}.

However it is worth recalling that unlike for the pure wino, the
splitting for the higgsino states can more easily be affected by the
presence of heavier states,  which can generate $\Delta m \sim M_Z^2/M$ --- which could be comparable to the radiative splittings if the heavier electroweak states are near $M \sim 5$ TeV.
If these splittings are comparable, resulting in a reduction of the width by a factor of 2, the decay length increases by a factor of $\sim 10 - 30$, and the higgsino reach becomes comparable to that for winos as shown in the right panel (dashed contour) of Fig.~\ref{fig:higgsino}. %Fig.~
This could be extremely exciting --- not only discovering the higgsino, but giving direct evidence for new multi-TeV electroweak states needed to reduce the higgsino mass-splittings in order to account for its anomalously long lifetime.

\subsubsection{Mixed dark matter} %%%%%%%%%%%%%%%%%%%%%%%%%%%%%%%%%%%%%%%%%%%%%%%

In the case of mixed dark matter we can expect mass splittings of tens
of GeV, and so the search is dominated by looking for the soft leptons
from chargino decays via off-shell $W$'s and $Z$'s. This will give us
a more powerful reach than with the monojet alone. On the other hand,
with these splittings the decays are prompt and we lose the advantages
of the disappearing-tracks search. We will focus on two representative examples, with mass splittings of 20 GeV. The first is a bino/higgsino mixture and 
the second is a bino/wino(/higgsino) mixture, obtained by dialing all three of $|M_1|,|M_2|,|\mu|$ close to each other.
The mass reach for these scenarios is shown in Fig.~\ref{fig:softlepton}.
\begin{figure}[h!]
  \centering
  \includegraphics[scale=0.32]{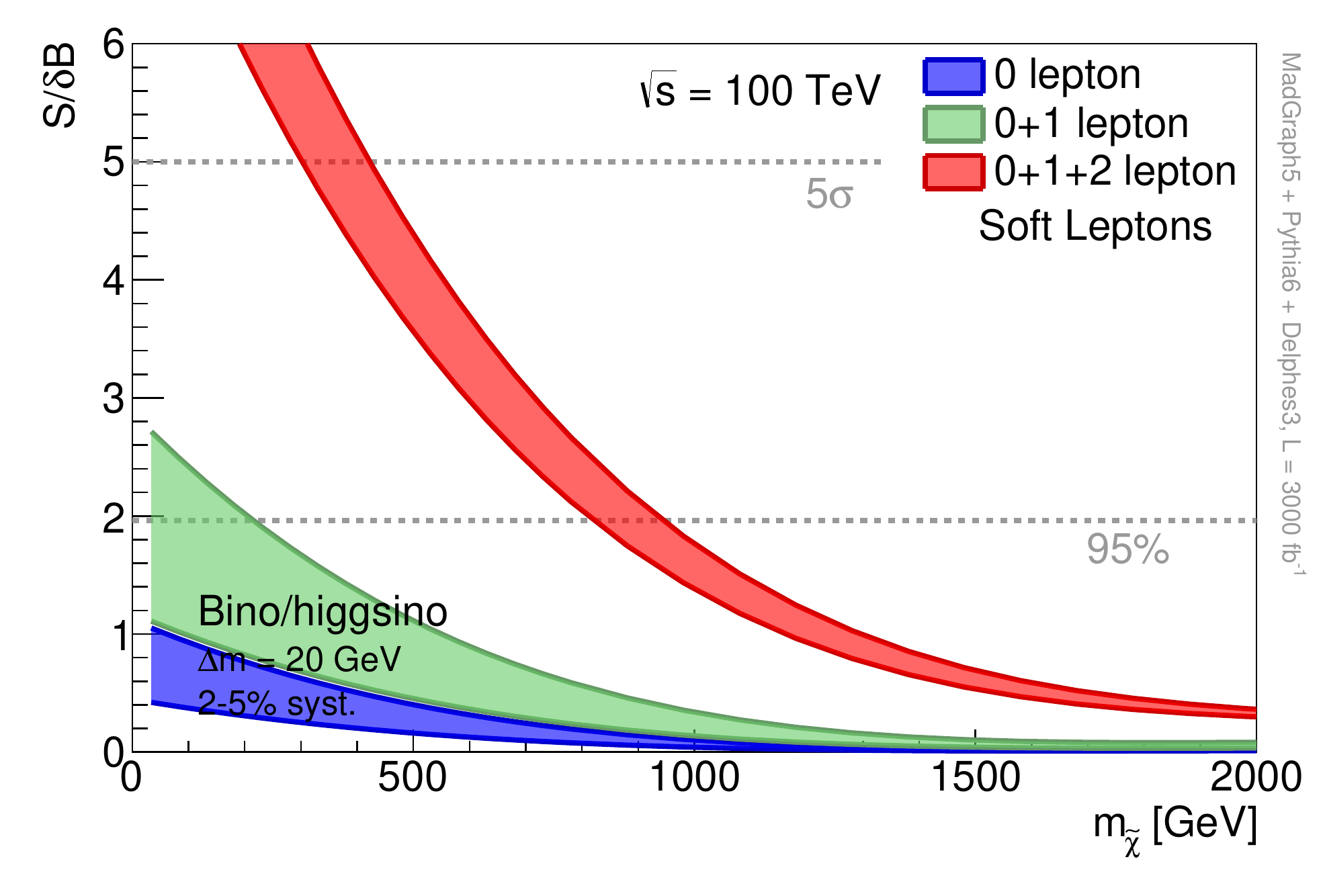}
  \includegraphics[scale=0.32]{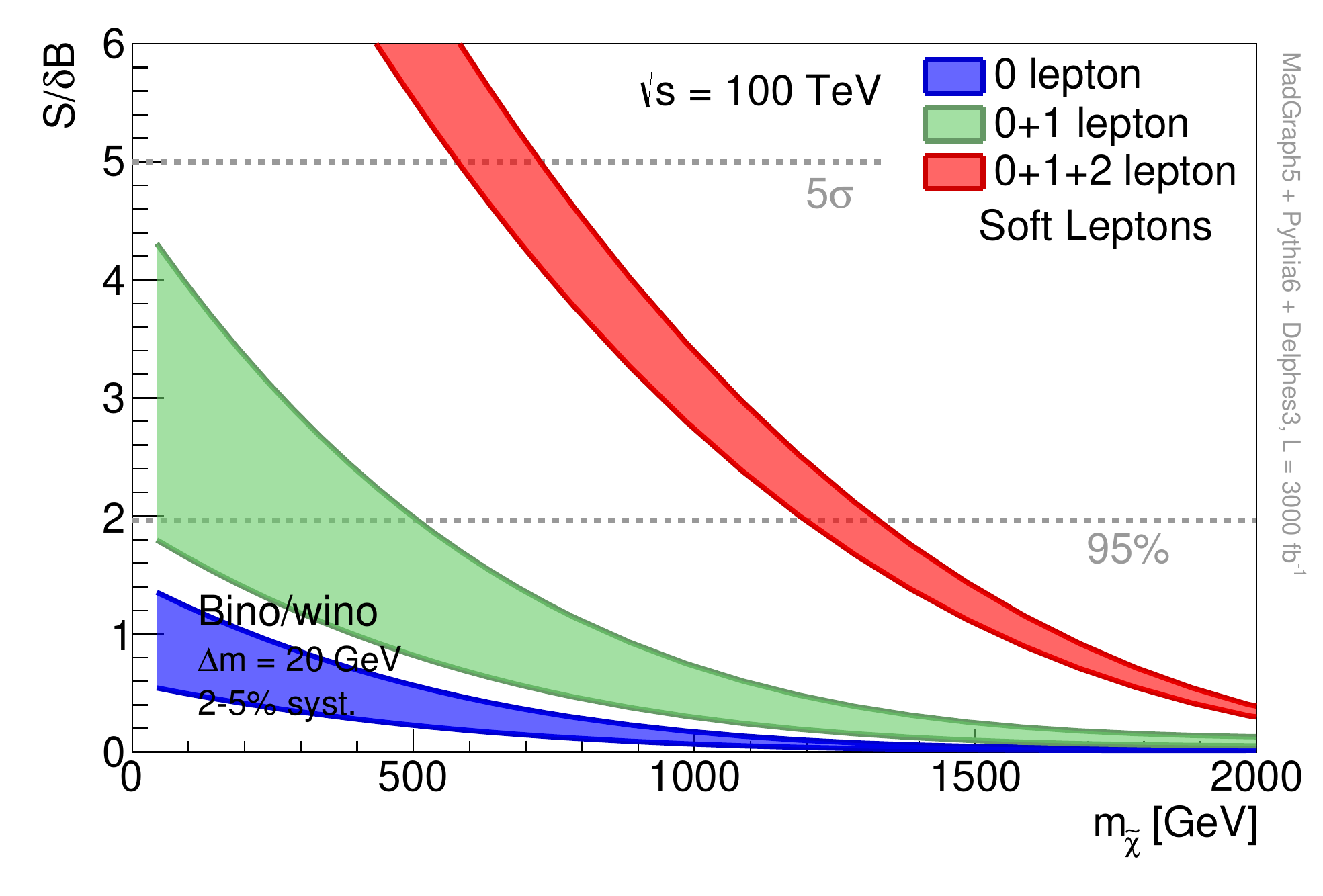}
  \caption{The mass reach in the mixed bino/higgsino ($\Delta = $ 20
    GeV) scenario in the soft lepton channel at 100 TeV with
    $\mathcal{L}=3~\rm{ab}^{-1}$, looking for 0 leptons
    (blue), 0 or 1 leptons (green), and 0, 1, or 2 leptons (red).  The
    bands are generated by varying the background systematics between
    $2-5 \%$ and the signal systematic uncertainty is set to $10
    \%$~\cite{Low:2014cba}.}
  \label{fig:softlepton}
\end{figure}

Note that in all cases the tagging of soft leptons is very important to maximize the mass reach. The 2-lepton bin is most important: while the 0 and 1 leÄpton bin backgrounds are dominated by single gauge-bosons, the 2-lepton backgrounds are controlled by diboson production with a much smaller cross section.
%process, which has a much smaller cross section.
We find exclusions reaching up to $\sim$ 1 TeV masses, and discovery up to several hundred GeV.

\subsubsection{Electroweak cascades} %%%%%%%%%%%%%%%%%%%%%%%%%%%%%%%%%%%%%%%%%%%%%%%

We have so far focused on the most difficult cases for dark matter
production, where the lightest electroweak states are produced and
their decays contain only soft particles. The mass reach can be
considerably higher if there is an electroweak spectrum with sizable
splittings. If the heavier states can be  produced, they will decay to
the dark matter state, emitting hard $W$'s, $Z$'s, and Higgses. This
leads to the familiar signals of multi-lepton plus missing energy, and
searches for events with leptons, such as 4 leptons, opposite- and
same-sign di-leptons.  A study of the reach at 100~TeV for electroweak cascades has recently been carried out in \cite{Gori:2014oua}, for four representative cases of the production of NLSP's decaying to the LSP:
\begin{itemize}
  \item Wino NLSP and higgsino LSP ($M_1 \gg M_2 > \mu$)
  \item Higgsino NLSP and wino LSP ($M_1 \gg \mu > M_2$)
  \item Higgsino NLSP and bino LSP ($M_2 \gg \mu > M_1$)
  \item Wino NLSP and bino LSP ($\mu \gg M_2 > M_1$)
\end{itemize}
The heaviest electroweakino in all cases is fixed to 5 TeV.  Bino NLSP's have too small a production cross section to be relevant, so they are never considered as the NLSP.

The reach for the final case, with wino NLSP and bino LSP, depends importantly on the wino branching ratios: for very heavy higgsinos, the decay
$\tilde{W} \to \tilde{B} h$ dominates, but it is also possible to have sizable branching ratios for emitting $W's,Z's$ as well.
The 100~TeV reach for these four scenarios is summarized in Fig.~\ref{fig:cascade}.
\begin{figure}[h!]
  \centering
  \includegraphics[width=0.49\textwidth]{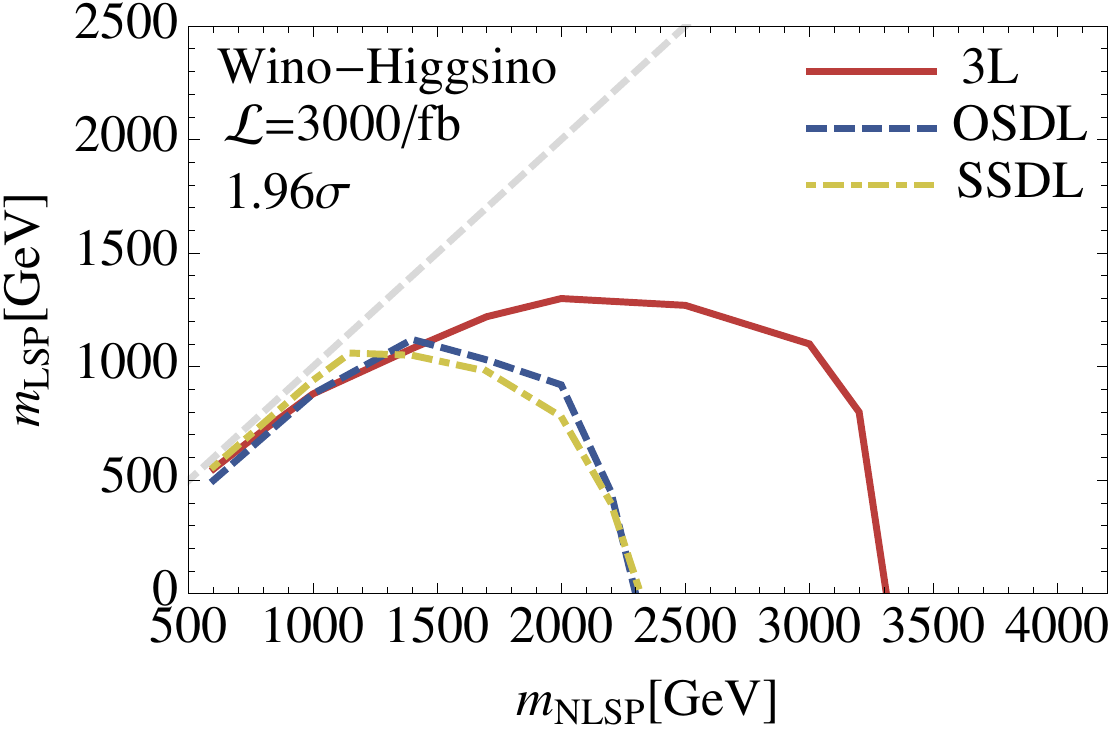}
  \includegraphics[width=0.49\textwidth]{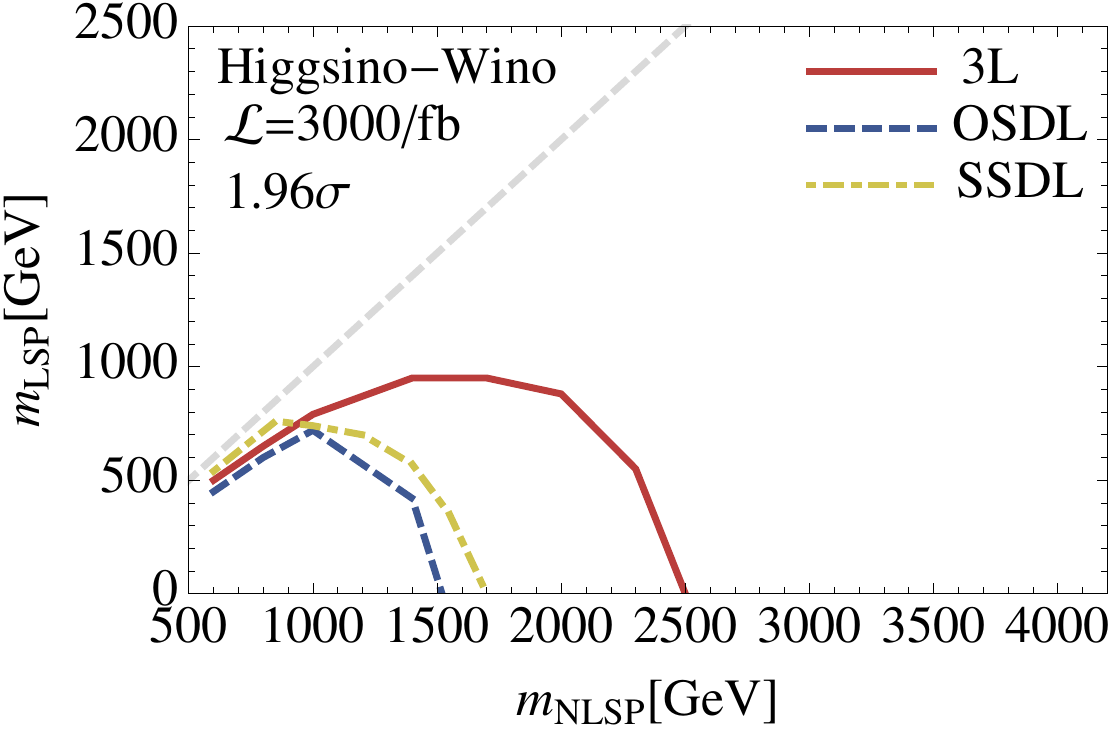} \\
  \includegraphics[width=0.49\textwidth]{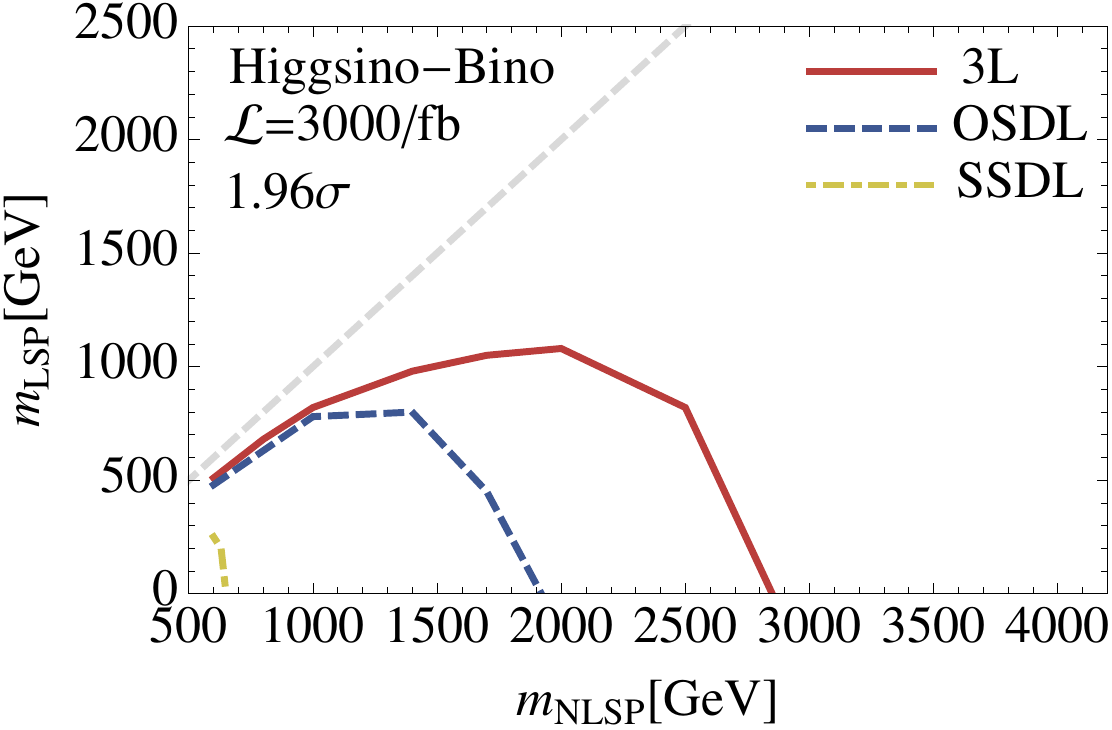}
  \includegraphics[width=0.49\textwidth]{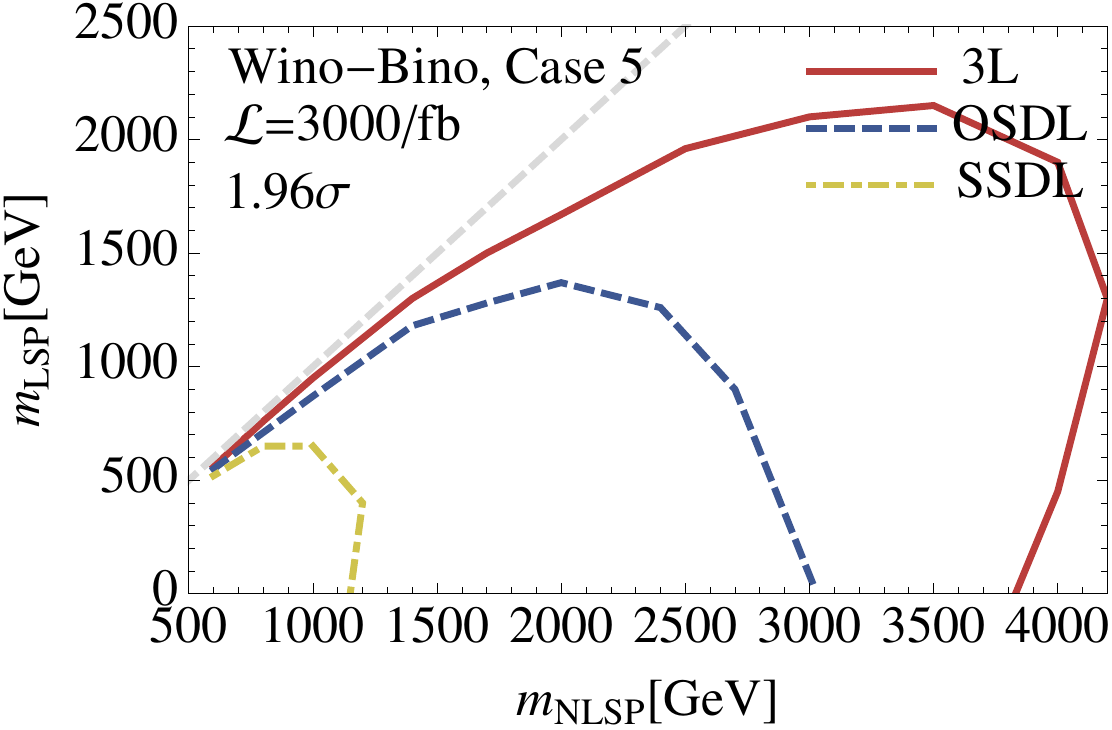}
  \caption{95 \% CL limits for wino-NLSP and higgsino-LSP (top left),
    higgsino-NLSP and wino-LSP (top right), higgsino-NLSP and bino LSP
    (bottom left) and wino-NLSP and bino-LSP (bottom
    right)~\cite{Gori:2014oua}.}
  \label{fig:cascade}
\end{figure}

This represents a major gain over the reach of the LHC. Most notably, the entire interesting range for higgsino masses can be probed in this way, provided the wino is lighter than $3$ TeV, and not too degenerate with the higgsino.

\subsubsection{Co-annihilation with Bino dark matter} %%%%%%%%%%%%%%%%%%%%%%%%%%%%%%%%%%%%%%%%%%%%%%%

So far we have only briefly considered the case of bino dark matter.  Due to its small couplings, the bino does not annihilate efficiently as it freezes out, and typically overcloses the universe unless it is extremely light.  Bino dark matter can be made viable in a supersymmetric context, if there are other superpartners with a mass nearly degenerate with it. Their presence can enhance the bino annihilation rate and give the correct relic abundance for heavier bino masses. If the co-annihilators are gluinos, stops, or squarks, the bino masses giving the correct relic abundance are in the multi TeV region, $\sim 7$ TeV for gluino co-annihilation, and $\sim 2$ TeV for stop or squark co-annihilation.  Since the colored states are very close in mass to the bino, they can have large production rates at a 100 TeV collider.  They will then decay to the bino and soft SM particles, resulting in the monojet signal.  Due to the colored production, however, these rates will be much higher than with electroweakino monojet signals.

\begin{figure}[h!]
  \centering
  \includegraphics[scale=0.30]{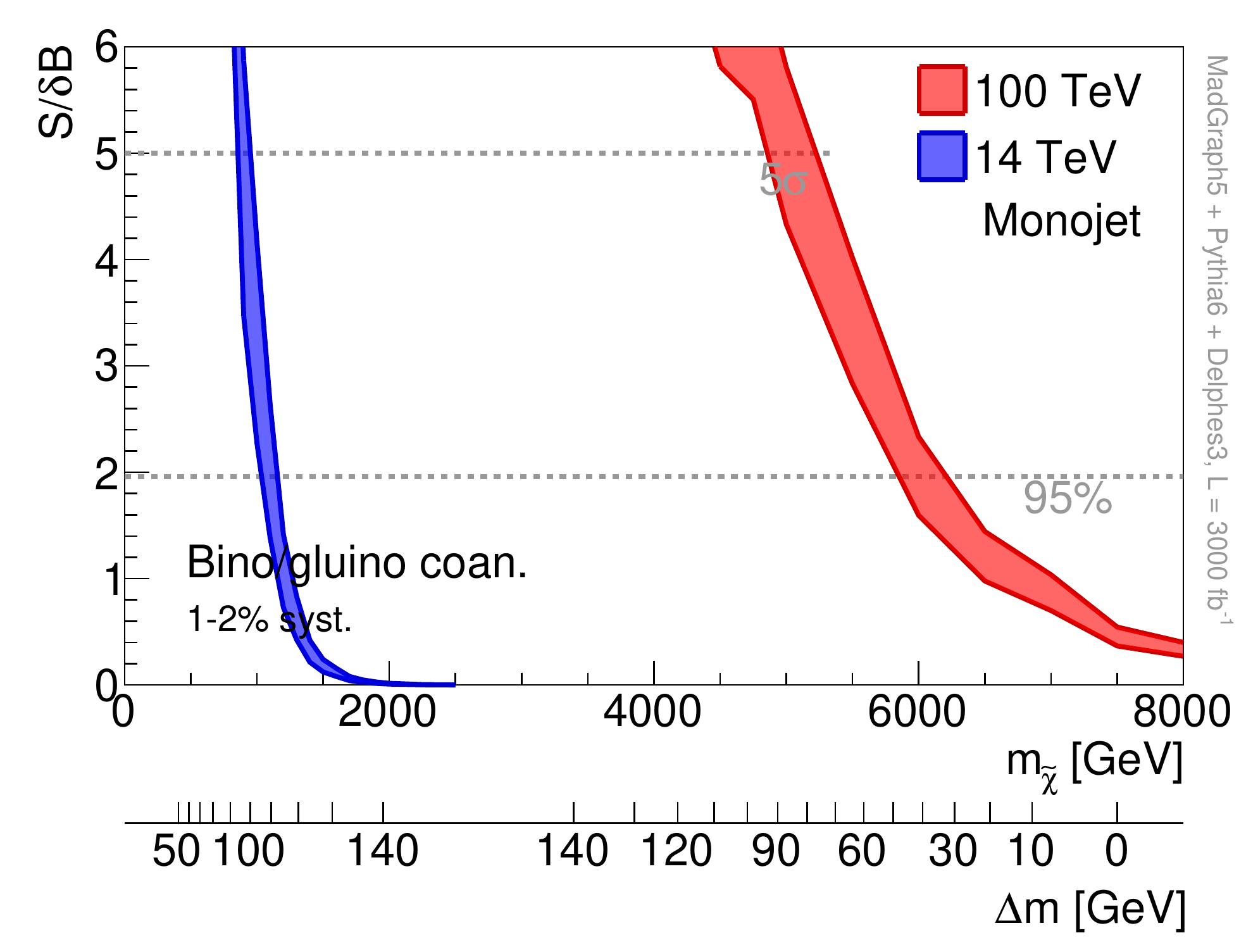}
    \includegraphics[scale=0.30]{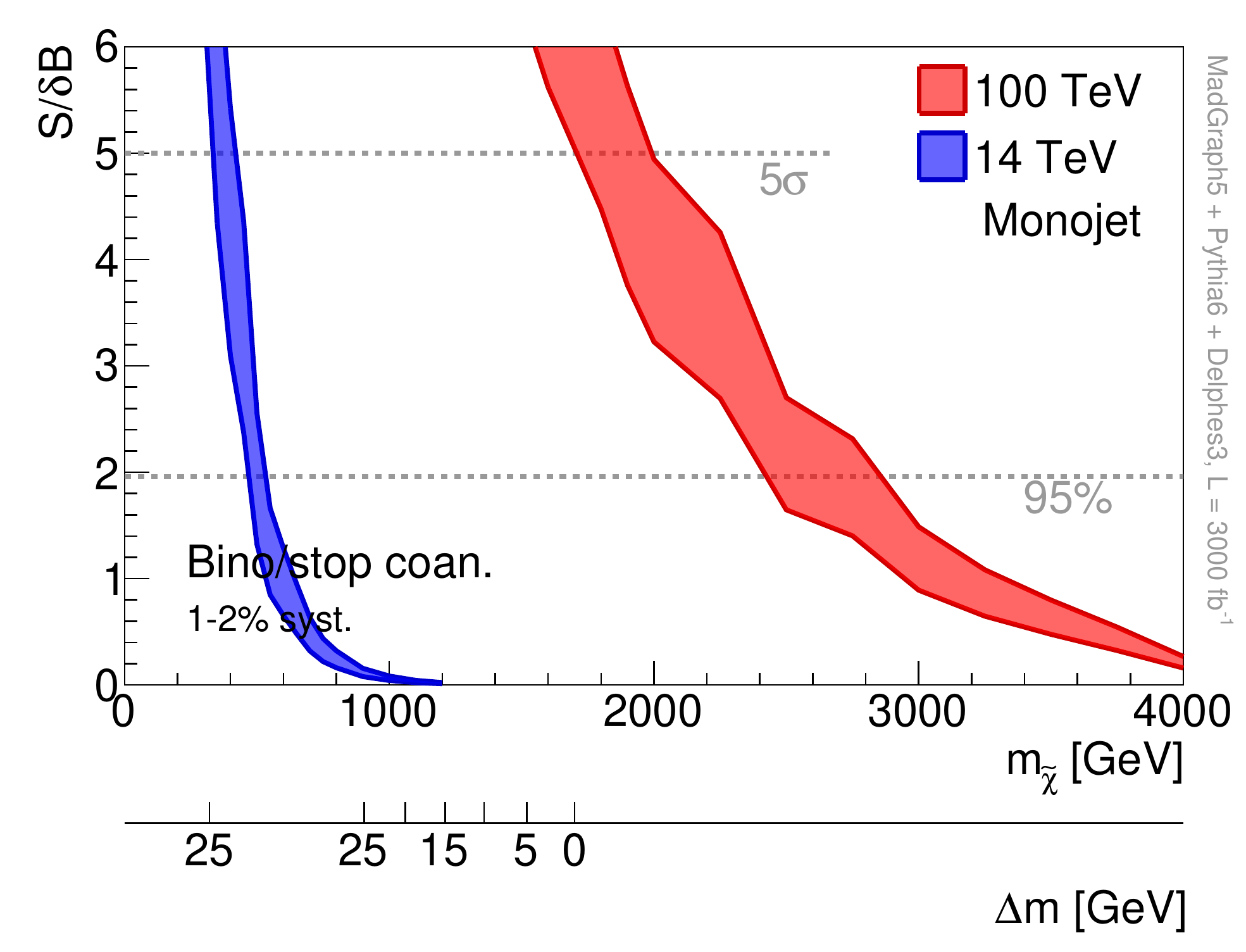}

  \caption{Left: The mass reach in the gluino coannihilation scenario
    in the monojet channel with $\mathcal{L}=3~\rm{ab}^{-1}$ for
    the 14 TeV LHC (blue) and at 100~TeV (red).  The bands are
    generated by varying the background systematics between $1-2 \%$
    and the signal systematic uncertainty is set to $10 \%$.  The
    lower $x$-axis displays the gluino-bino mass splitting $\Delta m$
    for a given bino mass that is required to saturate the relic
    density~\cite{Harigaya:2014dwa,deSimone:2014pda}.  A tick is
    placed every 10 GeV with the exception of the consecutive $\Delta
    m=140$ GeV ticks~\cite{Low:2014cba}. Right: The mass reach in the
    stop coannihilation scenario in the monojet channel with
    $\mathcal{L}=3~\rm{fb}^{-1}$ for the 14 TeV LHC (blue) and at
    100~TeV (red).  The bands are generated by varying the background
    systematics between $1-2 \%$ and the signal systematic uncertainty
    is set to $10 \%$.  The lower $x$-axis displays the stop-bino mass
    splitting $\Delta m$ for a given bino mass that is required to
    satisfy the relic density~\cite{deSimone:2014pda}.  A tick is
    placed every 5 GeV with the exception of the consecutive $\Delta
    m=25$ GeV ticks~\cite{Low:2014cba}.}
  \label{fig:coan}
\end{figure}

The  reach at 100~TeV for gluino and stop annihilations, as obtained
by \cite{Low:2014cba}, is shown in the left and right panels of
Figs.~\ref{fig:coan}, respectively.  For gluino co-annihilation the
gluino-bino splitting required to get the right relic abundance is
shown on the bottom $x$-axis of the left panel of Fig.~\ref{fig:coan}.
We see that a 100 TeV collider covers most of this parameter space.
It is also worth recalling, that we have presented the most conservative search as we assume that whatever accompanies the LSP from the co-annihilator decay is undetectable.  In practice, the searches can be augmented by looking for the possibly soft decay products.

The mass splitting for the correct relic abundance in stop co-annihilation has also been computed and is displayed on the bottom $x$-axis of the right panel of Fig.~\ref{fig:coan}.  Here, a 100 TeV collider can make strong statements about this spectrum. Both exclusion and discovery are possible even in the degenerate stop-bino limit.

\subsection*{Summary} %%%%%%%%%%%%%%%%%%%%%%%%%%%%%%%%%%%%%%%%%%%%%%%

A broad summary of the dark matter reaches we have discussed is given in Fig.~\ref{fig:DMsummary}.
\begin{figure}[h!]
  \centering
  \includegraphics[scale=0.325]{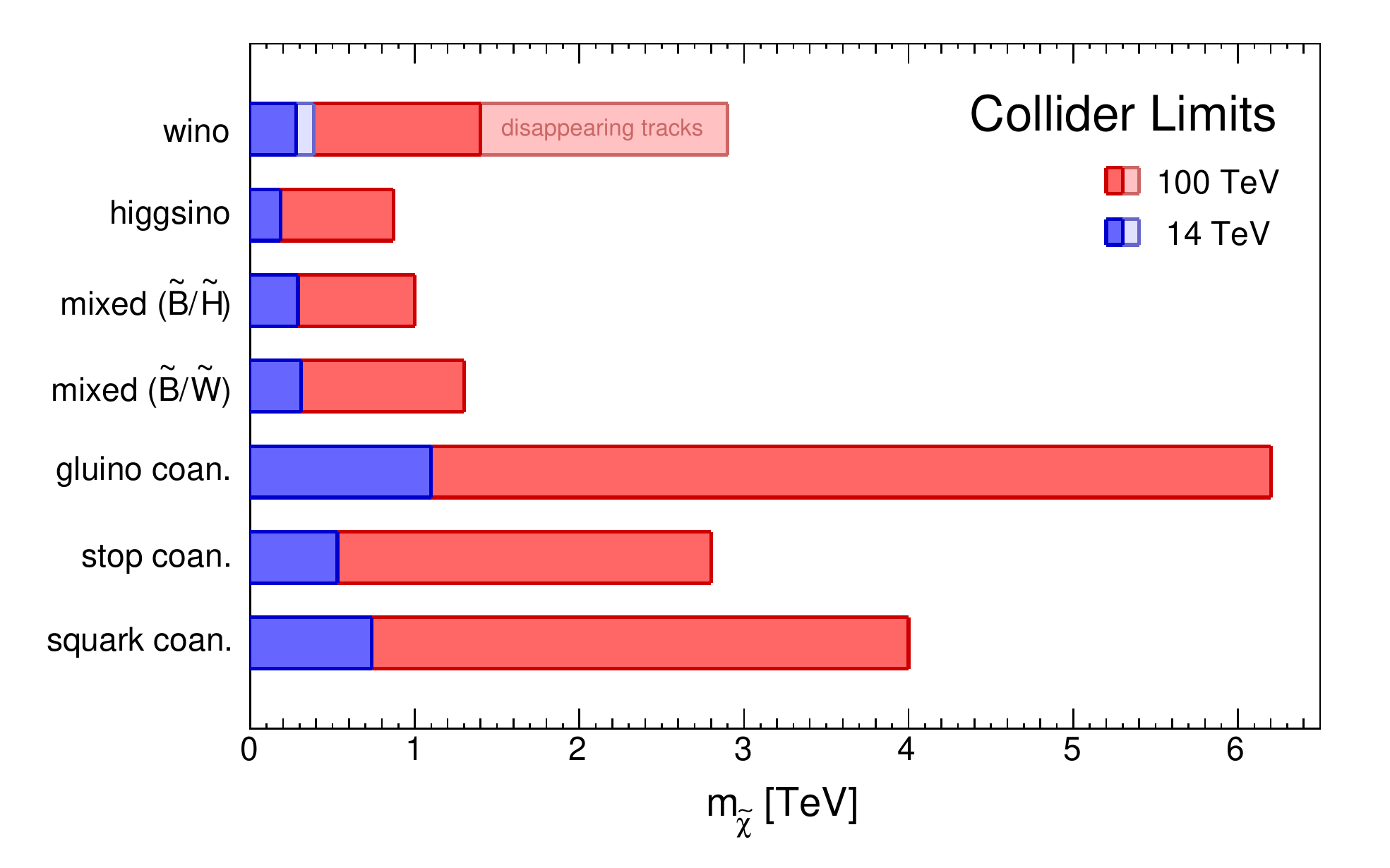}
  \includegraphics[scale=0.325]{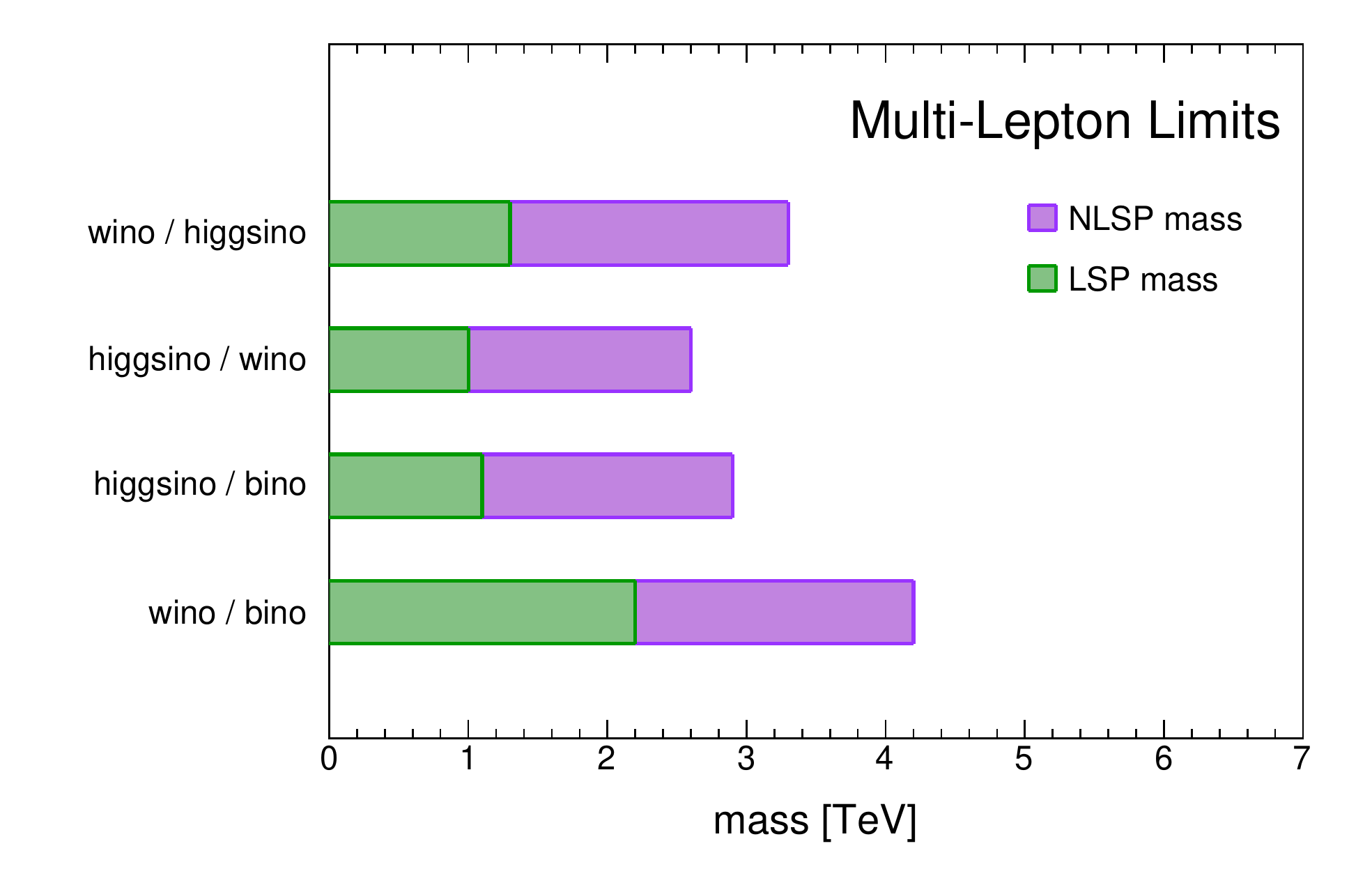}
  \caption{Summary of colliders' reach for neutralino dark
    matter~\cite{Low:2014cba} and in electroweakino
    cascades~\cite{Gori:2014oua}.}
  \label{fig:DMsummary}
\end{figure}
While the LHC can look for electroweak states up to a few hundred GeV, it will not probe the TeV mass range that is most natural for thermally saturating dark matter. By contrast, the jump to 100 TeV extends the LSP mass reach from the LHC roughly by a factor of 5, and thus allows us to go deep into this territory, with a great potential to discover WIMP dark matter.

\section{Other New Physics Searches}
\label{sec:NP}
As the next exploration facility at the energy frontier, the 100 TeV
$pp$ collider will lead us into completely new territory. 
In this section, we present the projections of a variety of new particles and phenomena that could show up. 
We show the cross section
increases with respect to the LHC, and provide qualitative estimates of
the observability in experiments at 100~TeV.
%

%%%%%%%%%%%%%%%%%%%%%%
\subsection{New Color Resonances}
A high energy hadron collider is a QCD machine. Any new states with
QCD interactions would be copiously produced via quark and gluon
partons. Some such exotic states have been systematically classified in
Ref.~\cite{Han:2010rf}, and the LHC experiments have been actively
searching for them \cite{Aad:2014aqa,Chatrchyan:2013qha}. The
non-observation at the LHC sets  bounds on their mass, bounds that
will extend well beyond a few TeV after the LHC energy increase to
$13-14$ TeV. 
This mass reach would be substantially extended by the 100 TeV collider. 

\begin{figure}[h!]
\centering
\includegraphics[scale=0.60]{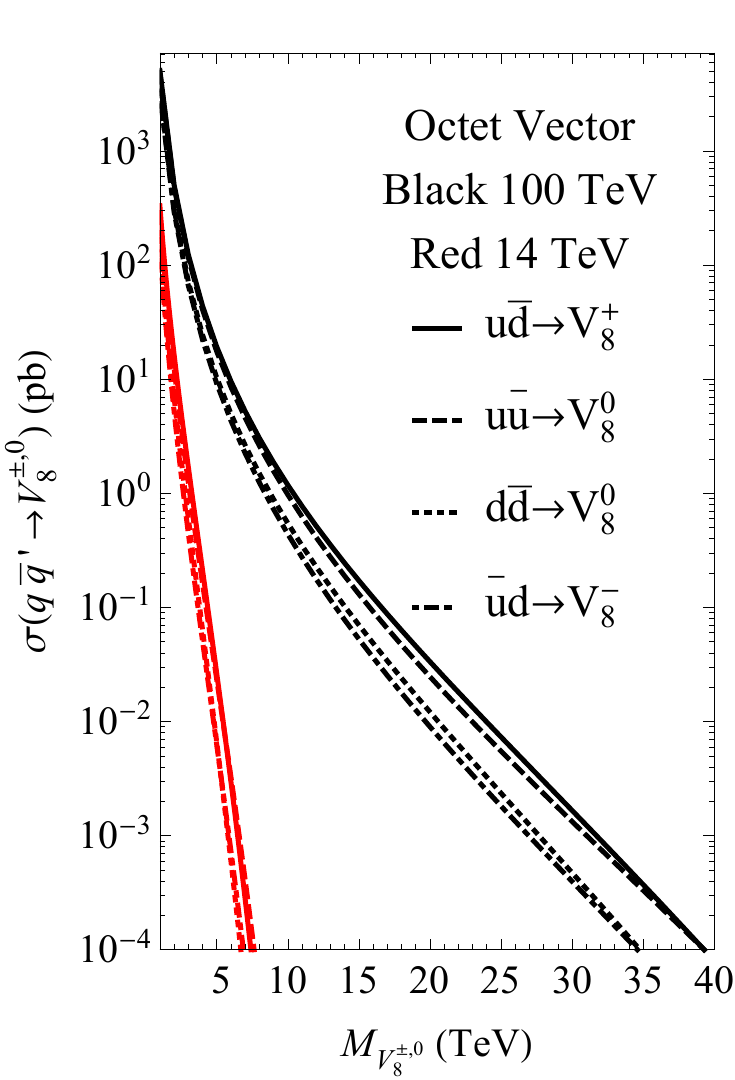}
\includegraphics[scale=0.60]{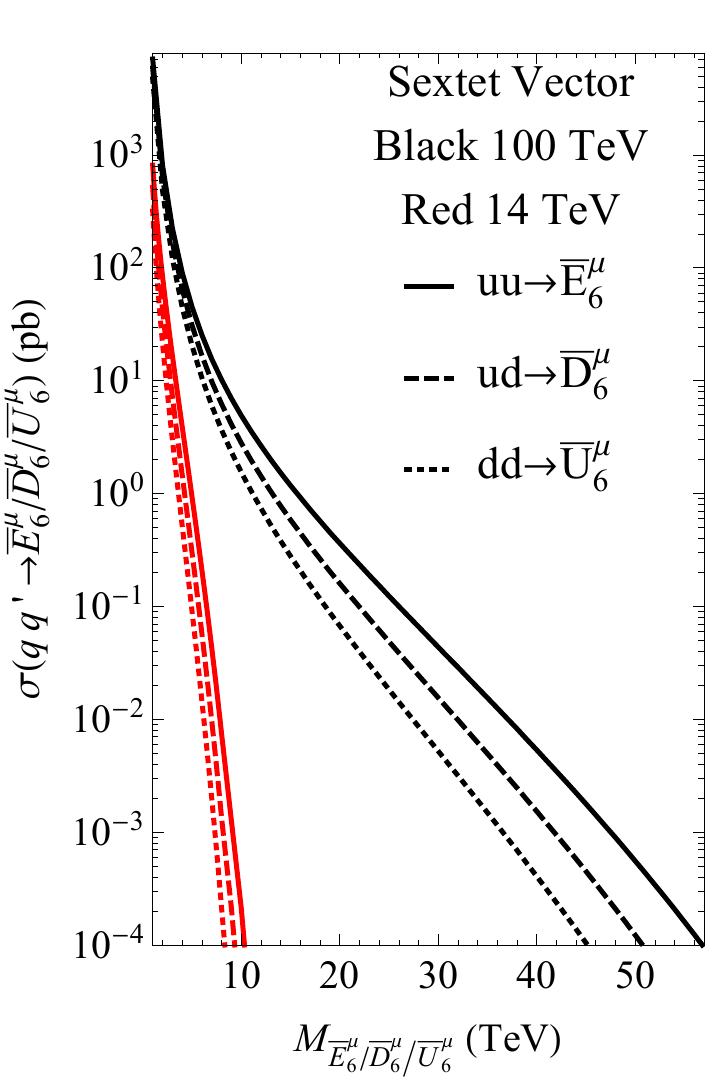}
\includegraphics[scale=0.60]{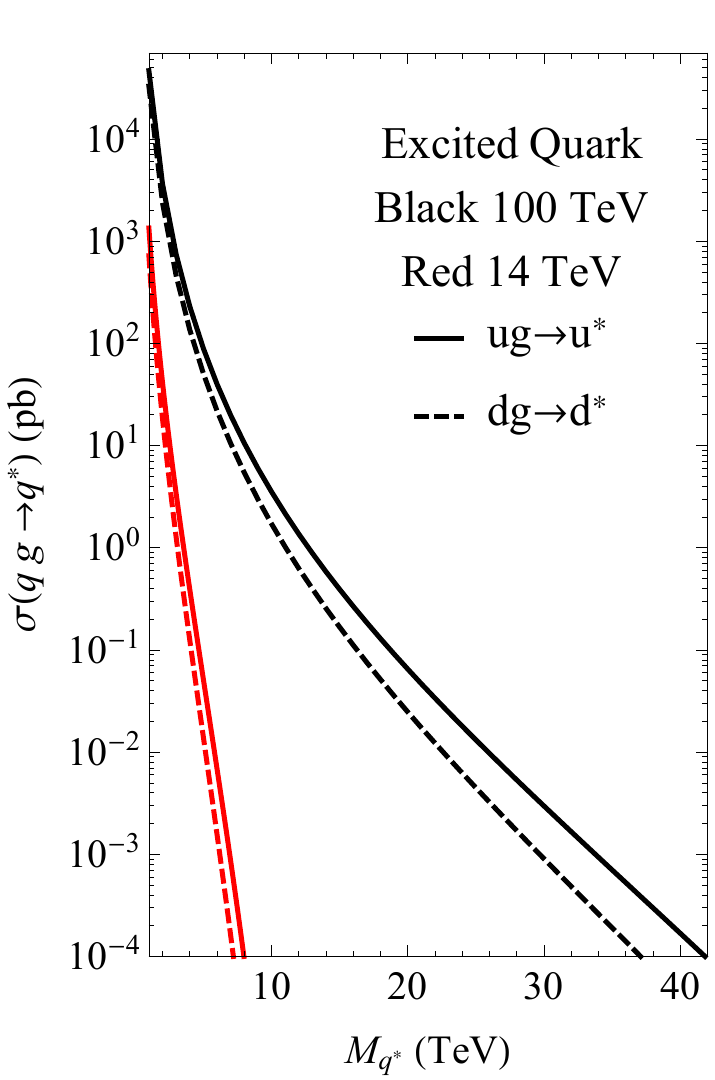}
\caption{Production cross sections for exotic colored resonances at 14 TeV and 
  100~TeV, for (a) charged and neutral color-octet vector states, (b)
  fractionally charged color-sextet vector states (di-quark-like), and
  (c) spin-3/2 excited quark states.} 
\label{fig:colored}
\end{figure}

We show the total production cross sections for the a representative
set of new resonant states in Fig.~{\ref{fig:colored}}, considering
(a) charged or neutral color-octet vector states (techni-$\rho$-like),
(b) fractionally charged color-sextet vector states (di-quark-like),
and (c) excited quark states (spin-$3/2$ for quark compositeness).
We see that the cross sections\footnote{There is a model-dependent dimensionless coupling constant for each of the couplings. We have set it to be unity for illustration \cite{Han:2010rf}.}  for these exotic colored states' production
can reach 0.1$-$1 fb for the mass range of 25$-$55 TeV, a rate which is
expected to be easily observable given the planned several \iab. The color-octet vector states in
Fig.~{\ref{fig:colored}}(a) are produced via the Drell-Yan process from  $q\bar q$ annihilation, which results in a lower reach by
about 10 TeV than the colored di-quark states in
Fig.~{\ref{fig:colored}}(b), produced through the valence quark pair
annihilation. In contrast, the excited quark states in
Fig.~{\ref{fig:colored}}(c) are produced via dimension-5 operators\footnote{We set the cutoff scale to be equal to the resonance mass for simplicity \cite{Han:2010rf}.}
with $qg$ fusion and have a sensitivity reach in between the above
two.
The exotic colored states will typically decay back to two jets, leading to di-jet resonances. One will thus expect
that the 100~TeV experiments would be able to significantly extend the LHC coverage of the exotic colored states, reach a broad mass range beyond about 25$-$55 TeV.

%%%%%%%%%%%%%%%%%%%%%%
\subsection{New Gauge Bosons and Vector Resonances}
One of the most striking signals would be the new electroweak gauge
boson resonant production with the subsequent decay to leptonic final
states --- the typical Drell-Yan mechanism. New 
charged $W'$ and neutral $Z'$ gauge bosons exist in many theories with gauge
extensions beyond the SM.
We illustrate the typical cross sections for $W'$ and $Z'$ production
for various well-motivated models
\cite{Langacker:2008yv,Agashe:2014kda} in Fig.~\ref{fig:WpZp} at both
14 and 100~TeV. As expected, the LHC may be able to
uncover a $W',Z'$ signal up to a mass of about 5 TeV with a cross section of the order 0.1 fb. At 100~TeV, one will extend the mass reach to about 25 TeV for a $(B-L)\ Z'$ (the smallest in rate), and to about 35 TeV for a left-right symmetric model $W'$ (the largest in rate). Somewhere in
between, a sequential SM $Z'$ may be observable to about 30 TeV.
Similarly, the production rate of a color-singlet $\rho$-like vector state in the minimal version of composite Higgs models is shown in Fig.~\ref{fig:vector_resonance_reach}. The production rate is roughly comparable to that of the $(B-L)\ Z'$.

\begin{figure}[h!]
\begin{center}
  \includegraphics[scale=0.4]{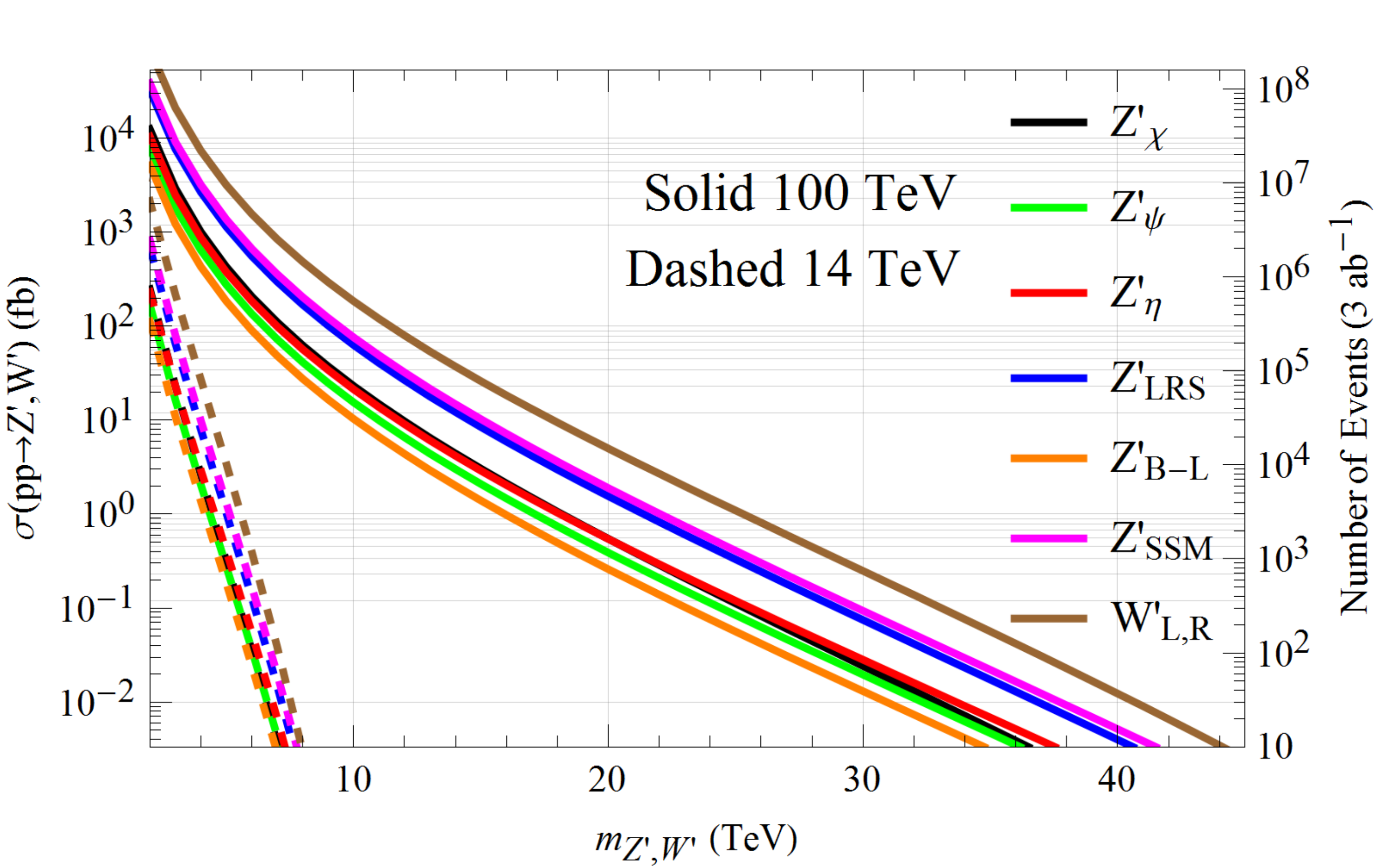}
\end{center}
\caption{Production cross section of new heavy electroweak gauge bosons 
$W'$ and $Z'$ in various models \cite{Langacker:2008yv,Agashe:2014kda}
  at 14 and 100~TeV.} 
\label{fig:WpZp}
\end{figure}

\begin{figure}[h!]
\centering
\includegraphics[scale=1.1]{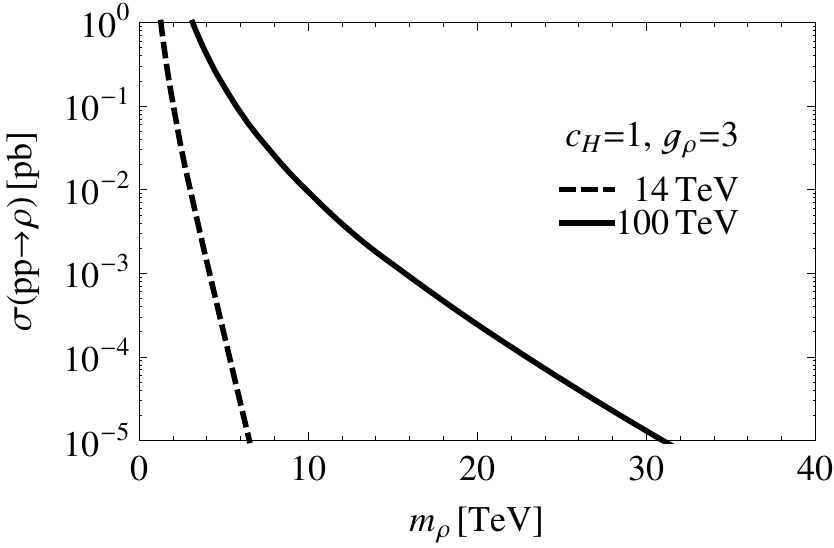}
\caption{Production cross section of color-singlet vector resonances
  $\rho$ at 14 and 
100~TeV.}
\label{fig:vector_resonance_reach}
\end{figure}

%%%%%%%%%%%%%%%%%%%%%%%%%%%%%%%%
\subsection{Heavy Higgs Bosons in Doublet and Triplet Models}
Many theories beyond the SM need the extension of the Higgs sector,
resulting in the prediction of new Higgs bosons, some of the commonly considered examples are denoted as $H^0$, $A^0$, $H^\pm$,
and $H^{\pm\pm}$. Searching for the heavy Higgs bosons will be extremely important from the point of view of both understanding the full electroweak sector, and exploring the naturalness paradigm.  Indeed, the mere existence of additional Higgs bosons at
the TeV scale would unambiguously reveal new principles in the construction of the EW sector, and presenting new challenges in comprehending the naturalness
problem with multiple scalars. However, it would be very challenging to discover those
states at the LHC because of the rather small production cross section
and the large SM backgrounds to their decay products in the final state, perhaps limited to a mass scale around 1 TeV \cite{Craig:2015jba,Hajer:2015gka}. It is thus expected that the significant increase of the CM
energy to 100~TeV would allow us to extend the coverage for heavy
Higgs boson searches.

\begin{figure}[h!]
\begin{center}
  \includegraphics[width=0.8\textwidth]{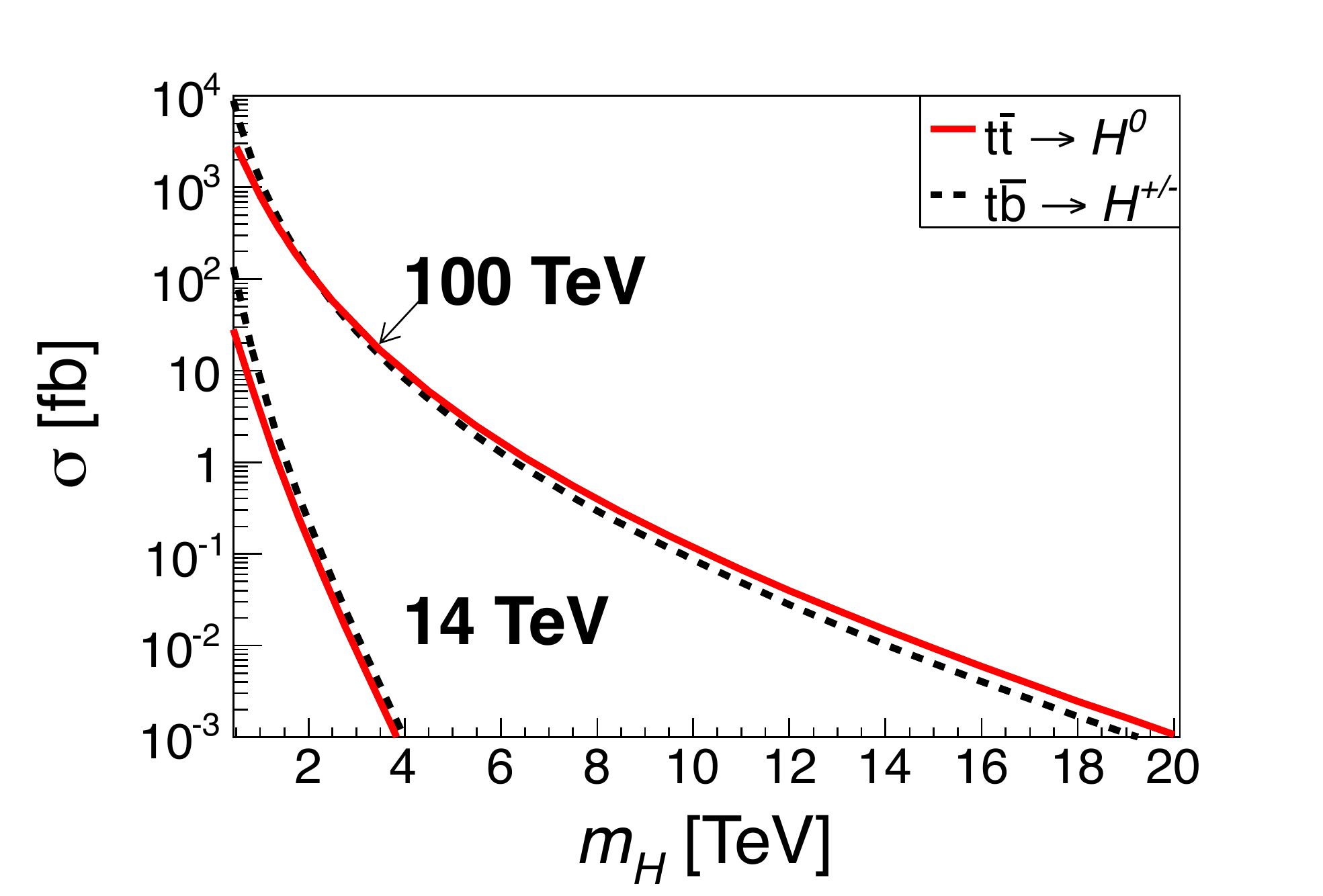}
\end{center}
\caption{Single heavy Higgs boson production ($H^0, A^0$ and $H^\pm$)
  associated with a heavy quark ($t$ and $b$) at 14 and 100~TeV for $\tan\beta=10$. The
  calculations include the gluon initial as well as the heavy
  quark initial processes with proper collinear subtraction \cite{Han:2014nja} in the
  ACOT scheme. }
\label{fig:hq}
\end{figure}

The leading production channels for heavy Higgs bosons are the single
Higgs boson associated with heavy quarks ($b$ and $t$). Figure
\ref{fig:hq} shows the total cross section for $H^0\ (A^0)$ and
$H^\pm$ processes.  The calculations include $gg\to t\bar t
H^0,\ t\bar b H^\pm$ and $gt\to t H^0$, $gb\to t H^\pm$ and with
proper treatment for the collinear subtraction of the massive quark partons in the ACOT scheme \cite{Aivazis:1993pi,Han:2014nja}, for $\tan\beta=10$.  We see the orders of magnitude increase of the cross section going from 14 TeV to 100 TeV, and the relative enhancement becomes more substantial at higher masses. 
The typical experimental signatures of the heavy Higgs doublet are the
decays of the Higgs bosons to heavy fermions $H^\pm \to
tb,\ \tau\nu;\ H^0, A^0\to t\bar t,\ b\bar b, \tau\tau$. While having
to face the challenge of the highly boosted objects from the heavy
Higgs decays, one can expect the 100~TeV collider to extend the LHC
coverage substantially. 
Depending on the value of $\tan\beta$, it is conceivable to reach the
heavy Higgs mass up to 10 TeV \cite{Hajer:2015gka} with a cross section of the order 0.1 fb at the 100 TeV collider.

It is important to note that the cross sections for Higgs pair
production $\gamma^*, Z^*\to H^+H^-,\ H^{++}H^{--}$ and $W^{\pm *}\to
H^\pm A^0,\ H^{++}H^{-}$ depend only on the electroweak gauge
interactions. In contrast, the complementary processes 
$Z^*\to A^0H^0,\ A^0 h^0$ and $W^{\pm *}\to H^\pm H^0,\ H^\pm h^0$, are sensitive to the model parameter
of the neutral scalar mixing $\cos(\beta-\alpha)$ (here assumed to be
close to 1).  In Fig.~\ref{fig:2Higgs}, we present those cross
sections for (a) a generic two-Higgs doublet model (2HDM), and (b) a
$Y=1$ triplet model.
The results are shown for the Higgs pair production at 14 and 100~TeV,
versus the heavy Higgs mass (assumed to be degenerate). Because of the
heavy mass and the electroweak coupling, the pair production cross
sections are lower than the heavy-quark associated production by about three 
 orders of magnitude. It is nevertheless conceivable to reach the sensitivity of the Higgs pair production with a mass scale of about a few TeV. 

\begin{figure}[h!]
\begin{center}
  \includegraphics[width=0.45\textwidth]{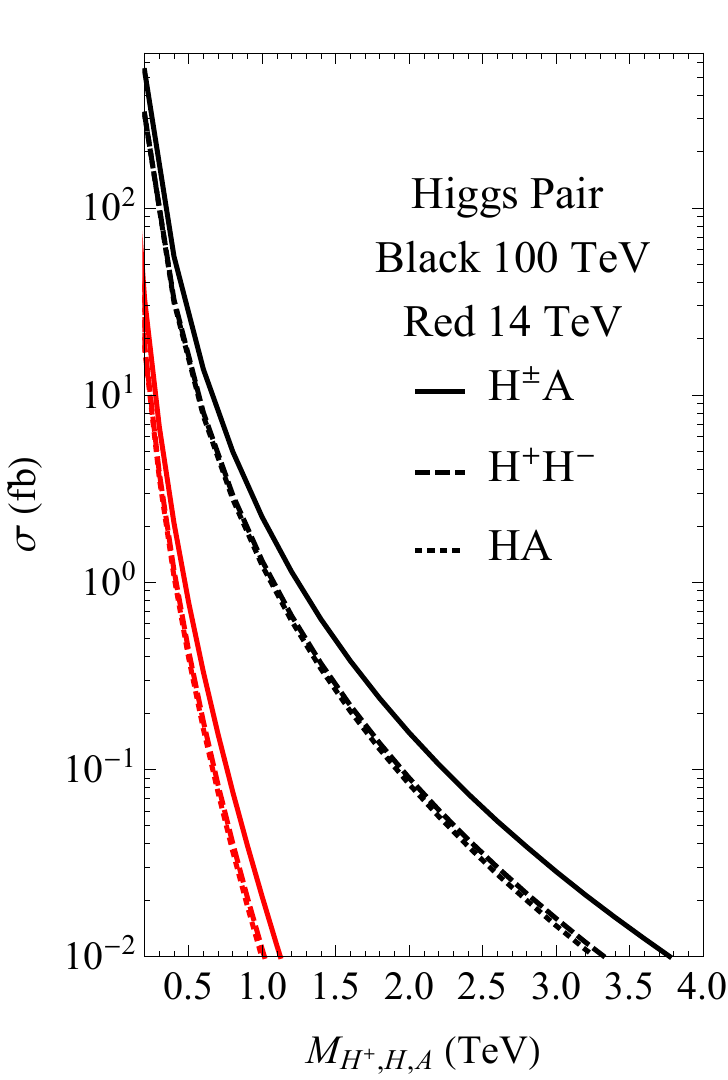}~
  \includegraphics[width=0.45\textwidth]{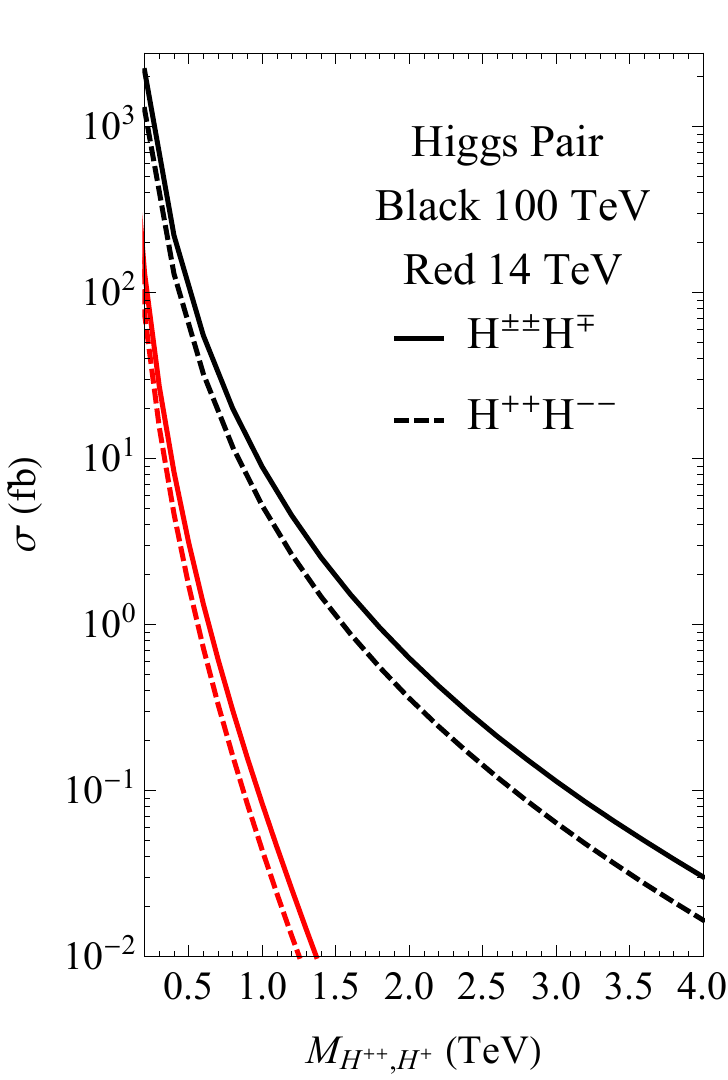}
\end{center}
\caption{Pair production of heavy Higgs bosons via pure EW gauge interactions at 14 and 100~TeV.}
\label{fig:2Higgs}
\end{figure}

The striking feature for the $Y=1$ triplet Higgs $(\Phi)$ is the existence of the doubly-charged Higgs boson $H^{\pm\pm}$. 
The severe constraint from the SU(2) custodial symmetry bounds the
triplet vev to be smaller than about a GeV.
The co-existence of the couplings to the lepton doublet and to the
Higgs doublet
$$-y_\nu \bar L \Phi i\sigma_2 L^c + \mu H^T i\sigma_2 \Phi H$$
features the Type-II seesaw mechanism for the Majorana neutrino mass generation
\cite{Cheng:1980qt,Mohapatra:1980yp}.  The model breaks the lepton
number by two units and leads to a neutrino Majorana mass $m_\nu \sim
y_\nu v' \sim y_\nu \mu v^2/M_\Phi^2$.  An interesting borderline is $v' \sim
10^{-4}$ GeV, above which $H^{\pm\pm}\to
W^\pm W^\pm$ dominates, and  below which $H^{\pm\pm}\to
\ell^\pm_i\ell^\pm_j$ takes over.  This is an extremely attractive
scenario because not only it leads to very clean like-sign di-lepton
signals at hadron colliders with unambiguous lepton-number violation, but also the flavor combinations of the lepton pairs $\ell_i \ell_j$ would correlate with the low-energy
neutrino mixing patterns and thus could help probe the neutrino mass
hierarchy \cite{Perez:2008ha}. The LHC will probe the doubly charged
Higgs to about a TeV in mass and a 100~TeV collider would be able to
extend the coverage to about 5 TeV.

\begin{figure}[h!]
\begin{center}
  \includegraphics[scale=0.25]{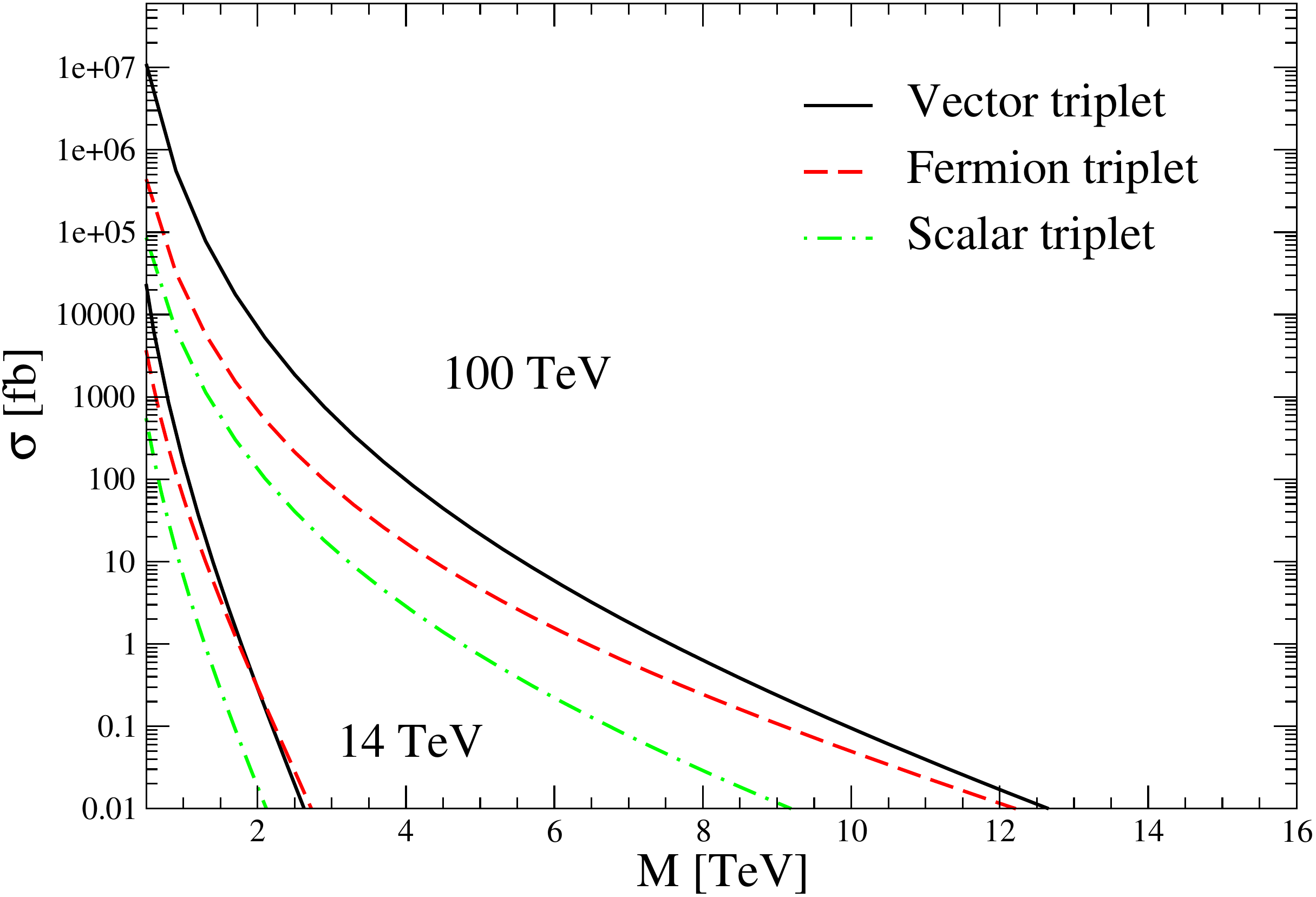}
    \includegraphics[scale=0.25]{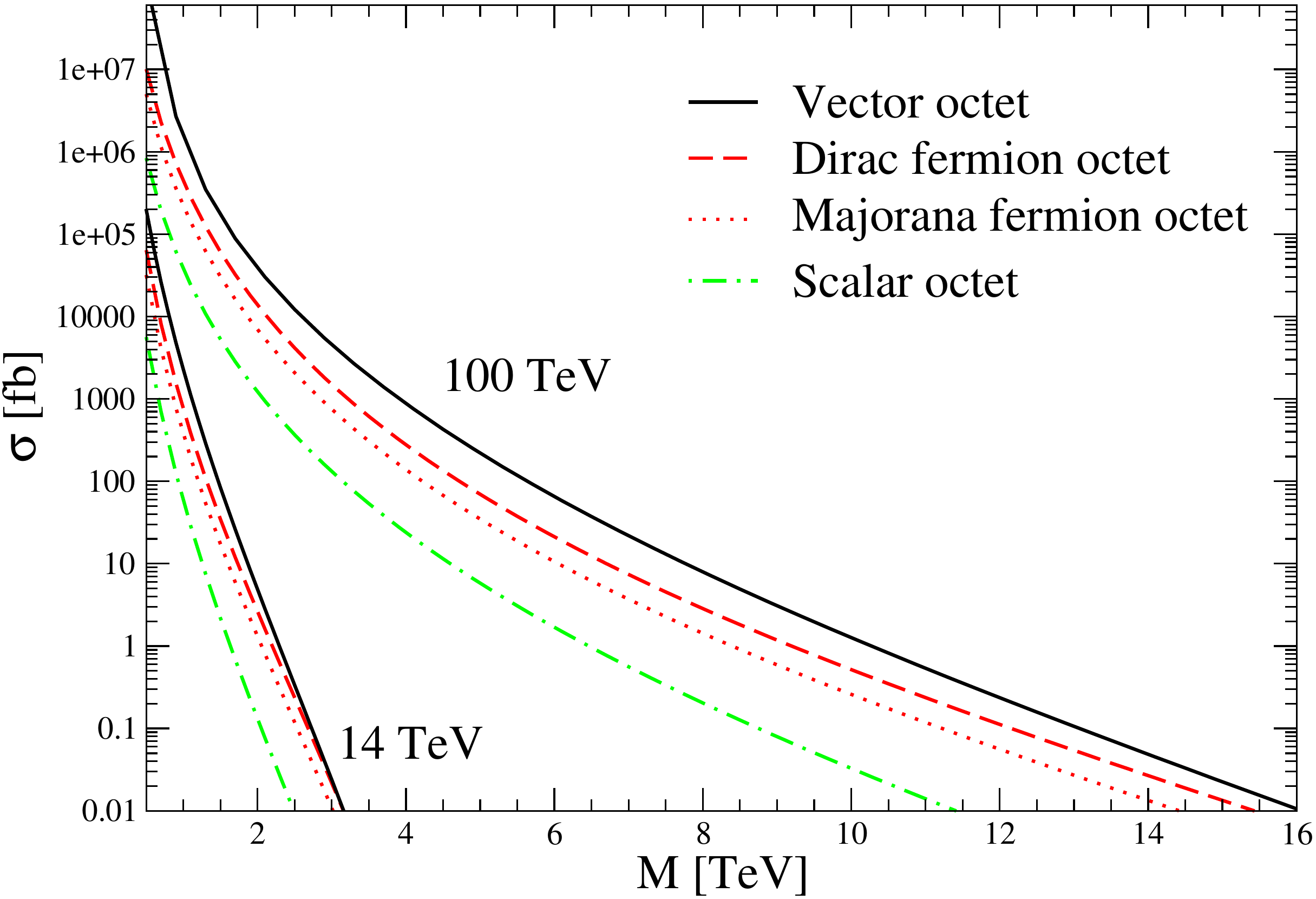}
\end{center}
\caption{Pair production of new states of color-triplets (left panel) and color-octets (right panel) for spin-1 (solid curves), $1/2$ (dashed), and 0 (dot-dashed) at 14 and 100 TeV.}
\label{fig:Tp}
\end{figure}

%%%%%%%%%%%%%%%%%%%%%%%%%%%%%%%%
\subsection{Pair Production of Exotic Color States}
Because of the large top-quark Yukawa coupling, the key ingredient for
 naturalness requires the existence of a top-quark partner at or below the TeV scale, most commonly a scalar partner in SUSY (stop $\tilde t$) or a fermionic partner in composite 
theories ($T'$). They are arguably the ``most wanted'' new particles in a natural theory of the Higgs sector, 
and the discovery sensitivity of them has been presented in an earlier section Sec.~\ref{sec:natural}.
On the other hand, 
there are other possible colored states that may be directly or indirectly associated with the top partners, such a the gluinos, massive gluons, and even spin-1 top partners \cite{Cai:2008ss}. As a QCD machine, a 100 TeV collider would certain open up a new perspective for the discovery of the exotic colored states. 

In Fig.~\ref{fig:Tp}, we show the typical production cross sections at 14 and 100 TeV, for color-triplets \cite{Chen:2012uw} (upper panel) and color-octets \cite{Chen:2014haa} (lower panel), for the possible states of spin-0, $1/2$, and 1. Among those color-triplet states, a scalar is obviously stop-like (dot-dashed lines), a fermion is $T'$-like (dashed lines), and a color-triplet vector can be a spin-1 top-quark partner in an extended gauge theory \cite{Cai:2008ss}. 
Among the color-octets, a spin-0 state could be a colored Higgs or a techni-meson (dot-dashed), a fermion is obvious gluino-like, that could be either a Dirac or Majorana state (two close-by dotted lines), and a color-octet vector can be KK-gluon-like (solid line). 

Once the color and spin quantum numbers are specified, the pair production cross sections are completely determined by the QCD dynamics. As expected, the production rate for the fermionic states is larger than that of the scalar by about a factor of 8, largely due to the spin-state counting and the threshold behavior. For the same reason, the vector states yield even a larger cross section. Because of the color factors, the octet states lead to a higher production cross section than that of the triplets.
Given the substantial production rates as seen in Fig.~\ref{fig:Tp}, as long as the decay channels are not too disfavored for the signal identification, it would be quite conceivable that the mass coverage for those states would be extended to about 5, 10 and 15 TeV, respectively.

%%%%%%%%%%%%%%%%%%%%%%%%%%%%%%%
\subsection{Pair Production of Heavy Leptons}
To complete our overview of new particle production at the future hadron
colliders, we discuss the case of new heavy leptons. While the
existence of a purely sequential SM-like $4^{th}$ generation is
disfavored by existing data on the production of the 125~GeV Higgs
boson, 
there are good motivations to consider new heavy leptons, both neutral
and charged, for example in connection with possible models of 
neutrino mass generation. 
\begin{figure}[h!]
\begin{center}
  \includegraphics[scale=0.4]{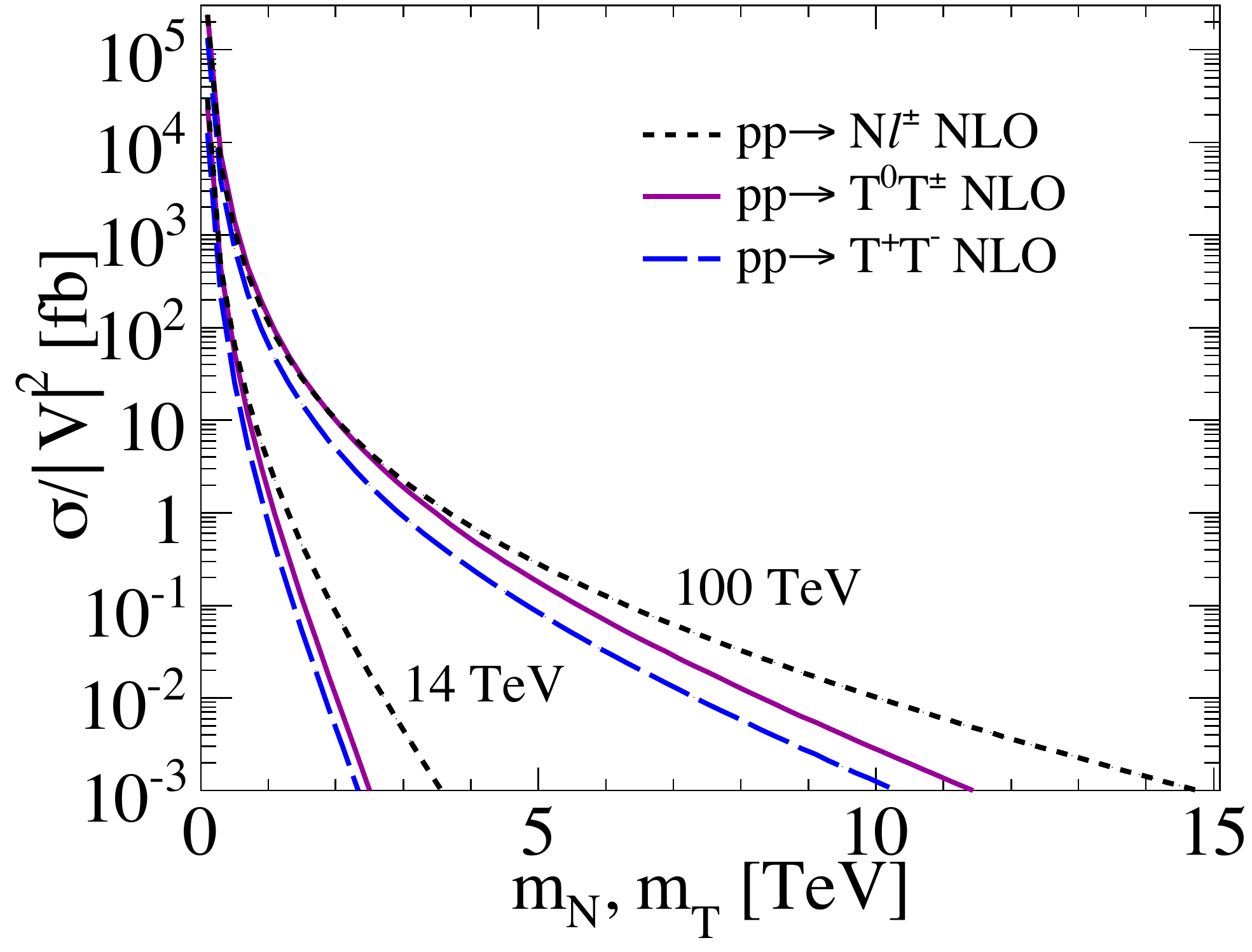}
\end{center}
\caption{Pair production of new heavy leptons at 14 and 100~TeV, for an SU(2) triplet ($T^{\pm,0}$)  and for a singlet state $N \ell^\pm$ via mixing.}
\label{fig:Seesaw}
\end{figure}

We present the production cross sections for heavy lepton pairs at 14 and 100~TeV in Fig.~\ref{fig:Seesaw}. The states $T^\pm, T^0$ form an SU(2) triplet (representative in the Type-III seesaw model \cite{Foot:1988aq}) and the production cross sections,  $q\bar q' \to W^* \to T^0 T^\pm$ (solid curves) and $q\bar q \to \gamma^*, Z^*\to T^+ T^-$ (dashed),
are fully determined by their SU(2) gauge quantum numbers \cite{Arhrib:2009mz,Li:2009mw}, as they should be the same as any other pure SU(2) triplet production. For heavy leptons of a TeV mass, the cross section increase from 14 TeV to 100 TeV may reach a factor $4-5$. 
The triplet components are nearly degenerate in mass. 
Although their mixing with the SM leptons may be small, 
the leading decay channels will still be $T^\pm \to W^\pm\nu, Z\ell^\pm,
h\ell^\pm$ and $T^0 \to W^\pm\ell^\mp, Z\nu, h\nu$, leading to
distinctive and reconstructable final states. It is conceivable that a
100~TeV collider will be able to extend the heavy lepton mass coverage to about
$6-8$ TeV.

For illustration, we have also included the cross section for the production of a heavy neutral lepton $N$ (a heavy neutrino) in association with a SM charged lepton $\ell^\pm$. This production rate, however, is governed by the mixing matrix $V$ between $N$ and the charged leptons $\ell^\pm$. 
This may be a representative for variations of Type-I seesaw models. The dotted curves are for the $N\ell^\pm$ production normalized to $V^2=1$, which would correspond
to the production of an SU(2) fermion doublet. 

\section{Benchmark Standard Model processes}

\newcommand{\metslash}{E\!\!\!\!/_T} 
\def\iab{ab$^{-1}$}
\def\ifb{fb$^{-1}$}
\def\ipb{pb$^{-1}$}
\def \gsim{\mathrel{\vcenter
     {\hbox{$>$}\nointerlineskip\hbox{$\sim$}}}}
\def \lsim{\mathrel{\vcenter
     {\hbox{$<$}\nointerlineskip\hbox{$\sim$}}}}

Standard Model particles play multiple roles in the 100~TeV collider
environment. In the context of BSM phenomena, and for most scenarios,
new BSM particles eventually decay to the lighter SM states, which
therefore provide the signatures for their production. BSM
interactions, furthermore, can influence the production properties of
SM particles, and the observation of SM final states can probe the
existence of an underlying
BSM dynamics. SM processes therefore provide both signatures and
potential backgrounds for any exploration of BSM phenomena. SM
backgrounds have an impact on BSM studies in different ways: on one
side they dilute, and can hide, potential BSM signals; on the other,
SM processes influence the trigger strategies, since they determine
the irreducible contributions to trigger rates and may affect the
ability to record data samples of interest to the BSM searches.

The observation of SM processes has also an interest per se. The huge
rates available at 100~TeV allow, in principle, to push to new limits
the exploration of rare phenomena (e.g. rare decays of top quarks or
Higgs bosons), the precision in the determination of SM parameters,
and the test of possible deviations from SM dynamics. The
extremely high energy kinematical configurations probe the shortest
distances, and provide an independent sensitivity to such deviations. 

Finally, SM processes provide a necessary reference to benchmark the
performance of the detectors, whether in the context of SM
measurements, or in the context of background mitigation for the BSM
searches. 

In this Chapter we review the key properties of SM processes at
100~TeV, having in mind the above considerations. This will serve as a
reference for future studies, and to stimulate new ideas on how to
best exploit the immense potential of this collider. We shall focus on
the production of key SM objects, such as jets, heavy quarks, gauge
bosons and the Higgs boson.  We shall not address issues like the
current or expected precision relative to given processes. On one
side, and with some well understood exceptions notwithstanding,
leading-order calculations are typically sufficient to give a reliable
estimate of the production rates, and assess possible implications for
trigger rates, background contributions, and detector spefications. On
the other, any statement about the precision of theoretical
calculations today will be totally obsolete by the time this collider
will operate, and assumptions about the accuracy reach cannot but be
overly conservative.

\begin{figure}[h!]
\centering
\includegraphics[width=0.5\textwidth]{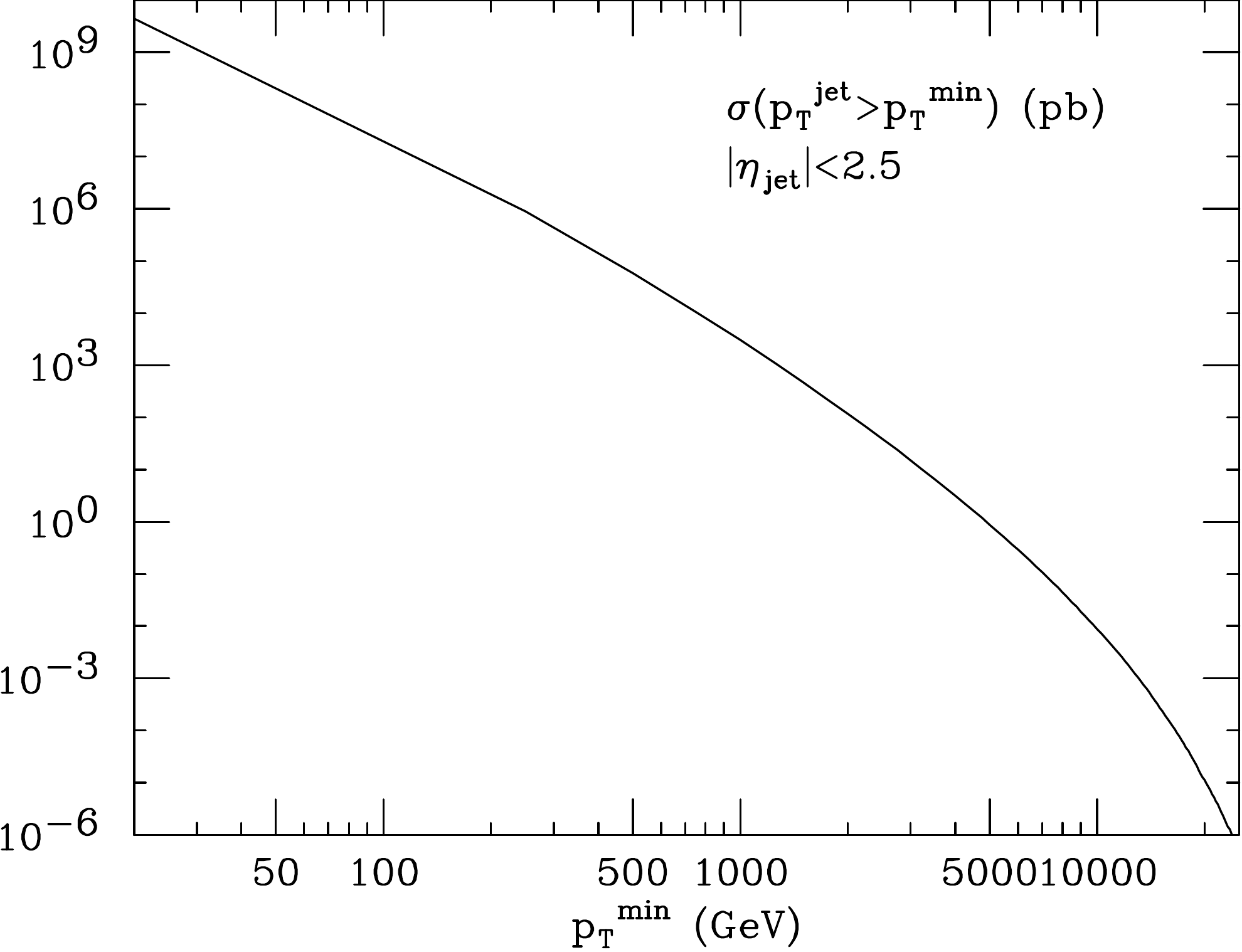}
\caption{Rates of 1-jet inclusive events with $\vert \eta\vert<2.5$ and
  $p_T>p_{T}^{min}$.} 
\label{fig:jetpt}
\end{figure}
\subsection{Jets}
The production of jets is the process that by far dominates, at all
distance scales, the final states emerging from hard collisions among the
proton consituents. 
Figure~\ref{fig:jetpt} shows the integrated rates for the
production of events with at least one jet of transverse momentum
$p_T$ larger than a given
threshold. The distribution refers to jets with pseudorapidity $\eta$
in the range $\vert \eta \vert < 2.5$. Figure~\ref{fig:etaj} shows the
probability that events with jets above certain $p_T$ threshold be
contained inside certain $\eta$ ranges. Notice the huge $\eta$
extension, even for jets with $p_T$ in the TeV range. Assuming
integrated luminosities in excess of 1~\iab, the reach in
$p_T$ extends well above 20~TeV. Fully containing and accurately
measuring these jet energies 
sets important constraints on the design of calorimeters,
e.g. requiring big depth and therefore large transverse size, with a big
impact on the overall dimensions and weight of the detectors. 

\begin{figure}[h!]
\centering
\includegraphics[width=0.8\textwidth]{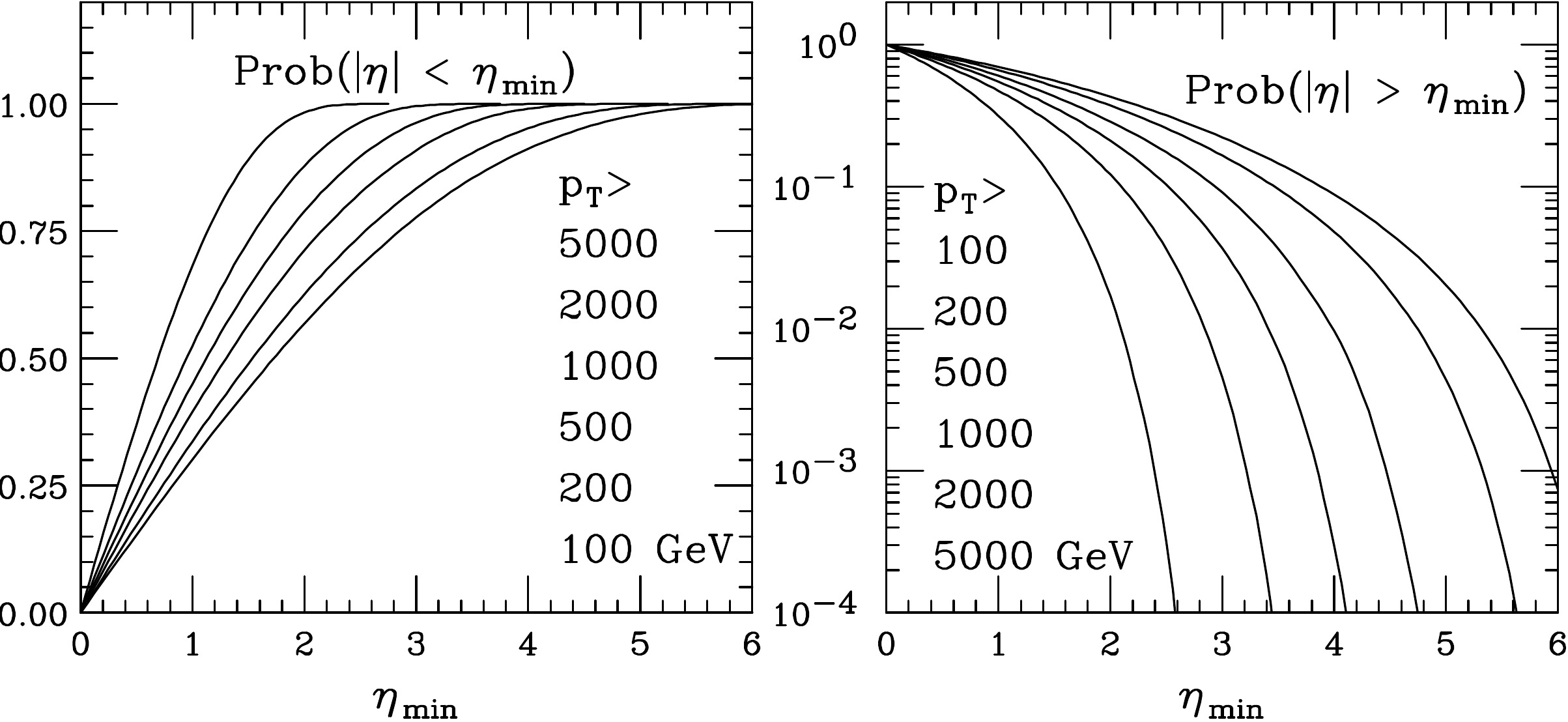}
\caption{Left: acceptance, for jets above various $p_T$ thresholds, to be
  contained within $\vert\eta_j\vert<\eta_{min}$. Right: probability
  to be outside the $\eta_{min}$ acceptance.}
\label{fig:etaj}
\end{figure}
\begin{figure}[h!]
\centering
\includegraphics[width=0.4\textwidth]{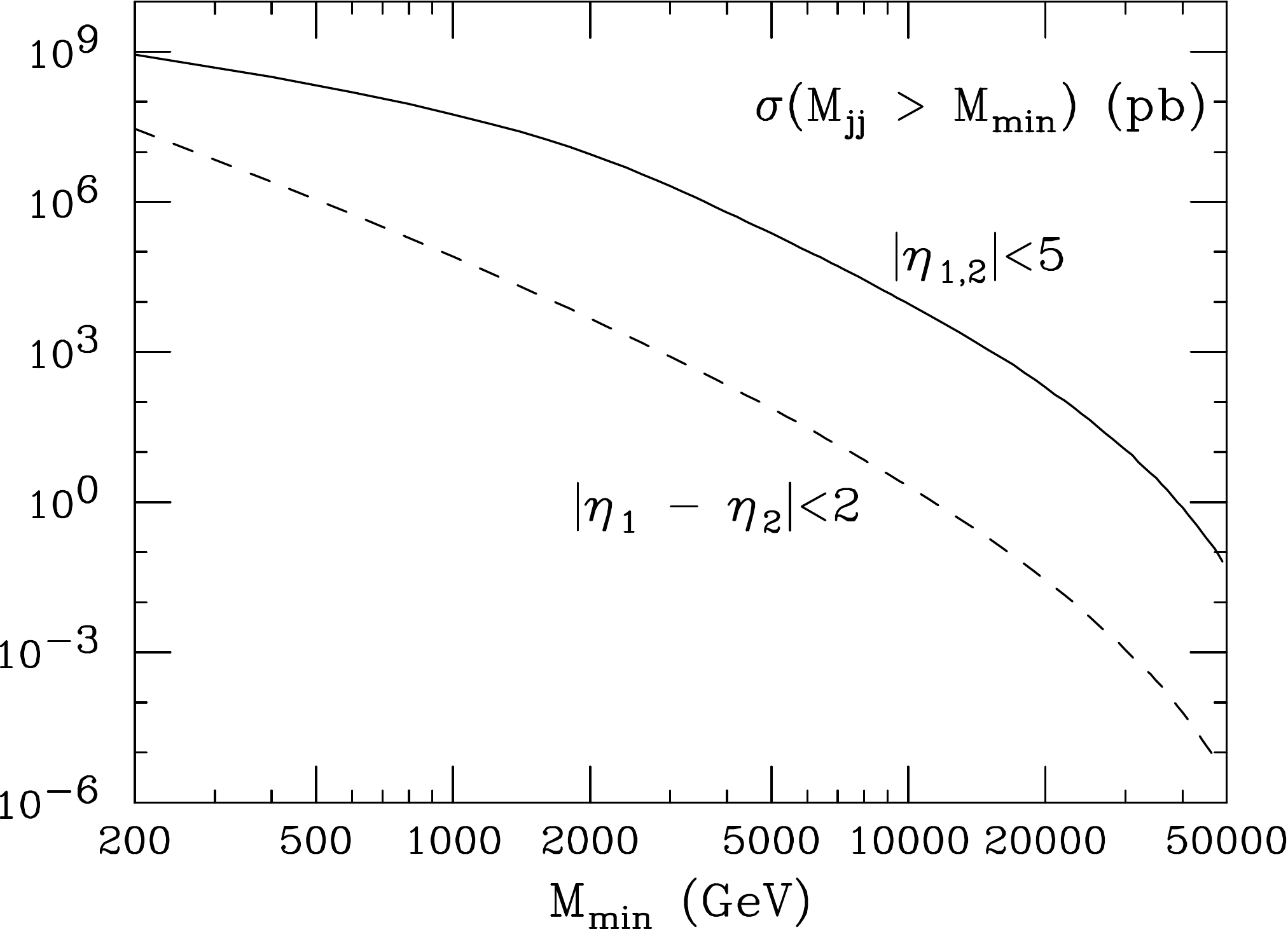} \hfil
\includegraphics[width=0.4\textwidth]{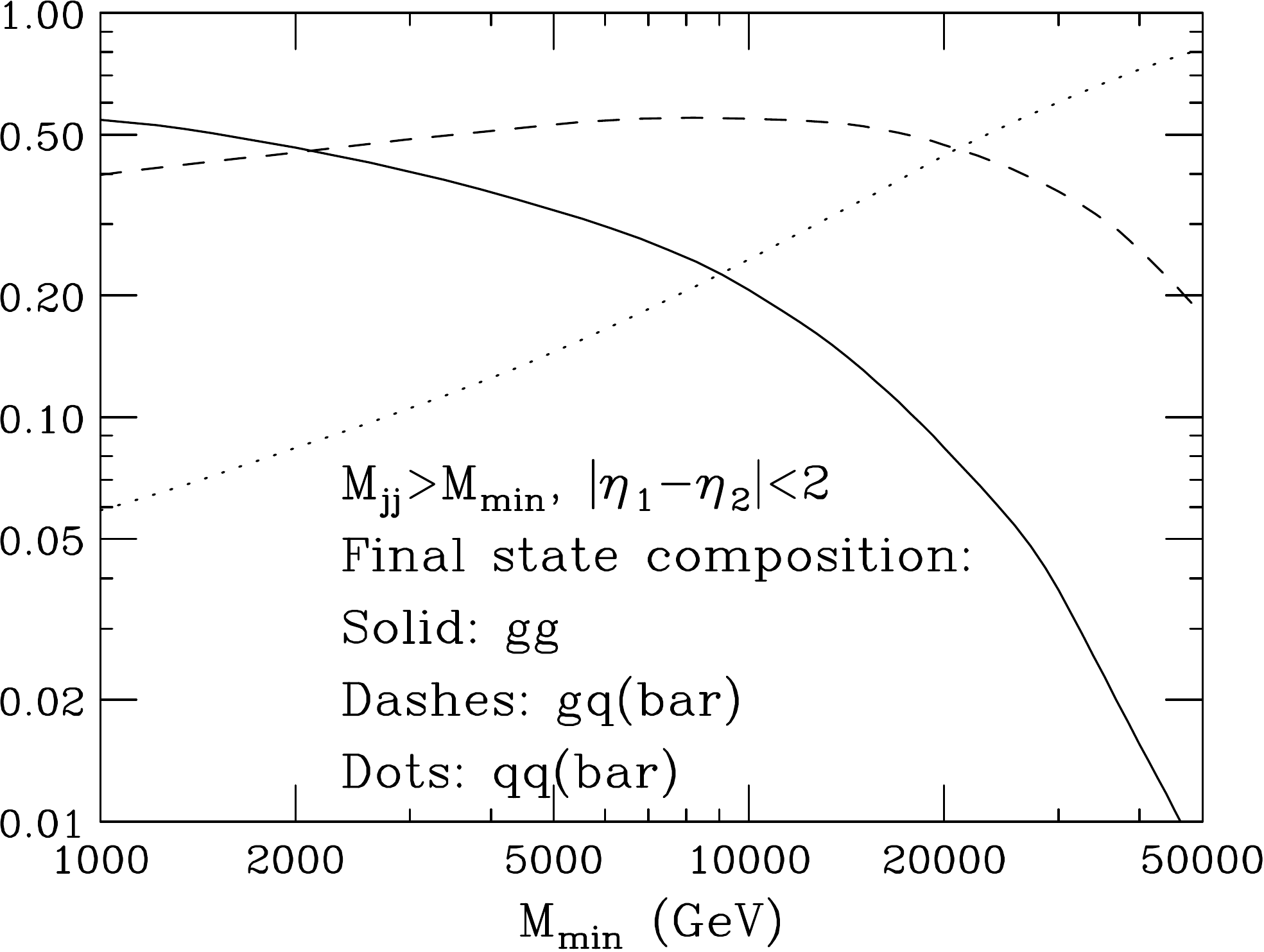}
\caption{Left: dijet mass spectra, for different $\eta$
  constraints. Right: partonic composition of dijet final states, as a
function of the dijet mass. }
\label{fig:mjj100}
\end{figure}

These choices become particularly relevant in the context of searches
for high-mass resonances in dijet final states, where the separation
from the continuum background of possibly narrow states requires good
energy resolution.  Figure~\ref{fig:mjj100} shows the rates for QCD
production of final states with a dijet of invariant mass above a
given threshold. We consider two cases: the dijet mass spectrum of all
pairs with jets within $\vert \eta\vert < 5$, and the spectrum limited
to jets produced at large angle in the dijet center of mass ($\vert
\eta_1 - \eta_2 \vert<2$), a configuration which is more typical of
the production and decay of a possible resonance. Notice that,
particularly at the largest masses, the former rates are several
orders of magnitude larger than the latter ones. This is because one
is dominated there by the low-angle scattering. But even for central
production we have rates in excess of 1 event/\iab\ for masses above
50~TeV. The relative partonic composition of central dijet events, as
a function of the dijet mass, is shown in the right plot of
Fig.~\ref{fig:mjj100}. In the region 2~TeV$ \lsim M_{jj} \lsim$20~TeV
the final states are dominated by $qg$ pairs. Above 20~TeV, we find
mostly $qq$ pairs (the $q\bar{q}$ component is greatly suppressed
throughout).

The ability to tag the nature of the partons that originate the jets
at these energies could be crucial to understand the properties of
possible signals of new physics, such as a decaying resonance with a
multi-TeV mass. Some general features of multi-TeV jets from the QCD
background processes, or from the evolution and hadronic decay of
bottom, top or $W$ bosons, are shown in
Figs.~\ref{fig:multiplicity}-\ref{fig:mass}.

\begin{figure}[h!]
\centering
\includegraphics[width=0.8\textwidth]{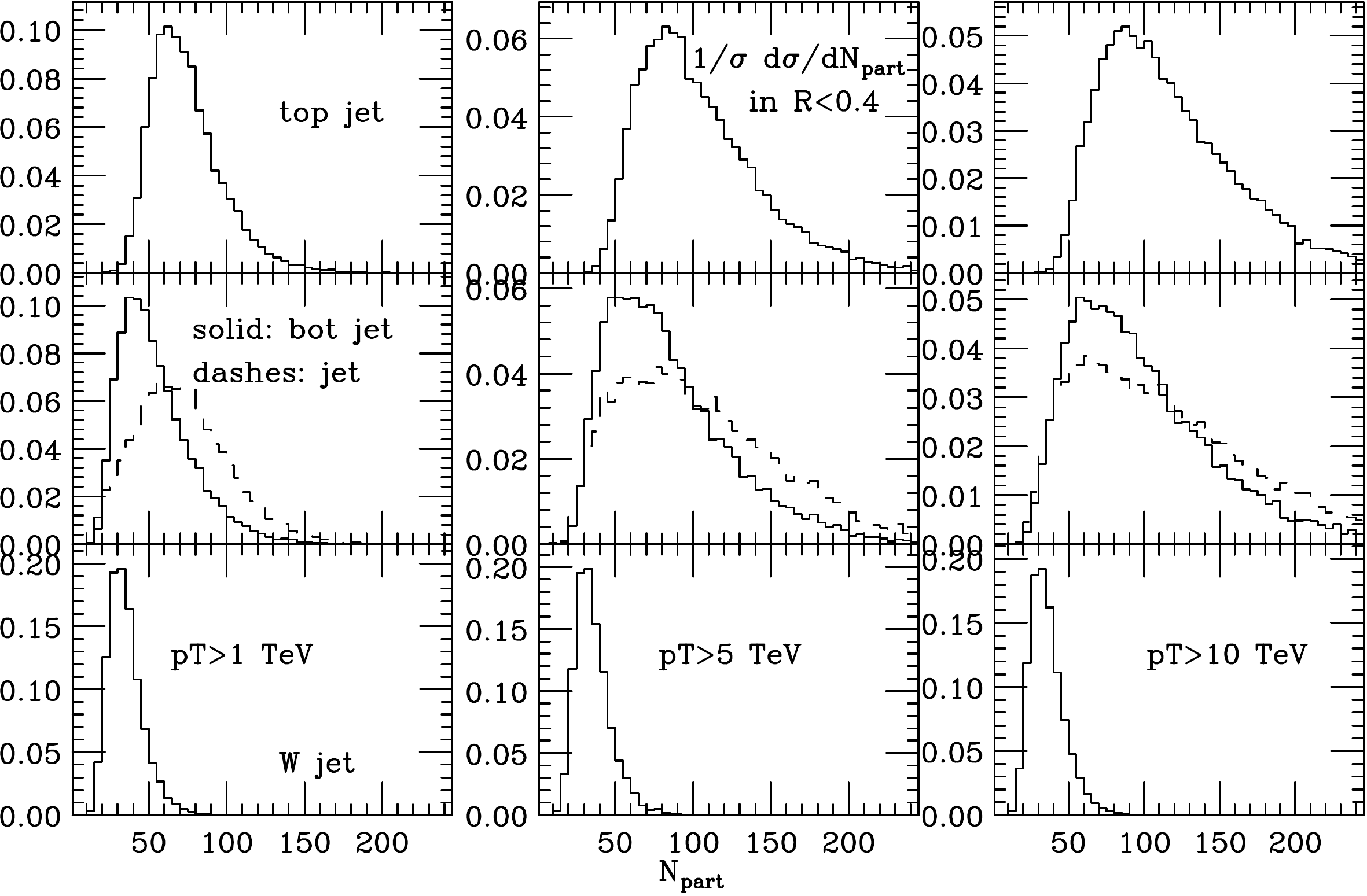}
\caption{Multiplicity distribution in high-$p_T$ jets orginating from
  hadronically decaying top quarks (upper rows), bottom quarks and
  light partons (central rows) and hadronic decays of $W$ bosons
  (lower rows). }
\label{fig:multiplicity}
\end{figure}
\begin{figure}[h!]
\centering
\includegraphics[width=0.8\textwidth]{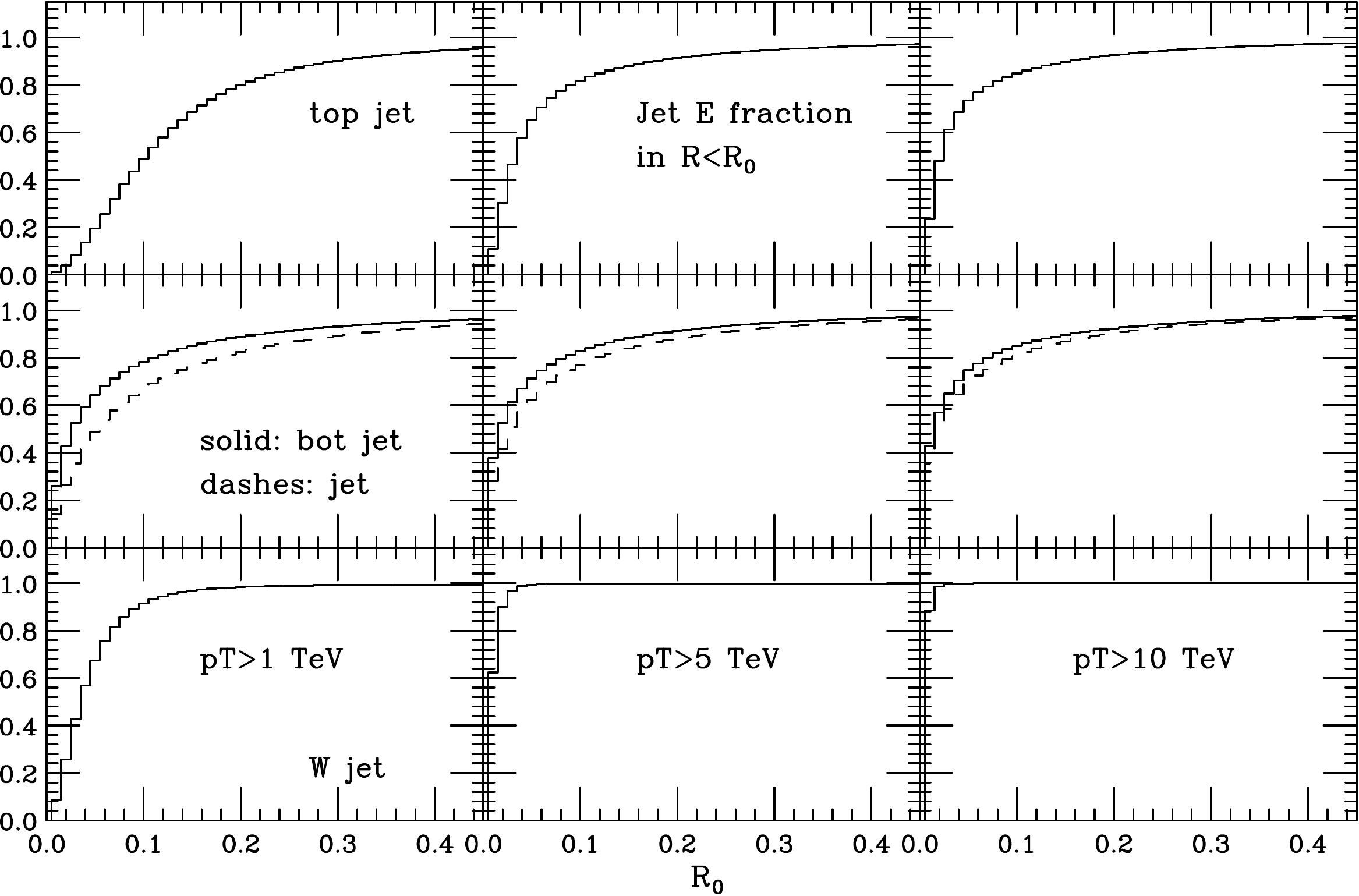}
\caption{Fraction of the total energy for $R=1$ jets, contained within
smaller radii $R_0$. }
\label{fig:Efrac}
\end{figure}
\begin{figure}[h!]
\centering
\includegraphics[width=0.8\textwidth]{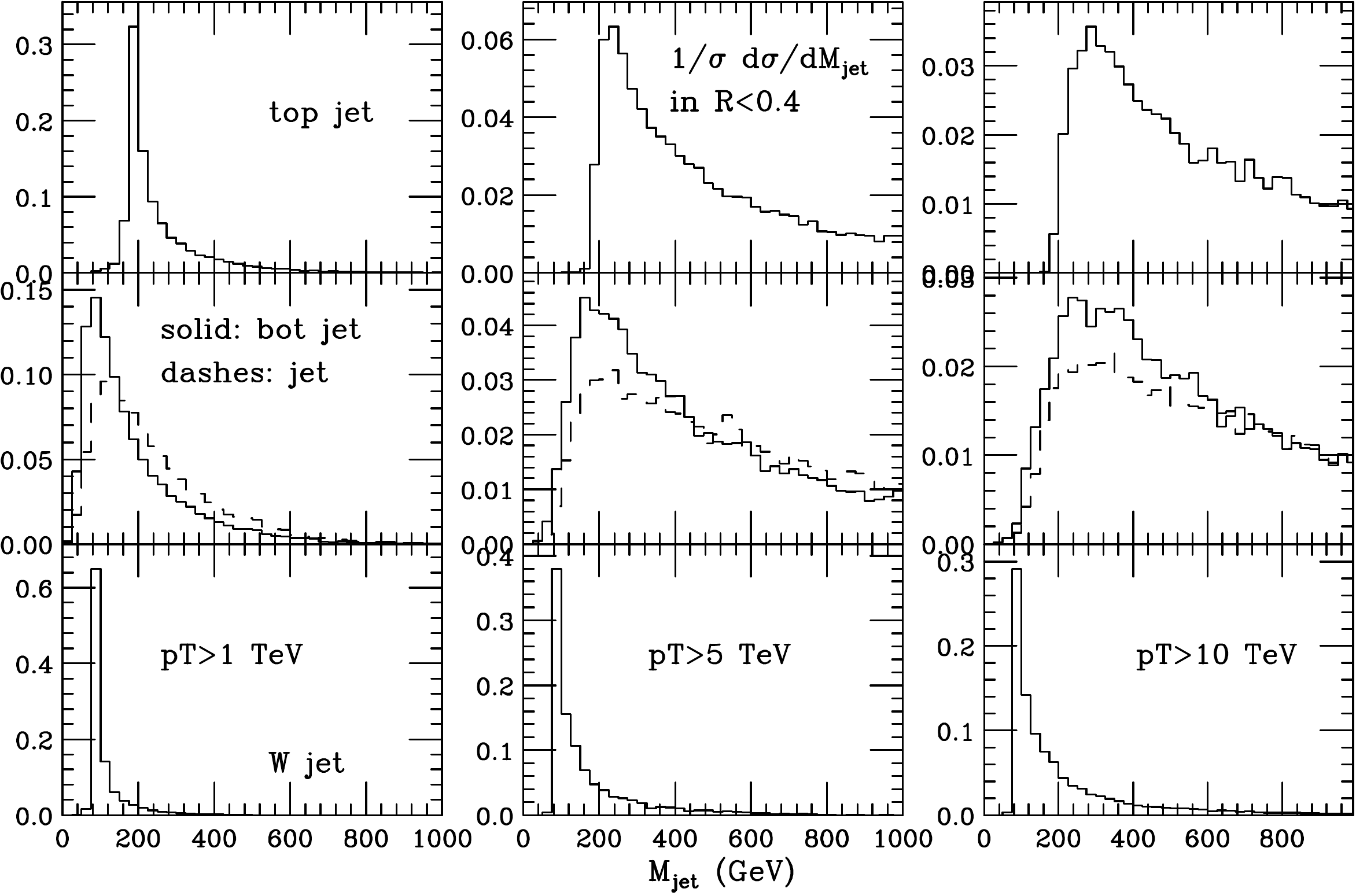}
\caption{Distribution of the jet invariant mass contained with a cone
  of radius $R=0.4$ around the jet axis.}
\label{fig:mass}
\end{figure}

Figure~\ref{fig:multiplicity}
plots the multiplicity distribution of particles (both
charged and neutral, assuming stable $\pi^0$'s) contained within a
cone of radius $R=0.4$ around the jet axis. The three columns of plots
refer to jets of $p_T>1$, 5 and 10~TeV, respectively (jets are defined
here by the anti-$k_T$ algorithm~\cite{Cacciari:2008gp}, 
with a wide cone of $R=1$). The three rows
contain the distributions relative to hadronically-decaying top jets
(upper row), bottom jets and inclusive jets (light quarks and gluons,
according to the QCD-predicted fraction), and hadronically-decaying
$W$ bosons (lower row). For $W$ bosons, the multiplicity is
practically independent of $p_T$, and reflects the multiplicity of a
$W$ decay at rest, with a negligible contamination from initial-state
radiation. The other objects show a clear evolution with $p_T$, and
tend asymptotically to very similar spectra, as expected since at
large $p_T$ the differences induced by the bottom and top masses, and
by the top decay products, are reduced. Notice of course that the
multiplicity of top jets has a sharp onset at about $N_{part}\sim 40$,
because of the presence of the $W$ decay products. Similar features
are observed in the distribution of the jet energy fraction contained
in a subcone of radius $R_0$, shown in Fig.~\ref{fig:Efrac}, and in
the distribution of the jet mass within a cone of $R=0.4$, shown in
Fig.~\ref{fig:mass}. Figure~\ref{fig:Efrac}, in particular, shows that
practically all the energy from a 10~TeV $W$ jet is contained within a cone
of radius $R\lsim 0.02$; this means an average of 40 particles all
inside this tiny radius, making their individual reconstruction
experimentally very challenging.  Efforts are ongoing to exploit
the small differences observed in distributions such as those shown
here, in order to statistically separate with good efficiency objects
such as top quarks of gauge bosons from each other, and from lights
jets. Such techniques, developed for the $p_T\sim$TeV range of relevance to
LHC physics, are being extended to the more challenging multi-TeV
regime relevant to the future physics of a 100~TeV collider. See for
example the study in Ref.~\cite{Larkoski:2015yqa}, dedicated to top quarks.  

\subsection{$W/Z$ Production}
The production of $W$ and $Z$ bosons is a valuable probe of both EW
and QCD dynamics. The production properties are known today up to
next-to-next-to-leading order (NNLO) in QCD, leading to a precision of
the order of the percent. A detailed discussion of the implications of
this precision, and of the possible measurements possible with $W$ and
$Z$ final states at 100~TeV, is outside the scope of this review, also
because the LHC has only started exploiting the full
potential of what can be done with them. 
We shall therefore focus here on documenting some basic
distributions, to show the extreme kinematical configurations that may
be accessed at 100~TeV, and to highlight some of the novel features of
EW interactions that will emerge at these energies.

\begin{figure}[h!]
\centering
\includegraphics[width=0.8\textwidth]{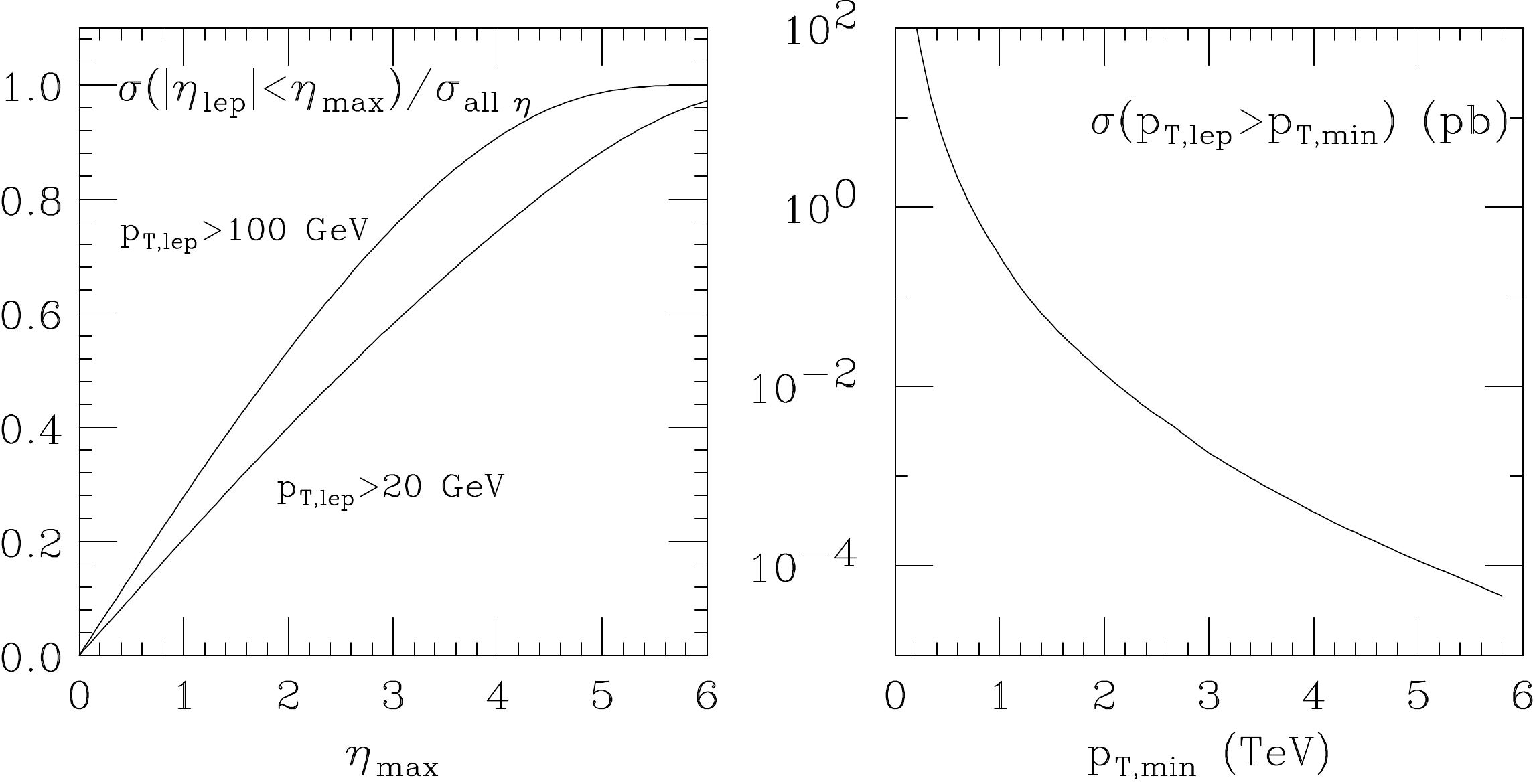}
\caption{Left: rapidity acceptance for leptons from inclusive $W$
  production and decay, for $p_T$ thresholds of 20 and 100~GeV. Right:
inclusive lepton $p_T$ spectrum.}
\label{fig:wleppt}
\end{figure}

The total production rate of $W^\pm$ ($Z^0$) bosons at 100~TeV is about
1.3 (0.4)~$\mu$b. This corresponds to samples of $O(10^{11})$ leptonic ($e,\mu$)
decays per \iab. At 100~TeV, gauge bosons will have a rather broad
rapidity distribution and, as shown in the left plot of
Fig.~\ref{fig:wleppt}, more than 50\% of the leptons with $p_{T}>20$~GeV will be
produced at $\vert \eta \vert>2.5$ (w.r.t. $\sim 30\%$ at
14~TeV). Even leptons with $p_{T}>100$~GeV will have a large forward
rate, with about 40\% of them at $\vert \eta \vert>2.5$ ($\sim 10\%$ at
14~TeV). Their $p_T$ spectrum will also extend to large values, as
shown in the right plot of Fig.~\ref{fig:wleppt}. The largest fraction
of these high-$p_T$ leptons will arise from $W$'s produced at large
$p_T$, in association with jets.

\subsubsection{Multiple gauge boson production}
\begin{table}
\begin{center}
\def\arraystretch{1.5}
\begin{tabular}{ l | c | c | c | c }
Proc &  $WWW$ & $WWZ$ & $WZZ$ & $ZZZ$ 
\\   \hline
$\sigma$(fb) & $4.3\times 10^3$ & $4.0\times10^3$ & $1.4\times10^3$ &
$2.6\times 10^2$ 
\\
\end{tabular}
\\[0.5cm]
\def\arraystretch{1.5}
\begin{tabular}{ l | c | c | c | c | c }
Proc &  $WWWW$ & $WWWZ$ & $WWZZ$ & $WZZZ$ & $ZZZZ$ 
\\   \hline
$\sigma$(fb) & 41 & 60  & 33 & 7.1 & 0.8
\\
\end{tabular}
\end{center}
\caption{NLO cross sections for production of multiple gauge bosons, 
at 100 TeV~\cite{Torrielli:2014rqa}.}
\label{tab:multiVB}
\end{table}
Table~\ref{tab:multiVB} shows the rates of associated production of
multiple gauge bosons from full NLO calculations \cite{Alwall:2014hca,Torrielli:2014rqa}. Production of each additional EW gauge boson brings the cross section down roughly by the order of $\alpha$, as naively expected in perturbation theory.
Even including the branching ratios for the best visible leptonic decays,
the rates are sufficient in principle to observe the production of up
to four gauge bosons. This will lead to unprecendeted precision in the
measurement of anomalous triple gauge couplings, and to the detection
of quartic couplings, furthermore providing a probe of anomalous
higher-dimension operators involving multiple gauge bosons.

\subsubsection{FSR effects of the gauge bosons and initial state partons}
\label{sec:WZatHighE}
The left plot in Fig.~\ref{fig:wpt} shows the integrated $p_T$
spectrum of $W$ bosons.\footnote{This calculation only includes the QCD
  effects. For $p_T$ beyond the TeV scale, the effects of virtual EW
  corrections are known to lead to important
  corrections~\cite{Denner:2011vu}.}  With luminosities in excess of
1~\iab, data will extend well beyond 15~TeV. For processes involving
gauge bosons and jets at such large energies, however, a very
interesting new phenomenon emerges, namely the growth of the gauge
boson emission probability from high-$p_T$ jets. If we ask what is the
most likely mechanism to produce gauge bosons in final states with at
least one multi-TeV jet, it turns out that this is not the LO QCD
process where the gauge boson simply recoils against the jet, but the
higher-order process where it is a second jet that absorbs the leading
jet recoil, and the gauge boson is radiated off some of the
quarks~\cite{Rubin:2010xp}, the effect of ``final state radiation'' (FSR).  In
other words, the parton-level scattering $q q \to qq V$ dominates over
$qg \to qV$ (for simplicity, we do not show explicitly the possibly
different quark flavour types involved in the processes). The emission
probability of gauge bosons in this case is enhanced by large
logarithms of $p_{T,jet}/M_V$, and can reach values in the range of
10\% and more, as shown in the right plot of Fig.~\ref{fig:wpt}. This
gives the emission probability for one or more $W$ bosons in events in
which there is at least one jet above a given $p_T$ threshold. The
kinematical properties of these events are illustrated in
Fig.~\ref{fig:WiniHighEtjet},  in the case of final states with a jet above
1~TeV, and above 10~TeV, to highlight the kinematical evolution with
jet $p_T$. In the case of largest $p_T$, we see the dominance of
events in which the two jets balance each other in transverse
momentum, while the $W$ carries a very small fraction of the leading
jet momentum. One third of the $W$'s are emitted within $\Delta R<1$
from the subleading jet, with a large tail of emission at larger
angles, due in part to $W$ radiation from the initial state.  

\begin{figure}[h!]
\centering
\includegraphics[width=0.46\textwidth]{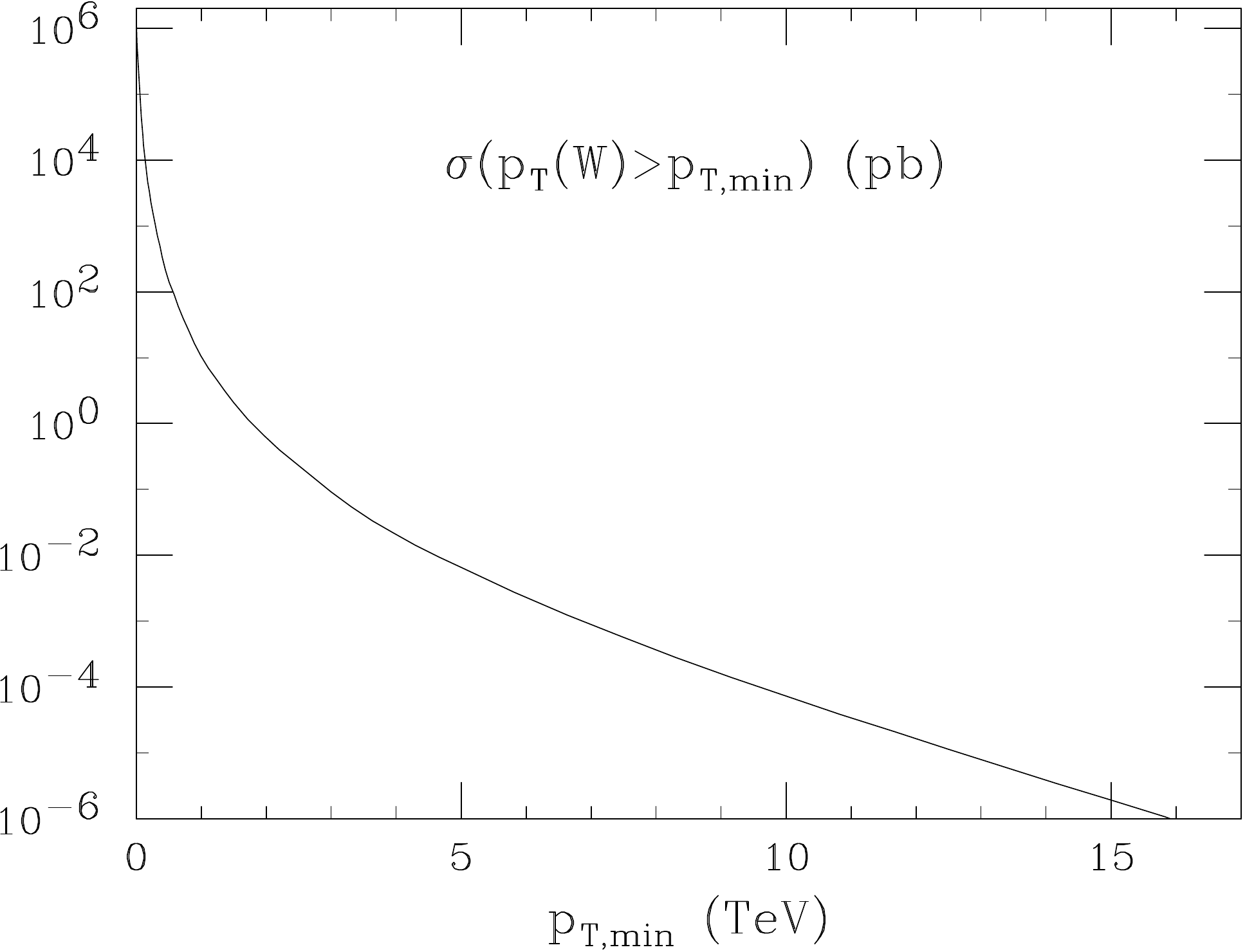}
\hfill
\includegraphics[width=0.45\textwidth]{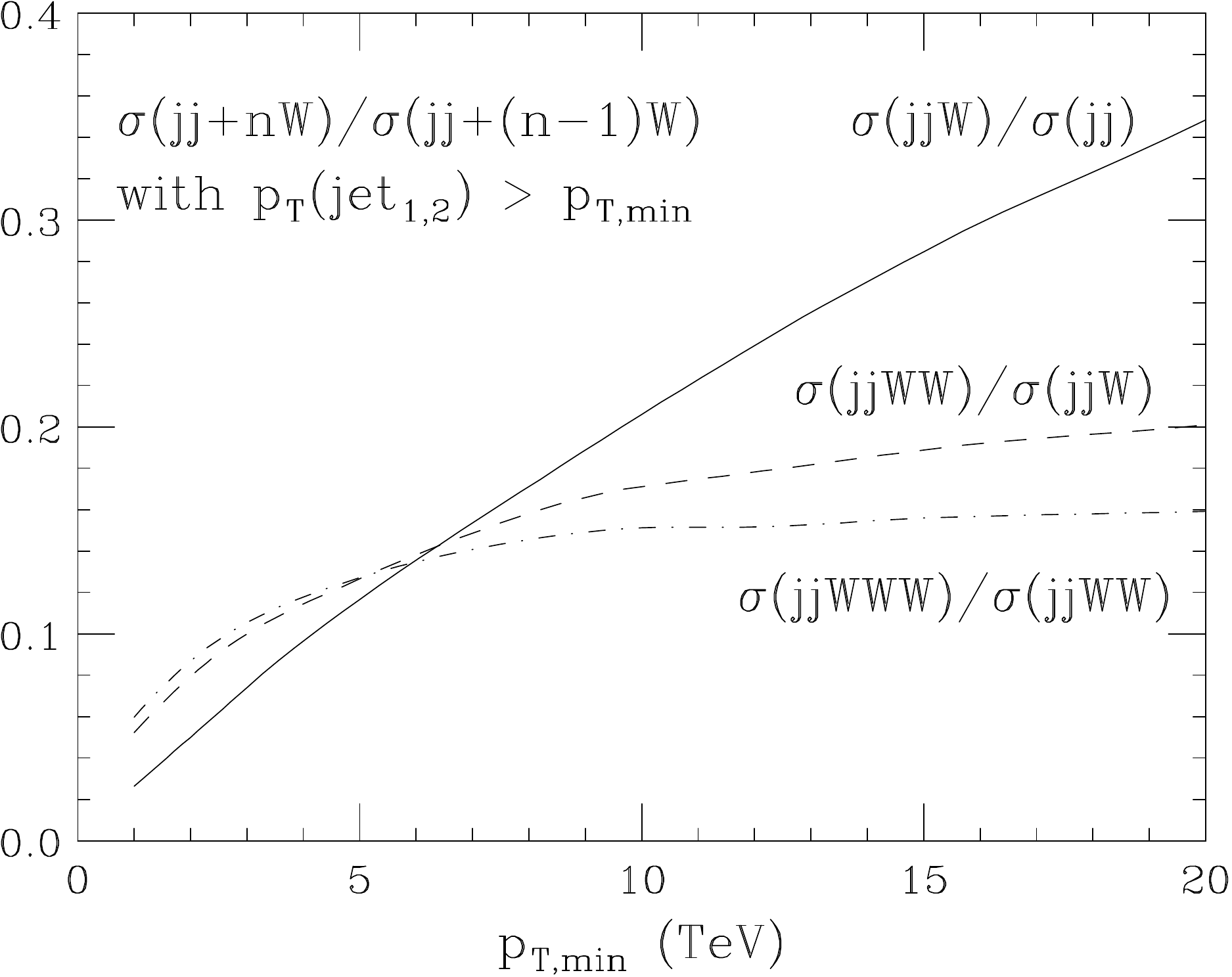}
\caption{Left: inclusive $p_T$ spectrum of $W$ bosons. Right: emission
probability for additional $W$ bosons in dijet events at large $p_T$. }
\label{fig:wpt}
\end{figure}

\begin{figure}[h!]
\centering
\includegraphics[width=0.9\textwidth]{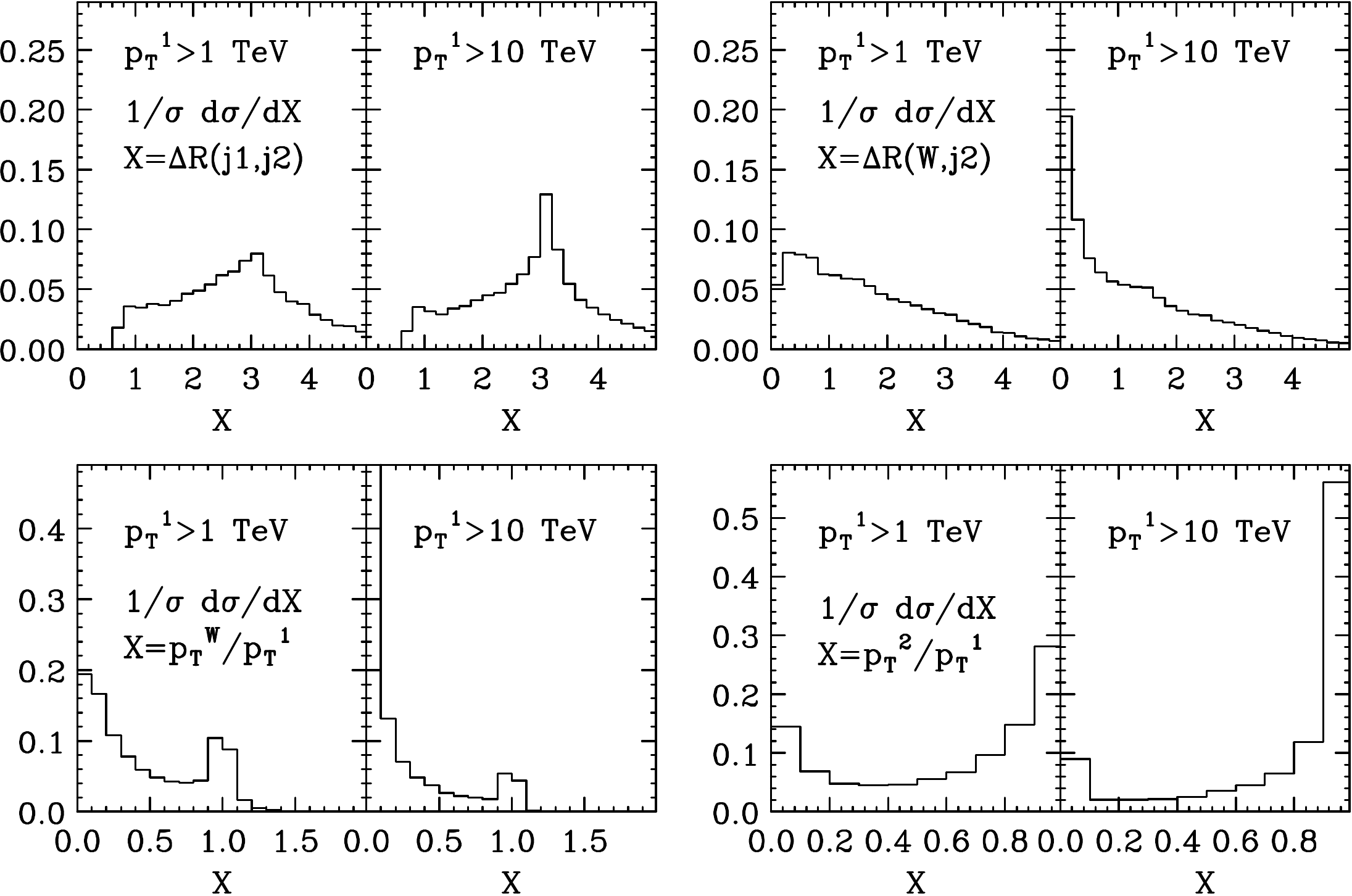}
\caption{Kinematical correlations in high-$p_T$ jet events with $W$
  radiation, for values of the leading jet $p_T>1$ and 10~TeV.}
\label{fig:WiniHighEtjet}
\end{figure}

The process considered above is just one manifestation of the general
fact that, in hard electroweak interactions at multi-TeV energies, the
soft/collinear structure of almost {\it any} multi-TeV process can
become significantly altered, as the logarithmic enhancements familiar
from QED and QCD will become active for electroweak emissions (see,
e.g.,~\cite{Moretti:2006ea,Bell:2010gi,Dittmaier:2012kx,Christiansen:2014kba,Hook:2014rka}).
Obtaining correct descriptions of the complete event structure when
$\sqrt{\hat{s}} \gg M_W$ can be then greatly facilitated by
incorporating factorization and resummation, such as that provided by
parton showering and parton distribution functions.  In effect, we
will begin to see weak bosons (including the Higgs boson) behaving as
nearly-massless partons, in stark contrast to the conventional
perspective in which they are viewed as ``heavy'' particles.  Jets,
whether initiated by QCD processes, electroweak process, or new
physics processes, will be found to contain electroweak splittings
with probabilities at the $O(10\%)$ level.  Similarly, weak bosons can
usefully be thought of as collinear components of the protons, at the
same level as gluons and photons.

To develop some intuition of the collinear splitting behavior of
electroweak ``partons,'' it is useful to first consider a conceptual
limit with an unbroken $SU(2) \times U(1)$ gauge symmetry with
massless gauge bosons and fermions, supplemented by a massless scalar
doublet field $\phi$ without a VEV (the would-be Higgs doublet).  In
this limit, many processes are direct analogs of those in QED and QCD.
Fermions with appropriate quantum numbers may emit (transverse)
$SU(2)$ and $U(1)$ gauge bosons with both soft and collinear
enhancements.  The $SU(2)$ bosons couple to one another via their
non-abelian gauge interactions, and undergo soft/collinear splittings
of the schematic form $W\to WW$, similar to $g\to gg$.  All of the
electroweak gauge bosons may also undergo collinear-enhanced
splittings into fermion pairs, similar to $g\to q\bar q$ or $\gamma
\to f \bar f$.  Beyond these, the major novelty is the introduction of
the scalar degrees of freedom.  First, the scalars may themselves
radiate $SU(2)$ and $U(1)$ gauge bosons, with soft/collinear limits
identical to their counterparts with fermionic sources.  Second, the
electroweak gauge bosons can split into a pair of scalars, again in
close analog with splittings to fermion pairs.  Third, fermions with
appreciable Yukawa couplings to the scalar doublet can emit a scalar
and undergo a chirality flip.  Finally, the scalars can split into
collinear fermion pairs.

In the realistic case of spontaneously-broken symmetry, several
important changes take place.  Primarily, all of the soft and
collinear divergences associated with the above splittings become
physically regulated, effectively shutting off at $p_T \lsim M_W$ (or
$m_H$, $m_t$ where appropriate).  Roughly speaking, $M_W$ plays a role
similar to $\Lambda_{\rm QCD}$ in the QCD parton shower, albeit with
far less ambiguity of the detailed IR structure since this regulation
occurs at weak coupling.  Another major difference is the mixing of
the scalar doublet's Goldstone degrees of freedom into the $W$ and $Z$
gauge bosons, allowing for the appearance of longitudinal modes.  In
many cases, the longitudinal gauge bosons behave identically to the
original scalars, as dictated by the Goldstone equivalence
theorem~\cite{Lee:1977eg,Chanowitz:1985hj}.  For example the splitting
$W_T^+ \to W_{L}^+ Z_{L}$ is, up to finite mass effects, an exact
analog of $W_T^+ \to \phi^+ {\rm Im}(\phi^0)$ in the unbroken theory.
Similarly for longitudinal gauge boson emissions from heavy fermions,
such as the equivalence between $t_L \to Z_{L} t_R$ and $t_L \to {\rm
  Im}(\phi^0) t_R$.

But important exceptional cases now also occur for emissions near $p_T
\sim M_W$.  Most well known, even a massless fermion exhibits a kind
of soft/collinear-enhanced emission of $W_{L}$ and
$Z_{L}$~\cite{Kane:1984bb,Dawson:1984gx}.  These emissions have no
Goldstone equivalent analog, and are highly power-suppressed for $p_T
\gsim M_W$.  But the overall population of emissions at the boundary
between ``broken'' and ``unbroken'' behavior nonetheless grows
logarithmically with the fermion energy.  This is formally subdominant
to the double-logarithmic growth of transverse emissions, but remains
numerically important at multi-TeV energy scales.  Emissions from
massless quarks also cause the energetic initial-state protons to act
as sources of longitudinal boson beams, allowing for studies of the
high-energy interactions of the effective Goldstone bosons through
weak boson scattering (discussed further below).  Similar types of
emissions occur in the splittings of transverse bosons, such as $W_T^+
\to Z_{L} W_T^+ \, / \, Z_T W_{L}^+$.

\begin{table}
\begin{center}
\begin{tabular}{ l | c | c | c }
Process \ &  ${\mathcal P}(p_T)$  & \ ${\mathcal P}(1~{\rm TeV})$ \  & \ ${\mathcal P}(10~{\rm TeV})$ \  \\   \hline\hline
$f \to V_Tf$ \ &  $(5\times10^{-3}) \log^2\frac{p_T}{m_{\rm EW}}$ &  3\% &  12\%    \\
$f \to V_{L}f$ \ &  $ (2\times10^{-3})\log\frac{p_T}{m_{\rm EW}}$ &  0.6\% &  1\%   \\ \hline
$V_T \to V_T V_T$ \ & $(0.01) \log^2\frac{p_T}{m_{\rm EW}}$  &  6\% &  25\%   \\
$V_T \to V_{L}V_T$ \ & $(0.01)\log\frac{p_T}{m_{\rm EW}}$  &  2\% &  4\%   \\
$V_T \to f\bar f$ \ & $(0.01)\log\frac{p_T}{m_{\rm EW}}$  &  2\% &  4\%   \\
$V_T \to V_{L} h$ \ & $(2\times10^{-3})\log\frac{p_T}{m_{\rm EW}}$  &  0.6\% &  1\%   \\ \hline
$V_{L} \to V_T h$ \ & $(5\times10^{-3}) \log^2\frac{p_T}{m_{\rm EW}} $  &  3\% &  12\%   \\
$V_{L} \to V_L h$ \ & $(2\times10^{-3}) \log\frac{p_T}{m_{\rm EW}} $  &  0.6\% &  1\%   \\
\hline\hline
\end{tabular}
\end{center}
\caption{An illustrative set of approximate total electroweak
  splitting rates in final-state showers at two representative energies \cite{EWshower}.}
\label{table:splittingRates}
\end{table}

Table~\ref{table:splittingRates} provides a few estimates for total
splitting rates of individual final-state particles \cite{EWshower}, including
approximate numerical values for particles produced at $p_T = 1$~TeV
and 10~TeV.  The $SU(2)$ self-interactions amongst transverse gauge
bosons tend to give the largest rates, quickly exceeding 10\% as the
energy is raised above 1~TeV (these rates are slightly lower than
those extracted from Fig.~\ref{fig:wpt}, since there an important
contribution to $W$ emission came from initial state radiation).  This has significant impact on
processes with prompt transverse boson production such as
$W/Z/\gamma$+jets, and especially on multiboson production including
transverse boson scattering.  Generally, it is important to appreciate
that {\it any} particle in an event, whether initial-state or
final-state, or even itself produced inside of a parton shower, can
act as a potential electroweak radiator.  Consequently, the total rate
for finding one or more electroweak splittings within a given event
must be compounded, and can sometimes add up to $O(1)$.

\begin{figure*}[tp!]
\begin{center}
\includegraphics[width=0.6\textwidth]{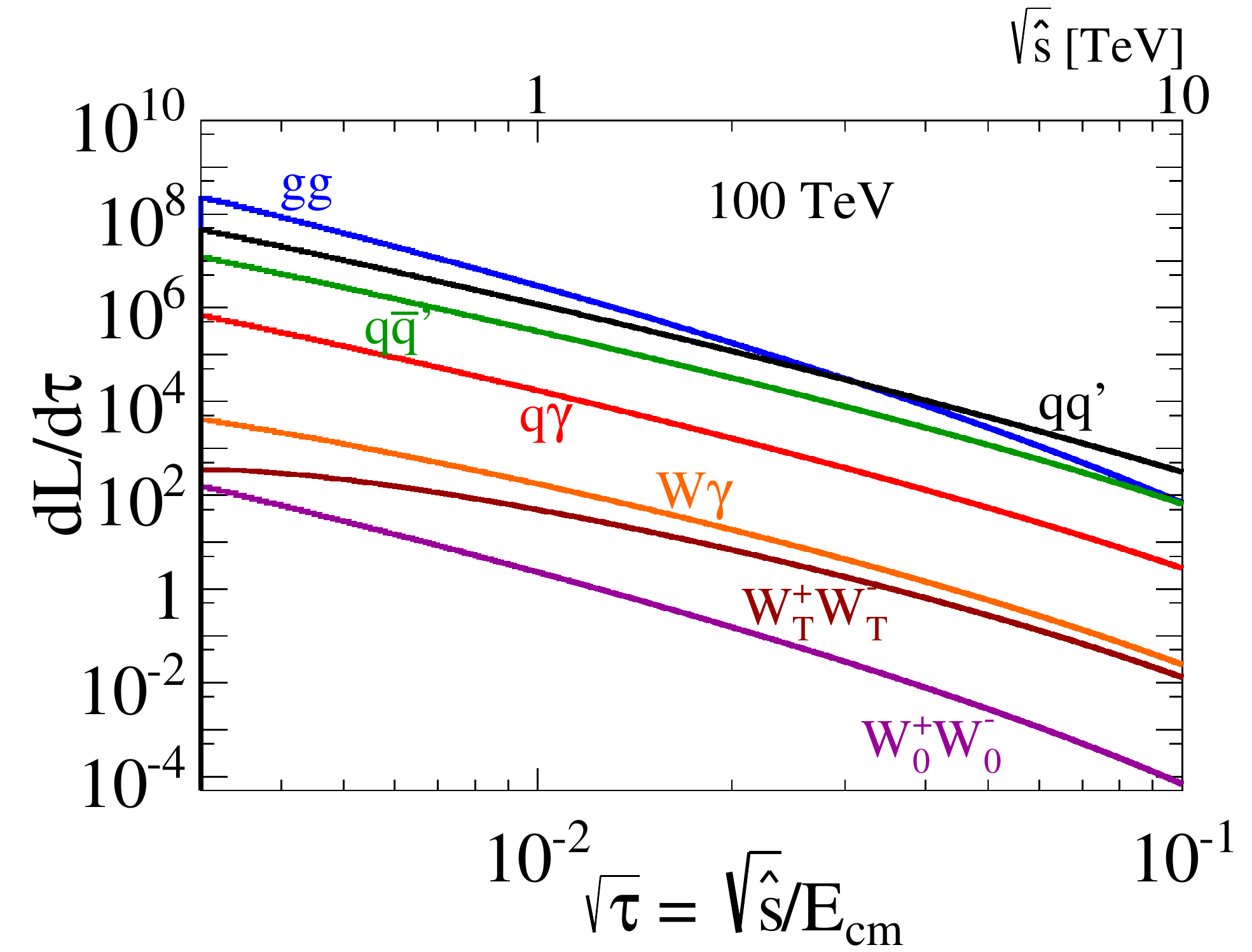}
\caption{Partonic luminosities at 100~TeV, illustrating the
  relative contributions from weak bosons when treated as partons in
  the proton \cite{EWshower}.}
\label{fig:ewPartonLumi100TeV}
\end{center}
\end{figure*}

In this regards, it would be interesting, both conceptually and technically, to consider the electroweak bosons as patrons in high-energy collisions. 
Fig.~\ref{fig:ewPartonLumi100TeV} summarizes the parton luminosities
when electroweak bosons are included in the PDFs.  One immediate
observation from comparing the $W_T\gamma$ and $W_TW_T$ luminosities
is that transverse weak bosons begin to appear on the same footing as
photons, as might have been anticipated.  Ultimately, they must be
folded into the full DGLAP evolution, though at 100~TeV energies the
running effects are not yet sizable.  The longitudinal bosons are
sourced from the quarks as described above at $p_T \sim M_W$, with
individual splitting rates $O$(3--10) times smaller than their
transverse counterparts at multi-TeV energies.  This leads to
$O$(10--100) times smaller luminosities.  For VBF process initiated by
the longitudinal bosons, the PDF approach effectively integrates out
the usual forward tagging jets, treating them as part of the ``beam.''
This of course becomes a progressively more justifiable approach, as
these jets with $p_T \sim M_W$ will appear at extremely high
rapidities, and may anyway become a less distinctive feature to
discriminate against backgrounds in the presence of copious QCD
initial-state radiation at similar $p_T$.  From a practical
perspective, the ability to treat VBF as a $2\to 2$ process rather
than $2\to 4$ would significantly reduce the computational burden for
event simulation.  The tagging jets can then be resolved using the
usual initial-state radiation machinery, appropriately adapted for
this unique electroweak splitting process.

 \begin{figure}[h!]
\centering
\includegraphics[width=0.44\linewidth]{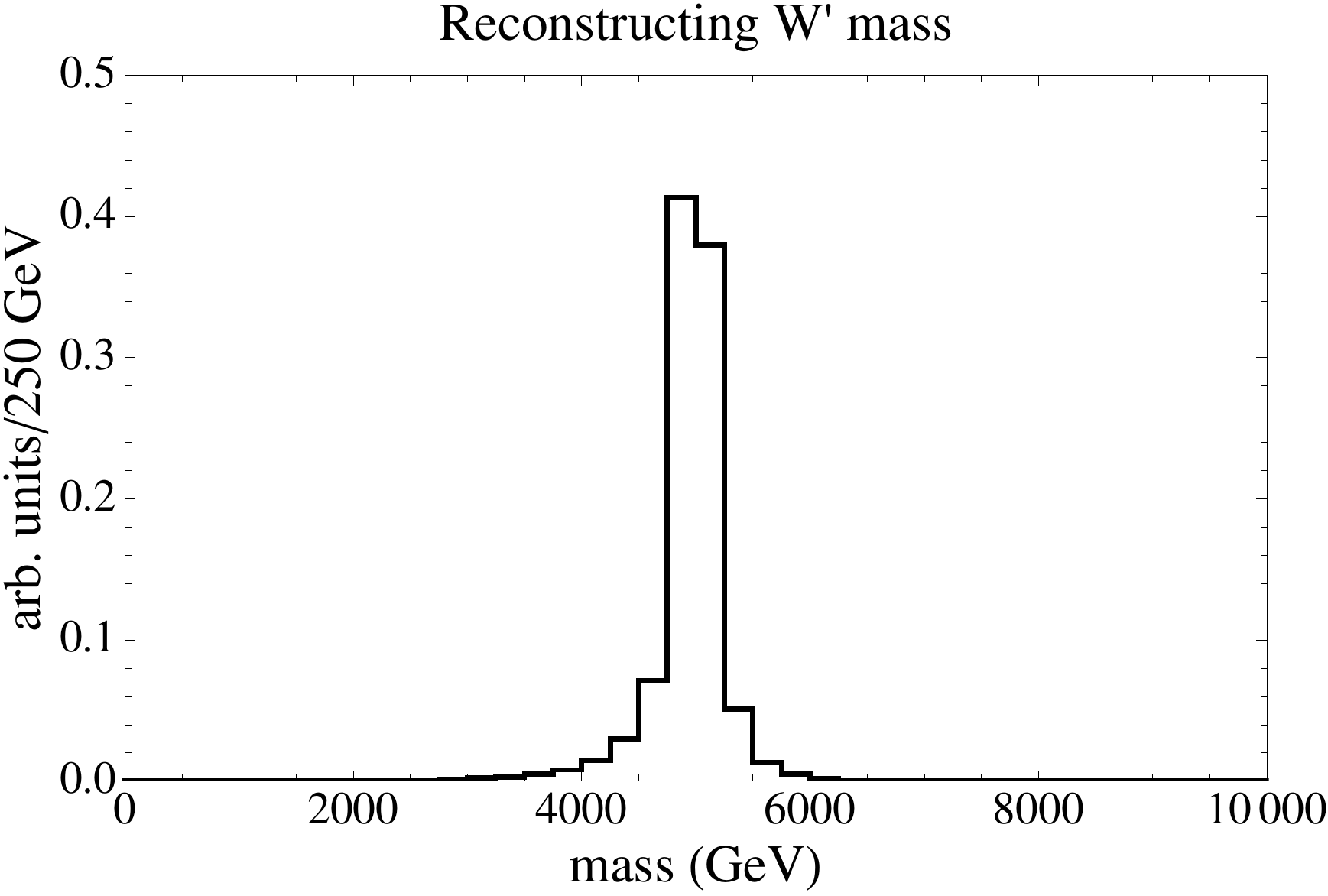}
  \includegraphics[width=0.44\linewidth]{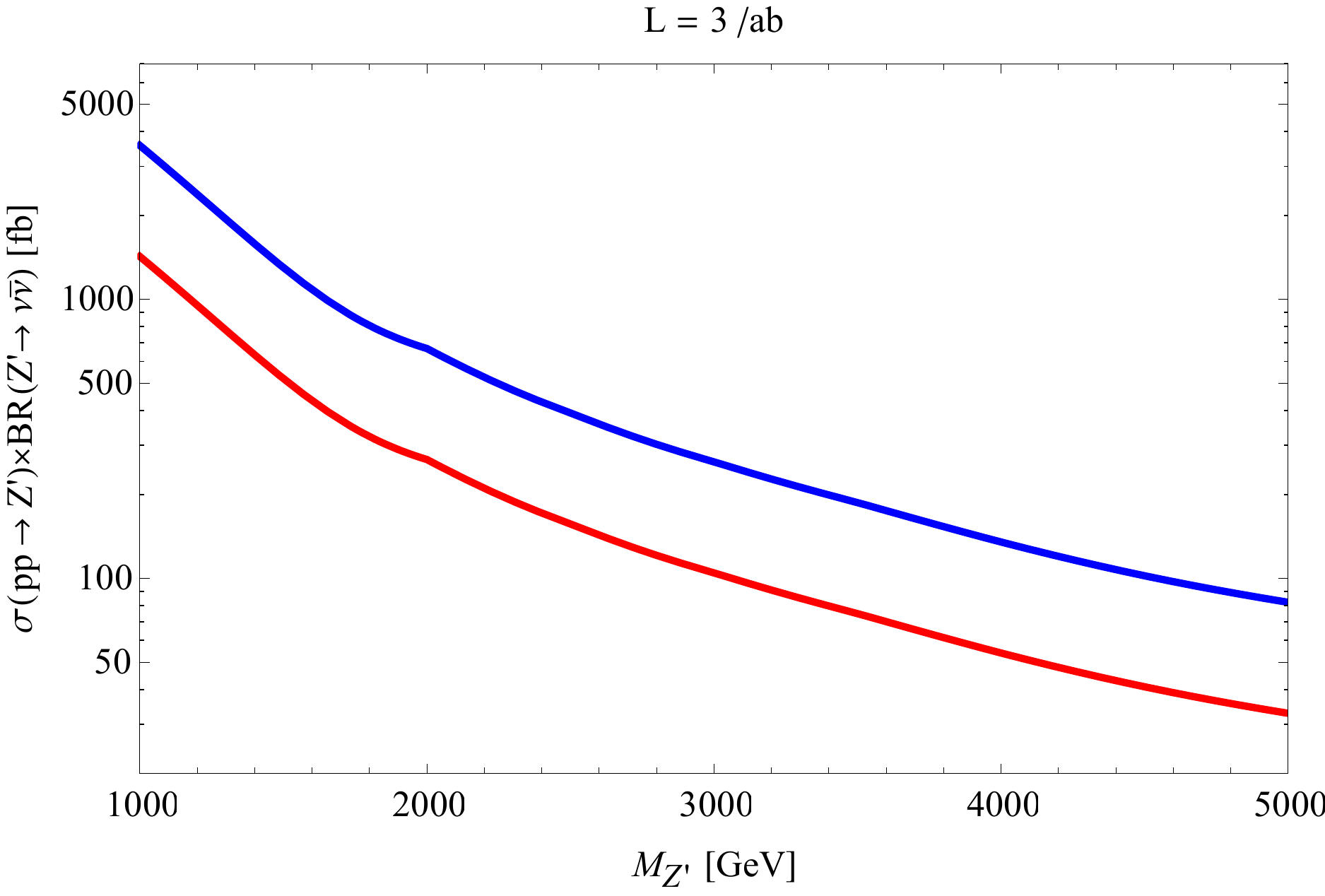}
\caption{Under  the assumption that the
  neutrino is collinear with the leptonic $Z$, the reconstructed
  neutrino allows one to guess the real missing energy in an event as
  well as reconstruct the full mass peak of a $W'$ particle (plot on
  the LHS). The mass resolution is smeared since the $Z$ is not always
  collinear with the neutrino, but the peak is very clearly at the
  $W'$ mass of 5~TeV. On the right hand side, we plot the reach of a 100 TeV collider to a $Z'$ decaying invisibly for
   a luminosity of 100 fb$^{-1}$ and 3000 fb$^{-1}$.  The blue and red
   lines are the 5 and 2 $\sigma$ results respectively. }  \label{Fig: wprime_zprime}
 \end{figure}

The enhanced $W$ and $Z$ radiation can also have interesting applications in new physics searches. We briefly mention 
a couple of examples here. 
The invisible and semi-invisible decays $Z' \to \nu \nu$ and $W' \to \ell \nu$ are
difficult to probe directly.  At large energies, neutrinos can emit
$W$ and $Z$ bosons which can help tagging these processes.\footnote{The
  importance of heavy $Z'$ three-body decays was first mentioned in
  Ref.~\cite{Cvetic:1991gk} in the context of SSC and later in
  Ref.~\cite{Rizzo:2014xma,Hook:2014rka} in context of a 100~TeV
  collider.}  The Sudakov enhancement of this process can make the
three-body decays of a $W'$ or $Z'$ significant if the leptons are
sufficiently boosted, e.g. $Z' \to \nu \bar \nu Z$ or $Z ' \to \nu l^-
W^+$. If a $Z$ boson is
radiated, the collinear enhancement results in a strong tendency for
the $Z$ boson to be emitted parallel to the neutrino.   This  allows one to
reconstruct approximately the neutrino momentum.  If a $W$ boson is radiated and reconstructed (most likely in a
hadronic decay mode), the small $\Delta R$ distance between it and the
lepton allows one to tag the lepton as originating from a neutrino.
 These effects at an 100 TeV $pp$ collider have been studied in Ref.~\cite{Hook:2014rka}. The analysis is at parton level and Madgraph5~\cite{Alwall:2011uj,Stelzer:1994ta,Maltoni:2002qb} was used to generate the events. The results are shown in Fig.~\ref{Fig: wprime_zprime}.

This approach can be pursued further and help determine
quantum numbers of new particles based on total EW gauge bosons
emission.  Particles which are not charged under $SU(2)_L \times
U(1)_Y$ do not radiate $W$ and $Z$ bosons and can thus be
distinguished from their charged counterparts.

We illustrate this effect in an example where we assume a ``natural
SUSY'' - like spectrum at the TeV scale, namely a stop as an NLSP
decaying into a neutralino LSP.  SUSY with light third generation
squarks is a well motivated~\cite{Dimopoulos:1995mi,Cohen:1996vb} and
well studied scenario~\cite{Brust:2011tb,Essig:2011qg,Papucci:2011wy}.
The left and right handed stops have different couplings to the $Z$.
Due to electroweak symmetry breaking, they mix so that the NLSP is an
admixture of the two.  
At large masses, the chirality of the stops can be measured by the
additional radiation of a $Z$ or $W$ in the event.  The Sudakov
enhancement for the radiation of $Z$s and $W$s makes this measurement
feasible at a 100~TeV machine.  Note however that the radiation of the
EW gauge bosons from the stop is only single log enhanced because the
collinear singularity in this case is cut off by the mass of the
emitting particle (the stop) and effectively does not lead to any
enhancement.  Meanwhile, both ISR and FSR have a Sudakov double log
enhancement.  Because both the decay products of the stop and the
initial state quarks have the same chirality as the stop, the
radiation strength provides a good measure of the chirality of the
stop regardless of where the radiation came from.

 \begin{figure}
 \centering
 \includegraphics[width=0.44\linewidth]{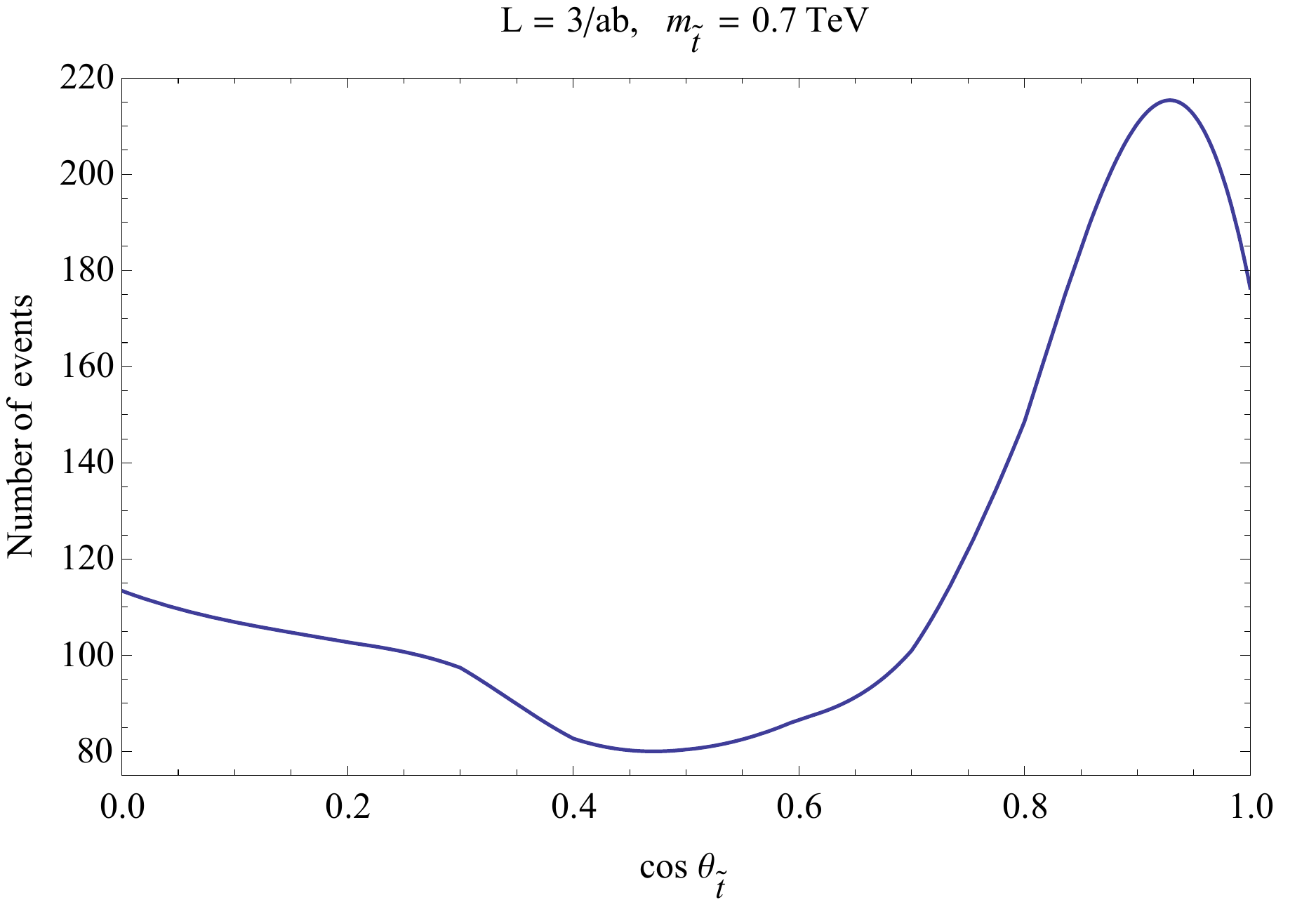}
  \includegraphics[width=0.44\linewidth]{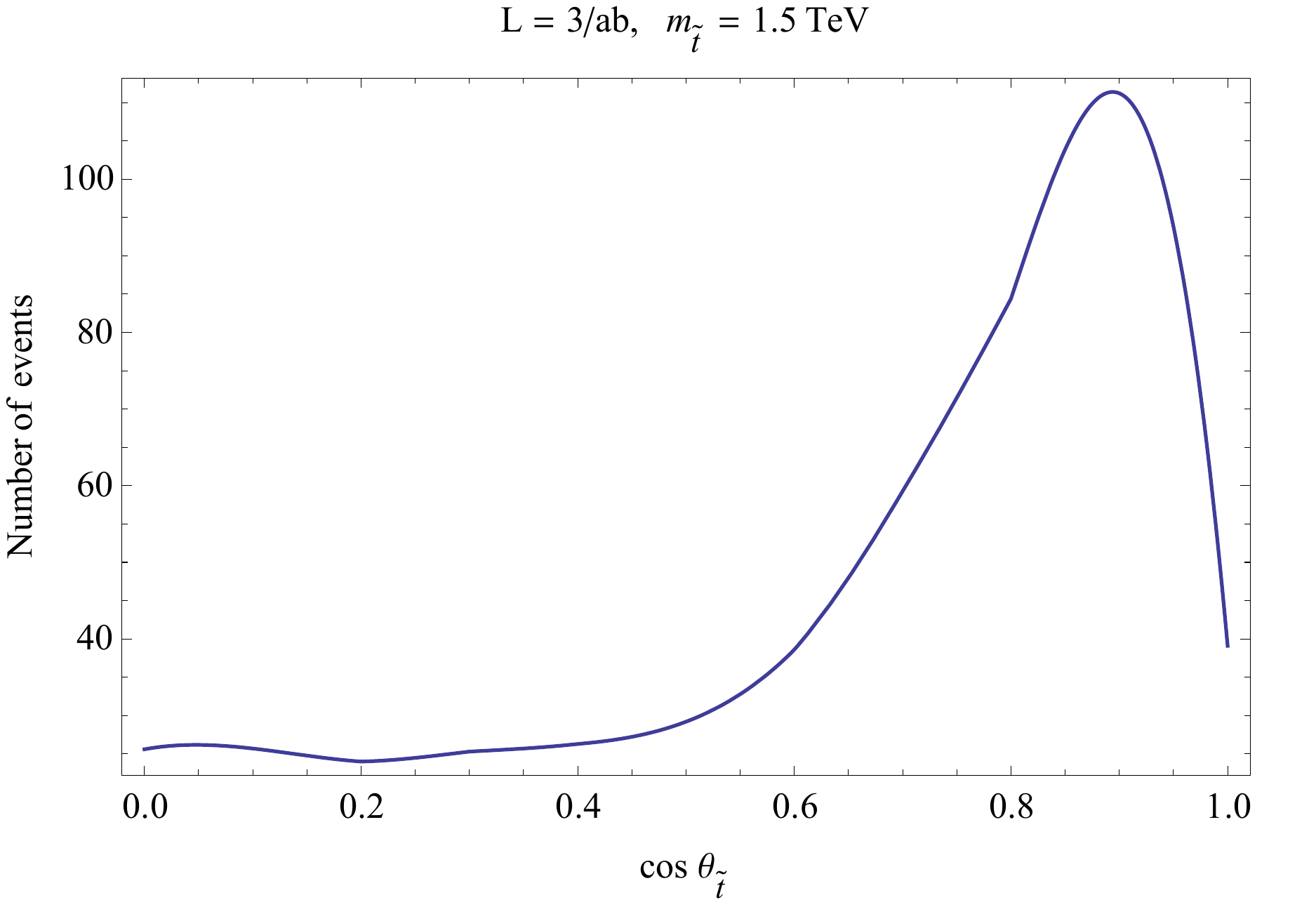}
 \caption{ Number of signal events after imposing all selection cuts, detailed in Ref.~\cite{Hook:2014rka},  as a function of $\cos\theta_{\tilde t}$ for three mass points.  $\cos\theta_{\tilde t} = 0$ is a right handed stop. }
\label{Fig: stops}
 \end{figure}
Fig.~\ref{Fig: stops} demonstrated such a measurement with two benchmark stop masses:  $m_{\tilde t}
= 0.7$~TeV and $m_{\tilde t} = 1.5$~TeV, all decaying into a massless
bino-like neutralino.  
Note that the first benchmark point can be
easily discovered by the LHC while the second one is inaccessible even
for the LHC14.  There is a clear difference between $\cos\theta_{\tilde
  t} = 0$ and $1$.  Thus purely left and purely right handed stops can
be distinguished.

\subsection{Heavy Quarks}
\subsubsection{Inclusive bottom production}
Inclusive production of $b$ hadrons in hadronic collisions offers
unlimited opportunities for flavour studies in the $b$ sector, as
shown very well by the Tevatron and LHC experiments. 
\begin{figure}[h!]
\centering
\includegraphics[width=0.44\textwidth]{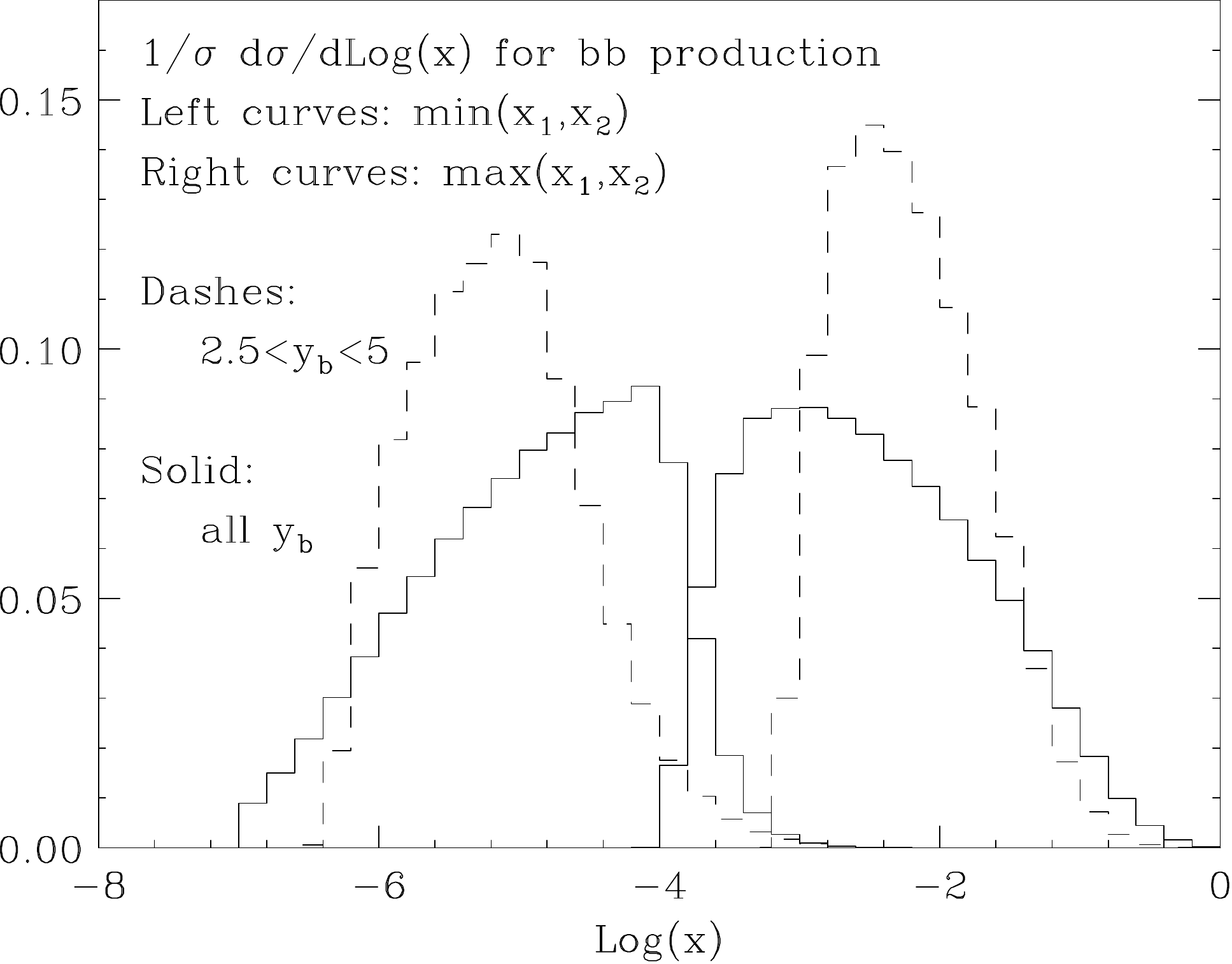}
\includegraphics[width=0.55\textwidth]{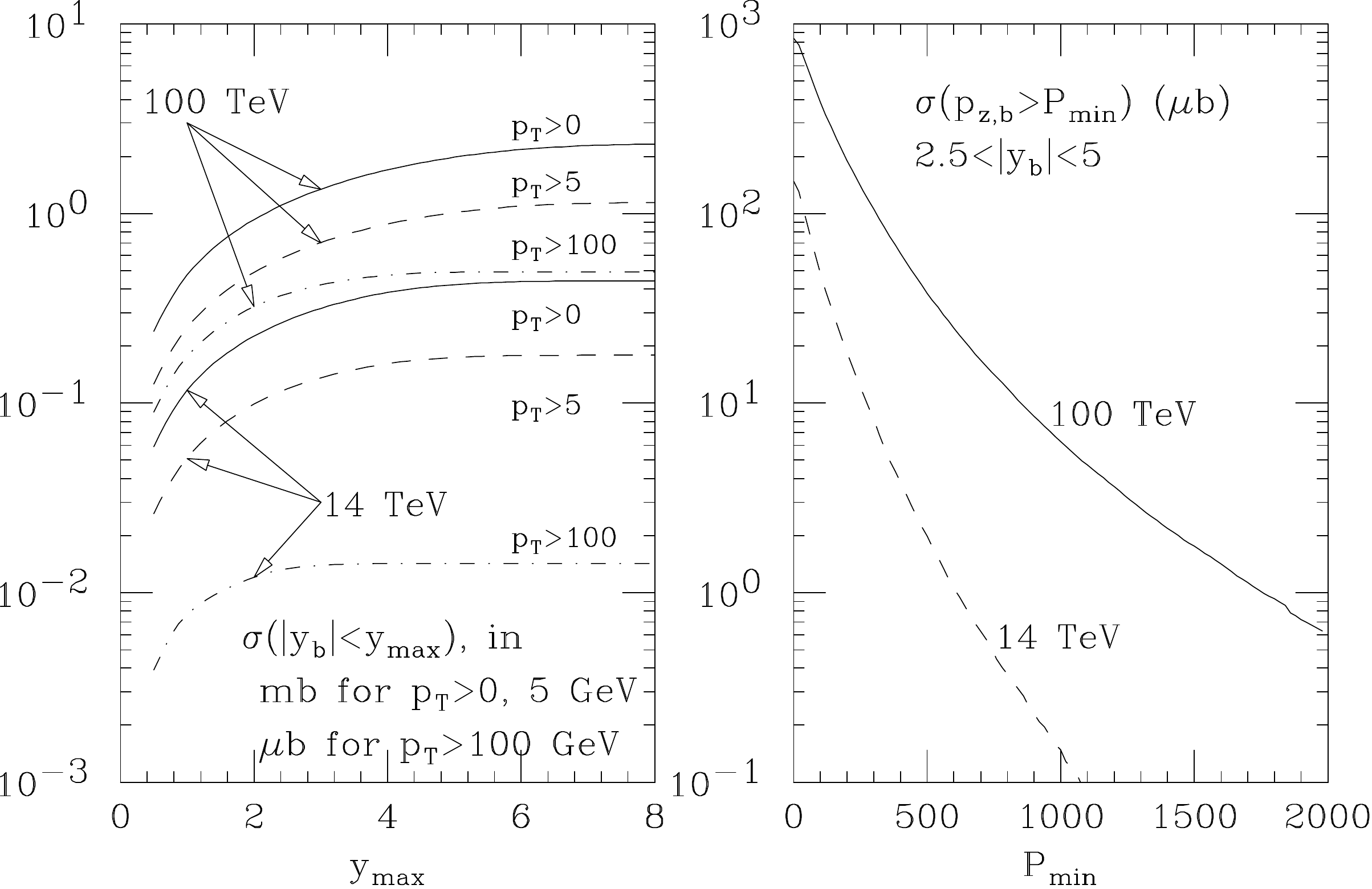}
\caption{Left: distribution of the smaller and larger values of the
  initial partons momentum fractions in inclusive $b\bar{b}$ events
  (solid) and in events with at least one $b$ in the rapidity range
  $2.5<\vert y \vert < 5$ (dashes). Center: production rates for $b$
  quarks as a function of detection acceptance in $y$, for various
  $p_T$ thresholds (rates in $\mu$b for $p_T>100$~GeV, in mb
  otherwise). Right: forward $b$ production rates, as a function of
  the $b$ longitudinal momentum. }
\label{fig:bot100x}
\end{figure}
The long-term interest in these studies will depend on what future
LHCb and Belle2 data will tell us, and on the flavour implications of
possible LHC discoveries in the high-$Q^2$ region. But it is likely
that heavy flavour studies will remain a pillar of the physics
programme at 100~TeV.  The total $b\bar{b}$ production cross section
at 100~TeV is about 2.5mb, and increase of $\sim 5$ relative to the
LHC, and it is more than a 1\% fraction of the total $pp$ cross
section. This rate comes with a large uncertainty, due to the
contribution of gluons at very small $x$ values, where the knowledge
of PDFs is today extremely poor and mostly dictated, at best, by
reasonably guessed extrapolations.  The left plot of
Fig.~\ref{fig:bot100x} shows that, for a detector like LHCb, covering
the rapidity region $2.5<y<5$, about 50\% of the $b$ events would
originate from gluons with momentum $x<10^{-5}$, i.e. in a domain
totally unexplored so far! The following two plots of
Fig.~\ref{fig:bot100x} provide the rapidity distributions for $b$
quarks produced above some thresholds of $p_T$ and, for $b$ quarks
produced in the region $2.5<\vert y \vert < 5$, the integrated
spectrum in longitudinal momentum $p_z$, comparing results at 14 and
100 TeV. We note that, while the total production rate grows only by a
factor of $\sim 5$ from 14 to 100~TeV, the rate increase can be much
greater once kinematic cuts are imposed on the final state. For
example, at 100~TeV $b$ quarks are produced in the forward region
$2.5<\vert y \vert < 5$ with $p_z>1$~TeV at the astounding rate of
$10\mu$b, 100 times more than at the LHC. To which extent this opens
concrete opportunities for new interesting mesurements, to be
exploited by the future generation of detectors, remains to be
studied.

\subsubsection{Inclusive top production}
Table~\ref{tab:topXS} shows the NLO cross sections for the inclusive
production of top quark pairs, and for production in association with
one and two gauge bosons. The $\sim 30$~nb inclusive rate is more than
30 times larger than at 14~TeV. For the planned total integrated
luminosity, two experiments would produce of the order of $10^{12}$
(anti)top quarks.  The possible applications emerging from this huge
statistics have yet to be explored in detail. It would be interesting
to consider the potential of experiments capable of recording all
these events (only a small fraction of top quarks produced at the LHC
survives for the analyses). Triggering on one of the tops, would allow
for unbiased studies of the properties of the other top and of its
decay products: studies of inclusive $W$
decays~\cite{Mangano:2014xta}
(which are impossible
using the $W$'s produced via the Drell-Yan process), of charm and
$\tau$ leptons produced from those $W$ decays, of flavour-tagged $b$'s
from the top decay itself~\cite{Gedalia:2012sx}.

\begin{figure}[h!]
\centering
\includegraphics[width=0.47\textwidth]{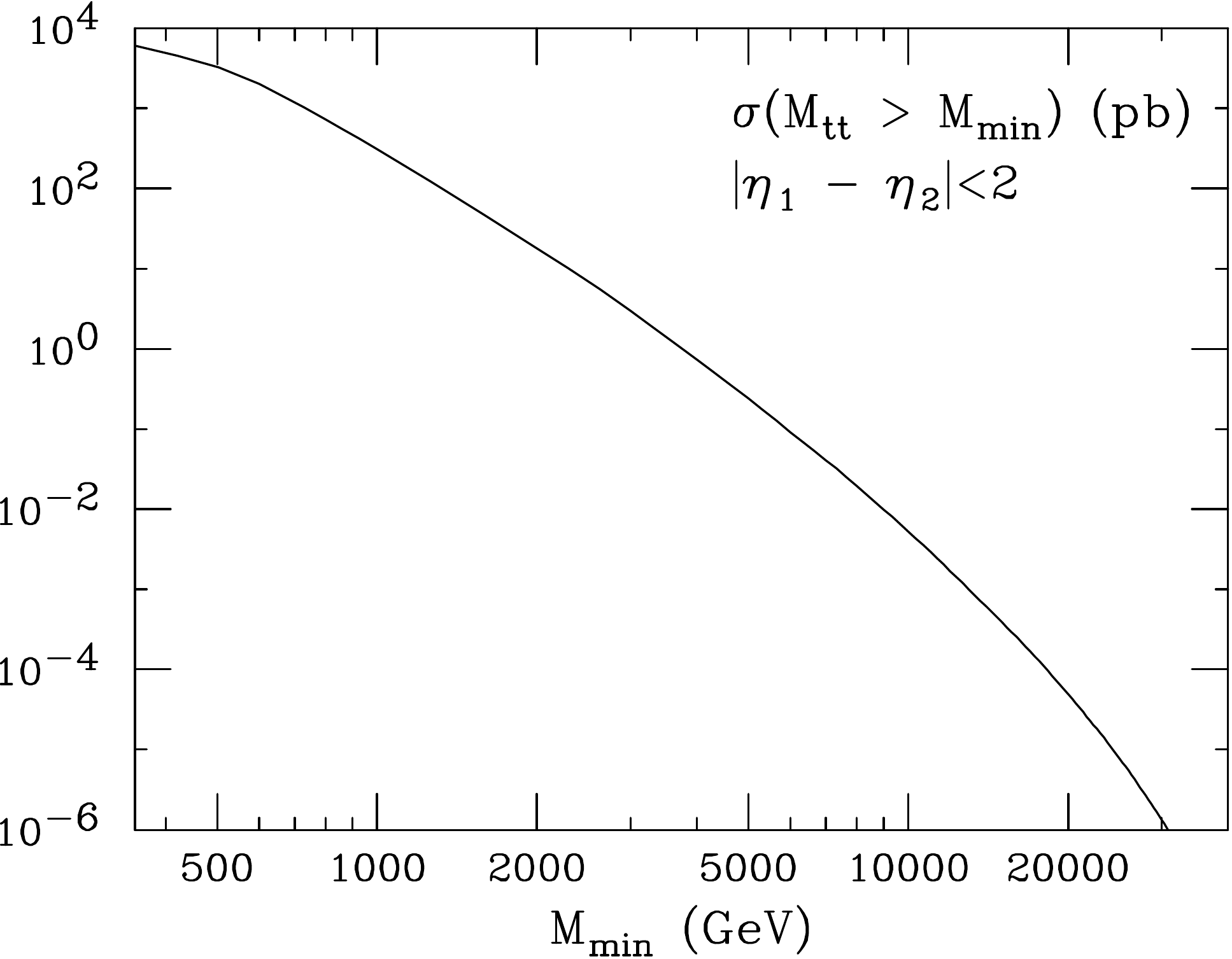}
\includegraphics[width=0.47\textwidth]{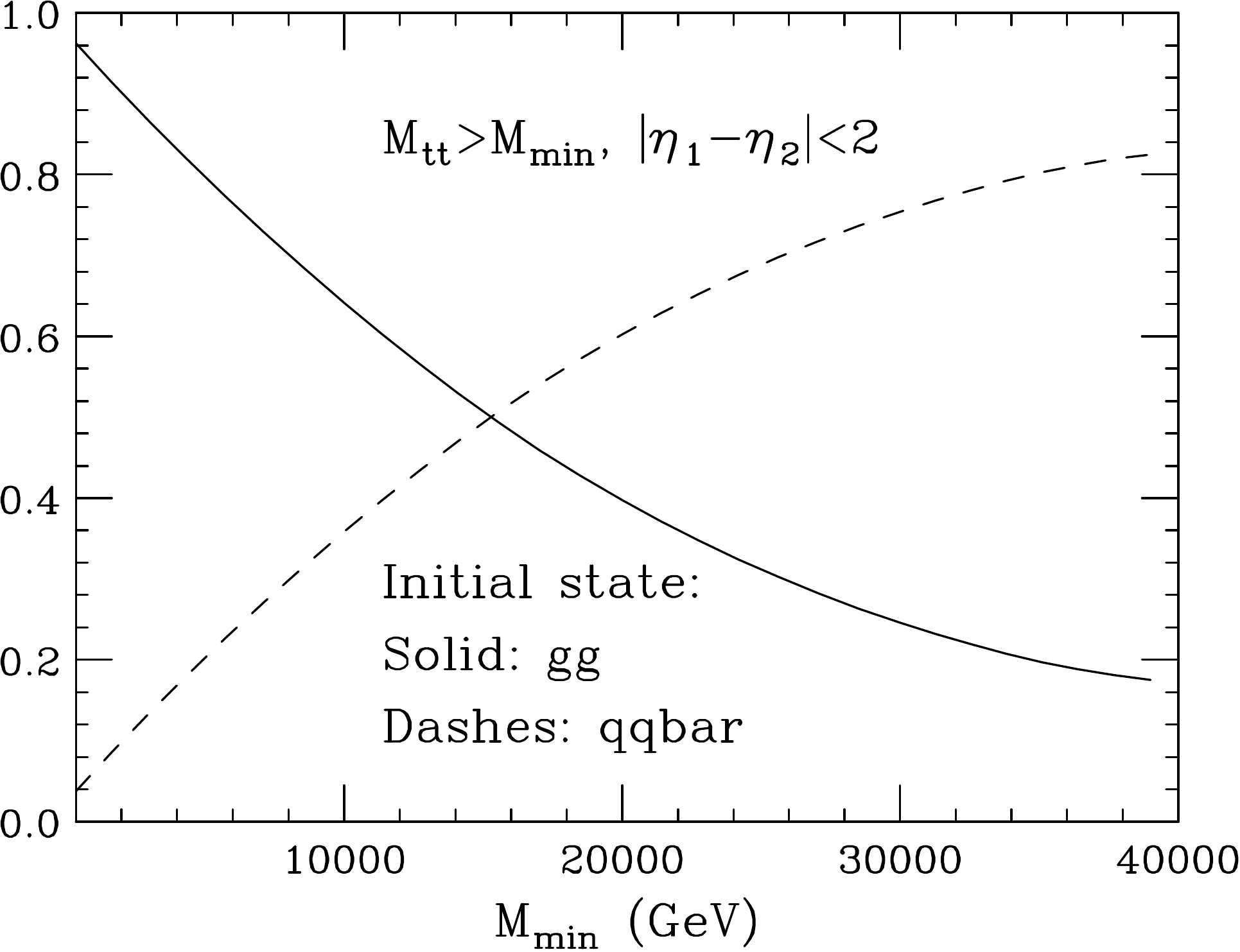}
\caption{Left: integrated invariant mass distribution for production
  of central $t\bar{t}$ quark pairs. Right: initial state composition
  as a function of the $t\bar{t}$ invariant mass.}
\label{fig:mtt}
\end{figure}

Comparing the rates for associated production, in
Table~\ref{tab:topXS}, with those in Table~\ref{tab:multiVB} for
multiple gauge boson production, and considering that each top quark
gives rise to a $W$ through its decay, we remark that top quark
processes at 100~TeV will provide the dominant source of final states
with multiple $W$ bosons, and thus with multiple leptons. This will
have important implications for the search of new physics signals
characterized by the presence of many gauge bosons or leptons from the
decay of the new heavy particles.

Notice also that $t\bar{t}Z^0$ production is more abundant than
$t\bar{t}W^\pm$, contrary to the usual rule that $W$ bosons are
produced more frequently than $Z^0$'s in hadronic collisions. This is
because the $t\bar{t}Z^0$ process is driven by the $gg$ initial state,
which for these values of $\hat{s}/s$ has a much larger luminosity
than the $q\bar{q}'$ initial state that produces
$t\bar{t}W$. This also implies that studies of top production via
initial state light quarks (e.g. in the context of
 $t$ vs $\bar{t}$ production asymmetries) will benefit from a higher
purity of the $q\bar{q}$ initial state w.r.t. $gg$ if one requires the presence
of a $W$ boson (see e.g. Ref.~\cite{Maltoni:2014zpa}).

\begin{table}
\begin{center}
\def\arraystretch{1.5}
\begin{tabular}{ l | c | c | c | c | c| c | c }
 &  $t\bar{t}$ & $t\bar{t}t\bar{t}$ &$t\bar{t}\,W^\pm $ & $t\bar{t}\,Z^0$ & $t\bar{t}\,WW$ &
  $t\bar{t}\,W^\pm Z$ & $t\bar{t}\,ZZ$  
\\   \hline
$\sigma$(pb) & $3.2 \cdot 10^4$ & 4.9 & 16.8 & 56.3 & 1.1 & 0.17 & 0.16
\\
\end{tabular}
\end{center}
\caption{NLO cross sections for associated production of (multiple)
  top quark pairs and gauge bosons~\cite{Torrielli:2014rqa,Maltoni:2015ena}.}
\label{tab:topXS}
\end{table}

\subsubsection{Bottom and top production at large $Q^2$}
Production of bottom and top quarks at large $Q^2$ is characterized by
two regimes. On one side we have final states where the heavy quark
and antiquark ($Q$ and $\bar{Q}$)
give rise to separate jets, with a very large dijet invariant mass
$M_{QQ}$. These are the configurations of relevance when, for example,
we search for the $Q\bar{Q}$ decay of massive resonances.  In the case
of top quarks, the left-hand side of Fig.~\ref{fig:mtt} shows the
production rate for central $t\bar{t}$ pairs above a given invariant
mass threshold.  
At 100~TeV there will be events well above
$M_{tt}>30$~TeV. The right plot in Fig.~\ref{fig:mtt} furthermore
shows that, due to the absence at LO of
contributions from $qq$ or $qg$ initial states, $gg$ initial states
remain dominant up to very large mass, $M_{tt}\sim 15$~TeV. Well above
$M_{QQ}\sim$~TeV, the results for $b\bar{b}$ pair production are
similar to those of the top.

The second regime occurs when we request only one jet to be tagged as
containing a heavy quark. This could be of interest, for example, in
the context of high-$p_T$ studies of single top production. 
In this regime, 
configurations in which the heavy quark pair arises from the splitting
of a large-$p_T$ gluon are enhanced. 
The final state will then contain a jet
formed by the heavy-quark pair, recoiling against a gluon jet. An
example of the role of these processes is shown in
Fig.~\ref{fig:bbjet}, where we compare the $p_T$ spectrum of $b$ jets
in events where the $b\bar{b}$ pair is produced back to back (as in
the first case we discussed above), and the
spectrum of jets containing the $b$ pair (here jets are defined by a
cone size $R=0.4$). The latter is larger by approximately one order of
magnitude at the highest $p_T$ values, leading to rates in excess of 1
event/\iab\ for $p_T>15$~TeV. Similar considerations apply to
the case of top quark production in this multi-TeV regime, as shown in
the right plot of 
Fig.~\ref{fig:bbjet}. In this case the rate for $t\bar{t}$ jets is
only slightly larger than that for single-top jets, due to the much
larger mass of the top quark, which leads to a smaller probability of
$g\to t\bar{t}$ splitting. 
\begin{figure}[h!]
\centering
\includegraphics[width=0.47\textwidth]{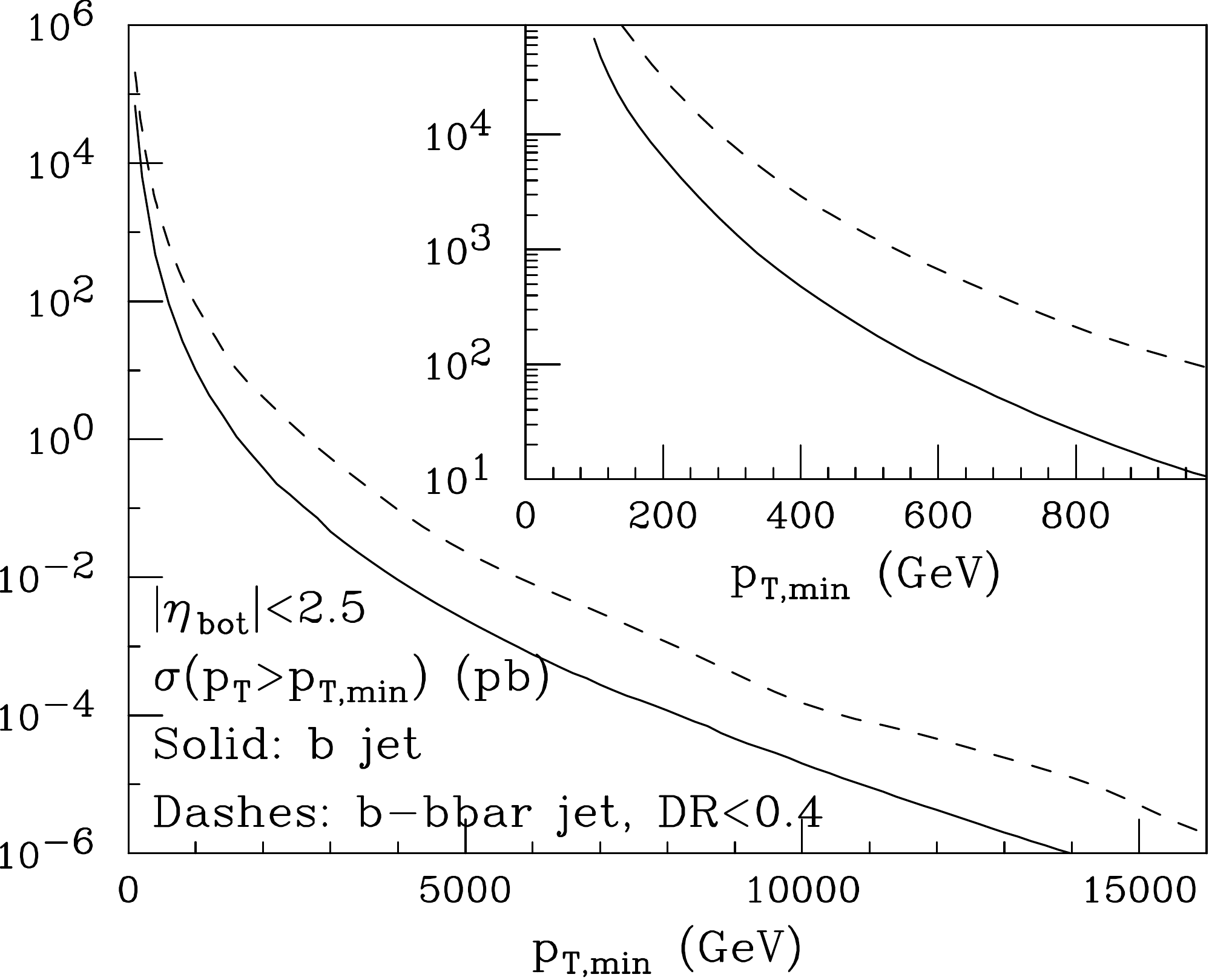}
\includegraphics[width=0.47\textwidth]{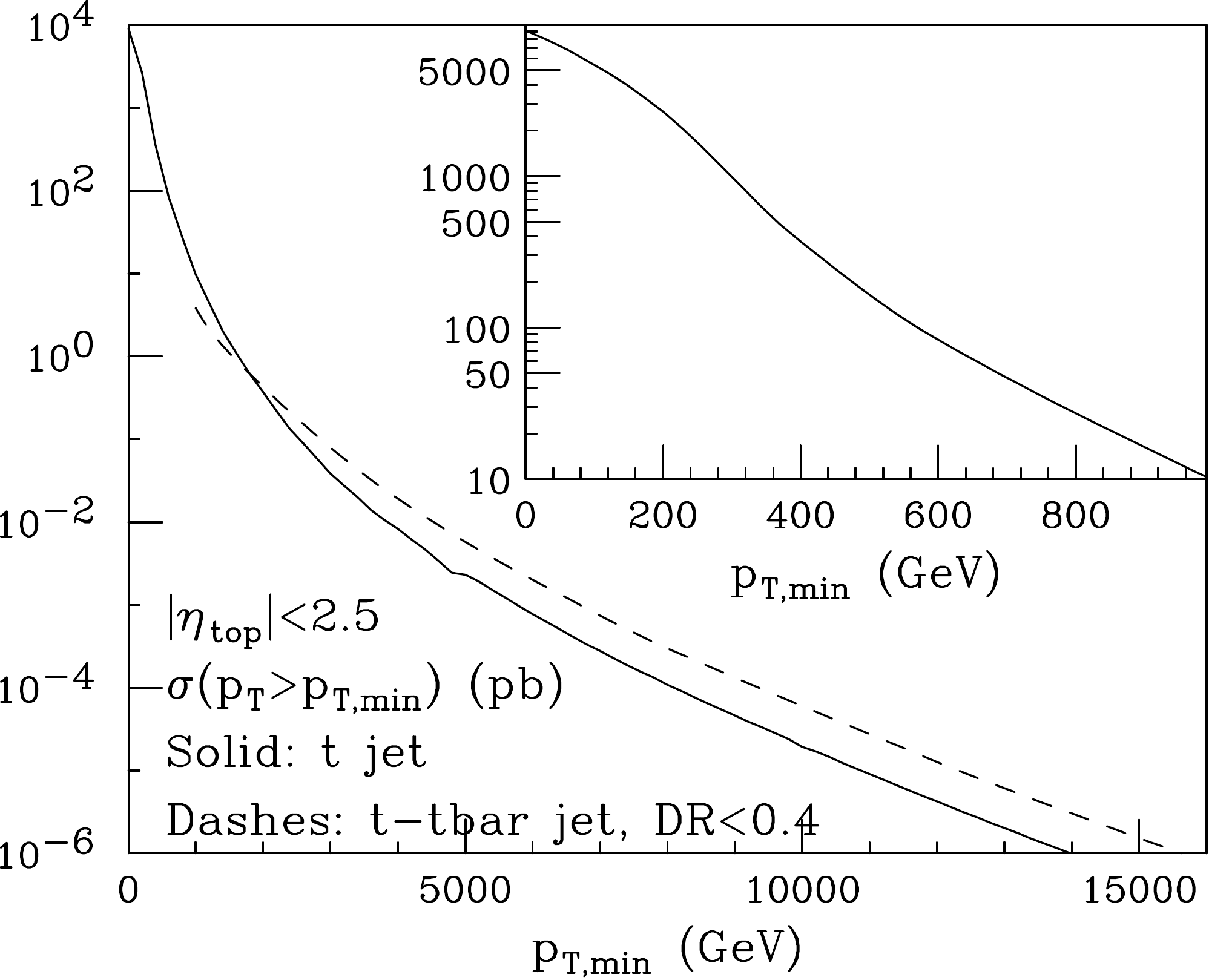}
\caption{Left: production rates for $b$ jets (solid), 
and for jets containing a $b\bar{b}$ pair within $\Delta R<0.4$
(dashes). Right: same, for top-quark jets (top treated as stable).  }
\label{fig:bbjet}
\end{figure}

%%%%%%%%%%%%%%%%%%%%%%%%

\subsubsection{Heavy quark partons}
%%%%%% top PDF, Tao Han et al
\label{sec:toppdf}
At 100~TeV, particles with masses around the electroweak scale appear as light as the bottom
quark at the Tevatron collision energy of $\sqrt{s}=2 \ \text{TeV}$. 
When a very heavy scale is involved in the process, the gluon splitting into a top-antitop pair may present a large logarithmic enhancement.
For $Q\sim 10 \ \text{TeV}$, for instance, $\alpha_s(Q)\log(Q^2/m_t^2)\sim 0.6$, which makes a
perturbative expansion of the hard process questionable. Defining a
parton distribution function (PDF) for the top-quark inside the proton
allows us to resum large collinear logarithms
$\alpha_s^n(Q)\log^n(Q^2/m_t^2)$ to all orders  in perturbation
theory. Initial heavy quarks have been studied in detail in the
context of bottom-initiated processes~\cite{Barnett:1987jw,Olness:1987ep}, and the main concepts
can be adopted for the top-quark. The NNPDF collaboration has released
a top-quark PDF as part of their NNPDF2.3 set~\cite{Ball:2012cx},
which facilitates the implementation. 
\begin{figure}[h]
\begin{center}
\includegraphics[width=9cm]{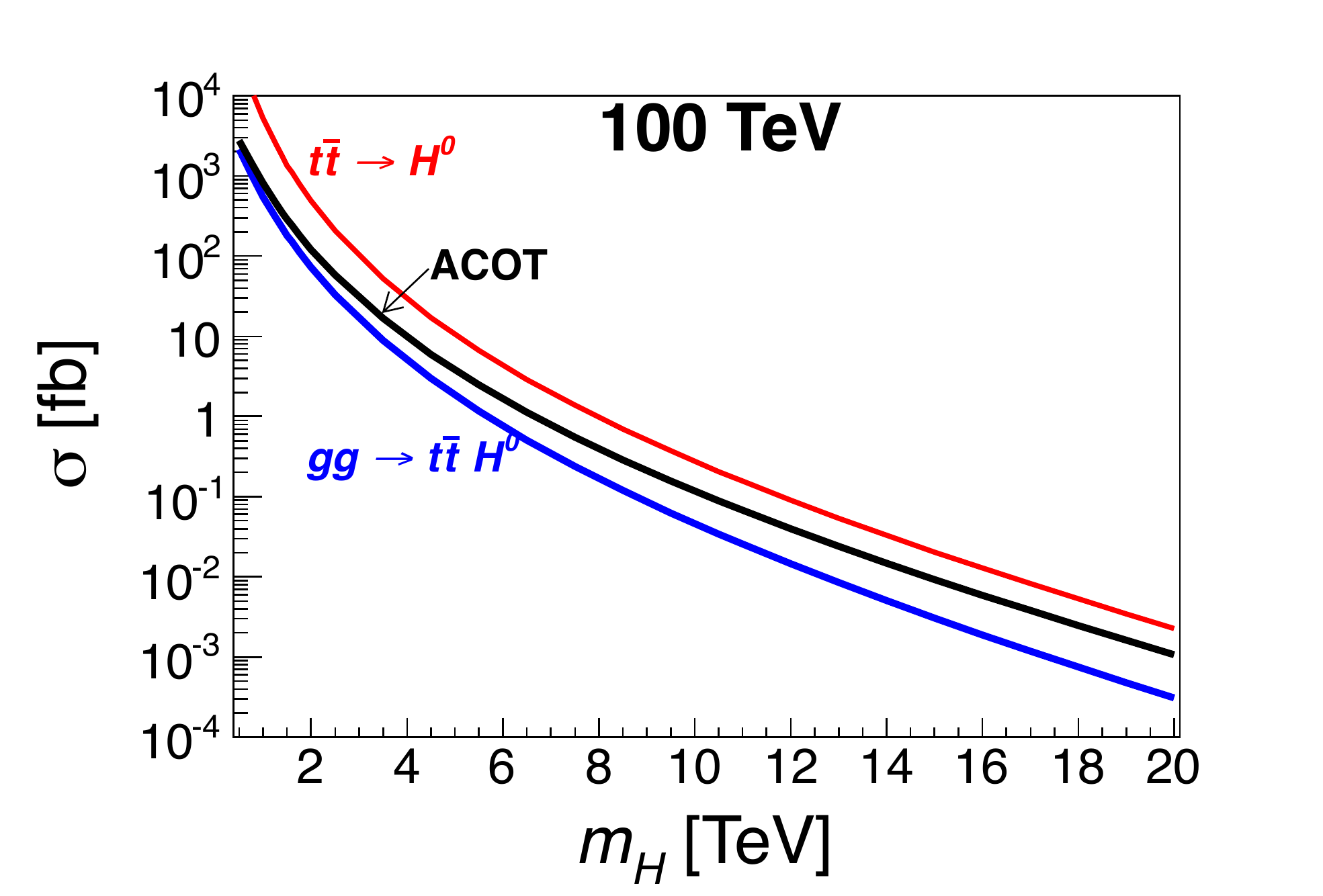}
\end{center}
\caption{Inclusive cross section for a heavy scalar $H^0$ production with Yukawa
coupling $y=1$ at $100$ TeV versus its mass $m_{H}$, in the 5-flavor scheme (bottom blue), the 6-flavor scheme (upper red), and the ACOT scheme with proper subtraction (middle black). }
\label{fig:schemes1}
\end{figure}
%%%%%%%%%%%%

Figure~\ref{fig:schemes1} shows a comparison of calculations in the
5-flavor, massless 6-flavor, and ACOT schemes \cite{Aivazis:1993pi,Collins:1998rz} for the inclusive production of a hypothetic heavy scalar ($H^0$) at a $100$ TeV $pp$ collider \cite{Han:2014nja}. The ACOT scheme with proper treatment of collinear subtraction shows the desired behavior of interpolating between the region
near the top threshold and the very high energy limit. 
We point out that the simplest LO 6-flavor
calculation is unreliable for masses below $10$~TeV, indicating that
the minimum scale above which a parton interpretation for the top quark
becomes justified is much larger than the top mass itself.

\subsection{Higgs Production Rates}
We collect here, for reference, the production rates at 100 TeV of SM
Higgs bosons, including both the canonical production channels, as
well as more rare channels of associated production. Associated
production of Higgs bosons with other objects could allow independent
tests of the Higgs boson properties, and might provide channels with
improved signal over background, with possibly reduced sistematic
uncertainties. 

\begin{table}[h!]
\begin{center}
\def\arraystretch{1.5}
\begin{tabular}{ l | c | c | c | c| c }
 &  $gg\to H$ & $VBF$ & $HW^\pm$ & $HZ$ & $t\bar{t}H$
\\   \hline
$\sigma$(pb) & 740 & 82 & 15.9 & 11.3 & 37.9
\\
$\sigma$(100 TeV)/$\sigma$(14 TeV) &  14.7 & 18.6  & 9.7 & 12.5 & 61
\\
\end{tabular}
\end{center}
\caption{Upper row: cross sections~\cite{HXSWG} for production of a SM Higgs
  boson in $gg$ fusion, vector boson fusion, associated production
  with $W$ and $Z$ bosons, and associated production with a $t\bar{t}$
  pair. Lower row: rate
  increase relative to 14~TeV. All results are NNLO, except $ttH$
  (NLO), with the central PDF from the MSTW2008(N)NLO set.}
\label{table:HiggsXS}
\end{table}
Table~\ref{table:HiggsXS}, extracted from the compilation produced by
the LHC Higgs Cross Section working group~\cite{HXSWG}, shows the
rates for channels that will already be accessible and used at the
LHC.  The rates are typically a factor of 10-20 larger than at the
LHC, except for the associate $t\bar{t}H$ production, where the $gg$
initial state and the large mass of the final state benefit more
significantly from the higher energy, leading to a rate growth by a
factor of 60. The samples obtained with a luminosity of 10~\iab\ will
therefore be a factor of 30-200 larger than what available after the
completion of the HL-LHC progamme. The statistical uncertainties for
the extraction of the Higgs couplings to the third generation
fermions, to the charm and the muon, and to the EW gauge bosons, will
become smaller than the percent level. It is difficult today to
estimate how the theoretical progress will improve the theoretical
systematics, and the determination of experimental systematics will
require detailed simulation studies, based on realistic detector
concepts. The large statistics for both signals and backgrounds will
certainly help in improving the modeling systematics, which in many
cases are a limitation to the precision foreseen for the HL-LHC. It is
therefore not excluded that the final uncertainties, at least in some
channels, may reach the percent level. 

An example is the extraction of the top Yukawa coupling $y_{top}$ from
the $t\bar{t}H$ process~\cite{Plehn:2015cta}. The large cross
section at 100~TeV allows to consider boosted topologies for the
hadronic decays of both the top quarks and the Higgs boson ($H\to
b\bar{b}$), placing tight cuts on the emerging jets, and drastically
reducing the various sources of backgrounds, while maintaining a
statistical sensitivity on the production rate at the percent level.
This matches the theoretical systematics, which, already today, is at
the percent level~\cite{Plehn:2015cta}, if one considers the ratio
$\sigma(t\bar{t}H)/\sigma(t\bar{t}Z)$, which is very stable with
respect to PDF and scale uncertainties. The branching ratio for the
$H\to b\bar{b}$ decay, needed to extract the top Yukawa coupling from
this measurement, will be known with sufficient accuracy if an
$e^+e^-$ Higgs factory (at a linear or circular collider) will be
operating. Otherwise, a percent-level measurement of
$y_{top}*\mathrm{BR}(H\to b\bar{b})$ will still be one of the most
precise determinations of a combination of Higgs couplings, with
direct sensitivity on $y_{top}$.

Studies are also
available~\cite{Yao:2013ika,He:2015spf,Barr:2014sga,Azatov:2015oxa}
of the determination of the Higgs self-coupling in the $HH\to
b\bar{b}\gamma\gamma$ decay channel\footnote{For a study of more rare
  decay modes, see Ref.~\cite{Papaefstathiou:2015iba}.}, with a
projected uncertainty on the measurement of the SM coupling in the
range of $5-10\%$ with a total of 30~\iab.

\begin{table}
\begin{center}
\def\arraystretch{1.5}
\begin{tabular}{ l | c | c | c | c| c | c}
 &  $HH$ & $HHjj$ (VBF) & $HHW^\pm$ & $HHZ$ & $HHt\bar{t}$ & $HHtj$
\\   \hline
$\sigma$(fb) & $1.2\cdot 10^3$ & 81 & 8.1 & 5.5 & 86 & 4.6
\\
\end{tabular}
\end{center}
\caption{NLO cross sections for production of a SM Higgs
  boson pair, including associated production channels, 
  at 100 TeV~\cite{Alwall:2014hca}.}
\label{table:Higgspair}
\end{table}
Table~\ref{table:Higgspair}, extracted from the NLO results of
Ref.~\cite{Alwall:2014hca}, reports the rates for
SM Higgs pair production, including channels of associated production
with jets, gauge bosons and top quarks. Once again, the possible
implications of the measurement and study of these exotic Higgs
production channels are under study. 

Table~\ref{table:Higgsrare}, extracted from the NLO results of the aMC@NLO
code~\cite{Alwall:2014hca,Torrielli:2014rqa}, reports the rates for
associated production of a SM Higgs with gauge boson
pairs. Theoretical systematics, including scale and PDF
uncertainties, are typically below 10\%.  

\begin{table}
\begin{center}
\def\arraystretch{1.5}
\begin{tabular}{ l | c | c | c | c| c }
 &  $HW^+W^-$ & $HW^{\pm}Z$ & $HZZ$ & $HW^{\pm}\gamma$ &
$HZ\gamma$  
\\   \hline
$\sigma$(fb) & 170 & 100 & 42 & 78 & 43
\\
\end{tabular}
\end{center}
\caption{NLO cross sections for associated production of a SM Higgs
  boson with multiple gauge bosons~\cite{Alwall:2014hca}.}
\label{table:Higgsrare}
\end{table}

\subsection{Sources of Missing Transverse Energy}
\begin{figure}[h!]
\centering
\includegraphics[width=0.49\textwidth]{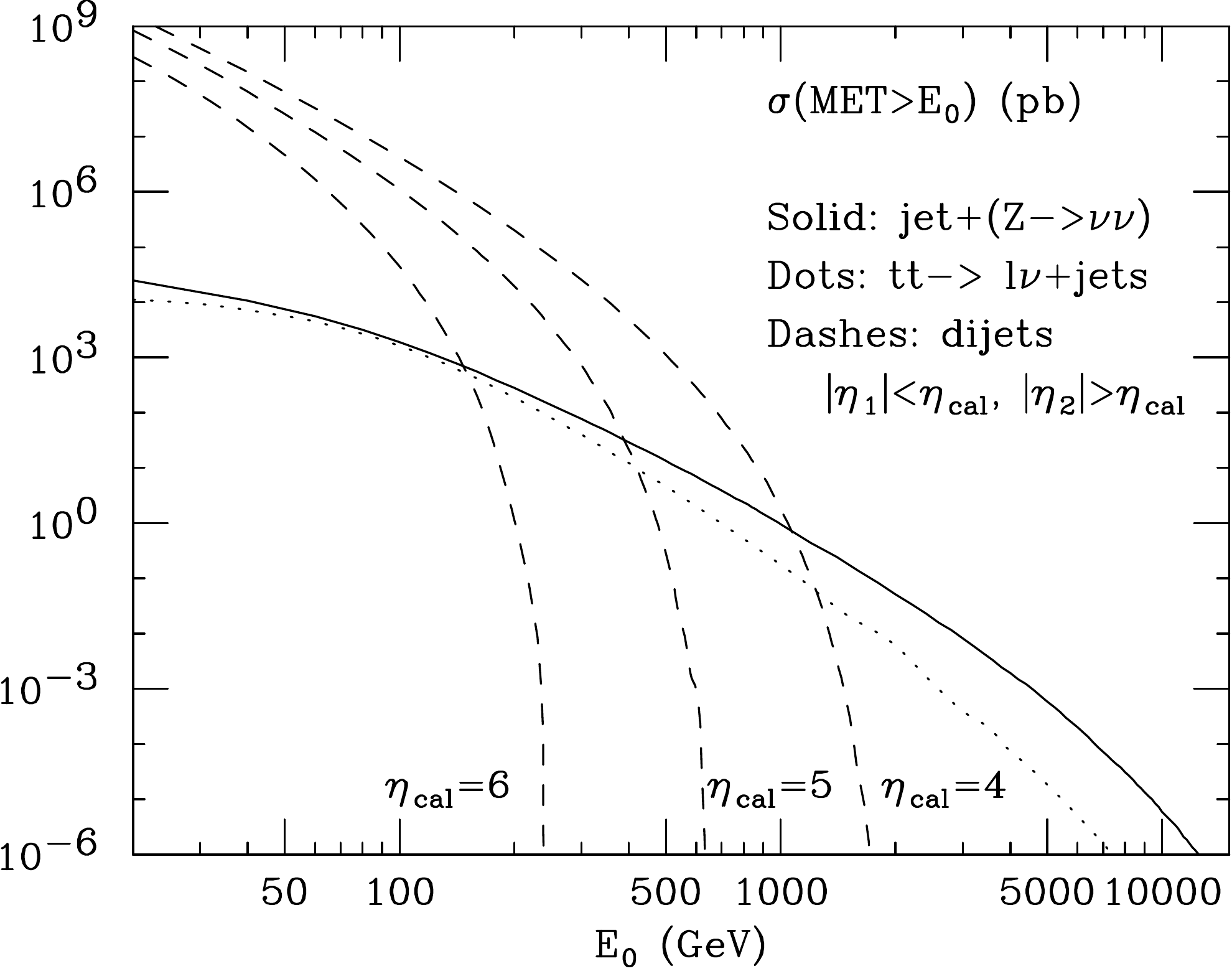}
\includegraphics[width=0.49\textwidth]{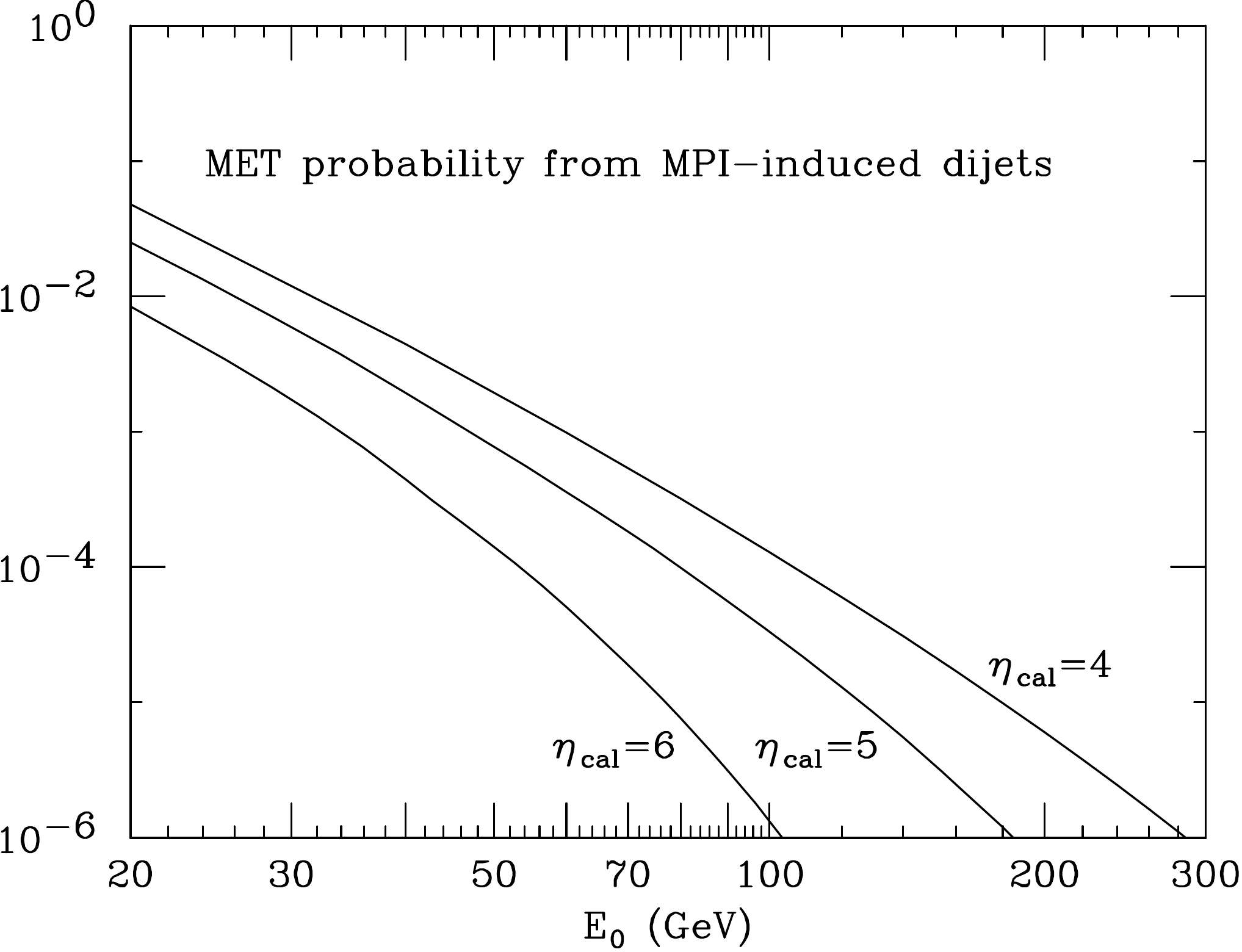}
\caption{Left: Missing transverse energy rates, from jet+$(Z\to
  \nu\bar{\nu})$ events and from dijets, with a jet escaping
  undetected at large rapidity. Right: Missing transverse energy
  probability induced by multiple-parton interactions, for different
  values of the jet rapidity acceptance. }
\label{fig:met}
\end{figure}
Missing transverse energy ($\metslash$) is an important signature for
many BSM processes. At 100~TeV, SM sources of $\metslash$ can
contribute with very large rates of irreducible backgrounds. We
consider here, for illustration, the effect of three of the leading sources
of irreducible $\metslash$: the associated production of jets and a $Z^0$
boson decaying to neutrinos, the semileptonic decay of top quarks, and the production of jets
outside the calorimeter acceptance. The latter channel is important,
since the high energy available in the CM allows for the production of
large $p_T$ jets at very forward rapidities. This is shown in
Fig.~\ref{fig:met}, where the dashed lines correspond to the rate of
dijet events in which one jet is within the calorimeter acceptance
(defined by the $\eta_{cal}$ label), and the other is outside. With the
standard LHC calorimeter coverage, $\eta_{cal}=5$,
dijets would give a $\metslash$ signal larger than $Z$+jets 
for $\metslash$ up to $\sim
400$~GeV. This is reduced to $\sim 150$~GeV with a calorimeter
extending out to $\eta_{cal}=6$. 

It must be noticed that the limited calorimeter acceptance can induce
a $\metslash$ signal in any hard process, due to the finite
probability of the coincidence of a multiparton
interaction. Multiparton interactions are hard scatterings taking
place among the partons not engaged in the primary hard process, and
cannot be separated experimentally since the resulting particles
emerge from exactly the same vertex as the primary scattering. The
probability that a multiparton interaction leads to a secondary hard
process $X$ in addition to the primary one is parameterized as
$\sigma(X)/\sigma_0$, where $\sigma_0$ is a process-independent
parameter. The right plot of Fig.~\ref{fig:met} shows the probability of
multiparton interactions leading to dijet final states, with one jet
inside the calorimeter and the other outside. For this example we
chose $\sigma_0=30$~mb, a number consistent with the direct
experimental determinations from Tevatron and LHC data. $\metslash$
signals in the range of 30-70~GeV are induced with probability of
about $10^{-3}$ if $\eta_{cal}$ is in the range 4 to 6, stressing
once again the need to instrument the detectors with a calorimetric
coverage more extended than at the LHC.

\section{Directions for Further Exploration}
\label{sec:future}

Many years will go by before a 100~TeV $pp$ collider becomes a
reality. The technological, financial and political challenges--which 
we have not even alluded to in this report--are immense, and far
from being trivially surpassable. But the physics opportunities are even more compelling, and have motivated 
ongoing efforts taking place worldwide
to define more precisely the tasks required to give substance to this
dream. 

In this report we just scratched the surface of the vast array
of contributions that the 100 TeV collider could give to physics. New
ideas and proposals appear in the literature almost on a daily basis,
and the overall picture will continue to evolve, not least in view of what
will emerge from the future LHC runs. In this report we presented some
examples of concrete and central theoretical issues that are unlikely to be
settled conclusively by the LHC, and will require exploration at much
higher energies. Discoveries at the LHC may of course change the
theoretical priorities, and give higher weight to alternative future
avenues. For example, discoveries at the edge of the LHC mass reach
can be studied in detail with good statistics even by just doubling
the LHC energy, to 28~TeV. The discovery of new weakly interacting
particles well below the TeV would strengthen the case of a lepton
collider in the TeV range. It is clear, therefore, that, while the
physics case of a 100~TeV collider is strong and clear as a long-term goal for
the field--no other proposed or foreseeable project can have direct
sensitivity to such large mass scales--the
precise route used to to get there must take account of the fuller picture to emerge from the LHC as well as other current and future experiments in areas ranging from flavour physics to dark matter searches.

In the meantime, several directions are open for continued studies of
the physics capabilities  of the 100~TeV collider. To start with, it is
essential to assess the progress that can be anticipated in detector and data acquisition
technologies, and how these might impact the ultimate experimental
performance and systematics. This is critical to give more
realistic estimates of the physics reach. The prospects in areas such
as precise measurements of the Higgs properties, or searches and
studies of signatures that today are elusive because of trigger or
background limitations, will be strongly affected by the detector
performance. In this report, we gave some examples of projected
performance for Higgs studies where the impact of a 100~TeV collider
will certainly be comparable, if not superior, to any sub-TeV $e^+e^-$
collider (the measurement of the Higgs self-coupling and of the top
Yukawa coupling). But it may happen that future studies, in
the context of ambitious detector concepts, could expose important
complementarity, if not superiority, also in areas that typically
represent golden channels for the $e^+e^-$ colliders.  Future
theoretical improvements in the precision of calculation and modeling
of hard processes in $pp$ collisions will also play a key
role. Important inputs for these calculations include the
determination of $\alpha_S$ and of the PDFs. Whether the LHC, or the
100~TeV collider itself, can provide sufficient inputs to reach the
needed precision, is a further relevant question. Scrupulous studies
of the complementarity and synergy of the $pp$ and $e^+e^-$
approaches, as well as the role of a possible future $ep$ collider,
are therefore a pressing priority for future work.

We discussed here some landmark issues that might lie in the exclusive
domain of the 100~TeV collider, such as the understanding of the
nature of the EW phase transition and the observation of dark matter,
should this be due to a WIMP thermal relic. More work is needed to
explore all possible scenarios, and to support more firmly the
statement that this facility could give conclusive answers to those
questions.  With these clear goals in mind, phenomenological analyses
of the most important processes will ideally provide useful benchmarks
for the detector performance.

The infrastructure of a 100~TeV collider enables in principle a
broader range of studies than we discussed: from higher-energy and
higher-intensity fixed target experiments, to higher-energy heavy ion
collisions, to dedicated collider experiments focused on flavour
physics or further explorations of the properties of SM particles (the
top quark, gauge bosons, etc). In the case of  flavour physics,  for example, the importance
of these studies may be enhanced by future findings, at the LHC, at Belle2, in
rare kaon decay experiments, or in the lepton sector. Thus, even in these
areas, more comprehensive studies  of the
physics opportunties and of the complementarity of a 100 TeV collider with other
experimental approaches is desirable. 

The process of consolidating the physics case of a 100~TeV collider
will certainly be rich and fruitful, no matter what: it will prompt
theorists to consider possible new ideas -- which may not emerge if
one just focused on what is measurable by today's facilities -- and
will challenge experimentalists and detector experts to push further
their creativity -- opening new avenues to approach measurements
considered impossible before. This progress on all fronts will likely
bear fruits already during the LHC era, and will hopefully give
confidence to the high-energy community to unite forces behind this
fantastic project, which will push our knowledge of the fundamental
laws of nature well beyond our current limited view of physics at the TeV
scale.

\vskip 0.5cm
\noindent
{\bf Acknowledgments}\\
%\vskip -0.5cm
We would like to thank the Institute for High Energy Physics in Beijing for hospitality during our visits. TH would like to thank Chien-Yi Chen, Junmou Chen, Zhen Liu, Richard Ruiz, Josh Sayre, Brock Tweedie, and Susanne Westhoff for collaboration and for providing some figures. 
LTW would like to thank Matthew Low for supplying some of the figures and results. 
The work of TH is supported in part by the U.S.~Department of Energy under grant 
No.~DE-FG02-95ER40896 and in part by PITT PACC. 
LTW's research is supported in part by a DOE grant DE-SC0013642.
The work of MLM is supported by the ERC grant 291377
\textsl{LHCtheory: Theoretical predictions and analyses of LHC physics:
advancing the precision frontier}.

\newpage

\newpage

\bibliographystyle{elsarticle-num}
\bibliography{pp100TeV}

\end{document}
\endinput
%%
%% End of file `elsarticle-template-num.tex'.